\documentclass[a4paper,fleqn]{cas-sc}

\usepackage[authoryear,longnamesfirst]{natbib}

\usepackage{braket}
\usepackage{slashed}
\usepackage{extarrows}
\usepackage{enumerate}
\usepackage{orcidlink}
\usepackage[normalem]{ulem}
\usepackage{graphicx}

\usepackage{graphicx}
\usepackage{bm}

\newcommand{\tr}{\text{tr}}
\newcommand{\Eq}[1]{Eq.~(\ref{#1})}
\newcommand{\Eqs}[1]{Eqs.~(\ref{#1})}

\newcommand{\Sec}[1]{Sec.~\ref{#1}}

\def\tsc#1{\csdef{#1}{\textsc{\lowercase{#1}}\xspace}}
\tsc{WGM}
\tsc{QE}

\begin{document}
\let\WriteBookmarks\relax
\def\floatpagepagefraction{1}
\def\textpagefraction{.001}

\shorttitle{Perspectives on QCD, Topology and the Strong CP Problem}    

\shortauthors{Anthony G. Williams}  

\title [mode = title]{
Perspectives on QCD, Topology and the Strong CP Problem
}  



%

\author[1]{Anthony G. Williams}[orcid=0000-0002-1472-1592]
\cormark[1]


\ead{anthony.williams@adelaide.edu.au}



\affiliation[1]{organization={ARC Centre of Excellence for Dark Matter Particle Physics and CSSM},
            addressline={Department of Physics, Adelaide University}, 
            city={Adelaide},
            postcode={5005}, 
            state={South Australia},
            country={Australia}}







\cortext[1]{Corresponding author}



\begin{abstract}
The strong $CP$ problem occupies a unique position in modern
particle physics, connecting nonperturbative quantum
chromodynamics, topology, anomalies and the global structure
of gauge theories. It has motivated extensive theoretical
developments, including instantons, the $\theta$ vacuum,
lattice studies of topological charge, and the Peccei-Quinn
mechanism and its associated axion phenomenology. At the
same time, the relationship between local quantum field
theory, global topology and the conventional construction
of the $\theta$ vacuum continues to raise important
conceptual and mathematical questions.

This review presents a pedagogical overview of the strong $CP$
problem from the perspectives of its historical development,
physical motivation and mathematical formulation. Beginning
with the conventional formulation of the problem and the local
structure of QCD, we review the roles of the axial anomaly,
topological charge density, topological susceptibility,
instantons, lattice gauge theory, and the Hamiltonian and
Euclidean formulations of nonperturbative QCD. A recurring
theme is the distinction between local structures that
follow directly from relativistic quantum field theory
and additional global constructions commonly employed in
the conventional treatment of topological sectors and
$\theta$ vacua.

The review clarifies which aspects of the conventional
framework are direct consequences of established local
quantum field theory and which rely on additional mathematical
assumptions concerning global topology. In particular, standard
nonperturbative results involving local correlation functions
and topological susceptibility can be formulated without
requiring the full global construction of topological sectors
and the $\theta$ vacuum. Alternative viewpoints and outstanding
conceptual questions are discussed in order to provide a
balanced assessment of the present theoretical understanding
of the strong $CP$ problem.
\end{abstract}

\begin{highlights}
\item Clarifies assumptions underlying the $\theta$ term in QCD
\item Distinguishes local QFT structure from global topological
assumptions
\item Shows a local $\theta=0$ formulation is consistent with
established QCD phenomenology
\item Identifies additional assumptions required for a physical
$\theta$ parameter
\item Shows how local correlators reproduce standard nonperturbative relations
\end{highlights}

\begin{highlights}
\item Clarifies assumptions underlying the theta term in QCD
\item Distinguishes local structure from global topological assumptions
\item Shows theta=0 is consistent with a minimal local QFT formulation
\item Identifies assumptions required for a physical theta parameter
\item Local correlators reproduce standard nonperturbative relations
\end{highlights}


\maketitle

\tableofcontents

\section{Introduction}
\label{Sec:Intro}

Quantum Chromodynamics (QCD), the nonabelian gauge theory
describing the strong interaction, is one of the central 
pillars of the Standard Model of particle physics. Since its
establishment in the early 1970s, it has provided a remarkably
successful description of the interactions of quarks and gluons,
accounting for phenomena ranging from asymptotic freedom at
high energies \citep{Gross:1973id,Politzer:1973fx} to
confinement, dynamical chiral symmetry breaking, and the rich
spectrum of hadronic bound states at low energies. Lattice
gauge theory has now established QCD as a quantitatively 
successful nonperturbative theory of the strong interaction,
reproducing hadron masses, matrix elements and many other
observables directly from the underlying Lagrangian
\citep{Wilson:1974sk,Creutz1983,MontvayMunster,Neuberger:1998,
Smit,GattringerLang,Rothe2012}. At the same time, QCD has provided
the setting for many of the deepest developments in modern
quantum field theory, including anomalies, topology, 
instantons, nonperturbative dynamics, large-$N_c$
methods, and the vacuum structure of gauge theories.

Among the conceptual questions arising within QCD, the 
strong $CP$ problem occupies a unique position. 
The most general local, Lorentz-invariant, gauge-invariant 
and renormalizable QCD Lagrangian allows the additional term
\begin{align}
\mathcal{L}_\theta &= \theta \frac{g^2}{32\pi^2}F_{\mu\nu}^a
    \tilde F^{a\mu\nu}=\theta\frac{g^2}{16\pi^2}
    \tr\left(F_{\mu\nu}\tilde F^{\mu\nu}\right),
\end{align}
where $F_{\mu\nu}^a$ denotes the gluon field strength,
$\tilde F^{a\mu\nu}$ its dual, $g$ is the strong coupling constant, and
$\theta$ is a real dimensionless parameter. Repeated color indices $a$ 
are summed over. This operator is odd under parity ($P$), time reversal ($T$),
and the combined charge-conjugation and parity transformation ($CP$).
Measurements of the neutron electric dipole
moment imply the remarkably stringent bound
\begin{align}
|\bar\theta| \lesssim 10^{-10},
\end{align}
where
\begin{align}
\bar\theta = \theta + \arg\det M
\end{align}
is the physically relevant parameter after allowing for complex phases
in the quark mass matrix $M$. Current experimental limits therefore imply
that strong interactions conserve $CP$ to extraordinarily high precision
\citep{Baker2006ts,Abel2020gbr}. Explaining the apparent absence of
strong $CP$ violation, despite the absence of an obvious symmetry requiring
$\bar\theta$ to vanish, has become known as the strong $CP$ problem
\citep{Crewther1979,Witten:1979vv,Vafa:1984xg,Abel2020gbr}.

The development of the strong $CP$ problem has closely paralleled the
development of modern nonperturbative quantum field theory itself. Many of
the concepts that are now standard in gauge theory were introduced, or
acquired their modern significance, through attempts to understand the role
of topology in QCD. Consequently, the strong $CP$ problem has become much
more than a question concerning the numerical value of a single parameter. It
provides a window into the interplay between gauge symmetry, topology,
anomalies, nonperturbative dynamics and the mathematical foundations of
quantum field theory.

A major breakthrough came with the discovery of finite-action instanton
solutions by Belavin, Polyakov, Schwarz and Tyupkin
\citep{Belavin:1975fg}. These classical Euclidean solutions demonstrated that
nonabelian gauge theories possess a rich semiclassical topological structure absent in
Abelian gauge theories. They showed that gauge fields with different
topological properties can contribute to the semiclassical expansion of the
Euclidean functional integral and thereby influence the quantum theory in
ways that are invisible in perturbation theory. Shortly afterwards,
't~Hooft demonstrated how instantons describe nonperturbative tunnelling 
between the distinct classical vacuum configurations of the conventional 
semiclassical construction and derived the corresponding effective
fermion interactions \citep{tHooft:1976snw,tHooft:1976rip}. These
developments fundamentally changed our understanding of the nonperturbative
vacuum.

Closely related work by Jackiw and Rebbi \citep{Jackiw:1976pf} and by
Callan, Dashen and Gross \citep{Callan:1976je} led to the conventional
Hamiltonian picture in which gauge fields are classified into topological
sectors labelled by an integer winding number and the physical vacuum is
constructed as a coherent superposition of these sectors, the
$\theta$ vacuum. The Adler-Bell-Jackiw
anomaly \citep{Adler:1969gk,Bell:1969ts} and Fujikawa's functional-integral
derivation of the anomaly \citep{Fujikawa:1979ay,Fujikawa:1980eg}
demonstrated the intimate connection between the axial anomaly, gauge fields and the quantum structure of the fermion measure.
These developments revealed one of the central lessons of modern quantum
field theory: the quantum theory may possess properties that cannot be
anticipated from the classical equations of motion alone.

Further important advances followed. The large-$N_c$ analyses of Witten and
Veneziano \citep{Witten:1979vv,Veneziano:1979ec} established a remarkable
connection between the topological susceptibility of pure Yang-Mills theory
and the unexpectedly large mass of the $\eta'$ meson. Closely related
developments by Leutwyler and Smilga
\citep{Leutwyler:1992yt,Smilga:1998dh} clarified the relationship between the
topological susceptibility, chiral symmetry breaking and the light-quark
mass dependence of QCD. These results demonstrated that correlations
of the topological charge density are not merely of formal interest
but play an essential role in hadron phenomenology.

During the same period lattice gauge theory matured into a powerful
nonperturbative framework for studying QCD directly from first principles.
Beginning with Wilson's lattice formulation \citep{Wilson:1974sk}, enormous
progress has been made in understanding confinement, chiral symmetry
breaking, hadron spectroscopy and topological properties of the QCD vacuum.
Modern lattice calculations now determine quantities such as the topological
susceptibility with impressive precision and provide one of the principal
sources of nonperturbative information concerning the role of topology in
QCD.

The strong $CP$ problem also became one of the principal motivations for
physics beyond the Standard Model. The Peccei-Quinn mechanism
\citep{Peccei:1977ur,Peccei:1977hh} proposed a dynamical explanation for the
smallness of the effective vacuum angle through the spontaneous breaking of
a global chiral symmetry. This mechanism predicts the existence
of a light pseudoscalar particle, the axion, independently proposed by
Weinberg and Wilczek
\citep{Weinberg:1977ma,Wilczek:1977pj}. Subsequent developments led to the
invisible axion models of Kim, Shifman, Vainshtein and Zakharov, and
Dine, Fischler and Srednicki
\citep{Kim:1979if,Shifman:1979if,Dine:1981rt,Zhitnitsky:1980tq}, while more
recent work has considerably broadened the phenomenology to include
axion-like particles arising in many extensions of the Standard Model.

Alternative approaches seek instead to explain the smallness of
$\bar\theta$ through exact discrete symmetries that are broken
spontaneously. The best-known examples are the Nelson-Barr mechanism
\citep{Nelson:1983zb,Barr:1984qx}, in which $CP$ is an exact symmetry of the
underlying theory that is spontaneously broken while maintaining a naturally
small value of $\bar\theta$, and left-right symmetric models in which parity
plays a central role
\citep{Mohapatra:1978fy,Senjanovic:1978ev}. Together with the Peccei-Quinn
mechanism, these ideas have shaped much of the theoretical exploration of
the strong $CP$ problem over the past four decades
\citep{Peccei:2006as,Strumia2025}.

Consequently, the strong $CP$ problem now connects particle physics, 
astrophysics and cosmology, with ongoing experimental searches spanning
laboratory experiments, stellar evolution, supernovae, dark matter and
the early Universe. It lies at the intersection of particle physics, 
quantum field theory, mathematical physics, lattice gauge theory and 
cosmology. It has stimulated many of the
major developments in our understanding of nonperturbative gauge theories,
while continuing to motivate new theoretical ideas and increasingly
sensitive experimental searches. Equally importantly, it has raised deep
conceptual questions concerning the relationship between local quantum field
theory, global topology and the mathematical formulation of gauge theories.
These questions continue to attract considerable attention and form one of
the principal themes of the present review.

Although the phenomenology associated with the strong $CP$ problem is well
established, its formulation involves a number of logically distinct
ingredients that are often presented together. These include the local
topological charge density, the axial anomaly, instantons, topological
susceptibility, the Hamiltonian and Euclidean formulations of QCD, lattice
definitions of topological charge, and the construction of the
$\theta$ vacuum. Each of these subjects possesses its own mathematical and
physical foundations, while the logical relationships between them are not
always made explicit. As a consequence, readers entering the field often
encounter a collection of closely related ideas whose respective roles in
the conventional formulation of the strong $CP$ problem are not always easy
to disentangle.

A recurring theme throughout this review is the distinction between three
closely related, but conceptually distinct, aspects of the conventional
framework. The first consists of the local structure of quantum field
theory, including local gauge invariance, causal locality, the axial
anomaly, local composite operators, Ward identities and correlation
functions. The second concerns the additional global constructions that
appear in conventional nonperturbative treatments, including finite-action
boundary conditions, topological sectors, winding numbers, large gauge
transformations and the $\theta$ vacuum. The third comprises the practical
realizations of these ideas in semiclassical analyses, lattice gauge theory,
gradient flow, overlap fermions and related nonperturbative techniques.
Keeping these three levels conceptually distinct helps clarify both the
logical structure of the conventional formulation and the assumptions upon
which it rests.

The distinction between local and global structures is particularly
important. Physical observables in quantum field theory are expressed
through correlation functions of local gauge-invariant operators,
schematically
\begin{align}
\braket{\hat O_1(x_1)\cdots\hat O_n(x_n)}
    &\equiv \braket{\Omega|T[\hat O_1(x_1)\cdots\hat O_n(x_n)]|\Omega}
    =\frac{1}{Z}\textstyle \int \!\mathcal D\Phi\,
    \mathcal O_1(x_1)\cdots\mathcal O_n(x_n)\,e^{iS},
\end{align}
where $\Phi$ collectively denotes the dynamical fields. Such observables
are associated with measurements performed within finite spacetime regions.
By contrast, the conventional construction of the $\theta$ vacuum invokes
additional assumptions concerning the behavior of gauge fields at infinity,
the topology of gauge transformations and the global organization of the
space of gauge-field configurations. One of the principal objectives of
this review is therefore to distinguish carefully between those aspects of
the theory that follow directly from local quantum field theory and those
that rely upon additional global mathematical structure.

A related question concerns the nature of the functional integral itself.
The semiclassical description of nonperturbative phenomena is formulated in
terms of smooth classical solutions, such as instantons, around which one
constructs a saddle-point expansion. On the other hand, constructive
quantum field theory and lattice gauge theory indicate that typical
quantum-field configurations are highly irregular and need not resemble
the smooth classical configurations that underlie semiclassical analysis
\citep{Glimm:1987ylb,Simon:2015spl,colella1973sample}.
Understanding the relationship between
these two perspectives is one of the recurring conceptual themes of modern
quantum field theory and plays an important role in the discussion of
topology developed later in this review.

It should be emphasized at the outset that the purpose of the present review
is not to challenge the internal consistency of QCD, nor the 
well-established role played by instantons, topological susceptibility,
large-$N_c$ arguments, lattice gauge theory or the extensive body of
phenomenological work that has grown from these developments. Rather, our
aim is to examine carefully the logical structure of the conventional
formulation of the strong $CP$ problem, to distinguish between established
results and the assumptions upon which they depend, and to clarify the
relationships between the various theoretical ingredients that have emerged
during the historical development of the subject.

The strong $CP$ problem also remains an active area of research.
Considerable recent effort has been devoted both to sharpening its
conceptual foundations and to exploring possible solutions. These include
developments in axion physics, gauged discrete symmetries, higher-form
gauge theories, lattice calculations of topology, and alternative
approaches to the vacuum structure of QCD
\citep{Benabou2025,Kaplan:2025JHEP050,Dvali2025,Strumia2025,
Nakamura:2021meh,Schierholz:2024var,
Gattringer:2020mbf,Iida:2024irv,Kotov_2025,
bonanno2025strongcpproblemtheta,Sannino:2026wgx,Williams:2026tuw}.
Recent work has also reconsidered aspects of the $\theta$ parameter from
the viewpoints of canonical quantization, the physical Hilbert space,
large gauge transformations, the ordering of infinite-volume and
topological-sector limits, and simpler quantum-mechanical analogues,
with differing conclusions concerning their implications for the strong
$CP$ problem \citep{Ai:2020ptm,Ai:2024vfa,Benabou2025,Strocchi:2024tis,
Albandea:2024fui,Khoze:2025auv,Bhattacharya:2025qsk}.
These continuing developments illustrate that the strong $CP$ problem is
not merely a question concerning the numerical value of a single parameter,
but continues to provide a fertile testing ground for our understanding of
nonperturbative gauge theories, quantum topology and the mathematical
foundations of quantum field theory.

The primary purpose of this review is to provide a pedagogical overview of
the strong $CP$ problem from the perspectives of its historical
development, physical motivation and mathematical formulation. Rather than
concentrating on a particular proposed solution, we seek to clarify the
logical structure underlying the conventional formulation of the problem and
the relationships between its various theoretical ingredients. Throughout
the review we distinguish carefully between results that follow directly
from established local quantum field theory and those that rely upon
additional assumptions concerning the global structure of gauge fields.
Alternative viewpoints are discussed where appropriate, not with the aim of
advocating a particular resolution, but rather to illuminate the logical
connections between the different formulations and to identify the
conceptual and mathematical questions that remain open. Some of the 
conceptual issues discussed here have also been explored in a
recent research article by the present author \citep{Williams:2026tuw}.
The central aim of this review, however, is not to advocate any particular
perspective, but to place the underlying issues within the broader historical
and theoretical development of the subject.

The review is organized to follow, as far as possible, both the historical
development of the subject and the logical progression from established local
quantum field theory to the additional global structures that enter the
conventional formulation of the strong $CP$ problem.

Section~\ref{Sec:QCD} introduces the conventional formulation of the problem. We 
begin with the QCD Lagrangian and the $\theta$ term, discuss their transformation
properties under the discrete symmetries, and summarize the present
experimental limits on strong $CP$ violation from searches for the neutron
electric dipole moment. This establishes the conventional statement of the
strong $CP$ problem and the phenomenological motivation for the remainder of
the review.

Sections~\ref{Sec:local_QCD}--\ref{Sec:top_suscept} develop the local foundations 
of the subject. Section~\ref{Sec:local_QCD} reviews the local structure of QCD, 
including gauge invariance, local observables, the functional-integral 
formulation, the classical and semiclassical limits, and the axial anomaly. Section~\ref{Sec:thetabar} explains how the physically observable parameter
$\bar\theta$ arises through the interplay between chiral rotations, the quark
mass matrix and the anomaly. Section~\ref{Sec:top_suscept} discusses the 
topological susceptibility, its derivation from local correlation functions,
and its role in hadron phenomenology through the Witten-Veneziano and
Leutwyler-Smilga relations. Together these sections establish the local
nonperturbative framework that underlies all subsequent discussions.

Sections~\ref{Sec:global_top}--\ref{Sec:H_vs_L} then examine the conventional 
global formulation of nonperturbative QCD. Section~\ref{Sec:global_top} reviews 
finite-action gauge fields, fibre bundles, homotopy, topological sectors, 
instantons and the conventional construction of the $\theta$ vacuum. 
Section~\ref{Sec:FI_beyond} discusses the functional integral beyond the 
regime of the semiclassical approximation, emphasizing the distinction between 
smooth and rough gauge-field configurations, gauge orbits, the role of global 
versus local structure, and the circumstances under which the $\theta$ term 
contributes. Section~\ref{Sec:top_latt_QCD} reviews topology in lattice gauge
theory, including modern definitions of topological charge based on gluonic
operators, gradient flow and overlap fermions, together with their
interpretation in the continuum limit. Section~\ref{Sec:H_vs_L} compares the 
Hamiltonian and Euclidean formulations of nonperturbative QCD, reviewing 
canonical quantization, the Gauss-law constraint, large gauge transformations 
and the correspondence between the two formulations.

Having established both the local and global aspects of the conventional
framework, Section~\ref{Sec:foundations} discusses a number of broader conceptual 
and mathematical issues. These include locality and causality, global structure,
boundary conditions, spacetime topology, the mathematical status of the
continuum functional integral, constructive quantum field theory, and the
rigorous definition of continuum Yang-Mills theory.

Section~\ref{Sec:Perspectives} then draws together the principal conceptual 
perspectives on the strong $CP$ problem itself, distinguishing between the 
conventional viewpoint, alternative interpretations, the respective roles of 
local and global assumptions, and the present state of theoretical understanding.

Finally, Section~\ref{Sec:Outlook} summarizes the principal conclusions of 
the review and discusses several outstanding conceptual and mathematical 
questions that continue to motivate current research. Particular emphasis
is placed on distinguishing consequences that follow directly from established
local quantum field theory from those that rely upon additional assumptions
concerning the global organization of gauge-field configuration space.
There is also an instructive comparison with quantum
electrodynamics, where no analogous strong $CP$ problem conventionally arises. 
By comparing the Abelian and nonabelian theories, we examine the roles of the 
axial anomaly, chiral rotations, the local operator $F_{\mu\nu}\tilde F^{\mu\nu}$,
homotopy, instantons and global topology. This comparison provides a useful
framework for identifying which ingredients are common to both theories and
which are specific to nonabelian gauge theories.
We conclude by discussing possible future directions in both the mathematical
foundations of nonperturbative gauge theory and the continuing search for the
origin of the apparent absence of strong $CP$ violation.

Although the review necessarily touches upon many of the central ideas of
modern quantum field theory, our emphasis throughout is on the logical
relationships between them. By distinguishing carefully between local quantum
field theory, semiclassical methods, lattice gauge theory, global topology and
the mathematical formulation of the functional integral, we hope to provide a
coherent framework within which the diverse perspectives on the strong $CP$
problem can be understood and compared. In this sense, the strong $CP$
problem serves not only as a long-standing puzzle concerning the smallness of
a single parameter, but also as a valuable case study in the interplay
between locality, topology and the mathematical foundations of quantum field
theory.


\section{Quantum Chromodynamics and the Origin of the Strong $CP$ Problem}
\label{Sec:QCD}

We begin our discussion with an overview of quantum chromodynamics, the $\theta$ term and the strong $CP$ problem.

\subsection{QCD Lagrangian density}
\label{Sec:QCD_Lagrangian}

Quantum chromodynamics (QCD) is the nonabelian gauge theory describing the
strong interaction. It is based on the local gauge symmetry group
$SU(3)_c$ with $c$ denoting color, under which the quark fields transform in the fundamental
representation and the gluon fields transform in the adjoint
representation. One of the most remarkable features of QCD is that the
simultaneous requirements of Lorentz invariance, locality, local gauge
invariance and power-counting renormalizability almost uniquely determine
the form of the theory
\citep{Yang:1954ek,Gross:1973id,Politzer:1973fx}.

In four spacetime dimensions the canonical mass dimensions of the
fundamental fields are
\begin{align}
[A_\mu]=1, \qquad [\psi]=\textstyle \frac32, \qquad [F_{\mu\nu}]=2.
\end{align}
Consequently, power-counting renormalizability restricts the Lagrangian
density to local operators of mass dimension four or less. The gauge-invariant
operators constructed from the quark and gluon fields are therefore highly
constrained. In addition to the Yang-Mills kinetic term and the quark
kinetic terms, the most general theory permits a complex quark-mass matrix,
corresponding in general to both scalar and pseudoscalar mass terms. It also
permits the independent dimension-four pseudoscalar gluonic operator
$\tr(F_{\mu\nu}\tilde F^{\mu\nu})$. The relationship between phases in the
quark-mass matrix and the coefficient of this gluonic operator will be
examined below in connection with the axial anomaly and the physical
parameter $\bar\theta$.

Working in natural units, $\hbar=c=1$, the Minkowski-space action is
\begin{align}\label{Eq:QCDAction}
S_M = \textstyle \int \! d^4x\, \mathcal L_M ,
\end{align}
where the Lagrangian density is
\begin{align}\label{Eq:QCDLag}
\mathcal L_M = \textstyle
-\frac14F_{\mu\nu}^aF^{a\mu\nu}
+ \bar\psi (i\gamma^\mu D_\mu-M) \psi
= -\frac12 \tr(F_{\mu\nu}F^{\mu\nu})
+ \bar\psi (i\slashed D-M) \psi .
\end{align}
The covariant derivative is
\begin{align}\label{Eq:CovDeriv}
D_\mu = \partial_\mu - igA_\mu^aT^a = \partial_\mu - igA_\mu ,
\end{align}
where
\begin{align}\label{Eq:GaugeField}
A_\mu \equiv A_\mu^aT^a ,
\end{align}
and the nonabelian field-strength tensor is defined by
\begin{align}\label{Eq:Fmunu}
F_{\mu\nu} = (i/g)  [D_\mu,D_\nu] = \partial_\mu A_\nu
    - \partial_\nu A_\mu - ig[A_\mu,A_\nu] .
\end{align}
In component form this becomes
\begin{align}\label{Eq:FmunuComp}
F_{\mu\nu}^a = \partial_\mu A_\nu^a
    - \partial_\nu A_\mu^a + gf^{abc}A_\mu^bA_\nu^c ,
\end{align}
where the generators satisfy
\begin{align}\label{Eq:Generators}
[T^a,T^b] = if^{abc}T^c, \qquad
\tr(T^aT^b) = \textstyle \frac12\delta^{ab}.
\end{align}
The dual field-strength tensor is defined by
\begin{align}
\tilde F^{\mu\nu} = \textstyle \frac12 \epsilon^{\mu\nu\rho\sigma}
F_{\rho\sigma},
\qquad\mbox{where}\qquad \epsilon^{0123}=+1.
\end{align}
The quark field $\psi$ contains the six quark flavours and the quark-mass
matrix is
\begin{align}\label{Eq:MassMatrix}
M = \mathrm{diag} (m_u,m_d,m_s,m_c,m_b,m_t),
\end{align}
where for now we are assuming a purely real diagonal mass matrix. We
will return to this issue later in the review. Throughout this review 
we employ the notation and conventions of \cite{Williams:2022bzq}.

The defining symmetry of QCD is invariance under local gauge
transformations,
\begin{align}
U(x)\in SU(3)_c ,
\end{align}
under which the quark fields transform according to
\begin{align}
\psi(x) \rightarrow \psi'(x) = U(x)\psi(x).
\end{align}
Rather than simply quoting the corresponding transformation law for the
gauge field, it is instructive to derive it from the requirement that the
covariant derivative transform covariantly,
\begin{align}
D'_\mu\psi' = UD_\mu\psi .
\end{align}
Since $D_\mu$ is a differential operator this implies
\begin{align}
D'_\mu = UD_\mu U^\dagger ,
\end{align}
from which one immediately obtains
\begin{align}\label{Eq:gaugetransf_Amu}
A_\mu' &= UA_\mu U^\dagger + (i/g) U\partial_\mu U^\dagger 
    = UA_\mu U^\dagger - (i/g) (\partial_\mu U) U^\dagger ,
\end{align}
where the two forms are equivalent because
$\partial_\mu(UU^\dagger)= \partial_\mu I=0$.

Using the definition of the field-strength tensor we find
\begin{align}
[D_\mu,D_\nu]\psi = -igF_{\mu\nu}\psi ,
\end{align}
which immediately implies
\begin{align}
[D'_\mu,D'_\nu] = U[D_\mu,D_\nu]U^\dagger ,
\end{align}
and hence
\begin{align}\label{Eq:Fmunu_transf}
F_{\mu\nu}' = UF_{\mu\nu}U^\dagger ,
\qquad \tilde F_{\mu\nu}' = U\tilde F_{\mu\nu}U^\dagger .
\end{align}
Gauge invariance of the gluon kinetic term then follows directly,
\begin{align}\label{Eq:FF_gauge_inv}
\tr(F_{\mu\nu}'F^{\prime\mu\nu}) = \tr(UF_{\mu\nu}F^{\mu\nu}U^\dagger)
= \tr(F_{\mu\nu}F^{\mu\nu}),
\end{align}
and the same argument shows that
\begin{align}
\tr(F_{\mu\nu}\tilde F^{\mu\nu})
\end{align}
is likewise gauge invariant.

Unlike quantum electrodynamics, the commutator term in
Eq.~(\ref{Eq:Fmunu}) gives rise to three- and four-gluon interaction
vertices. These gluon self-interactions are responsible for many of the
qualitative features that distinguish QCD from Abelian gauge theories,
including asymptotic freedom, confinement, dynamical chiral symmetry
breaking and the rich nonperturbative structure of the QCD vacuum.

The classical Lagrangian density provides the starting point for quantization.
Throughout this review we shall employ both the operator and
functional-integral formulations of QCD. Since many of the topological
concepts discussed later are most naturally expressed in the
functional-integral formalism, we briefly review the essential ideas here.

The Minkowski-space generating functional is
\begin{align}\label{Eq:Z_M}
Z_M = \textstyle \int \! \mathcal DA_\mu\,
\mathcal D\bar\psi\, \mathcal D\psi\;
e^{\,iS_M[A,\psi,\bar\psi]} .
\end{align}
Strictly speaking, this expression is only a formal one. Gauge-equivalent
field configurations represent the same physical configuration, so that the
gauge redundancy must be treated appropriately. In perturbation theory one
introduces gauge fixing together with the corresponding Faddeev-Popov ghost
fields \citep{Faddeev:1967fc,Becchi:1975nq,Tyutin:1975qk}. Beyond perturbation
theory, however, the implementation of gauge fixing is considerably more
subtle owing to the existence of Gribov copies, namely distinct 
gauge-equivalent configurations satisfying the same gauge-fixing condition
\citep{Gribov:1977wm}. Consequently, the precise mathematical definition of
the continuum functional integral remains an important topic in the
nonperturbative formulation of Yang-Mills theory \citep{Williams:2022bzq}.  Since the present review is concerned primarily
with gauge-invariant aspects of the theory, we shall generally suppress the
explicit gauge-fixing and ghost contributions except where they play an
essential role.

Expectation values of gauge-invariant operators are formally obtained from
the functional integral according to
\begin{align}\label{Eq:QCD_observ}
\braket{\hat O}_M
\equiv \langle\Omega|\hat O|\Omega\rangle_M
= \frac{ \displaystyle \textstyle \int \!
\mathcal DA_\mu\,\mathcal D\bar\psi\,\mathcal D\psi\;
O[A,\bar\psi,\psi]\,e^{\,iS_M}}
{\displaystyle\textstyle\int\!
\mathcal DA_\mu\,\mathcal D\bar\psi\,\mathcal D\psi\;
e^{\,iS_M}},
\end{align}
where $\hat O$ denotes a gauge-invariant operator expressed in terms of the
quark and gluon fields and $|\Omega\rangle$ is the exact nonperturbative
vacuum. Correlation functions of this form provide the fundamental
connection between the quantum field theory and physical observables.

Although the Minkowski-space formulation most directly reflects the causal
structure of relativistic quantum field theory, many nonperturbative
developments are more naturally formulated after a Wick rotation,
\begin{align}
x^0 \rightarrow -ix_4 ,
\end{align}
which transforms the oscillatory factor
$e^{\,iS_M}$ into the exponentially damped factor
$e^{-S_E}$. The Euclidean action is then found to be
\begin{align}
S_E = \textstyle \int\! d^4x_E\,\mathcal L_E ,
\end{align}
where
\begin{align}\label{Eq:L_E}
\mathcal L_E = \textstyle \frac14
F_{\mu\nu}^aF_{\mu\nu}^a + \bar\psi
(\gamma_{E\mu}D_{E\mu}+M) \psi 
= \frac12 \tr(F_{\mu\nu}F_{\mu\nu})
+ \bar\psi(\slashed D_E+M) \psi .
\end{align}
Using the convention that 
$\gamma^5\equiv i\gamma^0\gamma^1\gamma^2\gamma^3$,
the Euclidean Dirac matrices are defined by
\begin{align}
\gamma_{E4} \equiv \gamma^0,\qquad\gamma_{Ei}\equiv -i\gamma^i, \qquad 
\gamma_{E5} \equiv -\gamma_{E1}\gamma_{E2}\gamma_{E3}\gamma_{E4} 
= \gamma^5 ,
\end{align}
and satisfy the Euclidean Clifford algebra
\begin{align}
\{\gamma_{E\mu}, \gamma_{E\nu} \} = 2\delta_{\mu\nu}.
\end{align}
Throughout this review we adopt the convention
\begin{align}
\epsilon_{1234}=+1 .
\end{align}
The Euclidean gamma matrices satisfy
\begin{align}
\gamma_{E\mu}^\dagger = \gamma_{E\mu},
\qquad \gamma_{E5}^\dagger = \gamma_{E5}.
\end{align}

The corresponding Euclidean generating functional is
\begin{align}\label{Eq:Z_E}
Z_E = \textstyle \int \! \mathcal DA_\mu\,
    \mathcal D\bar\psi\, \mathcal D\psi\; e^{-S_E[A,\psi,\bar\psi]},
\end{align}
and Euclidean expectation values are given by
\begin{align}\label{Eq:O_E}
\braket{\hat O}_E
\equiv \langle\Omega|\hat O|\Omega\rangle_E
=\frac{\displaystyle\textstyle \int \!
\mathcal DA_\mu\,\mathcal D\bar\psi\,
\mathcal D\psi\;O[A,\bar\psi,\psi]\,e^{-S_E}}
{\displaystyle\textstyle \int \!
\mathcal DA_\mu\,\mathcal D\bar\psi\,
\mathcal D\psi\;e^{-S_E}}.
\end{align}
The Euclidean formulation is considered to be better behaved mathematically
than the Minkowski formulation and provides the natural setting for lattice 
gauge theory, semiclassical analyses based on instantons, and many rigorous
developments in quantum field theory. Throughout this review we shall move
freely between the Minkowski and Euclidean formulations, taking care to
distinguish those results that depend only upon local gauge invariance from
those that rely upon additional assumptions concerning the global structure
of the gauge-field configuration space.

The Lagrangian density in Eq.~(\ref{Eq:QCDLag}) is often referred to as
``the'' QCD Lagrangian density. Strictly speaking, however, it is not the most
general local, Lorentz-invariant, gauge-invariant and renormalizable
Lagrangian density that can be constructed from the quark and gluon fields.
As discussed above, there exists an additional independent gauge-invariant
operator of mass dimension four, namely the pseudoscalar operator
\begin{align}\label{Eq:FFdual}
\tr(F_{\mu\nu}\tilde F^{\mu\nu}) .
\end{align}
Unlike the Yang-Mills kinetic term, this operator is odd under parity and
under the combined $CP$ transformation. Nevertheless, it is fully consistent
with locality, Lorentz invariance, gauge invariance and renormalizability.
Consequently, the most general QCD Lagrangian may also contain the additional
term
\begin{align}\label{Eq:L_theta}
\mathcal L_\theta(x)
= \theta q(x) = 
\theta\,
\frac{g^2}{16\pi^2}
\tr[F_{\mu\nu}(x)\tilde F^{\mu\nu}(x)]
=\theta \frac{g^2}{32\pi^2}
F^a_{\mu\nu}(x)\tilde F^{a \mu\nu}(x),
\qquad q(x) \equiv \frac{g^2}{16\pi^2}
\tr[F_{\mu\nu}(x)\tilde F^{\mu\nu}(x)], 
\end{align}
where $\theta$ is a real dimensionless parameter and the pseudoscalar
density $q(x)$ is referred to as the topological
charge density for reasons that will be discussed later in this review.
Note that $q(x)$ is a local quantity. The conventional
interpretation of the spacetime integral of this local density lies
at the heart of the strong $CP$ problem. It is possible to
generalize this through the inclusion of a pseudoscalar mass term
by allowing a complex mass matrix. However, as we will discuss
in \Sec{Sec:thetabar}, the axial anomaly allows the corresponding
$CP$-violating phase to be shifted between the gluonic $\theta$
term and the quark mass matrix.

An important special property of this operator is that, for a sufficiently
smooth gauge field and within a spacetime patch in which the gauge potential
is defined, the topological charge density may be written as a total
four-divergence,
\begin{align}\label{Eq:qTotalDivergence}
q(x) &= \partial_\mu K^\mu(x),
\end{align}
where
\begin{align}\label{Eq:ChernSimonsCurrent}
K^\mu(x)
&=
\frac{g^2}{8\pi^2}
\epsilon^{\mu\nu\rho\sigma}\mathrm{tr}
    \left[A_\nu\partial_\rho A_\sigma-\frac{2ig}{3}
    A_\nu A_\rho A_\sigma\right].
\end{align}
is the Chern-Simons current. This identity is local. Although $q(x)$ is
gauge invariant, $K^\mu(x)$ is not gauge invariant and need not possess a
single globally defined representation over the complete spacetime.
Consequently, the fact that $q(x)$ can locally be written as a total
four-divergence does not by itself imply that its spacetime integral
vanishes. Whether the integral can be converted into a surface term, and
what value that surface term may take, depends on additional smoothness,
boundary and global assumptions. These distinctions will be examined in
detail below.

At first sight the introduction of such an operator appears entirely natural.
It satisfies all of the usual criteria employed in constructing a local
quantum field theory, and no symmetry of the classical QCD Lagrangian forbids
its presence. However, as we shall see throughout this review, the physical
interpretation of this operator differs in important respects from that of the
other terms appearing in the Lagrangian. In particular, while 
$\tr(F_{\mu\nu}\tilde F^{\mu\nu})$ is a local gauge-invariant operator,
the conventional interpretation of its spacetime integral is closely
connected with the global properties of the gauge field, the topology of
gauge-field configuration space and the boundary conditions imposed upon
the gauge fields.

The purpose of the following sections is to examine how these additional
ingredients enter the conventional formulation of the strong $CP$ problem.
We begin by reviewing the standard introduction of the $\theta$ term and its
transformation properties under the discrete symmetries.

\subsection{$\theta$ term and discrete symmetries}
\label{Sec:ThetaDiscrete}

As discussed in the previous subsection, the QCD
Lagrangian is highly
constrained by the simultaneous requirements of
locality, Lorentz invariance, local gauge invariance
and renormalizability. Nevertheless, it is not the 
most general renormalizable Lagrangian density that
can be constructed from the quark and
gluon fields. One may also include the 
gauge-invariant pseudoscalar operator
$\tr(F_{\mu\nu}\tilde F^{\mu\nu})$, giving
\begin{align}\label{Eq:ThetaLag}
\mathcal{L}_{\rm QCD}
& = \mathcal L^{\rm min}_{\rm QCD} 
+ \mathcal L_\theta
= -{\textstyle \frac14} F_{\mu\nu}^aF^{a\mu\nu}
+\bar\psi(i\slashed D-M)\psi
+\theta\,\frac{g^2}{16\pi^2}
\tr(F_{\mu\nu}\tilde F^{\mu\nu}) ,
\end{align}
where $\mathcal L^{\rm min}_{\rm QCD}$ is that
defined in Eq.~(\ref{Eq:QCDLag}) and $\theta$ is a real dimensionless parameter. 
Since $\tr(F_{\mu\nu}\tilde F^{\mu\nu})$ is gauge invariant, Lorentz invariant
and has mass dimension four, the $\theta$ term satisfies the same fundamental
requirements as the remaining terms in the QCD Lagrangian. Consequently,
nothing appears to forbid its inclusion on the basis of locality, Lorentz
invariance or gauge invariance alone.

The chromoelectric and chromomagnetic fields are defined as
Lie-algebra-valued fields by
\begin{align}
\mathbf E &\equiv \mathbf E^aT^a,
\qquad \mathbf B \equiv \mathbf B^aT^a,
\qquad F^{0i} = -E^i , 
\qquad F^{ij} = -\epsilon^{ijk}B^k ,
\end{align}
where $E^i\equiv E^{ai}T^a$ and 
$B^i\equiv B^{ai}T^a$. With these conventions,
\begin{align}
F_{\mu\nu}^a\tilde F^{a\mu\nu}
&=
-4\,\mathbf E^a\cdot\mathbf B^a,
\end{align}
and, using $\tr(T^aT^b)=\frac12\delta^{ab}$,
\begin{align}
\tr(F_{\mu\nu}\tilde F^{\mu\nu})
&=
-4\,\tr(\mathbf E\cdot\mathbf B)
=
-2\,\mathbf E^a\cdot\mathbf B^a.
\end{align}
The physical significance of the $\theta$ term
follows from its transformation properties under
the discrete symmetries. 
One finds that under parity
\begin{align}\label{Eq:ParityFields}
\mathbf E(\mathbf x,t)
&\to -\mathbf E(-\mathbf x,t), 
\qquad
\mathbf B(\mathbf x,t)
\to  \mathbf B(-\mathbf x,t) ,
\end{align}
under charge conjugation
\begin{align}\label{Eq:ChargeFields}
\mathbf E &\to -\mathbf E^{\,T}, \qquad
\mathbf B \to -\mathbf B^{\,T} ,
\end{align}
and under time-reversal
\begin{align}
\mathbf E(\mathbf x,t)
&\to \mathbf E(\mathbf x,-t),
\qquad
\mathbf B(\mathbf x,t) \to
-\mathbf B(\mathbf x,-t).
\end{align}
Note that the color transpose is necessary under charge conjugation
for nonabelian fields.

It follows immediately that
\begin{align}\label{Eq:CPProperties}
C:\quad
\tr(F_{\mu\nu}\tilde F^{\mu\nu})
&\rightarrow
\phantom{-}\tr(F_{\mu\nu}\tilde F^{\mu\nu}), \nonumber\\
P:\quad
\tr(F_{\mu\nu}\tilde F^{\mu\nu})
&\rightarrow
-\tr(F_{\mu\nu}\tilde F^{\mu\nu}), \nonumber\\
T:\quad
\tr(F_{\mu\nu}\tilde F^{\mu\nu})
&\rightarrow
-\tr(F_{\mu\nu}\tilde F^{\mu\nu}), \nonumber\\
CP:\quad
\tr(F_{\mu\nu}\tilde F^{\mu\nu})
&\rightarrow
-\tr(F_{\mu\nu}\tilde F^{\mu\nu}) .
\end{align}
The operator $\tr(F_{\mu\nu}\tilde F^{\mu\nu})$ is therefore odd under parity,
time reversal and the combined $CP$ transformation, while remaining invariant
under charge conjugation. Consequently, a nonvanishing value of $\theta$
introduces explicit $P$, $T$ and $CP$ violation into the strong interaction,
while preserving $CPT$ invariance.

This immediately gives rise to the conventional formulation of the strong
$CP$ problem. Since the $\theta$ term appears to be allowed by all of the
fundamental principles used to construct the QCD Lagrangian, one might expect
the corresponding parameter $\theta$ to be of order unity. Experiment,
however, tells a very different story. 
Measurements of the neutron electric
dipole moment imply that the effective strong $CP$ parameter is extremely
small, currently satisfying $|\bar\theta|\lesssim10^{-10}$
\citep{Baker2006ts,Abel2020gbr}. Explaining this enormous suppression has
motivated decades of theoretical investigation and numerous proposed
solutions.

The remainder of this review examines the assumptions that underlie the
introduction of the $\theta$ term as a physical contribution to the QCD
action.

\subsection{Experimental constraints}
\label{Sec:Experimental}

The strong $CP$ problem is not merely a theoretical curiosity. It arises
because experiments indicate that, if the $\theta$ term is present in the QCD
Lagrangian, its coefficient must be extraordinarily small. The most stringent
constraint comes from measurements of the neutron electric dipole moment
(EDM), which is one of the most sensitive probes of $CP$ violation in the
strong interaction.

Classically, the interaction energy of a particle with an electric 
dipole moment $\mathbf d$ in an external electric field $\mathbf E$ is
\begin{align}
H_{\rm int}
=
-\mathbf d\cdot\mathbf E ,
\end{align}
where $\mathbf E$ denotes the applied electric field.

A fundamental requirement of relativistic quantum field theory is that the
description of a physical system be independent of its spacetime location
and of the inertial frame from which it is observed. The Poincar\'e group is
precisely the group relating all such descriptions. Elementary particles are
therefore classified by the irreducible representations of the Poincar\'e
group. A spin-$\frac12$ particle transforms according to the irreducible
spin-$\frac12$ representation and consequently possesses only one
independent intrinsic direction, namely its spin. Its intrinsic electric
dipole moment must therefore be aligned with the spin,
\begin{align}
\mathbf d
=
2d_n\,\mathbf S ,
\end{align}
where $\mathbf S$ is the neutron spin operator and $d_n$ is the neutron
electric dipole moment. The interaction Hamiltonian therefore takes the form
\begin{align}\label{Eq:EDM}
H_{\rm EDM}
=
-2d_n\,\mathbf S\!\cdot\!\mathbf E .
\end{align}

The electric field is a polar vector, whereas the spin is an axial vector.
Consequently, the scalar product $\mathbf S\!\cdot\!\mathbf E$ is odd under
parity. Furthermore, under time reversal the spin changes sign while the
electric field remains unchanged, so the interaction is also odd under time
reversal. Assuming the validity of the $CPT$ theorem, a permanent neutron
electric dipole moment therefore violates $P$, $T$ and $CP$ invariance.
Within the Standard Model, the dominant strong-interaction contribution to
the neutron electric dipole moment is proportional to the effective
parameter $\bar\theta$. Consequently, experimental searches for a nonzero
neutron electric dipole moment provide the most direct and stringent
experimental test of strong $CP$ violation.

Within the Standard Model, the $CP$ violation associated with the
Cabibbo-Kobayashi-Maskawa (CKM) matrix induces an exceedingly small neutron
EDM, many orders of magnitude below current experimental sensitivity
\citep{Pospelov:2005pr,Engel:2013lsa}. In contrast, a nonzero effective QCD
vacuum angle generates a neutron EDM approximately proportional to
$\bar\theta$,
\begin{align}\label{Eq:nEDMTheta}
d_n \sim 10^{-16}\,\bar\theta\; e\,{\rm cm} ,
\end{align}
where the numerical coefficient depends upon the nonperturbative treatment of
QCD and has been estimated using a variety of approaches, including chiral
effective field theory, QCD sum rules and lattice gauge theory
\citep{Crewther1979,Pospelov:2005pr,Guo:2015tla,Dragos:2019oxn}.

The current experimental limit on the neutron electric dipole moment is
\begin{align}\label{Eq:nEDMLimit}
|d_n|
<
1.8\times10^{-26}\;e\,{\rm cm}
\qquad
(90\%~{\rm C.L.}),
\end{align}
leading to the approximate constraint
\begin{align}\label{Eq:ThetaLimit}
|\bar\theta|
\lesssim
10^{-10} ,
\end{align}
where the precise numerical bound depends upon the theoretical relation
between $d_n$ and $\bar\theta$
\citep{Abel2020gbr,Dragos:2019oxn}.

This striking disparity between theoretical expectation and experimental
observation is the conventional strong $CP$ problem. If the $\theta$ term is
allowed by the fundamental principles used to construct the QCD Lagrangian density,
there appears to be no obvious reason why its coefficient should be so
extraordinarily small. Explaining the apparent suppression of $\bar\theta$ has
motivated numerous proposed solutions, including the Peccei-Quinn mechanism,
axion models, spontaneous $CP$ violation and parity-based extensions of the
Standard Model.

The discussion up to this point represents the conventional formulation of the
strong $CP$ problem. The remainder of this review examines the assumptions
that underlie this formulation, beginning with the role of chiral symmetry,
the axial anomaly and the origin of the physically observable parameter
$\bar\theta$.


\section{Properties of QCD from local structure}
\label{Sec:local_QCD}

In this section we focus on properties of QCD that
follow from its local structure and so these properties 
are independent of assumptions about the global 
structure of the theory.

\subsection{Gauge invariance and local observables}
\label{Sec:LocalGauge}

Local gauge invariance is the defining symmetry principle of QCD. It
determines the form of the interactions between quarks and gluons and ensures
that physical observables are independent of the choice of gauge. Although the
gauge fields themselves are gauge dependent, measurable quantities must be
constructed from gauge-invariant operators. This distinction between the
fundamental fields and the observable quantities plays a central role in the
modern formulation of quantum field theory.

Under a local $SU(3)$ gauge transformation,
\begin{align}\label{Eq:GaugeTransform}
\psi(x) &\rightarrow U(x)\psi(x), \qquad
A_\mu(x) \rightarrow U(x)A_\mu(x)U^\dagger(x)
+ (i/g)U(x)\partial_\mu U^\dagger(x),
\end{align}
where $U(x)\in SU(3)$ is an arbitrary smooth function of spacetime. The
Lagrangian density is invariant under these transformations, and consequently all
physical predictions of QCD must also be gauge invariant.

The elementary quark and gluon fields are therefore not themselves observable.
Instead, physical quantities are constructed from gauge-invariant operators.
Examples include the scalar gluon operator
$\tr(F_{\mu\nu}F^{\mu\nu})$,
the pseudoscalar topological charge density
\begin{align}\label{Eq:GaugeInvFFDual}
q(x)
=
\frac{g^2}{16\pi^2}
\tr(F_{\mu\nu}\tilde F^{\mu\nu}),
\end{align}
and hadronic interpolating fields constructed from color-singlet combinations
of quark and gluon fields. Since $F_{\mu\nu}$ transforms covariantly,
$F_{\mu\nu}\rightarrow UF_{\mu\nu}U^\dagger$,
both $\tr(F_{\mu\nu}F^{\mu\nu})$ and $q(x)$ are gauge invariant at each
spacetime point.

Gauge transformations relate different mathematical representatives of the
same physical gauge configuration. Formally, the \emph{gauge orbit} of a gauge
field $A_\mu$ is the set of configurations obtained by acting on it with
the all admissible gauge transformations $U$.
Physical observables and local gauge-invariant quantities, such 
as $\tr(F_{\mu\nu}F^{\mu\nu})$ and $\tr(F_{\mu\nu}\tilde F^{\mu\nu})$, 
are constant along such an orbit.

Observable quantities are obtained from correlation functions of such local
gauge-invariant operators. In the Euclidean formulation these are given by \Eq{Eq:O_E} and so
\begin{align}\label{Eq:Correlation}
\langle \hat O_1(x_1)\cdots \hat O_n(x_n)\rangle_E
= \frac{\displaystyle
\textstyle \int \mathcal \!D\Phi\;
O_1(x_1)\cdots O_n(x_n)\, e^{-S_E}}
{\displaystyle
\textstyle \int \!\mathcal D\Phi\; e^{-S_E}} ,
\end{align}
where $\mathcal D\Phi$ denotes integration over the quark and gluon fields.
These correlation functions encode hadron masses, matrix elements,
scattering amplitudes and other observable consequences of QCD. In
particular, they provide the basis for lattice gauge theory calculations and
for the nonperturbative definition of the theory.

An important example is the topological susceptibility,
\begin{align}\label{Eq:TopSuscept}
\chi = \textstyle \int\!d^4x\,
\langle \hat q(x) \hat q(0)\rangle_E ,
\end{align}
which is defined entirely in terms of the local gauge-invariant operator
$\hat q(x)$. As will be discussed in later sections, the existence of this local
correlation function does not by itself require any particular interpretation
of the spacetime integral of $\hat q(x)$ itself or any decomposition of the functional
integral into topological sectors. It is therefore useful to distinguish
carefully between local observables, which are directly defined through
correlation functions, and additional global structures that may be introduced
in the formulation of the theory.

This distinction forms the conceptual starting point for the discussion that
follows. We first examine the local operator $\hat q(x)$ and its mathematical
properties before turning to the conventional interpretation of its spacetime
integral as a global topological charge.

\subsection{Functional integral formulation}
\label{Sec:FunctionalIntegral}
 
The modern nonperturbative formulation of QCD is most naturally expressed in 
terms of the functional integral. In this approach, physical observables are 
obtained as expectation values of products of local gauge-invariant operators
computed by averaging over gauge and quark field configurations with weights
determined by the classical action.
This formulation provides the foundation for perturbation
theory, lattice gauge theory and most contemporary approaches to
nonperturbative QCD. As outlined above, for the
purposes of perturbation theory gauge fixing is
achieved through the introduction of ghost fields through
the Faddeev-Popov approach. 
In perturbation theory one expands about the trivial classical vacuum
$A_\mu=0$, where Gribov copies do not affect the perturbative expansion.
In lattice calculations of gauge-invariant observables no gauge fixing 
is required, so Gribov copies play no direct role. 
They become relevant only when gauge-dependent quantities are considered.

Formally, the Minkowski-space generating functional is
\begin{align}\label{Eq:ZM}
Z_M
= \int \mathcal D A_\mu \, \mathcal D\bar\psi \, \mathcal D\psi \;
e^{\,iS_M[A,\psi,\bar\psi]} ,
\end{align}
where the action is
\begin{align}\label{Eq:ActionM}
S_M = \textstyle \int \! d^4x \; \mathcal L_M .
\end{align}
Although this expression is formally useful, the oscillatory nature of the
weight $e^{iS_M}$ makes it difficult to interpret mathematically and unsuitable
for most nonperturbative calculations.

Instead, one usually performs a Wick rotation,
\begin{align}\label{Eq:WickRotation}
x^0 \rightarrow -ix_4 ,
\end{align}
leading to the Euclidean generating functional
\begin{align}\label{Eq:ZE}
Z_E
= \textstyle \int\! \mathcal D A_\mu \, \mathcal D\bar\psi \, \mathcal D\psi \;
e^{-S_E[A,\psi,\bar\psi]} ,
\end{align}
where
\begin{align}\label{Eq:ActionE}
S_E = \textstyle \int\! d^4x_E \; \mathcal L_E .
\end{align}
The Euclidean weight $e^{-S_E}$ is exponentially damped and therefore provides
the natural starting point for lattice gauge theory, semiclassical analyses and
constructive approaches to quantum field theory.

The functional measures
$\mathcal D A_\mu$, $\mathcal D\bar\psi$ and $\mathcal D\psi$ denote
integration over all gauge and quark field configurations, where the fermion 
field configurations are Grassmann valued. Formally, they may
be viewed as infinite-dimensional analogues of ordinary integration measures.
In practice, however, they require regularization before becoming
mathematically well defined. 
Vacuum expectation values of gauge-invariant operators are obtained from the
functional integral using
\begin{align}\label{Eq:VEV}
\langle \hat O \rangle_E
=
\frac{\displaystyle
\textstyle \int \!\mathcal D A_\mu \, \mathcal D\bar\psi \, \mathcal D\psi \;
O[A,\psi,\bar\psi]\,
e^{-S_E}}
{\displaystyle
\textstyle \int\! \mathcal D A_\mu \, \mathcal D\bar\psi \, \mathcal D\psi \;
e^{-S_E}} .
\end{align}
More generally, all physical observables are determined by correlation
functions of local gauge-invariant operators, from which one may extract
particle masses, decay constants, matrix elements and scattering amplitudes.
This functional-integral formulation is therefore the primary framework within
which nonperturbative QCD is understood.

On the lattice the measure becomes an ordinary
finite-dimensional integral over link variables and fermion fields, while in
the continuum it is usually understood through regularization and
renormalization. Although the precise mathematical definition of the continuum
functional measure remains a deep problem in constructive quantum field theory,
its regularized versions have proved extraordinarily successful in describing
the observed properties of the strong interaction \citep{Creutz1983, MontvayMunster,Neuberger:1998,Smit,GattringerLang,Rothe2012}.

In lattice QCD, since the fermion action is bilinear in the Grassmann fields, 
the fermionic functional integral can be performed analytically. Wick’s theorem
reduces products of fermion fields to sums of fermion propagator contractions,
leaving a functional integral over the gauge fields weighted by the fermion
determinant.
Consider the lattice estimate of some Euclidean Green's function
$\langle\Omega| (O[\hat A,\hat{\bar\psi},\hat\psi])|\Omega\rangle_E$, where 
$O[\hat A,\hat{\bar\psi},\hat\psi]$ is some gauge invariant quantity.
Upon discretizing spacetime, the gauge field is represented by link variables
$U_\mu(x)\in SU(3)$, and the continuum functional integral becomes an ordinary 
finite-dimensional integral over these variables.
Writing the observable 
$O[A,{\bar\psi},\psi]$ in terms of the link variables leads
to a  corresponding quantity $O[U,{\bar\psi},\psi]$ in the functional 
integral. We then find
\begin{align}\label{Eq:phys_observ_lattice_QCD}
&\langle\Omega| O[\hat A,\hat{\bar\psi},\hat\psi]|\Omega\rangle_E =
\frac{\int\!{\cal D}U(\prod_f {\cal D}\bar\psi^f{\cal D}\psi^f)\,
	O[U,\bar\psi,\psi] e^{-S_E[\bar\psi,\psi,U]}}
	{\int\!{\cal D}U(\prod_f {\cal D}\bar\psi^f{\cal D}\psi^f) \,
	e^{-S_E[\bar\psi,\psi,U]}} \\
&\!=\!\frac{\int\!{\cal D}U\,
	O[U,M^{-1}[U]] \,\{\prod_f \det D^f[U]\} \,e^{-S_E[U]}}
	{\int\!{\cal D}U \, \{\prod_f \det D^f[U] \}\, e^{-S_E[U]}} 	
	= \lim_{N_{\rm config}\to \infty}\!\!
		\frac{\sum_{i=1}^{N_{\rm config}}  O[U_i,M^{-1}[U_i]]}{N_{\rm config}} 
		, \nonumber
\end{align}
where $f$ is the quark flavor, $(D^f)^{-1}[U]$ is the quark propagator
for that $f$ in the background gauge field $U$ and
$O[U,D^{-1}[U]]$ is the observable after all of the fermion field
contractions have been carried out and summed over. Gauge configurations 
are then generated using Monte Carlo techniques with relative probability 
\begin{align}\label{Eq:unquenched_QCD_P[U]}
P[U']/P[U]&= \textstyle \{\prod_f \det(D^f[U'])\} e^{-S_E[U']}/ \{\prod_f \det(D^f[U])\} e^{-S_E[U]}
\end{align}
to generate the ensemble of lattice configurations,
$\{U_1,\ldots,U_{N_{\rm config}} \}$. 
The continuum limit ($a\to0$) and subsequently the infinite-volume limit
($V\to\infty$) are understood to be taken after the Monte Carlo average 
in \Eq{Eq:phys_observ_lattice_QCD} has been formed.
A continuum Dirac fermion at zero chemical potential and positive real 
mass has a positive determinant.
For two degenerate light flavors the product of their
fermion determinants is always positive even on the lattice,
since on the lattice $\det D^f[U]\in \mathbb{R}$ and so for
mass-degenerate fermion pairs $\prod_f \det(D^f[U'])>0$.
Modern $(2+1)$-flavor simulations incorporate the strange quark using
pseudofermion representations together with rational hybrid Monte Carlo
(RHMC) algorithms, so that the appropriate fermion determinant is
included in the probability measure used to generate the ensemble in
\Eq{Eq:phys_observ_lattice_QCD}.

Gauge-equivalent field configurations represent the same physical state, so
the physical configuration space of a gauge theory is the quotient of the
space of gauge fields by the gauge group. A functional-integral formulation
equivalent to the constrained Hamiltonian theory must therefore assign each
gauge orbit a single physical contribution with the appropriate measure and
must not introduce orbit-dependent multiplicities. Gauge fixing attempts to
implement this quotient by selecting a representative from each orbit. In
perturbation theory the Faddeev-Popov construction is adequate locally around
the trivial vacuum. Nonperturbatively, however, Gribov copies show that a
standard local gauge condition does not select a unique representative.
For example, while restriction to the first Gribov region removes 
infinitesimal copies it does not, in general, restrict the field to 
the fundamental modular region, which
contains the absolute minima of the gauge-fixing functional on each orbit.

For gauge-invariant lattice observables no gauge fixing is imposed in 
practice. A finite Monte Carlo ensemble is overwhelmingly unlikely to 
contain two representatives of the same gauge orbit; indeed, 
gauge-equivalent configurations must have identical Euclidean action, 
whereas independently generated configurations almost never have the 
same action to machine precision. It is therefore straightforward to 
verify that no two sampled configurations even satisfy this necessary 
condition for gauge equivalence. Operationally, the sampled ensemble 
therefore represents distinct physical gauge orbits without requiring 
the explicit construction of a global gauge condition.

No smoothing of the generated gauge ensemble is required in order to
calculate ordinary renormalization-group-invariant observables such as
hadron masses, decay constants or matrix elements directly from QCD. 
In particular, the gauge configurations generated in standard lattice 
simulations are not first transformed into smooth representatives 
satisfying the conditions required for an integer-valued gluonic 
topological charge. Such constructions are introduced only when 
topology itself is the quantity under investigation.
Smoothing procedures 
such as gradient flow are instead introduced when defining particular
renormalized operators, composite operators or topological observables.

It is important to distinguish between the definition of the functional 
integral itself and the later construction of global topological quantities. 
The lattice functional integral is defined directly as an integral over 
gauge-field configurations. It does not require the initial ensemble 
to be partitioned into configurations carrying integer-valued topological 
charge. When one wishes to define global topological quantities, 
additional constructions,
such as gradient flow, overlap fermions, or related techniques, may be 
introduced to define them. Whether such constructions should be regarded 
as revealing an intrinsic global decomposition of the original functional 
integral or as additional mathematical structures built upon it is a 
separate question that will be examined in later sections.
At this stage we require only the existence of the local functional
integral and the correlation functions that it defines.

\subsection{Classical and semiclassical limits}
\label{Sec:ClassicalLimit}

The functional integral defines quantum field theory as an average over 
all field configurations, not merely over solutions 
of the classical field equations. Consequently, the field configurations 
that dominate 
the functional integral are fundamentally quantum objects. They are 
typically highly irregular, fluctuating on arbitrarily short distance 
scales, and are more appropriately regarded as distribution-valued 
fields than as smooth classical functions 
\citep{Glimm:1987ylb,Simon:2015spl,colella1973sample}. This expectation is supported 
rigorously for free scalar fields, where typical configurations have 
been proved to be nowhere continuous with probability one, and 
asymptotic freedom strongly suggests that nonabelian gauge fields
exhibit analogous ultraviolet behavior \citep{colella1973sample}.
Nevertheless, the 
macroscopic world is accurately described by smooth classical and 
semiclassical fields. Understanding how these descriptions emerge 
from the underlying quantum theory is therefore of fundamental 
importance.

The key observation is that physical measurements are necessarily performed
over finite spacetime regions and therefore probe suitable averages of the
underlying quantum fields rather than their pointwise values. Consequently,
macroscopic observables are naturally represented by operator-valued
distributions smeared over spacetime regions large compared with the relevant
microscopic correlation lengths.
A simple example is that a quantity such as $\phi(x)^2$ may, at finite
resolution, be represented by the corresponding quantity constructed from
the coarse-grained field,
\begin{align}\label{Eq:CoarseObservable}
O[\phi]=\phi(x)^2 \to O_R[\phi]
= \left[(1/V_R)\textstyle \int_R d^4x\,\phi(x)\right]^2 ,
\end{align}
where $R$ denotes a spacetime region of volume $V_R$. Such an observable is
insensitive to ultraviolet fluctuations but probes the collective
behavior of the field throughout the region $R$.

In Euclidean quantum field theory, the principle of cluster decomposition
implies that fluctuations in sufficiently well-separated subregions of $R$
become approximately independent. Consequently, the coarse-grained field
entering $O_R$ is obtained by averaging a large number of nearly independent
contributions. The probability that such a coarse-grained observable differs
significantly from its typical value is therefore exponentially suppressed
with increasing spacetime volume. Schematically,
\begin{align}\label{Eq:LargeDeviation}
P(O_R)
\sim
\exp\!\left[-V_R I(O_R)\right] ,
\end{align}
where $I(O_R)\ge0$ is the corresponding large-deviation rate function
\citep{ZinnJustin2002,Varadhan1984}. As the averaging region
becomes macroscopic, the probability distribution becomes increasingly sharply
peaked around the typical value of the observable.

The typical macroscopic field configuration is determined by the stationary
point (or, in stable situations, the minimum) of the appropriate coarse-grained
effective action. In
this sense, classical field theory emerges naturally as the macroscopic limit
of the underlying quantum theory. The smooth classical fields familiar from
continuum physics should thus be regarded as effective descriptions of
coarse-grained quantum fields rather than as typical microscopic field
configurations contributing directly to the functional integral.

In fact, the classical and semiclassical descriptions both emerge from this 
coarse-graining 
process. After averaging over
sufficiently large spacetime regions, the probability distribution for the
coarse-grained fields becomes concentrated near stationary configurations of
an effective action appropriate to the corresponding resolution scale.
Schematically,
\begin{align}
P[\Phi_R]
\propto
\exp\!\left[
-\frac{1}{\hbar}
\Gamma_R[\Phi_R]
\right],
\end{align}
where $\Phi_R$ denotes the coarse-grained field and $\Gamma_R$ is the
corresponding effective action. 
This expression is schematic and is intended only to indicate the exponential
concentration of the probability distribution around the typical
coarse-grained field configurations.
The dominant coarse-grained configuration
$\bar\Phi_R$ satisfies
\begin{align}
\left.
\frac{\delta\Gamma_R[\Phi_R]}
{\delta\Phi_R(x)}
\right|_{\Phi_R=\bar\Phi_R}
=
0.
\end{align}
Retaining only this dominant configuration gives the classical
approximation, which is a smooth configuration. 
The semiclassical approximation includes the leading quantum
fluctuations about it. Since the effective action $\Gamma_R$ is already a
coarse-grained description appropriate to the resolution scale $R$, the
fluctuation field $\eta$ likewise represents a smooth coarse-grained
deviation rather than the microscopic ultraviolet fluctuations of the
underlying quantum fields. Writing
\begin{align}
\Phi_R
=
\bar\Phi_R+\eta,
\end{align}
one expands
\begin{align}
\Gamma_R[\bar\Phi_R+\eta]
&=
\Gamma_R[\bar\Phi_R]
+
\frac12
\int d^4x\,d^4y\,
\eta(x)
\Gamma_R^{(2)}(x,y)
\eta(y)
+
O(\eta^3).
\end{align}
The Gaussian functional integral over $\eta$ then produces the leading
quantum correction to the classical approximation.
In this sense, the semiclassical
approximation describes the leading quantum deviation from the classical
dominance that emerges after the underlying fields have been probed only at
a sufficiently coarse spacetime resolution.

This distinction will be central to the discussion that follows. Classical 
and semiclassical gauge fields provide invaluable analytical tools and 
underpin much of our understanding of nonperturbative gauge theory. 
However, they should not automatically be identified with the typical field 
configurations contributing to the full quantum functional integral, which 
are expected instead to be highly irregular distribution-valued fields.
Throughout this review it will therefore be important to distinguish
statements that depend only on the existence of the functional 
integral from those that rely upon semiclassical expansions about 
smooth gauge fields.

\subsection{Axial anomaly}
\label{Sec:AxialAnomaly}

Whenever a symmetry of a classical field theory fails 
to survive quantization, the symmetry is said to possess a quantum anomaly,
or simply an anomaly.
In the classical massless theory, the QCD Lagrangian possesses a global
chiral symmetry
\begin{align}\label{Eq:ChiralSymmetry}
SU(N_f)_L \times SU(N_f)_R \times U(1)_V \times U(1)_A ,
\end{align}
where $N_f$ is the number of quark flavours. 
The nonabelian chiral symmetries $SU(N_f)_L\times SU(N_f)_R$ remain exact 
symmetries of the quantum theory (up to explicit quark masses and 
spontaneous symmetry breaking), whereas the singlet axial symmetry 
$U(1)_A$ is broken by the anomaly. The vector symmetry
$U(1)_V$ corresponds to baryon-number conservation, 
while the non-Abelian chiral symmetry is spontaneously broken to the vector
subgroup, giving rise to Goldstone bosons which, for nonzero quark masses,
become the light pseudoscalar pseudo-Goldstone bosons.
Classically, the axial transformation
\begin{align}\label{Eq:AxialRotation}
\psi(x)
&\to
e^{\,i\alpha\gamma^5}\psi(x), \qquad
\bar\psi(x)
\to
\bar\psi(x)e^{\,i\alpha\gamma^5},
\end{align}
leaves the massless QCD Lagrangian invariant, implying conservation of the
axial current
\begin{align}\label{Eq:AxialCurrent}
J_5^\mu = \bar\psi\gamma^\mu\gamma^5\psi ,
\qquad
\partial_\mu J_5^\mu = 0 .
\end{align}
Writing
\begin{align}
\psi_L = \textstyle \frac12(1-\gamma^5)\psi,
\qquad
\psi_R = \textstyle \frac12(1+\gamma^5)\psi,
\end{align}
the nonabelian chiral symmetry acts independently on the left- and
right-handed quark fields according to
\begin{align}
\psi_L
&\rightarrow
U_L\psi_L,
\qquad
U_L\in SU(N_f)_L,\\
\psi_R
&\rightarrow
U_R\psi_R,
\qquad
U_R\in SU(N_f)_R.
\end{align}
Quantum mechanically, however, this conservation law is violated by the axial
anomaly. Although the classical action remains invariant under the chiral
transformation, the functional integration measure is not. The anomaly may be
derived perturbatively from the Adler-Bell-Jackiw triangle diagram
\citep{Adler:1969gk,Bell:1969ts}, or more generally through Fujikawa's elegant
analysis of the functional measure \citep{Fujikawa:1979ay,Fujikawa:1980eg}.
A derivation is provided here in Appendix~\ref{App:Fujikawa}.

Since the functional measure is defined as the product of the integration measures over the expansion coefficients, the Jacobian of the chiral transformation is obtained from the transformation of these coefficients in \Eq{Eq:chiral_Jacobian}.
In Fujikawa's approach the fermion fields are expanded in a complete set of
eigenfunctions of the Dirac operator. Under a chiral rotation the classical
action transforms in the expected way, but the fermionic functional measure
acquires a nontrivial Jacobian. After suitable gauge-invariant
regularization one has from \Eq{Eq:chiral_Jacobian} that
\begin{align}\label{Eq:FujikawaJacobian}
\mathcal D\bar\psi\,\mathcal D\psi \rightarrow
    \mathcal D\bar\psi\,\mathcal D\psi\,\exp\!\left(-2i\alpha N_f
    \textstyle \int \! d^4x\,q(x)\right) .
\end{align}
where
\begin{align}\label{Eq:TopDensity}
q(x) = \frac{g^2}{16\pi^2} \tr(F_{\mu\nu}\tilde F^{\mu\nu})
\end{align}
is the topological charge density introduced earlier in \Eq{Eq:L_theta}. 
Consequently, the divergence of the axial current becomes
\begin{align}\label{Eq:AxialAnomaly}
\partial_\mu J_5^\mu
=
2im\,\bar\psi\gamma^5\psi
+
2N_f\,q(x) ,
\end{align}
where, for simplicity, a common real quark mass has been assumed. For further
details see Appendix \ref{App:Fujikawa}.

The axial anomaly is an exact quantum effect. It is independent of
perturbative corrections beyond one loop through the Adler-Bardeen theorem
\citep{Adler:1969gk,Adler:1969er}, and plays a central role in many aspects of strong
interaction physics. Among its most important consequences are the
$\pi^0\rightarrow\gamma\gamma$ decay amplitude, the resolution of the
$U(1)_A$ problem, the large mass of the $\eta^\prime$ meson, and the
transformation properties of the QCD vacuum under chiral rotations.

For the purposes of the present review, the essential point is that the
anomaly is fundamentally a local statement. It relates the divergence of the
local axial current to the local gauge-invariant operator $q(x)$ and therefore
follows directly from the local structure of quantum field theory. The 
derivation of the anomaly
itself does not require the introduction of topological sectors, instantons or
a $\theta$ vacuum. Those additional concepts enter only when one considers the
spacetime integral of $q(x)$ and its interpretation, which will be discussed in
later sections.

The anomaly nevertheless has an important consequence for the strong $CP$
problem. Since chiral rotations modify the functional measure, they also alter
the coefficient multiplying the operator $q(x)$. This observation leads
naturally to the physically observable combination of the vacuum angle and the
phase of the quark mass matrix, which is the subject of the following
subsection.
This observation does not by itself require any global assumptions 
concerning the space of gauge fields. It follows directly from the 
local transformation properties of the functional measure. The relation 
between this local result and the conventional interpretation of the 
spacetime integral of $q(x)$ is one of the principal themes developed 
later in this review.

\section{How $\theta$ becomes $\bar\theta$ from the quark mass matrix}
\label{Sec:thetabar}

We now wish to understand how a complex phase of the quark mass matrix
contributes to the observable vacuum angle. As we shall see, the physically
relevant parameter is not $\theta$ itself, but the combination
$\bar\theta=\theta+\arg\det M$, where $M$ is the quark mass matrix.

The quark mass contribution to the QCD Lagrangian density is
\begin{align}\label{Eq:quark_L_m}
\mathcal{L}_m &=-\bar\psi_L M \psi_R +\mathrm{h.c.}
= -\bar\psi_L M \psi_R -\bar\psi_R M^\dagger \psi_L  ,
\end{align}
where $\psi$ is an $N_f$-component Dirac spinor in flavor space,
$M$ is a general complex $N_f\times N_f$ matrix, and
\begin{align}
\psi_{L,R} = P_{L,R}\psi,
\qquad
P_{L,R} = \textstyle\frac12(1\mp\gamma_5)
\end{align}
are the left- and right-handed chiral projections. This expression
generalizes the diagonal real mass matrix $M$ appearing in
\Eq{Eq:MassMatrix}.

It is convenient to parameterize the general complex mass matrix in terms of
two Hermitian matrices acting in flavor space,
\begin{align}
M &= M_S+iM_P,
\end{align}
where
\begin{align}
M_S &= \frac12 \left(M+M^\dagger\right),
\qquad
M_P=\frac{1}{2i}\left(M-M^\dagger\right),
\end{align}
so that
\begin{align}
M_S^\dagger &= M_S,
\qquad
M_P^\dagger=M_P,
\qquad
M^\dagger=M_S-iM_P.
\end{align}
The chiral projections satisfy
\begin{align}
\psi_L&=P_L\psi,
\qquad
\psi_R=P_R\psi,
\nonumber\\
\bar\psi_L&=\bar\psi P_R,
\qquad
\bar\psi_R= \bar\psi P_L,
\end{align}
where we have the spinor chiral projection matrices,
\begin{align}
P_L &= \textstyle\frac12 \left(I-\gamma_5\right),
\quad
P_R=\textstyle\frac12\left(I+\gamma_5\right) ,
\quad P_L^2 =P_L, \quad P_R^2=P_R,\quad  P_R+P_L = I,  \quad P_R-P_L = \gamma_5.
\end{align}
Since the flavor matrices commute with the chiral projectors 
the mass Lagrangian density becomes
\begin{align}
\mathcal L_m&=-\bar\psi_LM\psi_R-\bar\psi_RM^\dagger\psi_L
=-\bar\psi\left(MP_R+M^\dagger P_L\right)\psi
\nonumber\\
&=-\bar\psi\left[\left(M_S+iM_P\right)
P_R+\left(M_S-iM_P\right)P_L\right]\psi
\nonumber\\
&=
-\bar\psi
\left[M_S\left(P_L+P_R\right)
+iM_P\left(P_R-P_L\right)\right] \psi
\nonumber\\
&=-\bar\psi\left(M_S+i\gamma_5M_P\right)\psi.
\label{Eq:theta_as_mass}
\end{align}
Thus the most general Lorentz-invariant and Hermitian fermion mass term
contains both scalar and pseudoscalar couplings. The Hermitian matrix
$M_S$ appears in the scalar coupling $\bar\psi M_S\psi$, while $M_P$
appears in the Hermitian pseudoscalar coupling
$\bar\psi i\gamma_5M_P\psi$. This pseudoscalar mass term is $C$ even
and $P$ odd and so a nonvanishing $M_P$ therefore corresponds,
in the chosen flavor basis, to explicit $P$- and $CP$-violating terms in
the mass Lagrangian density. Whether these terms represent independent
physical $CP$ violation depends upon which phases can be removed by chiral
flavor transformations.

The appearance of the pseudoscalar coupling does not by itself imply the 
existence of an additional physical parameter. As we shall see below, 
suitable chiral rotations may be used to transfer phases between the quark 
mass matrix and the coefficient of the local operator $q(x)$. 
The axial anomaly implies that only the combination of the vacuum angle 
$\theta$ and the phase of the quark mass matrix is physically observable.

\subsection{Biunitary diagonalization, anomalies, and residual phase}

It is useful to separate an algebraic statement about general biunitary
diagonalization of complex matrices from the physical restrictions imposed 
by allowable field redefinitions in a quantum field theory.
An arbitrary complex $N_f\times N_f$ matrix $M$ has a biunitary 
(or singular value) decomposition,
\begin{align}
M = U_L\, D\, U_R^\dagger \qquad\mbox{or} \qquad D = U_L^\dagger M U_R  ,
\end{align}
for some $U_L\in U(N_f)_L$, $U_R\in U(N_f)_R$ and 
$D=\mathrm{diag}(m_1,m_2,\dots,m_{N_f})$ with  real and nonnegative $m_i$.
So for any complex matrix $M$  there is always a biunitary transformation
that brings $M$ into a real, diagonal, and nonnegative matrix with no 
residual complex phase. In this case there is no obstruction to removing 
\emph{all} phases of a complex mass matrix. 

Now define  $U_R=e^{i\phi_R/N_f}V_R$ and  $U_L\equiv e^{i\phi_L/N_f}V_L$ where 
$V_R\in SU(N_f)_R$ and $V_L\in SU(N_f)_L$.
Then 
\begin{align}
e^{i\Phi/N_f}D\equiv e^{i(\phi_L-\phi_R)/N_f} D 
=  e^{i(\phi_L-\phi_R)/N_f} U_L^\dagger M U_R = V_L^\dagger M V_R , 
\end{align}
which shows that we can always find a biunitary transformation with special 
unitary matrices that brings any complex $M$ to a real, diagonal and 
nonnegative matrix $D$ up to an overall phase $\Phi/N_f$. Since 
$\det D \in \mathbb{R}$, $\det V_R=\det V_L = 1$, then it follows that
\begin{align}
\arg\det M \!=\! \arg\left[\det V_L \det(e^{i\Phi/N_f}D)\det V_R^\dagger\right] \!=\! \arg\left[{e^{i\Phi}\det D}\right]=\Phi\!\!\!\pmod{2\pi}
\qquad\mbox{for}\;\;\det D \ne 0 .
\end{align}
Here $\Phi$ is defined only modulo $2\pi$, so one is free to choose any
convenient branch, for example $0\le\Phi<2\pi$ or $-\pi<\Phi\le\pi$.
If one or more of the $m_i$ vanish then $\det D=0$ and  then
$\arg\det M$ is undefined. This special case requires separate
treatment and will be discussed later.

From \Eq{Eq:UxU=etc} we have locally (i.e., near the identity),
\begin{align}
U(N_f)_L \times U(N_f)_R \cong SU(N_f)_L \times SU(N_f)_R 
    \times U(1)_V \times U(1)_A .
\end{align}
At the level of a classical field theory, a field redefinition is 
allowable if it is local, locally invertible and leaves the action 
invariant up to a total derivative. Thus a $U(N_f)_L\times U(N_f)_R$ 
rotation of fields is allowable at the classical level. At the level 
of the quantum theory one must also take account of the transformation 
of the functional measure.

The singlet axial subgroup $U(1)_A$ does not leave the fermion functional 
measure invariant. This failure of invariance is the axial anomaly.
Consequently, only the subgroup
$SU(N_f)_L\times SU(N_f)_R\times U(1)_V$ can be used without inducing
an anomalous change in the functional measure. Since $U(1)_V$ contributes
an identical phase to $\psi_L$ and $\psi_R$, it has no effect on the mass
term in \Eq{Eq:quark_L_m}. Thus only
$SU(N_f)_L\times SU(N_f)_R$ can be used to transform the mass matrix
without an anomalous contribution from the functional measure.
As we saw above this means that for an  arbitrary complex mass matrix 
$M$ we can only obtain a diagonal real mass matrix through
non-anomalous field redefinitions up to some overall  phase $\Phi/N_f$ 
so that $\det M = e^{i\Phi}\det D$ provided that $\det M\ne 0$.

Using  allowable chiral flavor transformations,
\begin{align}
\psi_L &\rightarrow V_L\psi_L, \qquad \psi_R \rightarrow V_R\psi_R,
    \qquad V_L,V_R\in SU(N_f),
\end{align}
the mass matrix may be diagonalized up to a common overall phase,
\begin{align}
V_L^\dagger M V_R = e^{i\Phi/N_f}D, \quad \mbox{where} \quad
    D = \mathrm{diag}(m_1,m_2,\ldots,m_{N_f}) , \;\; m_i>0
    \;\;\mbox{and}\quad \Phi = \arg\det M \!\!\! \pmod{2\pi}.
\end{align}
Recall that here we have restricted attention to the generic case in 
which all quark masses are nonzero ($\det M\ne 0$). The special case in which one or 
more quark masses vanish requires separate treatment and will be 
discussed below. 

In this flavor transformed basis, the quark mass 
contribution to the Lagrangian density is
\begin{align}
\mathcal L_m
&= -\bar\psi_L M \psi_R -\bar\psi_R M^\dagger \psi_L
= -e^{i\Phi/N_f}\bar\psi_L D\psi_R
-e^{-i\Phi/N_f}\bar\psi_R D\psi_L
\nonumber\\
&= \textstyle -\sum_{i=1}^{N_f} m_i
    \left[e^{i\Phi/N_f}\bar\psi_{Li}\psi_{Ri}+
    e^{-i\Phi/N_f}\bar\psi_{Ri}\psi_{Li} \right].
\end{align}
Using
\begin{align}
e^{\pm i\Phi/N_f} =
\cos(\Phi/N_f) \pm i\sin(\Phi/N_f),
\end{align}
together with
\begin{align}
\bar\psi_{Li}\psi_{Ri} + \bar\psi_{Ri}\psi_{Li}
    &= \bar \psi_i (P_R + P_L)\psi_i
    = \bar\psi_i\psi_i
\quad\mbox{and}\quad
\bar\psi_{Li}\psi_{Ri} - \bar\psi_{Ri}\psi_{Li}
    = \bar \psi_i( P_R - P_L)\psi_i
    = \bar\psi_i\gamma_5\psi_i,
\end{align}
we obtain
\begin{align}\label{Eq:mass_common_phase}
\mathcal L_m &= \textstyle -\sum_{i=1}^{N_f} m_i\,
    \bar\psi_i \left[\cos(\Phi/N_f) +
    i\gamma_5 \sin(\Phi/N_f) \right] \psi_i.
\end{align}
Thus all relative phases of the original complex mass matrix may be removed
by non-anomalous $SU(N_f)_L\times SU(N_f)_R$ flavor transformations. The
only remaining independent $CP$-odd phase is the common phase $\Phi/N_f$, or
equivalently the invariant quantity
\begin{align}
\Phi = \arg\det M \!\!\!\pmod{2\pi}.
\end{align}
It is this single residual phase that, through the axial anomaly, combines with the vacuum angle $\theta$ to produce the physically observable parameter $\bar\theta$.

\subsection{Phase of the quark mass matrix contributes like the $\theta$-term}

The axial anomaly provides the mechanism by which the overall phase of the
quark mass matrix, $\arg\det M$, enters the QCD functional integral in the
same manner as a vacuum angle $\theta$. For notational convenience we define
\begin{align}\label{Eq:Q_gluon}
Q_{\rm gluon} \equiv
[g^2/(32\pi^2)] \textstyle \int\! d^4x\, F_{\mu\nu}^a\tilde F^{a\mu\nu},
\end{align}
so that the $\theta$ term contributes 
$\theta Q_{\rm gluon}= \int d^4x \mathcal L_\theta$ to the action and hence a factor 
\begin{align}
\exp(i\theta Q_{\rm gluon})
\end{align}
to the QCD functional integral. 
At this stage $Q_{\rm gluon}$ is simply regarded as a real-valued functional of the
gauge field, defined as the spacetime integral of the local topological
charge density $q(x)$. No assumption is made that $Q_{\rm gluon}$ is integer valued or
that the functional integral is decomposed into topological sectors. These
issues will be examined in later sections.

Consider a global axial rotation acting on all $N_f$ quark flavors,
\begin{align}
\psi \to \psi' \equiv e^{i\alpha\gamma_5}\psi,
\qquad
\bar\psi \to \bar\psi' \equiv \bar\psi\,e^{i\alpha\gamma_5}.
\end{align}
Decomposing the Dirac fields into chiral components,
\begin{align}
\psi_{L,R} \equiv \tfrac12(I\mp\gamma_5)\psi,
\qquad
\bar\psi_{L,R} \equiv \bar\psi\,\tfrac12(I\pm\gamma_5),
\end{align}
one finds that the transformation acts as
\begin{align}
\psi_L \to \psi_L' = e^{-i\alpha}\psi_L,\;\,
\psi_R \to \psi_R' = e^{+i\alpha}\psi_R,\;\,
\bar\psi_L \to \bar\psi_L' = \bar\psi_L e^{+i\alpha},\;\,
\bar\psi_R \to \bar\psi_R' = \bar\psi_R e^{-i\alpha}.
\end{align}
At the classical level, the massless QCD action is invariant under this transformation. In the quantum theory, however, the fermion functional measure is not invariant due to the axial anomaly. Under the change of variables $(\bar\psi,\psi)\mapsto(\bar\psi',\psi')$ in the path integral, the measure transforms as
\begin{align}
\mathcal{D}\bar\psi\,\mathcal{D}\psi
\to
\mathcal{D}\bar\psi'\,\mathcal{D}\psi'
=
\mathcal{D}\bar\psi\,\mathcal{D}\psi\,
\exp\!\left(-i\,2N_f\alpha\,Q_{\rm gluon}\right) .
\end{align}
This Jacobian factor is equivalent to a shift of the coefficient of the $\theta$ term in the action,
\begin{align}
\theta \to \theta' = \theta - 2N_f\alpha.
\end{align}
Now include the quark mass term,
\begin{align}
\mathcal{L}_m = -\bar\psi_L M \psi_R -\bar\psi_R M^\dagger \psi_L,
\end{align}
where $M$ is a general complex $N_f\times N_f$ mass matrix in flavor
space. Under the axial rotation, the mass bilinears transform as
\begin{align}
\bar\psi_L M \psi_R &\to \bar\psi_L' M \psi_R'
    =(\bar\psi_L e^{+i\alpha})M(e^{+i\alpha}\psi_R)
    =e^{+2i\alpha}\bar\psi_LM\psi_R,
\\
\bar\psi_R M^\dagger \psi_L &\to\bar\psi_R' M^\dagger \psi_L'
    =(\bar\psi_R e^{-i\alpha})M^\dagger(e^{-i\alpha}\psi_L)
    =e^{-2i\alpha}\bar\psi_RM^\dagger\psi_L.
\end{align}
Thus, after the change of variables, the mass term in the transformed action reads
\begin{align}
\mathcal L_m \to \mathcal L_m' &=
    -e^{+2i\alpha}\bar\psi_LM\psi_R
    -e^{-2i\alpha}\bar\psi_RM^\dagger\psi_L.
\end{align}
It is convenient to rewrite $\mathcal{L}_m'$ in the same functional form as the original mass term by introducing a reparametrized mass matrix $M'$, defined by
\begin{align}
M' \equiv e^{+2i\alpha} M.
\end{align}
With this definition one has $(M')^\dagger = e^{-2i\alpha}M^\dagger$, and therefore
\begin{align}
\mathcal{L}_m' =
-\bar\psi_L M' \psi_R - \bar\psi_R (M')^\dagger \psi_L.
\end{align}
The reparametrization of the mass matrix implies
\begin{align}
\det M &\to \det M' = e^{+2iN_f\alpha}\,\det M, \\
\arg\det M &\to \arg\det M' = \arg\det M + 2N_f\alpha. \nonumber 
\end{align}
Combining this with the anomaly-induced shift of the vacuum angle,
\begin{align}
\theta \to \theta' = \theta - 2N_f\alpha,
\end{align}
one finds that the combination
\begin{align}
\bar\theta \equiv \theta + \arg\det M
\end{align}
is invariant under the axial rotation,
\begin{align}
\bar\theta' = \theta' + \arg\det M' 
=(\theta-2N_f\alpha) + (\arg\det M +2N_f\alpha) = \theta+\arg\det M
= \bar\theta.
\end{align}
For $\det M\neq0$, all strong-interaction $CP$ violation associated with
the vacuum angle and the quark mass matrix depends only on the invariant
parameter $\bar\theta$.

\emph{Observation:} 
In QCD with $N_f$ quark flavors and $\det M\neq0$, all relative phases of 
the complex quark mass matrix may be removed by non-anomalous 
$SU(N_f)_L\times SU(N_f)_R$ transformations. The remaining 
flavor-singlet pseudoscalar phase is proportional to $\arg\det M$. Through 
the axial anomaly this is equivalent to a shift of $\theta$, 
so that all $CP$ violation associated with the quark mass matrix and the
$\theta$ term is encoded in $\bar\theta=\theta+\arg\det M$.

The connection between the $\theta$ term and flavor-singlet 
pseudoscalar quark mass terms is entirely local and does not rely 
on any assumption about gauge-field topology. This connection arises 
through the axial anomaly, which relates the removal 
of a flavor-singlet pseudoscalar mass term to a shift in the 
coefficient of the local operator $F\tilde F$. It is important to 
emphasize that the axial anomaly itself is a \emph{local} statement, 
expressing the nonconservation of the axial current 
in terms of local composite operators. Any further connection to 
global topology enters only upon additional assumptions 
about boundary conditions and the global structure of gauge-field 
configurations.

The original estimate of the neutron electric dipole moment by 
Crewther {\it et al.} \citep{Crewther1979}
exploits the fact that, through the anomalous chiral rotation described
above, the $\theta$ term can be transferred entirely into a pseudoscalar
component of the quark mass matrix, whose form is displayed in
\Eq{Eq:theta_as_mass}.
As discussed earlier, the corresponding strong-$CP$ effects are not obtained
within ordinary perturbation theory about the trivial vacuum. Instead,
$CP$ violation is manifested through nonperturbative effects. Although the
flavor-diagonal operators $\bar\psi_i i\gamma_5\psi_i$ are local, their
relevant hadronic matrix elements are intrinsically nonperturbative: they
probe the chiral condensate, pion poles, and nucleon structure generated by
spontaneous chiral symmetry breaking. 
In the chiral effective theory, the $\theta$ dependence enters through 
the quark mass term in the vacuum energy, leading to a $CP$-odd 
realignment of the vacuum and inducing $CP$-violating pion--nucleon 
couplings. These couplings then generate a neutron EDM through 
long-distance pion dynamics and hadronic matrix elements. In this way, 
the Crewther {\it et al.} calculation avoids the perturbative 
vanishing of the $\theta$ term not by modifying perturbative Feynman 
rules, but by matching the $\theta$-induced deformation of the QCD 
vacuum onto nonperturbative low-energy degrees of freedom.

\subsection{Implications of massless quarks}

The discussion above implicitly assumes that all quark masses are nonzero, 
so that the quark mass matrix $M$ is invertible and $\arg\det M$ is well 
defined. If one or more quark masses vanish, the structure of $CP$-violating 
phases in QCD changes qualitatively. If $M$ has one or more zero eigenvalues, 
then
\begin{align}
\det M = 0,
\end{align}
and the phase $\arg\det M$ is undefined. In this case, there is no
invariant overall phase associated with the mass matrix.

Physically, the presence of a massless quark permits an anomalous chiral
change of variables that does not introduce a phase into the mass term.
For a quark flavor $i$ with vanishing mass, the axial rotation
\begin{align}
\psi_i \to e^{i\alpha\gamma_5} \psi_i
\end{align}
leaves the mass term invariant, since no mass term exists for that
flavor. The fermion functional measure is not invariant, however, and
the corresponding anomalous Jacobian shifts the vacuum angle according to
\begin{align}
\theta\to\theta-2\alpha .
\end{align}
Since this shift is not accompanied by any compensating phase in the
quark mass matrix, $\alpha$ may be chosen to remove the vacuum angle
entirely. Thus, if at least one quark is exactly massless, $\theta$ is
no longer a physical parameter.

\emph{Observation:} The strong $CP$ problem therefore disappears in
the presence of an exactly massless quark, since $CP$ violation
associated with the $\theta$ term can be eliminated by an anomalous
chiral change of variables that leaves the mass term unchanged.

\subsection{Topological charge density}
\label{Sec:TopologicalDensity}

The local gauge-invariant operator
\begin{align}\label{Eq:TopDensity}
q(x)
=
\frac{g^2}{16\pi^2}
\tr(F_{\mu\nu}\tilde F^{\mu\nu})
=
\frac{g^2}{32\pi^2}
F_{\mu\nu}^a\tilde F^{a\mu\nu}
\end{align}
plays a central role in the nonperturbative dynamics of QCD. As discussed in
the previous subsection, it appears naturally in the axial anomaly and is
therefore intimately connected with the chiral properties of the theory.
Because $F_{\mu\nu}$ transforms covariantly under local gauge
transformations, $F_{\mu\nu}\rightarrow UF_{\mu\nu}U^\dagger$,
the quantity $q(x)$ is gauge invariant pointwise throughout spacetime. It is
therefore a local gauge-invariant operator whose definition does not require
any assumption concerning global topology.

The local nature of $q(x)$ should be emphasized. Like any other local operator
in quantum field theory, it is defined at each spacetime point and enters
physical observables through its correlation functions. The operator itself
does not require the introduction of topological sectors or integer-valued
topological charges. Those concepts arise only when one considers the
properties of its spacetime integral under additional assumptions concerning
the global structure of the gauge fields.

One of the most important observables constructed from $q(x)$ is the
topological susceptibility,
\begin{align}\label{Eq:TopSusceptibility}
\chi
=
\int d^4x\,
\langle q(x)q(0)\rangle ,
\end{align}
which measures the fluctuations of the local topological charge density in the
QCD vacuum. Since it is defined entirely through a two-point correlation function 
of a local gauge-invariant operator, the susceptibility can be defined without
introducing a decomposition into topological sectors.
It has been calculated using lattice gauge
theory, chiral perturbation theory, large-$N_c$ methods and other
nonperturbative approaches, and plays a fundamental role in understanding the
vacuum structure of QCD.

\section{Topological susceptibility}
\label{Sec:top_suscept}

The topological susceptibility is an important quantity in QCD phenomenology
and warrants a careful discussion.

\subsection{Derivation of the topological susceptibility}
\label{Sec:chi_deriv}

In QCD the \emph{topological susceptibility} $\chi$
characterizes vacuum fluctuations associated with the local topological
charge density $q(x)$ given in \Eq{Eq:L_theta}.
The topological susceptibility is conventionally expressed as
the spacetime integral of the Euclidean two-point correlator of the
topological charge density, or equivalently, its zero-momentum 
limit when defined with the appropriate  contact-term prescription.

Time-ordering of the operators is unnecessary in Euclidean space
correlators and so there we write
\begin{align}\label{Eq:chi_defn}
\chi = \textstyle \int \! d^4 x_E \, \langle \Omega|
    \hat q_E(x) \hat q_E(0) |\Omega\rangle_E
	\equiv \textstyle \int \! d^4 x_E \,
	\braket{\hat q_E(x) \hat q_E(0)} ,
\end{align}
where this $\chi$ is understood to be defined in the absence of a
$\theta$-term ($\theta=0$).

In the continuum theory \Eq{Eq:chi_defn} requires some care because the
product of composite operators $\hat q_E(x)\hat q_E(0)$ has singularities
at $x=0$. Consequently, the spacetime integral cannot simply be regarded
as an ordinary integral of a well-defined two-point function without
specifying a nonperturbative prescription for the contact terms. Several
such constructions have been developed in lattice QCD, including
definitions based on chiral Ward identities and, more recently, gradient
flow~\cite{Luscher:2004fu,Luscher2010iy,Luscher:2021ygc}. We return to
this point below.

Consider a finite Euclidean region with periodic boundary
conditions and volume $V$. As we saw earlier, this is a four-torus with
four-volume $V$ and so is the continuum limit of a lattice in lattice QCD.
For the present discussion let us \emph{define} the operator $\hat Q$ as
the operator version of $Q_{\rm gluon}$ in \Eq{Eq:Q_gluon}. Assuming that
$\hat Q$ and the integrated product below are defined using a common
nonperturbative prescription, we may write
\begin{align}
\hat Q^2 &= \textstyle \left[ \int \!d^4x_E \, \hat q_E(x) \right]
    \left[ \int \!d^4y_E \, \hat q_E(y) \right]
     = \int d^4x_E \int d^4y_E \, \hat q_E(x)\, \hat q_E(y).
\end{align}
Taking the Euclidean vacuum expectation value at $\theta=0$ we have
\begin{align}
\braket{\hat Q^2}
&\equiv \braket{\Omega|\hat Q^2|\Omega}
= \textstyle \int \! d^4x_E \int \! d^4y_E \, \braket{\Omega|
    \hat q_E(x)\, \hat q_E(y) |\Omega}
= V \textstyle \int \! d^4x_E \, \braket{\Omega| \hat q_E(x)\,
    \hat q_E(0) |\Omega} = V\chi ,
	\nonumber \\
&\Rightarrow \quad \chi = \braket{\Omega|\hat Q^2|\Omega}/V
    \equiv \braket{\hat Q^2}/V
 = \textstyle \int\!d^4x_E\,
 \braket{\Omega| \hat q_E(x)\, \hat q_E(0) |\Omega}_E
	\nonumber ,
\end{align}
where we have used translation invariance. These
relations should therefore be understood within a prescription in which
the composite operators and their contact terms have been given a
consistent nonperturbative definition.

Gradient flow provides a particularly useful nonperturbative definition
of the composite operators appearing above. Given a gauge configuration
$A_\mu(x)$, or lattice links $U$, one constructs the flowed field
$A_\mu(t,x)$, or $U_t$, by solving the gradient-flow equation with the
original gauge field as the initial condition at $t=0$. One then defines
\begin{align}
\hat q_t(x) &\equiv [g^2/(32\pi^2)]\,
\hat F_{\mu\nu}^a(t,x)\,
\hat{\tilde F}_{\mu\nu}^a(t,x),
\qquad
\hat Q_t \equiv \textstyle \int \! d^4x\, \hat q_t(x) .
\end{align}
Thus $\hat q_t(x)$ is the usual $F\tilde F$ density evaluated on the
flowed gauge field. The flow does not modify the QCD action or the
functional measure used to generate the gauge-field ensemble. Rather,
it defines a smoothing map on each gauge configuration, with smoothing
radius of order $\sqrt{8t}$. At positive flow time the resulting
composite operators have improved ultraviolet behavior, allowing the
topological susceptibility and its moments to be defined
nonperturbatively. The relation of this construction to definitions
based on anomalous chiral Ward identities, and the universality of the
resulting continuum susceptibility, have been studied in
Refs.~\cite{Luscher:2004fu,Luscher2010iy,Luscher:2021ygc}.

At positive flow time $t>0$ the gauge field is smoothed over a radius
$r_{\rm sm}=\sqrt{8t}$, which suppresses ultraviolet fluctuations. In
particular, correlation functions constructed from the flowed density
$q_t(x)$ have well-controlled ultraviolet behavior. One works within a
``flow window'', $a\ll\sqrt{8t}\ll L$, where $a$ is the lattice spacing
and $L$ denotes a characteristic infrared scale. The left inequality
suppresses ultraviolet lattice artifacts, while the right inequality
ensures that the smoothing does not distort the infrared physics of
interest.

In the continuum theory the integrated flowed topological charge
$Q_t=\int d^4x\,q_t(x)$ is independent of the flow time under the usual
boundary conditions. At finite lattice spacing this relation is affected
by discretization effects, but the flow-time independence is recovered
in the continuum limit. Consequently,
\begin{align}
\chi = \lim_{a\to0}\frac{\langle Q_t^2\rangle}{V}, \qquad t>0,
\end{align}
provides a nonperturbative definition of the continuum topological
susceptibility. The resulting susceptibility is a renormalization-group
invariant observable.

It is also useful to recall how the physical content of the 
two-point function may be expressed through its spectral
representation. In Minkowski space we can define for example
\begin{align}
\Pi_{qq}(p^2)
&\equiv
i\int\!d^4x\,e^{ip\cdot x}
\braket{\Omega|T\{\hat q(x)\hat q(0)\}|\Omega}.
\end{align}
Insertion of a complete set of physical states gives the usual
Kallen--Lehmann representation,
\begin{align}
\Pi_{qq}(p^2)
&=
P(p^2)
+
\int_0^\infty\!ds\,
\frac{\rho_q(s)}{s-p^2-i0},
\label{Eq:q_spectral_rep}
\end{align}
where $P(p^2)$ denotes the subtraction polynomial required by 
the short-distance behavior of the composite-operator 
correlator. The spectral density has the schematic form
\begin{align}
\rho_q(s)
&=
\textstyle
\sum_n
(2\pi)\,
\delta(s-m_n^2)
\left|\braket{\Omega|\hat q(0)|n}\right|^2
+\rho_{\rm cont}(s),
\end{align}
where the sum runs over physical states with the quantum 
numbers of $q(x)$ and $\rho_{\rm cont}(s)$ denotes 
multiparticle continuum contributions. Since $q(x)$ is a
pseudoscalar, the relevant intermediate states lie in the
pseudoscalar channel. In particular, in full QCD the
flavor-singlet pseudoscalar sector, including the $\eta'$, 
plays an important role.

This representation makes explicit that the nontrivial physical
information contained in correlations of the local operator
$q(x)$ may be organized in terms of ordinary physical
intermediate states of the quantum theory. No classification
of individual gauge-field configurations by an integer
topological charge is required for this spectral
decomposition. At the same time, care is required at zero
momentum: because $\Pi_{qq}(p^2)$ requires subtractions, 
the susceptibility cannot in general be identified naively
with an unsubtracted positive spectral integral. The contact
and subtraction terms must be defined consistently with the
nonperturbative prescription used for the composite operator,
as discussed above. The pole structure, however, is 
unambiguous and will reappear in the discussion of the 
Witten--Veneziano relation in Sec.~\ref{Sec:app_top_suscept}
and Appendix~\ref{App:WV_relation}.

When $\theta=0$ the QCD action is $CP$ conserving and so 
$\ket{\Omega}\equiv\ket{\Omega_\theta}_{\theta=0}$ is  $CP$-invariant.
The topological charge density $q(x)$ is $CP$-odd and so we must have 
$\braket{\Omega|\hat q(x)|\Omega}\equiv \braket{\hat q(x)}=0$
and hence $\braket{\hat Q}=0$. This means that $\chi$ is the variance 
density of $Q$ since then
\begin{align}
\chi &= \braket{\hat Q^2}/V = \left[ \braket{\hat Q^2} 
    - \braket{\hat Q}^2 \right]/V .
\end{align}
More generally, for $\theta\ne0$ we define the topological susceptibility
as the connected second moment per unit volume,
\begin{align}
\chi_\theta &= \left[ \braket{\hat Q^2}_\theta 
    - \braket{\hat Q}_\theta^2 \right]/V ,
\end{align}
where the expectation value is evaluated in the presence of the
$\theta$-dependent source. For $\theta\ne0$, $CP$ is in general explicitly
broken, and the symmetry of the functional weight under $Q\to-Q$ is lost,
so that $\langle\hat Q\rangle_\theta$ need not vanish.

Consider the Euclidean QCD path integral with the $\theta$-term included,
which is $Z_E(\theta)$ of \Eq{Eq:QCD_genZ_theta}. We see that
\begin{align}\label{Eq:chi_theta_V} 
&Z_E(\theta)= \textstyle \int\!\mathcal{D}A_\mu \mathcal{D}\bar{\psi}  
    \mathcal{D}\psi \,e^{-S_E(\theta)} 
=  \textstyle \int\!\mathcal{D}A_\mu \mathcal{D}\bar{\psi} \mathcal{D}\psi 
	\,e^{-[ S_E -i \int\! d^4z_E \theta q_E]} , \\
&\frac{\partial\ln Z_E}{\partial \theta}\!=\! 
    \frac{1}{Z_E(\theta)}\frac{\partial Z_E(\theta)}{\partial \theta}
	\!=\! \frac{\textstyle \int\!\mathcal{D}A_\mu \mathcal{D}
    \bar{\psi} \mathcal{D}\psi 
	\, (i\int\!d^4x_E\,q_E) \,e^{-[ S_E -i \int\! d^4z_E \theta q_E]}}
	{\textstyle \int\!\mathcal{D}A_\mu \mathcal{D}\bar{\psi} 
    \mathcal{D}\psi
	\,e^{-[ S_E -i \int\! d^4z_E \theta q_E]}} \!=\! 
    i\braket{\hat Q}_\theta ,  \nonumber \\
&\frac{\partial^2\ln Z_E}{\partial \theta^2} 
= \frac{1}{Z_E(\theta)} \frac{\partial^2 Z_E(\theta)}{\partial \theta^2}
- \left(\frac{1}{Z_E(\theta)} \frac{\partial Z_E(\theta)}{\partial 
    \theta}\right)^2
	\!=\! - \braket{\hat Q^2}_\theta + (\braket{\hat Q}_\theta)^2 
    \!=\! - \chi_\theta V \nonumber . 
\end{align}
These identities depend only upon treating $\theta$ as the parameter
multiplying the source term coupled to the local operator $q(x)$. They are
therefore equally valid whether $\theta$ is interpreted as a physical
parameter of the QCD Lagrangian density or merely as an auxiliary source parameter
introduced in the generating functional in order to generate moments and
correlation functions involving $\hat Q=\int\!d^4x_E\,\hat q_E(x)$.
No assumption concerning the physical status of $\theta$ is required in
deriving Eq.~(\ref{Eq:chi_theta_V}).
Note that $\chi\ne 0$ does not imply that $\theta\ne 0$. 

It is important to distinguish several related facts in the Euclidean
formulation at real $\theta$. For each real Euclidean gauge-field
configuration, the quantity
\begin{align}
Q = \textstyle \int \! d^4x_E\,q_E(x)
\end{align}
is real. The $\theta$-dependent Euclidean functional weight
\begin{align}
e^{-S_E+i\theta Q}
\end{align}
is generally complex when $\theta\ne0$. The resulting partition function
$Z_E(\theta)$ is nevertheless real: configurations related by
$Q\rightarrow-Q$ contribute complex-conjugate phases, so that their sum is
proportional to $\cos(\theta Q)$. A demonstration that
$Z_E(\theta) \in \mathbb{R}$ is given in 
Appendix \ref{App:real_Z_E}. This reality of $Z_E(\theta)$ does not imply
that expectation values defined using the complex $\theta$-dependent weight
are themselves real. In particular,
\begin{align}
\braket{\hat Q}_\theta
=
\frac{\int\!\mathcal D\Phi\,
Q\,e^{-S_E+i\theta Q}}
{\int\!\mathcal D\Phi\,
e^{-S_E+i\theta Q}}
\end{align}
is generally purely imaginary. Indeed, pairing contributions with $Q$ and
$-Q$ gives a numerator proportional to
$Q[e^{i\theta Q}-e^{-i\theta Q}]
=2iQ\sin(\theta Q)$, while the denominator is real. Consequently,
\begin{align}
\frac{\partial\ln Z_E}{\partial\theta} = i\braket{\hat Q}_\theta
    \in \mathbb R 
\end{align}
as required. We see then that $i\braket{\hat Q}_\theta,
\braket{\hat Q^2}_\theta$, and $\braket{\hat Q}_\theta^2$ are all real.
Not that $\braket{\hat Q}_\theta^2$ can be nonpositive because
$\braket{\hat Q}_\theta$ is purely imaginary. Thus, for real
$\theta\ne0$, the connected second moment should not be interpreted as an
ordinary statistical variance with respect to a positive probability
measure. At $\theta=0$ the functional weight becomes real and nonnegative,
$\braket{\hat Q}\equiv\braket{\hat Q}_{\theta=0}=0$ by $CP$ symmetry,
and the connected second moment reduces to the ordinary variance
$\braket{\hat Q^2}\geq0$.
For QCD with nonzero physical quark masses the topological susceptibility
is nonzero, both in chiral effective theory and in lattice calculations,
so that $\braket{\hat Q^2}=V\chi>0$. Strict positivity is therefore a
physical property of QCD with nonzero quark masses rather than a consequence
of positivity of the Euclidean measure alone.

\emph{Corollary for one or more massless quarks:}
If at least one quark is exactly massless, the Euclidean generating
functional is independent of $\theta$. It then follows from the definition
of the topological susceptibility that it must vanish since
\begin{align}
\chi=-\frac{1}{V} \left.
\frac{\partial^2\ln Z_E(\theta)} {\partial\theta^2}\right|_{\theta=0} = 0 .
\end{align}

We can further relate these results to the vacuum energy density whenever
the $\theta$-dependent theory admits the usual Hamiltonian interpretation.
Consider a four-torus in Euclidean space with three-dimensional spatial
volume $V_3$ and temporal extent $T_E$, so that $V=V_3T_E$. For a quantum
system described by a time-independent Hamiltonian $\hat H$, 
the Euclidean partition function may be written as
\begin{align}
Z_E(T_E) = \mathrm{Tr}\!\left(e^{-\hat H T_E}\right)
    = \textstyle \sum_n e^{-E_nT_E},
\end{align}
where $E_n$ are the energy eigenvalues, with sums and integrals over
discrete and continuum states understood. The corresponding Minkowski
quantity is the trace of the evolution operator and is referred to as
the real-time partition function,
\begin{align}
Z_M(T) = \mathrm{Tr}\!\left(e^{-i\hat H T}\right).
\end{align}
Under the Wick rotation $T=-iT_E$ we have $Z_M(T) \to Z_E(T_E)$.
In the limit $T_E\to \infty$ we see that
\begin{align}
    Z_E &= \textstyle \sum_{n=0}^\infty e^{-E_n T_E} \to e^{-E_0 T_E} ,
\end{align}
where $E_0$  is the ground state energy. Then
$-\ln Z_E/T_E \to -\ln(e^{-E_0T_E})/T_E = E_0$. 

Thus, if the $\theta$-dependent Euclidean functional integral for QCD admits 
the usual Hamiltonian representation, then we can apply this and have
$Z_E(\theta)=\mbox{Tr}(e^{-\hat H_\theta T_E})
=\sum_{n=0}^\infty e^{-E_n(\theta) T_E}$,
which gives
\begin{align}\label{Eq:vareps_0}
-\frac{\ln Z_E(\theta)}{V} \xrightarrow{T_E\to\infty}
-\frac{\ln[e^{-E_0(\theta)T_E}]}{V_3T_E}
= \frac{E_0(\theta)}{V_3} \equiv \varepsilon_0(\theta).
\end{align}
where $\varepsilon_0(\theta)$ is the corresponding ground-state energy
density. Comparing with \Eq{Eq:chi_theta_V} 
it follows that in the limit $T_E\to \infty$ we can write
\begin{align}\label{Eq:chi_vareps}
	\chi_\theta &=  \frac{\partial^2 \varepsilon_0(\theta)}{\partial \theta^2} 
	= \frac{\braket{\hat Q^2}_\theta - (\braket{\hat Q}_\theta)^2 }{V} .
\end{align}
So in this limit the second derivative of the vacuum energy density with
respect to $\theta$ is the connected second moment of the topological charge
per unit volume. At $\theta=0$ this reduces to the variance density 
$\braket{\hat Q^2}/V$, which is
the usual topological susceptibility $\chi=\chi_{\theta=0}$.

At low energies, the dependence of the topological susceptibility
on the light quark masses is given by the
Leutwyler--Smilga relation. 
A derivation of this relation,
formulated entirely in terms of local operator relations and the
$\theta$-dependence of the vacuum energy, is presented in
Appendix~\ref{App:LS_relation}.

\paragraph*{Observation:} 
It is useful to note that the definition of the topological
susceptibility $\chi$ in Eq.~(\ref{Eq:chi_defn}) 
does not require a global classification of gauge fields 
and does not require integer values of $Q_{\rm gluon}$. Similarly, the results
in Eq.~(\ref{Eq:chi_theta_V}) hold with or without any
such global classification. The Leutwyler--Smilga relation similarly
does not require such a classification.

It is also important that none of these results requires us to decide
whether $\theta$ is regarded as a physical parameter of the QCD
Lagrangian density or simply as a source parameter introduced in the generating
functional. In the latter interpretation, $\theta$ is a source coupled
to the local operator $q(x)$, and derivatives with respect to $\theta$
generate the corresponding integrated correlation functions. The same
formal relations follow if $\theta$ is instead regarded as a physical
parameter of the theory. Thus the derivation and interpretation of the
topological susceptibility do not, by themselves, require either
interpretation of $\theta$.

We therefore conclude that the definition and physical interpretation 
of the topological susceptibility can be formulated entirely in terms 
of local operator correlation functions, without requiring a global 
classification of gauge field configurations into topological sectors
or any assumption concerning whether $\theta$ is a physical parameter
of QCD.

\subsection{Applications of the topological susceptibility}
\label{Sec:app_top_suscept}

The topological susceptibility also enters a number of important exact and
approximate relations. Through the anomalous axial Ward identities it is
connected with the chiral properties of the theory and with the dependence of
the vacuum energy upon the parameter $\theta$. 
In the large-$N_c$ limit, and to leading order in the chiral expansion,
the Witten-Veneziano relation connects the pure Yang-Mills topological
susceptibility $\chi_{\rm YM}$ to the large mass of the $\eta^\prime$ meson
\citep{Witten:1979vv,Veneziano:1979ec},
\begin{align}\label{Eq:WV}
m_{\eta'}^2 \simeq \frac{2N_f}{f_0^2}\chi_{\rm YM} ,
\end{align}
where $f_0$ is the flavor-singlet pseudoscalar decay constant, defined
more precisely in Appendix~\ref{App:WV_relation}. This relation provides
an elegant resolution of the long-standing $U(1)_A$ problem.
Appendix~\ref{App:WV_relation} provides a pedagogical derivation of this
important relation.

For full QCD with light quarks, the corresponding low-energy behavior is described
by the Leutwyler-Smilga relation
\citep{Leutwyler:1992yt},
\begin{align}\label{Eq:LS}
\chi=\textstyle \Sigma
\left(\sum_{i=1}^{N_f}(1/m_i)\right)^{-1} +\mathcal O(m_q^2) ,
\end{align}
where $m_i$ are the real quark masses,
$\Sigma\equiv-\langle\bar q q\rangle>0$ denotes the magnitude of the
chiral condensate in the chiral limit, and the topological susceptibility
is evaluated at $\theta=0$, i.e., $\chi=\chi_{\theta=0}$. This relation 
illustrates the close interplay between chiral symmetry 
breaking, the axial anomaly, and the fluctuations of the topological 
charge density encoded in the topological susceptibility.
Note that in the limit in which any one quark mass vanishes,
$\chi\to0$, as expected.
Since this relation plays an important conceptual role throughout 
the discussion, a pedagogical derivation is presented in 
Appendix \ref{App:LS_relation}. 

Taken together, the anomalous Ward identities, the Witten-Veneziano relation
and the Leutwyler-Smilga relation demonstrate that the operator $q(x)$
contains physically measurable information and plays an indispensable role in
nonperturbative QCD. The physical content of these relations can be formulated
in terms of the local operator $q(x)$, its correlation functions, and the
corresponding dependence of the vacuum energy on a source coupled to $q(x)$.
In particular, their physical consequences do not by themselves require that
the spacetime integral of $q(x)$ be interpreted as an integer-valued
topological charge. It is therefore useful to distinguish carefully between
the local physics encoded in $q(x)$ and the additional global structures that
are often introduced when discussing the $\theta$ vacuum.

\section{Global topology in gauge theories}
\label{Sec:global_top}

The previous sections concerned properties of QCD that follow directly from
its local gauge symmetry and the functional integral formulation. We now turn
to additional global properties of gauge fields that play an important role in
the conventional discussion of the strong $CP$ problem.

\subsection{Finite-action gauge fields}
\label{Sec:FiniteAction}

The standard construction of topological sectors begins by restricting attention
to gauge-field configurations that are sufficiently smooth that they
are square integrable and that have finite Euclidean action.
For pure Yang-Mills theory the Euclidean action is
\begin{align}\label{Eq:EuclideanActionYM}
S_{\rm YM} =
\textstyle \frac12 \int d^4x\, \tr(F_{\mu\nu}F_{\mu\nu}) .
\end{align}
Since the integrand is nonnegative, finite action requires
\begin{align}\label{Eq:FiniteActionCondition}
F_{\mu\nu}(x)
\rightarrow 0
\qquad
{\rm as}
\qquad
|x|\rightarrow\infty ,
\end{align}
sufficiently rapidly that the integral converges. This boundary condition
implies that the gauge field becomes asymptotically pure gauge for smooth
gauge fields.

Specifically, sufficiently close to the boundary the gauge field
approaches a pure gauge configuration,
\begin{align}\label{Eq:PureGauge}
A_\mu(x) = (i/g) U(x)\partial_\mu U^{-1}(x),
\end{align}
for some smooth gauge transformation
$U(x)\in SU(3)$. Since 
$F_{\mu\nu}=\partial_\mu A_\nu-\partial_\nu A_\mu-
ig[A_\mu,A_\nu]$,
it can be directly verified that
\begin{align}
F_{\mu\nu}=0
\end{align}
for every pure gauge configuration.

The asymptotic behavior of Eq.~(\ref{Eq:PureGauge}) has an important
geometrical interpretation. Euclidean spacetime is treated as
$\mathbb R^4$, and all points at infinity are identified. The boundary at
infinity therefore has the topology of a three-sphere, $S^3_\infty$.
The asymptotic pure-gauge field therefore determines a continuous mapping
\begin{align}\label{Eq:Map}
U: S^3_\infty \to SU(3).
\end{align}
Since
\begin{align}
\pi_3(SU(3))
=
\mathbb Z ,
\end{align}
such maps fall into homotopy classes labelled by an integer winding number.
This observation provides the starting point for the conventional
classification of smooth finite-action gauge fields into topological sectors.

It is important to distinguish carefully between the various assumptions 
that enter this construction. The requirement of finite Euclidean action 
implies that the field strength is square integrable and, for sufficiently 
smooth gauge fields, vanishes asymptotically. Under these assumptions the 
gauge field approaches a pure gauge configuration, leading to the boundary 
condition of Eq.~(\ref{Eq:PureGauge}). The subsequent interpretation of the 
asymptotic gauge transformation as a mapping $S^3_\infty\rightarrow SU(3)$, 
together with its classification by the homotopy group $\pi_3(SU(3))$, 
introduces additional global topological structure beyond that required 
by the local gauge symmetry and the local functional integral formulation 
of the theory.

These ideas play a central role in the semiclassical analysis of Yang-Mills
theory and underlie the conventional construction of instantons and
topological sectors. The following subsections examine this construction in
greater detail before examining the assumptions required to extend this 
semiclassical classification to the full quantum functional integral.

\subsection{Fiber bundles and topological charge}
\label{Sec:Bundles}

The boundary conditions discussed in the previous subsection admit a natural
geometrical interpretation in terms of principal fiber bundles. Although the
language of fiber bundles is not required for perturbative quantum field
theory, it provides the natural mathematical framework for describing global
properties of gauge fields and plays a central role in the conventional
discussion of topological charge.

It is useful to unpack the concept of fiber bundles here.
In simple terms, a fiber bundle consists of a base space, here
Euclidean spacetime with points $x\in\mathbb R^4$, together with
an internal space, called a fiber, attached to each point. For a
principal bundle with gauge group $G$, there is a copy of the full
group $G$ at each $x$; here
$G=SU(3)$. We may denote the fibre at $x$ as $G_x$. The fiber bundle 
is the space formed by the collection of all these fibers $G_x$.

The gauge field $A_\mu$ defines parallel transport between the
fibers at $x$ and $x'$ along a path $C$, specifying how an element
of $G_x$ is transported to an element of $G_{x'}$. This parallel
transport is in general path dependent, with the path dependence
determined by $A_\mu$. For this reason the gauge field $A_\mu$ is
said to define a connection on the fiber bundle. Different $A$
then define different connections on the same fibre bundle.

So in Yang-Mills theory the gauge field $A_\mu(x)$ may be regarded as a
connection on a principal $SU(3)$ bundle over spacetime, while the field
strength tensor $F_{\mu\nu}$ is the corresponding curvature. Local gauge
transformations correspond to changes of local trivialization of the bundle
and leave all gauge-invariant observables unchanged. Topological information
is therefore associated not with the local form of the gauge potential but
with the global structure of the bundle.

For smooth gauge fields satisfying the finite-action boundary conditions of
the previous subsection, the gauge field approaches a pure gauge at 
spacetime infinity, $A_\mu = (i/g) U\partial_\mu U^\dagger$,
where $U(x)\in SU(3)$. Since the sphere at infinity in four-dimensional 
Euclidean spacetime has the topology of $S^3$, the asymptotic gauge 
field defines a continuous mapping $U: S^3_\infty \to SU(3)$.
Two such maps are said to be homotopic if one can be continuously deformed
into the other while remaining a continuous map into $SU(3)$. The notation
$\pi_n(G)$ denotes the $n$th homotopy group of the group manifold $G$; for
present purposes, it classifies continuous maps from the $n$-sphere $S^n$
into the group manifold $G$ according to whether they can be continuously 
deformed  into one another. Maps belonging to different homotopy classes 
cannot be connected by such a continuous deformation. In the present case 
these maps are classified by the third homotopy group,
\begin{align}\label{Eq:Homotopy}
\pi_3(SU(3)) = \mathbb Z ,
\end{align}
which means that the distinct homotopy classes are labeled by an integer.
This integer is conventionally referred to as the winding number. Thus,
under these assumptions, each smooth finite-action gauge field $A_\mu$ is assigned
an integer winding number based on its pure gauge form $U$ at spacetime
infinity.

This integer may be written explicitly as
\begin{align}\label{Eq:WindingNumber}
n = \frac{1}{24\pi^2} \int_{S^3_\infty}\! d^3\xi\, \epsilon^{abc}\,
    \tr\left[(U^\dagger\partial_a U)(U^\dagger\partial_b U)
    (U^\dagger\partial_c U) \right] ,
\end{align}
where $\xi^a$, $a=1,2,3$, are local coordinates on the three-sphere at
infinity and $\partial_a\equiv\partial/\partial\xi^a$. Equivalently, in
differential-form notation,
\begin{align}
n = \textstyle -\frac{1}{24\pi^2} \int_{S^3_\infty}\!\tr\left[(U^\dagger dU)^3\right].
\end{align}
The integer $n$ is invariant under smooth deformations of $U$ and therefore
labels the homotopy classes of maps $S^3_\infty\rightarrow SU(3)$.

For smooth finite-action gauge fields satisfying the assumptions above, an 
equivalent characterization is obtained through the spacetime integral of
the local topological charge density, $q(x)$,
\begin{align}\label{Eq:TopCharge}
Q_{\rm gluon} = \textstyle \int\! d^4x\,q(x) 
    = [g^2/(16\pi^2)] \int\! d^4x\,
    \tr(F_{\mu\nu}\tilde F^{\mu\nu}) .
\end{align}
For sufficiently smooth gauge fields satisfying the above boundary
conditions, one finds
\begin{align}\label{Eq:QEqualsN}
Q_{\rm gluon} =n\in\mathbb Z ,
\end{align}
so that the spacetime integral of the topological charge density coincides 
with the winding number of the asymptotic gauge transformation. 
The local quantity $q(x)$ is also commonly referred to as the
Pontryagin density, since its spacetime integral is proportional to the
Pontryagin number for sufficiently smooth gauge fields satisfying the
appropriate global conditions. We will use the terms topological charge
density and Pontryagin density interchangeably..

Provided that we \emph{assume} that we only have smooth finite-action gauge 
fields in the functional integral, the equality (\ref{Eq:QEqualsN}) provides 
division into sectors labelled by an integer topological charge,
\begin{align}\label{Eq:SectorDecomp}
Z = \textstyle \sum_{Q_{\rm gluon}\in\mathbb Z} \,Z_{Q_{\rm gluon}} ,
\end{align}
where each sector contains gauge fields belonging to a fixed homotopy class.
This decomposition underlies the conventional construction of the
$\theta$-vacuum and much of the semiclassical analysis of nonabelian gauge
theories.

It is useful, however, to distinguish carefully between the various logical
steps entering this construction. The local operator $q(x)$ is defined
independently of any global assumptions, whereas the identification of its
spacetime integral with an integer winding number requires the gauge field to
define a sufficiently smooth connection on a globally defined principal
bundle satisfying the finite-action boundary conditions discussed above. The
extent to which these assumptions apply to the full quantum functional
integral will be examined in the following sections.

\subsection{Topological sectors}
\label{Sec:Top_sectors}

One may adopt a prescription that assigns an integer topological charge 
$Q\in \mathbb Z$ to all gauge field configurations, including rough 
configurations for which in general $Q\ne Q_{\rm gluon}$,
since $Q_{\rm gluon}$ may be ill-defined or real.
For example, this may be accomplished
by applying a smoothing procedure that  assigns an integer topological 
charge $Q$ to each configuration by equating $Q$ to the integer that 
$Q_{\rm gluon}\in \mathbb R$ approaches after sufficient 
smoothing has occurred. The details of possible integer 
assignments will be discussed in more detail later in 
\Sec{Sec:LatticeTopology}.

We may therefore collect the set of all of the gauge configurations with 
the same assigned value of $Q$ and denote that subspace as 
$\mathcal{C}_Q$. We can then express the entire gauge field configuration 
space $\mathcal{C}$ as the union of all of these 
$\mathcal{C}_Q$ subspaces,
\begin{align}\label{Eq:C=sumCQ}
   \mathcal{C} &= \textstyle \bigcup_{Q\in \mathcal{Z}} \, \mathcal{C}_Q .
\end{align}
The typical gauge field configurations
in each $\mathcal{C}_Q$ will be rough in the absence of additional
constraints on the gauge fields. 

Once such an assignment has been adopted, the space of gauge field 
configurations can be split into distinct \emph{topological sectors}, 
each labeled by an integer-valued topological charge $Q$. The QCD 
functional integral effectively becomes a sum over these sectors.
The gluonic definition of the topological charge in Minkowski and 
Euclidean spaces is
\begin{align}\label{Eq:EandM_top_Q}
 Q_{\rm gluon} & \!=\! \textstyle \int d^4x_M\; q_M(x)
 	\!=\! \textstyle  \frac{g^2}{16\pi^2} \int d^4x_M\; 
    \tr(F_{\mu\nu}\tilde F^{\mu\nu})
 	 \!=\! \textstyle \frac{g^2}{16\pi^2} \int d^4x_M \;
     \epsilon^{\mu\nu\rho\sigma} \, \tr(F_{\mu\nu} F_{\rho\sigma}) ,
 	\\
 Q_{\rm gluon} & \!=\!  \textstyle \int d^4x_E\; q_E(x)
	\!=\! \textstyle  \frac{g^2}{16\pi^2} \int d^4x_E\; 
    \tr(F_{\mu\nu}\tilde F_{\mu\nu})
	\!=\! \textstyle \frac{g^2}{16\pi^2} \int d^4x_E \;
    \epsilon_{\mu\nu\rho\sigma} \, \tr(F_{\mu\nu} F_{\rho\sigma}) ,
	\nonumber
\end{align}
respectively, where $q_M(x)$ and $q_E(x)$ are the topological charge 
densities in Minkowski and Euclidean space respectively.
The Levi-Civita tensors are completely antisymmetric under the 
pairwise exchange of indices and where $\epsilon^{0123}=+1$ and
$\epsilon_{1234}=+1$. In the usual way we understand that
$d^4x_M=dx^0dx^1dx^2dx^3$ and $d^4x_E=dx_1dx_2dx_3dx_4$ in Minkowski
and Euclidean spaces respectively. Note that \emph{for sufficiently 
smooth fields} $Q_{\rm gluon}$ is an integer and is the same quantity in both 
Euclidean and Minkowski space as explained in 
Appendix~\ref{App:Wick_rot}. If we restrict the functional integral to 
sufficiently smooth gauge fields, then $Q_{\rm gluon}$ is integer 
valued and we may use it as the assigned topological charge for the sector 
decomposition. In that case, we choose $Q=Q_{\rm gluon}$.

It follows from Eq.~(\ref{Eq:C=sumCQ}) 
that the functional integrals in Eqs.~(\ref{Eq:Z_M}) and (\ref{Eq:Z_E}) 
can be written as
\begin{align}\label{Eq:QCD_genZ_Q}
Z_M = \textstyle \sum_Q
	\int_{\mathcal{C}_Q} \! \mathcal{D}A_\mu \, \mathcal{D}\bar{\psi} \,
    \mathcal{D}\psi \; e^{iS_M[A,\psi,\bar{\psi}]}
	\, , \quad
Z_E = \textstyle  \sum_Q \int_{\mathcal{C}_Q}  \!\mathcal{D}A_\mu 
	\mathcal{D}\bar{\psi} \, \mathcal{D}\psi \; e^{-S_E[A,\psi,\bar{\psi}]} .
\end{align}
Since all of the $Q$-subspaces are separate we could in principle choose to 
include a relative weighting factor for each of them. 

For configurations supported in disjoint spacetime regions one has
\begin{align}
Q = Q_1 + Q_2,
\end{align}
and consistency of the functional integral requires that the $Q$-subspace 
weights factorize,
\begin{align}
w(Q_1+Q_2) = w(Q_1)\, w(Q_2).
\end{align}
The origin of this factorization condition can be seen by considering
two independent spacetime regions, or two regions whose separation is
taken sufficiently large for cluster decomposition to apply. Locality
implies that the action is additive and the functional measure factorizes,
\begin{align}
S[\Phi_1,\Phi_2] &= S_1[\Phi_1]+S_2[\Phi_2],
\qquad
{\cal D}\Phi = {\cal D}\Phi_1{\cal D}\Phi_2 .
\end{align}
The corresponding partition function must then have the independent
composition property
\begin{align}
Z_{12} &= Z_1Z_2 .
\end{align}
If an otherwise arbitrary relative weight $w(Q)$ is assigned according
to the total topological charge, the combined functional integral is
\begin{align}
&\textstyle 
\int\!{\cal D}\Phi_1{\cal D}\Phi_2\, e^{-S_1-S_2}\, w(Q_1+Q_2)
    = Z_{12}=Z_1 Z_2 
    = \left[ \int\!{\cal D}\Phi_1\,e^{-S_1}w(Q_1)\right]
    \left[\int\!{\cal D}\Phi_2\,e^{-S_2}w(Q_2)\right] ,
\end{align}
since the topological charge is additive, $Q=Q_1+Q_2$. 
For this composition property to hold generally, the
$Q$-dependent weights must therefore satisfy
$w(Q_1+Q_2) = w(Q_1)w(Q_2)$ as claimed.
The factorization condition fixes the relative weights up to the choice 
of $w(1)$. Since no topological sector is to be preferentially enhanced 
or suppressed solely by its integer label, the relative weights are 
taken to have equal magnitude. Equivalently, with the normalization 
$w(0)=1$, one requires $|w(Q)|=1$. It follows that $w(1)=e^{i\theta}$ 
and hence $w(Q)=e^{i\theta Q}$.

Thus, on closed manifolds the appearance
of the phase $e^{i\theta Q}$ follows from the assumption of a decomposition 
into integer-valued sectors together with locality and additivity.
Then we observe that
\begin{align}\label{Eq:Z_theta_phase}
\textstyle \int \mathcal{D}A_\mu (\cdots) =  \textstyle \sum_Q
	\int_{\mathcal{C}_Q} \! \mathcal{D}A_\mu (\cdots)
	\xrightarrow{\theta\ne 0} & \textstyle \sum_Qe^{i\theta Q}
    \int_{\mathcal{C}_Q} \! \mathcal{D}A_\mu (\cdots) .
\end{align}

With the introduction of the $\theta$ phase the Minkowski and Euclidean 
space functional integrals in \Eq{Eq:QCD_genZ_Q} become
\begin{align}\label{Eq:QCD_tildeZ}
\tilde Z_M(\theta) & = \textstyle \sum_Q e^{i\theta Q}\int_{\mathcal{C}_Q} 
    \! \!\mathcal{D}A_\mu 
	\mathcal{D}\bar{\psi}  \mathcal{D}\psi \,e^{iS_M[A,\psi,\bar{\psi}]}
	\equiv \textstyle  \sum_Q e^{i\theta Q} Z_{M}^Q
	, \\
\tilde Z_E(\theta) &= \textstyle  \sum_Q e^{i\theta Q} \int_{\mathcal{C}_Q} 
    \! \!\mathcal{D}A_\mu 
	\mathcal{D}\bar{\psi} \mathcal{D}\psi \,e^{-S_E[A,\psi,\bar{\psi}]} 
	\equiv \textstyle  \sum_Q e^{i\theta Q} Z_{E}^Q. \nonumber
\end{align}
Since $Q$ is constant on each sector $\mathcal C_Q$, the sector weight 
may equivalently be incorporated into the action. So we can now write
\begin{align}\label{Eq:QCD_tildeZ_S}
\tilde Z_M(\theta) & = \textstyle \sum_Q \int_{\mathcal{C}_Q} 
    \! \!\mathcal{D}A_\mu 
	\mathcal{D}\bar{\psi} \mathcal{D}\psi \,e^{iS_M[A,\psi,\bar{\psi}]
    +i\theta Q}
    = \textstyle \int  \!\mathcal{D}A_\mu 
	\mathcal{D}\bar{\psi}  \mathcal{D}\psi \,e^{i\tilde S_M(\theta)} , \\
\tilde Z_E(\theta) &= \textstyle \sum_Q  \int_{\mathcal{C}_Q} 
    \! \!\mathcal{D}A_\mu 
	\mathcal{D}\bar{\psi} \mathcal{D}\psi \,e^{-S_E[A,\psi,\bar{\psi}]+i\theta Q}
    = \textstyle \int  \!\mathcal{D}A_\mu 
	\mathcal{D}\bar{\psi} \mathcal{D}\psi \,e^{-\tilde S_E(\theta)} ,
    \nonumber
\end{align}
which has led to modified actions in both cases,
\begin{align}\label{Eq:theta_actions}
S_M &\to \tilde S_M(\theta)\equiv S_M+\theta Q  \quad\mbox{and}\quad 
	S_E \to \tilde S_E(\theta)\equiv S_E -i \theta Q \, .
\end{align}
Since $Q$ is an integer by construction then we see that, while the 
Minkowski space action remains real, the Euclidean space action with 
non-zero $\theta$ is generally complex. However, we show in 
Appendix \ref{App:real_Z_E} that $\tilde Z_E(\theta)$ remains real.
It is important to note that since 
$e^{i(\theta + 2n\pi)Q} = e^{i\theta Q}$ for all $n\in\mathbb{Z}$,
the functional integrals are periodic in $\theta$, 
\begin{align}\label{Eq:Z_periodic}
\tilde Z_M(\theta+2n\pi)=\tilde Z_M(\theta) 
	\quad\mbox{and}\quad 
\tilde Z_E(\theta+2n\pi)=\tilde Z_E(\theta) .
\end{align} 
This means that the $\theta$-parameter has a meaningful range of 
$2\pi$ such as $\theta\in[0,2\pi)$ or $\theta\in(-\pi,\pi]$, without 
loss of generality. All such choices are equivalent, differing only 
by a relabeling of the same set of physical theories. The point 
$\theta=0$ lies in the interior of any such domain.

Although the $\theta$ parameter is periodic with period $2\pi$, 
discrete symmetries impose additional structure on this circle. 
In particular, the operator $F_{\mu\nu}^a\tilde F^{a\mu\nu}$ is 
odd under the combined action of charge conjugation and parity (CP).
As a consequence, under a $CP$ transformation the vacuum angle 
changes sign, $\theta \to -\theta$. $CP$ invariance of the theory 
therefore requires that $\theta$ be equivalent to $-\theta$ modulo 
$2\pi$. Within a $2\pi$-periodic parameter space, this condition 
is satisfied only at the points $\theta = 0$ and $\theta = \pi$, 
which are mapped into themselves under $\theta \to -\theta$ up 
to an integer multiple of $2\pi$. These values are thus 
distinguished as the only $CP$-invariant points on the $\theta$ 
circle. Away from these special points, the $\theta$ term explicitly 
violates CP. The case $\theta=0$ corresponds to a CP-even vacuum 
continuously connected to the perturbative vacuum, while $\theta=\pi$
represents a distinct $CP$-invariant point that may exhibit nontrivial 
vacuum structure, such as spontaneous $CP$ violation, depending on 
the dynamics of the theory.

\subsection{Conventional construction}
\label{Sec:conventional}

Recall that for rough gauge fields the $Q_{\rm gluon}$ in \Eq{Eq:EandM_top_Q} 
may be ill-defined or non-integer. The conventional continuum construction 
proceeds by identifying the assigned sector label $Q$ appearing in 
\Eq{Eq:theta_actions} with the gluonic quantity $Q_{\rm gluon}$ of
\Eq{Eq:EandM_top_Q}, including the  rough gauge fields that dominate 
the functional integral.
This corresponds to replacing the $\theta$-actions in 
\Eq{Eq:theta_actions} with $S_M(\theta)$ and $S_E(\theta)$, where
\begin{align}\label{Eq:theta_actions_simeq}
\tilde S_M(\theta) &\to S_M(\theta) \equiv S_M+\theta Q_{\rm gluon}
    = \textstyle 
    \int \!d^4x_M\, \mathcal{L}_M(\theta) 
	\quad \mbox{and} \\ 
\tilde S_E(\theta) &\to S_E(\theta) \equiv S_E-i\theta Q_{\rm gluon}
    = \textstyle 
    \int \!d^4x_E\, \mathcal{L}_E(\theta) ,  
\end{align}
and where we have defined the Lagrangian densities including a 
$\theta$-term as 
\begin{align}\label{Eq:L_thetas}
\mathcal{L}_M(\theta) &\equiv  \mathcal{L}_M + \theta q_M
	= \textstyle -\frac{1}{2} \tr(F_{\mu\nu} F^{\mu\nu}) 
	+ \bar{\psi} (i \gamma^\mu D_\mu - m) \psi + 
    \theta(g^2/16\pi^2) \tr(F_{\mu\nu}\tilde F^{\mu\nu}) , \\
\mathcal{L}_E(\theta) &\equiv \mathcal{L}_E - i\theta q_E
	= \textstyle  \frac{1}{2} \tr(F_{\mu\nu} F_{\mu\nu}) 
	+ \bar{\psi} (\gamma_\mu D_\mu + M) \psi  - i 
    \theta(g^2/16\pi^2) \tr(F_{\mu\nu}\tilde F_{\mu\nu}) .
	\nonumber
\end{align}
Note that these equations and \Eq{Eq:theta_actions} coincide for sufficiently 
smooth fields with finite action.
In this way the $\theta$-term of QCD is associated with the global topological 
classification for that special class of gauge field configurations.

Of course, we could ignore topology and simply choose to 
\emph{define from the outset} the $\theta$-actions as $S_M(\theta)$ 
and $S_E(\theta)$ rather than $\tilde S_M(\theta)$ and 
$\tilde S_E(\theta)$, where the rationale is simply that Lorentz 
invariance and gauge invariance allow a $\theta$-term and so it 
should be included in the Lagrangian densities for that reason alone. 
This alternative definition does not directly connect global topological charge 
with the $\theta$-term for rough gauge field configurations. This alternative
definition leads to the functional integrals $\tilde Z_M(\theta)$ 
and $\tilde Z_E(\theta)$ being replaced by
\begin{align}\label{Eq:QCD_genZ_theta}
\tilde Z_M(\theta) \to Z_M(\theta) \equiv \textstyle \int\!
    \mathcal{D}A_\mu \mathcal{D}\bar{\psi}  \mathcal{D}\psi \,e^{iS_M(\theta)}
 \quad\mbox{and}\quad 
\tilde Z_E(\theta) \to Z_E(\theta) \equiv \int \!
    \mathcal{D}A_\mu \mathcal{D}\bar{\psi}  \mathcal{D}\psi \,e^{-S_E(\theta)} .
\end{align}
Alternatively, one might choose to formulate the functional integral by 
restricting it to sufficiently smooth square-integrable fiinite-action
gauge fields. Then the quantity 
$Q_{\rm gluon}$ in \Eq{Eq:Q_gluon} is always integer valued.
In that case the two constructions coincide,
\begin{align}
\tilde Z_M(\theta) = Z_M(\theta) \qquad\mbox{and}\qquad 
\tilde Z_E(\theta)= Z_E(\theta)
\end{align}
%

\subsection{Instantons}
\label{Sec:Instantons}

Instantons are among the most remarkable classical solutions of 
nonabelian gauge theories. They are smooth finite-action solutions 
of the Euclidean Yang-Mills field equations and play a central role
in the semiclassical analysis of QCD. Within the conventional smooth 
finite-action framework, their existence reflects the nontrivial global 
topology discussed in the previous subsection. 

The Euclidean Yang-Mills action is
\begin{align}\label{Eq:InstantonAction}
S_{\rm YM} = \textstyle \frac12 \int \! d^4x\, \tr(F_{\mu\nu}F_{\mu\nu}) .
\end{align}
Using
\begin{align}
\tilde F_{\mu\nu}\tilde F_{\mu\nu} = F_{\mu\nu}F_{\mu\nu}
\end{align}
in Euclidean space, we have
\begin{align}\label{Eq:SelfDualIdentity}
0 &\le \textstyle \frac14 \int \! d^4x\, \tr\left[
    (F_{\mu\nu}\mp\tilde F_{\mu\nu})
    (F_{\mu\nu}\mp\tilde F_{\mu\nu}) \right]
= \frac14 \int \! d^4x\, \tr\left[ F_{\mu\nu}F_{\mu\nu}
    + \tilde F_{\mu\nu}\tilde F_{\mu\nu} \mp
    2F_{\mu\nu}\tilde F_{\mu\nu} \right]
\nonumber\\
&= \textstyle \frac12 \int \! d^4x\, \tr(F_{\mu\nu}F_{\mu\nu})
    \mp \frac12 \int \! d^4x\, \tr(F_{\mu\nu}\tilde F_{\mu\nu})
    \nonumber\\
&=S_{\rm YM}\mp(8\pi^2/g^2)\, Q_{\rm gluon},
\end{align}
where
\begin{align}
Q_{\rm gluon} = \textstyle [g^2/(16\pi^2)]
    \int\! d^4x\, \tr(F_{\mu\nu}\tilde F_{\mu\nu}) .
\end{align}
Choosing the sign that makes the second term negative then gives
\begin{align}\label{Eq:Bogomolny}
S_{\rm YM} \ge (8\pi^2/g^2) |Q_{\rm gluon}| .
\end{align}
Self-dual and anti-self-dual fields correspond respectively to
\begin{align}\label{Eq:SelfDual}
F_{\mu\nu} = \pm\tilde F_{\mu\nu}\ne 0.
\end{align}
We see that the lower bound
$S_{\rm YM} = (8\pi^2/g^2) |Q_{\rm gluon}|$ is
saturated for self-dual ($Q_{\rm gluon}>0$), anti-self-dual fields
($Q_{\rm gluon}<0$), and pure gauge fields where
$F_{\mu\nu}=0$ ($Q_{\rm gluon}=0$).
For sufficiently smooth finite-action configurations
$Q_{\rm gluon}\in\mathbb Z$
and so the smallest nonvanishing action is
$S_{\rm YM}=8\pi^2/g^2$, where those solutions
with $Q_{\rm gluon}=+1$ are single-instanton gauge-field
configurations and those with $Q_{\rm gluon}=-1$ are single-anti-instanton
gauge-field configurations.

The simplest instanton solution, discovered by Belavin, Polyakov,
Schwartz and Tyupkin (BPST)~\citep{Belavin:1975fg}, is most simply
written by embedding an $SU(2)$ solution in the QCD gauge group.
We take this $SU(2)$ subgroup to be generated by the first three
generators, so that the color index $a=1,2,3$ in the expressions below
refers to this embedded subgroup, with the remaining $SU(3)$ gauge-field
components vanishing.

It is useful first to define the 't Hooft symbols that enter the
solution. With Euclidean indices $\mu,\nu=1,2,3,4$, $SU(2)$ indices
$a=1,2,3$, and $\epsilon_{1234}=+1$, the self-dual and anti-self-dual
't Hooft symbols are defined by
\begin{align}
\eta_{a\mu\nu} &= \epsilon_{a\mu\nu4}+\delta_{a\mu}\delta_{\nu4}
    -\delta_{a\nu}\delta_{\mu4}
    \qquad\mbox{and}\qquad
    \bar\eta_{a\mu\nu}=\epsilon_{a\mu\nu4}-\delta_{a\mu}\delta_{\nu4}
    +\delta_{a\nu}\delta_{\mu4}.
\end{align}
They are antisymmetric in their Euclidean indices and satisfy
\begin{align}
\textstyle \frac12\epsilon_{\mu\nu\rho\sigma}\eta_{a\rho\sigma}
    &= \eta_{a\mu\nu}
\qquad\mbox{and}\qquad
\textstyle \frac12\epsilon_{\mu\nu\rho\sigma}\bar\eta_{a\rho\sigma}
    = -\bar\eta_{a\mu\nu}.
\end{align}
Thus $\eta_{a\mu\nu}$ and $\bar\eta_{a\mu\nu}$ provide convenient
constant tensors for constructing self-dual and anti-self-dual gauge
fields, respectively. 

For an instanton centered at $x_0$ with size $\rho$, 
the gauge potential may be written in regular gauge as
\begin{align}\label{Eq:BPST_A}
A_\mu^a(x)
&=
\frac{2}{g}\,
\frac{\eta_{a\mu\nu}(x-x_0)_\nu}
{(x-x_0)^2+\rho^2}.
\end{align}
The term ``regular gauge'' refers to this choice of gauge representative,
for which the gauge potential is nonsingular at the instanton center
$x=x_0$. At large $|x-x_0|$ the potential falls as $1/|x-x_0|$ and
approaches a pure-gauge form, while its field strength falls more rapidly.
A gauge-equivalent representation, conventionally called singular gauge,
can instead be chosen in which the gauge potential falls more rapidly at
large distance but is singular at the instanton center. 
The singularity in that representation is a gauge singularity and does
not make gauge-invariant quantities constructed from the field strength,
such as the action density, singular.

The corresponding field strength in regular gauge is
\begin{align}\label{Eq:BPST_F}
F_{\mu\nu}^a(x)
&=
-\frac{4\rho^2}{g}\,
\frac{\eta_{a\mu\nu}}
{\left[(x-x_0)^2+\rho^2\right]^2},
\end{align}
and is self-dual, $F_{\mu\nu}^a=\tilde F_{\mu\nu}^a$.
The anti-instanton is obtained by replacing the self-dual 't Hooft
symbol $\eta_{a\mu\nu}$ by the anti-self-dual symbol
$\bar\eta_{a\mu\nu}$ and satisfies
$F_{\mu\nu}^a=-\tilde F_{\mu\nu}^a$.

For the unit instanton, Eq.~(\ref{Eq:BPST_F}) gives
\begin{align}
F_{\mu\nu}^a(x)F_{\mu\nu}^a(x)
&=
\frac{192\rho^4}{g^2
\left[(x-x_0)^2+\rho^2\right]^4}.
\end{align}
Since
\begin{align}
q_E(x)
&\equiv
\frac{g^2}{32\pi^2}
F_{\mu\nu}^a(x)\tilde F_{\mu\nu}^a(x),
\end{align}
self-duality therefore gives
\begin{align}\label{Eq:BPST_q}
q_E(x)
&=
\frac{6}{\pi^2}\,
\frac{\rho^4}
{\left[(x-x_0)^2+\rho^2\right]^4}.
\end{align}
The topological charge can then be evaluated explicitly. Setting
$r=|x-x_0|$ and using $d^4x=2\pi^2r^3dr$ gives
\begin{align}\label{Eq:BPST_Q}
Q_{\rm gluon} &= \textstyle \int\! d^4x\,q_E(x)
    =12\rho^4 \int_0^\infty dr\, \frac{r^3}{(r^2+\rho^2)^4}
    = 1.
\end{align}
Similarly, the charge contained within a four-dimensional ball of
radius $R$ centered on the instanton is
\begin{align}\label{Eq:BPST_QR}
Q(R) &\equiv  \int_{|x-x_0|\le R}d^4x\,q_E(x)
    = 12\rho^4 \int_0^R \! dr\, \frac{r^3}{(r^2+\rho^2)^4}
    = 1- \frac{3\rho^4}{(R^2+\rho^2)^2}
    + \frac{2\rho^6}{(R^2+\rho^2)^3},
\end{align}
so that $Q(0)=0$ and $Q(R)\to1$ as $R\to\infty$. Thus the BPST
instanton provides an explicit example in which the local topological
charge density is smoothly distributed over a region of characteristic
size $\rho$, while its integral over all Euclidean space gives the
integer topological charge $Q_{\rm gluon}=1$.

\begin{figure}
\centering
\includegraphics[width=0.8\columnwidth]{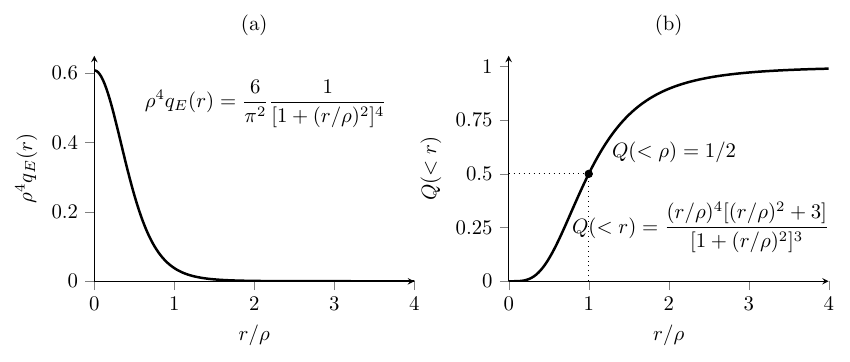}
\caption{Topological charge distribution of a BPST instanton of size
$\rho$, centered at the origin. The left panel shows the dimensionless
local density $\rho^4 q_E(r)$ as a function of $r/\rho$, using
Eq.~(\ref{Eq:BPST_q}). The right panel shows the topological charge
$Q(R)$ contained within a four-dimensional ball of radius $R$, using
Eq.~(\ref{Eq:BPST_QR}). The charge density is localized on the scale
$\rho$, whereas the integrated charge approaches the unit value
$Q_{\rm gluon}=1$ as $R/\rho\to\infty$. This illustrates explicitly
the distinction between the local density $q_E(x)$ and the global
quantity obtained by integrating it over all Euclidean spacetime for
this smooth configuration.}
\label{Fig:BPST_instanton}
\end{figure}

Figure~\ref{Fig:BPST_instanton} illustrates these two complementary
features of the solution: the localization of the topological charge
density and the accumulation of the integrated charge with increasing
radius.
The BPST solution possesses collective coordinates corresponding to its
position, size and overall gauge orientation. More general self-dual
multi-instanton solutions carrying arbitrary positive integer topological
charge, together with the corresponding anti-self-dual solutions of negative
charge, can be constructed using the ADHM construction~\citep{Atiyah:1978ri}.

Schematically, within the semiclassical approximation, the functional integral is 
expanded about these classical saddle points,
\begin{align}\label{Eq:SemiclassicalExpansion}
Z
\simeq
\textstyle \sum_{\rm saddles}\,
e^{-S_{\rm cl}}\,
{\cal N}_{\rm coll}\,
(\det' \mathcal{M}_{\text{fluct}})^{-1/2}
\cdots ,
\end{align}
where $S_{\rm cl}$ denotes the classical action of the saddle point,
$\det'$ indicates that zero modes are omitted from the fluctuation
determinant, and ${\cal N}_{\rm coll}$ schematically represents the
integration over the associated collective coordinates, together with the
appropriate gauge-fixing, ghost and fermionic factors.
Since the single-instanton action is
\begin{align}
S_{\rm inst} = 8\pi^2/g^2,
\end{align}
their contributions are exponentially suppressed by
\begin{align}
e^{-8\pi^2/g^2},
\end{align}
illustrating their intrinsically nonperturbative character.

In the conventional canonical interpretation, instantons describe tunnelling 
between classical Minkowski vacua characterized by different integer 
Chern-Simons numbers. The Euclidean instanton interpolates between the 
corresponding asymptotic configurations as Euclidean time runs from 
$-\infty$ to $+\infty$, changing the Chern-Simons number by one unit 
for a unit-charge instanton. This interpretation provides the
conventional physical picture underlying the $\theta$-vacuum construction and
many semiclassical treatments of nonabelian gauge theories.

Instantons have had a profound influence on the development of semiclassical
descriptions in modern quantum field theory. They provide important insights
into a variety of nonperturbative phenomena, including tunnelling between
classical vacua, fermion zero modes and the associated 't Hooft interaction,
and the axial $U(1)_A$ problem. Ensembles of instantons have also been
investigated as a possible mechanism contributing to spontaneous chiral
symmetry breaking. They also provide one of the principal historical and
semiclassical motivations for organizing the functional integral in terms of
topological sectors. A discussion of chiral symmetry breaking, the $U(1)_A$
problem and instantons is given in Appendix \ref{App:chiral_symm_U(1)}.

For the purposes of the present review, however, it is important to recall
the discussion of Sec.~\ref{Sec:ClassicalLimit}. Instantons are smooth
classical solutions contributing to the semiclassical approximation obtained
by expanding the functional integral about stationary points of the classical
action. The full quantum functional integral, by contrast, samples generic
quantum gauge-field configurations rather than only smooth classical
solutions. As discussed in Sec.~\ref{Sec:ClassicalLimit}, such configurations
are expected to be distributional or rough at arbitrarily short distances,
rather than smooth in the classical sense. Instantons therefore play
a central role in the semiclassical analysis of QCD, while the extent
to which the associated topological classification extends to the complete
quantum theory is a separate question that will be examined in the following
sections.

\subsection{The $\theta$ vacuum}
\label{Sec:ThetaVacuum}

The conventional interpretation of instantons is closely connected with the
construction of the $\theta$ vacuum. This construction is most naturally
formulated in the Hamiltonian approach, which will be examined in detail in
Sec.~\ref{Sec:CanonicalQuantization}. Here we summarize only the elements required to
establish the conventional connection between instantons, topological sectors,
and the $\theta$ vacuum.

In the Hamiltonian formulation one considers gauge fields on a spatial
hypersurface, for example a fixed-time spatial slice of Minkowski
spacetime. In a chosen inertial frame such a spatial hypersurface may
be identified with $\mathbb R^3$. The conventional topological
classification can then be understood by first considering the behavior
of the gauge field at spatial infinity.

For the sufficiently smooth finite-energy gauge fields considered in the
conventional construction, the field strength is required to vanish
sufficiently rapidly at spatial infinity,
\begin{align}
F_{ij}(\mathbf{x})\to 0
\qquad\text{as}\qquad
|\mathbf{x}|\to\infty .
\end{align}
The gauge field is therefore asymptotically flat and may locally be
represented as a pure gauge. The set of spatial directions at infinity
has the topology $S^2_\infty$. Since
\begin{align}
\pi_2(SU(3))=0,
\end{align}
there is no nontrivial homotopy classification associated with maps
from this asymptotic two-sphere into the gauge group.

The conventional Hamiltonian construction now introduces an additional
asymptotic condition on the admissible gauge transformations. They are
required to approach the same group element in every direction at spatial
infinity, which we denote by $U_\infty$, so that
\begin{align}
U(\mathbf{x})\to U_\infty
\qquad\text{as}\qquad
|\mathbf{x}|\to \infty .
\end{align}
This condition should be distinguished from the finite-energy condition
itself. The latter constrains the asymptotic field strength, whereas the
former is an \emph{additional} global boundary condition specifying the class
of gauge transformations admitted in the conventional
Hamiltonian formulation.

Once this boundary condition has been imposed, spatial infinity has a
unique gauge value and the spatial manifold may be one-point compactified
for the purpose of classifying the admissible gauge transformations,
\begin{align}
\mathbb R^3\cup\{\infty\}\simeq S^3 .
\end{align}
A familiar two-dimensional example helps to illustrate this construction.
Consider the plane $\mathbb R^2$ and functions that approach the same value
in every direction at infinity. One may then regard spatial infinity as
a single point at infinity and regard such functions as functions on the 
one-point compactification of the plane, which is topologically the surface 
of an ordinary sphere,
$S^2$. Similarly, the condition above permits all directions at spatial
infinity in $\mathbb R^3$ to be represented by a single added point.
Since $U_\infty$ is a constant group element, it may conventionally be
chosen to be the identity by replacing $U(\mathbf{x})$ with
$U_\infty^{-1}U(\mathbf{x})$. An admissible gauge transformation may
therefore be represented as a based map
\begin{align}
U:S^3_{\rm space}\to SU(3),
\qquad
U(\infty)=I,
\end{align}
and such maps are classified by
\begin{align}
\pi_3(SU(3))=\mathbb Z .
\end{align}
Thus the familiar integer classification does not follow directly from
the finite-energy condition alone.

It is useful to emphasize the sequence of assumptions entering this
construction. First sufficiently smooth gauge fields are assumed. Then
finite energy requires the field strength to vanish
sufficiently rapidly at spatial infinity and hence makes the gauge
field asymptotically flat. The fact that $\pi_2(SU(3))=0$ means that
there is no nontrivial homotopy classification associated with the
$S^2_\infty$ of asymptotic spatial directions. The conventional
Hamiltonian construction then imposes the \emph{additional} global boundary
condition that admissible gauge transformations approach the same
group element in every direction at spatial infinity. Only after this
additional condition has been imposed do the gauge transformations
extend continuously to the one-point compactification
$\mathbb R^3\cup\{\infty\}\simeq S^3$. It is then the nontrivial result
$\pi_3(SU(3))=\mathbb Z$ that produces their integer classification.
Thus the familiar integer winding number does not arise from winding
on the physical $S^2_\infty$ itself, but from maps of the one-point
compactified spatial hypersurface into the gauge group after the
additional asymptotic boundary condition has been imposed.

Consequently, the admissible gauge transformations fall into homotopy
classes labeled by an integer winding number $m$. Transformations in the
class $m=0$ can be continuously deformed to the identity while preserving
the asymptotic condition and are conventionally called \emph{small gauge
transformations}. Those with $m\ne0$ are conventionally called
\emph{large gauge transformations}.

Associated with a sufficiently smooth spatial gauge field $A_i(\mathbf{x})$
is the Chern-Simons number, defined by
\begin{align}\label{Eq:CS_number}
N_{\rm CS}[A] = \frac{g^2}{8\pi^2} \int\!d^3x\,
\epsilon_{ijk}\, \tr\left( A_i\partial_jA_k
-\frac{2ig}{3}A_iA_jA_k \right).
\end{align}
For a general gauge-field configuration $N_{\rm CS}[A]$ need not be an
integer. For gauge transformations satisfying the asymptotic condition
specified above, a small gauge transformation leaves $N_{\rm CS}[A]$
unchanged, whereas under a large gauge transformation $U_m$ of winding
number $m$,
\begin{align}
N_{\rm CS}[A^{U_m}]=N_{\rm CS}[A]+m,
\end{align}
where the sign of $m$ depends on the convention adopted for the winding
number.

On a fixed-time spatial hypersurface the four-dimensional gauge connection
$A_\mu$ restricts to the spatial connection $A_i$. The temporal component
$A_0$ is not being set to zero; rather, it does not form part of the
connection intrinsic to the spatial hypersurface. In the Hamiltonian
formulation $A_i$ are the gauge-field configuration variables, while $A_0$
acts as a Lagrange multiplier enforcing the Gauss-law constraint. The
Chern-Simons number is therefore naturally a functional of the spatial
connection $A_i$.

For a classical vacuum the chromoelectric and chromomagnetic fields
vanish,
\begin{align}
E_i=0,
\qquad
F_{ij}=0.
\end{align}
The latter condition implies that the spatial gauge field is pure gauge,
\begin{align}
A_i(\mathbf{x})=(i/g)U(\mathbf{x})\partial_iU^\dagger(\mathbf{x}).
\end{align}
Since the Chern-Simons number is a functional of the spatial connection
$A_i$, it is the condition $F_{ij}=0$ that is relevant for the following
topological classification.

For such a pure-gauge configuration the Chern-Simons number reduces to
the winding number of the map $U:S^3\to SU(3)$ and is therefore an integer,
\begin{align}
N_{\rm CS}[A]=n\in\mathbb Z .
\end{align}
Within this conventional description, the classical pure-gauge
configurations may consequently be labeled by this integer $n$.

\begin{figure}
\centering
\includegraphics[width=0.95\columnwidth]{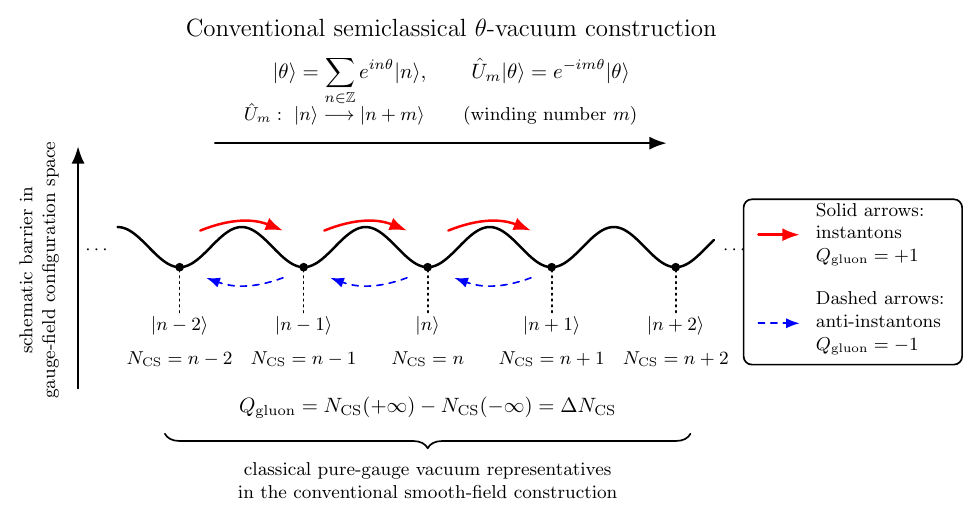}
\caption{Schematic representation of the conventional semiclassical
construction of the $\theta$ vacuum. Classical pure-gauge configurations
are represented by minima labelled by their integer Chern-Simons numbers
$N_{\rm CS}=n$. A large spatial gauge transformation of winding number
$m$ relates the corresponding vacuum-type states according to
$\hat U_m|n\rangle=|n+m\rangle$. Smooth Euclidean instanton and
anti-instanton configurations interpolate between neighboring asymptotic
vacua, with
$Q_{\rm gluon}=\Delta N_{\rm CS}=\pm1$.
The conventional $\theta$ vacuum is the coherent superposition
$|\theta\rangle=\sum_n e^{in\theta}|n\rangle$ and transforms by the phase
$e^{-im\theta}$ under a large gauge transformation of winding number $m$.
The one-dimensional barrier shown here is only a schematic representation
of the corresponding structure in the infinite-dimensional space of
gauge-field configurations.}
\label{Fig:theta_vacuum}
\end{figure}

We can now introduce the conventional vacuum-type states $\ket n$,
understood semiclassically as states localized about classical pure-gauge
vacua of integer Chern-Simons number $n$. These are analogous to states
localized about individual minima of a periodic potential in quantum
mechanics, where the role of the discrete translation operator is played
here by large gauge transformations. The periodic-potential picture is
only an analogy: the Yang-Mills configuration space is
infinite-dimensional, and the relevant interpolation occurs through
gauge-field configurations rather than along a single mechanical
coordinate. This conventional semiclassical picture is illustrated
schematically in Fig.~\ref{Fig:theta_vacuum}.

A large gauge transformation $\hat U_m$ having winding number $m$ changes
the Chern-Simons number by an integer and acts according to
\begin{align}
\hat U_m|n\rangle=|n+m\rangle ,
\qquad
m\in\mathbb Z .
\end{align}
Here it is useful to distinguish explicitly the two integers appearing
in this construction: $n$ labels the Chern-Simons number of the classical
pure-gauge configuration about which the state $|n\rangle$ is localized,
whereas $m$ is the winding number of the large gauge transformation.

The conventional $\theta$ vacuum is then defined as the coherent
superposition
\begin{align}\label{Eq:ThetaVacuum}
|\theta\rangle = \textstyle \sum_{n=-\infty}^{\infty}
e^{in\theta}|n\rangle .
\end{align}
Under a large gauge transformation one obtains
\begin{align}
\hat U_m|\theta\rangle
&= \textstyle
\sum_{n=-\infty}^{\infty} e^{in\theta}|n+m\rangle
= e^{-im\theta} \sum_{n=-\infty}^{\infty} e^{in\theta}|n\rangle
= e^{-im\theta}|\theta\rangle .
\end{align}
Thus $|\theta\rangle$ transforms only by an overall phase under large gauge
transformations and furnishes a one-dimensional representation of this
group. The parameter $\theta$ is an angular variable,
\begin{align}
\theta\equiv\theta+2\pi .
\end{align}

It is important to distinguish the integers $m$ and $n$ appearing in
this Hamiltonian construction from the four-dimensional Euclidean gluonic
topological charge $Q_{\rm gluon}$ introduced in the previous subsections.
The integer $m$ labels the homotopy class of an admissible spatial gauge
transformation, while $n$ labels the Chern-Simons number of a classical
pure-gauge configuration on a spatial hypersurface. By contrast,
$Q_{\rm gluon}$ characterizes sufficiently smooth finite-action gauge
configurations in four Euclidean dimensions. The connection between these
quantities is provided semiclassically by instantons.

For an instanton approaching pure-gauge configurations at Euclidean
times $x_4\rightarrow-\infty$ and $x_4\rightarrow+\infty$, one finds
\begin{align}\label{Eq:QDeltaN}
Q_{\rm gluon}
=
N_{\rm CS}(+\infty)-N_{\rm CS}(-\infty)
=
n_+-n_-
=
Q .
\end{align}
For these smooth instanton configurations $Q_{\rm gluon}$ coincides
with the integer sector label $Q$ introduced above. Thus a unit-charge
instanton interpolates between asymptotic classical vacua whose integer
Chern-Simons numbers differ by one unit.

The same conventional construction has a corresponding representation in
the Euclidean functional integral. Decomposing the functional integral into
integer-valued topological sectors gives
\begin{align}\label{Eq:SectorSum}
\tilde Z_E(\theta)
=
\textstyle \sum_{Q\in\mathbb Z} e^{i\theta Q}Z_E^Q ,
\end{align}
where $Z_E^Q$ denotes the functional integral restricted to the sector
labeled by $Q$. The relation in Eq.~(\ref{Eq:QDeltaN}) provides the
semiclassical connection between the relative phase associated with
transitions between the Hamiltonian vacua $|n\rangle$ and the factor
$e^{i\theta Q}$ weighting the four-dimensional Euclidean sectors.

One of the conventional motivations for selecting a definite $\theta$
vacuum is cluster decomposition. Restricting the Euclidean theory to a
single fixed value of the total topological charge imposes a global
constraint and can produce correlations between widely separated
spacetime regions. In the conventional construction, the Fourier
combination of fixed-$Q$ sectors with weights $e^{i\theta Q}$ selects
a state of definite $\theta$. It is these definite-$\theta$ theories,
rather than a theory restricted to fixed total $Q$, that provide the
standard framework in which cluster decomposition is satisfied in the
infinite-volume limit.

The $\theta$ vacuum provides the standard framework for discussing
instanton effects, anomalous chiral symmetry, and the dependence of the
vacuum energy on $\theta$. A discussion of the relative contributions of
the different topological sectors to the $\theta$ vacuum is given in
Appendix~\ref{Sec:ZQ_contrib}.

For the purposes of the present review, however, the construction above
should be regarded as a summary of the conventional connection between the
Hamiltonian $\theta$ vacuum and the Euclidean topological-sector
formulation. In the Hamiltonian formulation it relies on assumptions
concerning the global classification of gauge transformations and classical
vacua on compactified spatial slices, while in the Euclidean formulation it
relies on the classification of sufficiently smooth finite-action
four-dimensional gauge fields. The foundations and physical status of the
Hamiltonian construction, including the role of large gauge transformations,
will be examined separately in Sec.~\ref{Sec:CanonicalQuantization}.


\section{Functional integrals beyond the semiclassical approximation}
\label{Sec:FI_beyond}

The conventional discussion of topological sectors and the
$\theta$ vacuum is based upon smooth classical gauge fields
satisfying finite-action boundary conditions. As discussed 
in the previous section, these assumptions lead naturally
to topological classifications of gauge configurations and provide the foundation for the semiclassical theory of instantons.
The full quantum functional integral, however, is not 
restricted to smooth classical gauge-field configurations.
It is therefore important to examine the nature of
the configurations that dominate this integral and to 
compare them with the smooth classical fields appearing
in the semiclassical approximation.

\subsection{Smooth and rough gauge fields}
\label{Sec:SmoothRough}

As discussed in Sec.~\ref{Sec:ClassicalLimit}, the functional integral is not
dominated by smooth classical solutions. This statement can be made precise
in quantum field theories for which the functional measure has been
constructed rigorously. In four dimensions, for example, typical
configurations of a free scalar field have been shown to be nowhere
continuous with probability one
\citep{colella1973sample}. They are therefore
not ordinary differentiable classical fields, but are naturally described
as generalized fields or distributions. More generally, in Euclidean
quantum field theories for which rigorous functional-integral constructions
are available, the measure is typically supported on spaces of generalized
fields rather than smooth classical configurations
\citep{Glimm:1987ylb,Simon:2015spl}.

Although no corresponding rigorous construction is presently available for
four-dimensional QCD, asymptotic freedom provides strong grounds for
expecting analogous ultraviolet behavior. At sufficiently short distances
the running coupling becomes small, so that the ultraviolet fluctuations
increasingly approach those of a free quantum field. Weak coupling does not
imply that individual field configurations become smooth: the free-field
limit itself is supported on highly irregular configurations. Asymptotic
freedom therefore makes the short-distance behavior increasingly
perturbatively controlled without converting the configurations contributing
to the functional integral into differentiable classical gauge fields.

This roughness should not be regarded merely as an artifact of an ultraviolet
regulator. As the regulator is removed, fluctuations on progressively shorter
distance scales are included. Although their interactions become increasingly
weak, the fluctuations themselves are not thereby eliminated. Typical gauge
configurations should therefore be expected to remain rough in the continuum
limit rather than converge to smooth classical gauge fields. The continuum
quantum theory should accordingly not be regarded as a functional integral
over smooth fields, and its typical configurations should not be expected
to possess the differentiability properties assumed in semiclassical
analyses.

The functional integral itself may be written schematically as
\begin{align}\label{Eq:FullFunctionalIntegral}
Z_E
=
\textstyle \int\!
\mathcal D A_\mu\,
\mathcal D\bar\psi\,
\mathcal D\psi\;
e^{-S_E[A,\psi,\bar\psi]},
\end{align}
where the gauge-field integration is understood formally, with gauge
redundancy treated by an appropriate gauge-fixing or gauge-invariant
regularization. The measure
$\mathcal D A_\mu$ is not defined by first classifying smooth gauge fields
according to their topology. Rather, it is introduced through suitable
regularizations, such as lattice gauge theory, and subsequently interpreted
through the continuum limit.
This point is particularly transparent in lattice gauge theory. The
fundamental integration variables are group-valued link variables associated
with the edges of the lattice, and the functional measure is the product of
Haar measures over these variables. The definition of the lattice functional
integral does not require the link variables to arise from the discretization
of a smooth continuum gauge potential. In particular, generic configurations
included in the lattice measure need not possess any direct interpretation as
smooth continuum gauge fields.
Smooth continuum fields may be represented by special classes of lattice 
configurations and play an important role in semiclassical analyses, but 
typical configurations need not become smooth as the continuum limit is 
taken. Indeed, as the lattice spacing is reduced, fluctuations on 
progressively shorter physical length scales appear; asymptotic 
freedom makes their interactions increasingly weak but does not eliminate
the ultraviolet fluctuations themselves. Thus approaching the continuum limit 
does not imply that typical lattice configurations converge configuration 
by configuration to smooth classical gauge fields.

The distinction between smooth and rough gauge fields should therefore be kept
clearly in mind. Semiclassical analyses begin with smooth classical
configurations and study quantum fluctuations about them. The complete quantum
functional integral, by contrast, includes generic rough configurations that
are not small perturbations of any classical solution. Consequently, it is not
immediately evident that global topological constructions formulated for
smooth gauge fields extend without modification to the 
configurations relevant to the full quantum functional integral.

This observation does not invalidate the semiclassical approximation, nor does
it diminish the importance of instantons and related topological methods in
the study of nonperturbative QCD. Rather, it motivates a careful examination
of which properties of smooth finite-action gauge fields continue to hold for
the generic configurations relevant to the full quantum functional integral.
This question forms the starting point for the discussion in the following
subsections.

\subsection{Global versus local structure}
\label{Sec:GlobalLocal}

The preceding sections have described two complementary aspects of
nonabelian gauge theory. On the one hand, QCD is fundamentally a local
quantum field theory, whose dynamics is specified locally and in which
many physical quantities are obtained from correlation functions of local
gauge-invariant operators. On the other hand, the conventional discussion
of the strong $CP$ problem invokes global structures and classifications
involving principal bundles, homotopy classes, topological sectors and the
$\theta$ vacuum. It is therefore useful to
distinguish clearly between these local and global aspects of the theory and
to examine the relationship between them.

A basic set of quantities in quantum field theory is provided by correlation
functions of local gauge-invariant operators. For example, if
$\hat O_i(x_i)$ are local gauge-invariant operators, then their vacuum correlation
functions are given by
\begin{align}\label{Eq:LocalCorr}
\langle \Omega|  \hat O_1(x_1)\cdots \hat O_n(x_n) |\Omega \rangle_E
&=
(1/Z_E) \textstyle \int \! \mathcal D A_\mu\,\mathcal D\bar\psi\,
\mathcal D\psi\; O_1(x_1)\cdots O_n(x_n) \, e^{-S_E}.
\end{align}
Many physical observables can be obtained from correlation functions of
local gauge-invariant operators. Examples include hadron masses, scattering
amplitudes, decay rates and correlation lengths. Gauge theories also possess
important nonlocal gauge-invariant observables, such as Wilson loops.
The topological susceptibility is formally constructed from the correlation
function of the local density
$q(x)=g^2\tr(F_{\mu\nu}\tilde F^{\mu\nu})/(16\pi^2)$,
although its nonperturbative definition requires additional care because of
the short-distance singularities of the corresponding two-point function.
Ultimately, physical predictions are encoded in gauge-invariant quantities
that can, at least in principle, be related to measurements performed within
finite regions of spacetime.

The global topological classification discussed in the previous sections is
of a different character from the construction of local gauge-invariant
operators. The latter follows from the local gauge structure of QCD and does
not itself require a classification of gauge-field configuration space into
topological sectors. By contrast, for smooth gauge fields satisfying the
appropriate regularity, boundary, and global conditions, additional
topological constructions can classify entire configurations by
integer-valued quantities and, within the conventional framework, lead
ultimately to the $\theta$-vacuum construction. The relationship between
these local and global structures is therefore a nontrivial question rather
than an identity.

This distinction is familiar in differential geometry, where curvature is
defined locally at each point of a manifold whereas topology concerns the
manifold as a whole. Likewise, at the classical level the gauge-invariant
density $q(x)$ is a local quantity constructed from the field strength,
whereas its spacetime integral acquires the interpretation of an integer
topological invariant only under the additional regularity and global
conditions discussed in the previous sections. In the quantum theory the
definition of $q(x)$ as a composite operator introduces further ultraviolet
issues, which must be treated separately.

The purpose of emphasizing this distinction is not to question the internal
consistency or usefulness of the conventional topological formulation.
Indeed, it has proved extraordinarily successful in the semiclassical
analysis of nonabelian gauge theories. Rather, the objective is to identify
precisely which aspects of the theory follow directly from locality and gauge
invariance, and which additionally rely upon assumptions concerning the
global structure of gauge-field configuration space. This distinction
provides the conceptual framework for the remainder of the present review.

\subsection{Classical irrelevance of the $\theta$-term}
\label{Sec:theta_noclass}

The fact that the $\theta$ term does not contribute to the classical equations of motion can be 
understood in a purely local manner, without invoking boundary conditions at infinity or assumptions 
about the global structure of spacetime. Classical equations of motion are obtained by requiring 
stationarity of the action under arbitrary infinitesimal variations of the fields with compact support. 
That is, one considers variations $\delta A_\mu^a(x)$ which vanish outside a finite spacetime region, 
so that the variational principle probes only local properties of the action functional.
We have explained that the topological charge density $q(x)$ can always be written as a total 
derivative, $q(x) = \partial_\mu K^\mu(x)$,
for some $K^\mu$ in a given spacetime patch. We refer to $K^\mu$
as the Chern-Simons current.
Under an infinitesimal variation of the gauge field, the variation of  the 
$\theta$-dependent part of the action is
\begin{align}
\delta S_\theta
= \textstyle \theta \int \! d^4x \, \delta q(x)
= \theta\! \int\! d^4x \, \delta[\partial_\mu  K^\mu(x)]
= \theta\! \int\! d^4x \, \partial_\mu [\delta K^\mu](x).
\end{align}
Since $\delta K^\mu(x)$ is a local expression in the background gauge
field together with $\delta A_\mu^a(x)$ and its derivatives, every term
in $\delta K^\mu(x)$ contains either $\delta A_\mu^a(x)$ or one of its
derivatives. Consequently, if $\delta A_\mu^a(x)$ has compact support,
then $\delta K^\mu(x)$ also has compact support.
The spacetime integral of the divergence of a compactly supported function 
vanishes identically, so that
\begin{align}\label{Eq:int_dmuK}
\delta S_\theta = \textstyle \theta\! \int\! d^4x \, \partial_\mu [\delta K^\mu](x) 
	= 0.
\end{align}
This is a purely local statement, independent of any boundary conditions or 
assumptions about the behavior of the fields at infinity. 
This familiar argument uses a locally defined Chern-Simons current.

The result can also be seen directly without making use of the 
Chern-Simons current. Under an infinitesimal variation of the gauge field,
\begin{align}
\delta F_{\mu\nu}
=
D_\mu\delta A_\nu-D_\nu\delta A_\mu .
\end{align}
The variation of the topological density is therefore proportional to
\begin{align}
\delta\,\tr(F_{\mu\nu}\tilde F^{\mu\nu})
&=
2\,\tr(\delta F_{\mu\nu}\tilde F^{\mu\nu})
    = 4\,\tr[(D_\mu\delta A_\nu)\tilde F^{\mu\nu}]
    = 4\,\partial_\mu\tr(\delta A_\nu\tilde F^{\mu\nu})
    - 4\,\tr(\delta A_\nu D_\mu\tilde F^{\mu\nu}).
\end{align}
The second term on the right-hand side vanishes identically by the nonabelian 
Bianchi identity,
\begin{align}
D_\mu\tilde F^{\mu\nu}=0.
\end{align}
Consequently,
\begin{align}
\delta\,\tr(F_{\mu\nu}\tilde F^{\mu\nu}) =
    4\,\partial_\mu\tr(\delta A_\nu\tilde F^{\mu\nu}),
\end{align}
and hence
\begin{align}\label{Eq:delta_theta_local}
\delta S_\theta &= \theta\frac{g^2}{16\pi^2}
    \int\!d^4x\, \delta\,\tr(F_{\mu\nu}\tilde F^{\mu\nu})
    = \theta\frac{g^2}{4\pi^2} \int\!d^4x\, 
    \partial_\mu\tr(\delta A_\nu\tilde F^{\mu\nu}) =0,
\end{align}
because $\delta A_\mu$ has compact support.

This is a local variational statement and requires no assumption about the
behavior of the gauge field at spacetime infinity. Consequently, the
$\theta$ term makes no contribution to the local Euler-Lagrange equations
of motion. This result follows directly from the locality of the
variational procedure itself: classical dynamics is determined by the
response of the action to compactly supported field variations, and the
variation of the $\theta$ term is a total derivative for such variations.
This sharply distinguishes the classical role of the $\theta$ term from
its quantum-mechanical role.

\subsection{Absence of $\theta$-dependence in perturbation theory} 
\label{Sec:theta_perturb}

Perturbation theory is defined by expanding about the trivial vacuum,
which may be represented by $A_\mu=0$, and treating the gauge field as
a perturbative fluctuation about this background. With the usual boundary
or falloff conditions, the spacetime integral of the total derivative
associated with $F_{\mu\nu}\tilde F^{\mu\nu}$ vanishes order by order in
this perturbative expansion. Consequently, the $\theta$ term does not
contribute to the perturbative action. In particular, it changes neither
the quadratic action that determines the propagators nor the interaction
vertices, and therefore does not appear in the perturbative Feynman rules.

Perturbative correlation functions are consequently insensitive to
$\theta$ to all orders in the perturbative expansion. Any sensitivity
to $\theta$ must therefore arise from physics lying outside perturbation
theory about the trivial vacuum. In the conventional formulation this
sensitivity is associated with nonperturbative gauge-field configurations
for which the spacetime-integrated topological density can be nonzero,
\begin{align}
Q_{\rm gluon} = [g^2/(16\pi^2)]
\textstyle \int\!d^4x\,\tr(F_{\mu\nu}\tilde F^{\mu\nu}) \ne 0 .
\end{align}
Thus any physical sensitivity to $\theta$, if present, is intrinsically
nonperturbative.

The same observation applies to the contribution of the $\theta$ term
to the neutron electric dipole moment. A $\theta$-dependent contribution
is absent to all orders in perturbation theory about the trivial QCD
vacuum. In the conventional formulation, sensitivity to $\theta$ arises
through nonperturbative QCD dynamics, in which the functional integral
is sensitive to the spacetime-integrated topological density. The
resulting $CP$-odd structure then enters nonperturbative hadronic matrix
elements from which the $\theta$-dependent neutron electric dipole moment
is obtained. The neutron EDM induced by $\theta$ should therefore be
viewed as a probe of nonperturbative $CP$ violation in the strong
interactions, rather than as a perturbative consequence of the
$\theta$ term.

\section{Topology in lattice QCD}
\label{Sec:top_latt_QCD}

Lattice gauge theory provides a mathematically well-defined nonperturbative
regularization of QCD that is suitable for numerical calculation. It replaces
continuous spacetime by a four-dimensional Euclidean lattice with spacing
$a$, thereby rendering the regularized functional integral mathematically
well defined while preserving local gauge invariance exactly. 

More specifically, in lattice gauge theory
\citep{Creutz1983,Rothe2012,MontvayMunster,GattringerLang,Smit}
Euclidean spacetime is approximated by a finite hypercubic lattice
with lattice spacing $a$. With the conventional periodic boundary
conditions for the gauge fields, the finite lattice provides a discrete
approximation to a four-dimensional torus, $T^4$. Fermionic fields are
usually taken to be antiperiodic in the Euclidean time direction.
Other boundary conditions, including open temporal boundary conditions,
may also be employed for special purposes. The physical limit is typically 
approached by first taking the continuum limit, $a\to 0$, at fixed physical 
volume, followed by the infinite-volume limit. Gauge fields are represented
by link variables $U_\mu(x)\in SU(3)$, associated with oriented links between
neighboring lattice sites. These link variables $U_\mu(x)$ should not be
confused with the gauge transformations that we will now write as
$\Omega(x)$ to avoid confusion.

\subsection{Lattice gauge theory}
\label{Sec:Lattice}

In lattice QCD the fundamental gauge variables are not the continuum
gauge fields $A_\mu(x)$, but rather group-valued link variables
\begin{align}
U_\mu(x)\in SU(3),
\end{align}
associated with the directed links joining neighboring lattice sites.
For a sufficiently smooth continuum gauge field, the corresponding link
variable is the parallel transporter,
\begin{align}\label{Eq:LinkVariable}
U_\mu(x) &= \textstyle {\cal P}\exp\left[ig\int_x^{x+a\hat\mu} 
    dz^\nu A_\nu(z)\right]
    = \exp\left[ igaA_\mu(x)+{\cal O}(a^2) \right] ,
\end{align}
where this is the finite path-dependent parallel transporter connecting 
the $SU(3)$ fibers at neighboring lattice sites.

It is important, however, that the lattice functional integral is defined
directly in terms of the group-valued variables $U_\mu(x)$ and does not
require the existence of an underlying smooth continuum gauge field.
Under a local gauge transformation $\Omega(x)$,
\begin{align}\label{Eq:LinkGauge}
U_\mu(x) &\to \Omega(x) U_\mu(x) \Omega^\dagger(x+a\hat\mu),
\end{align}
where $\Omega(x)\in SU(3)$. To avoid confusion with the links
we use $\Omega(x)$ here rather than $U(x)$ for the $SU(3)$ gauge 
transformation. Local gauge invariance is therefore preserved
exactly at finite lattice spacing.

The elementary gauge-invariant quantity entering the Wilson gauge action is
the trace of the plaquette, where the plaquette variable is
\begin{align}\label{Eq:Plaquette}
U_{\mu\nu}(x) &= U_\mu(x) U_\nu(x+a\hat\mu)
    U_\mu^\dagger(x+a\hat\nu) U_\nu^\dagger(x).
\end{align}
The Wilson gauge action is
\begin{align}\label{Eq:WilsonAction}
S_W &= \textstyle \beta \sum_P \left[1-\frac13\text{Re}\,\tr(U_P)\right],
\end{align}
where $\beta=6/g^2$ for $SU(3)$ gauge theory and the sum extends over all
plaquettes.

For link variables corresponding to a sufficiently smooth continuum gauge
field, expansion at small lattice spacing gives
\begin{align}
U_\mu(x)
&=
1+igaA_\mu(x)+{\cal O}(a^2),
\end{align}
and the corresponding expansion of the plaquette yields the continuum
field strength. One then finds
\begin{align}
S_W
&=
\textstyle \frac12\int\!d^4x\,
\tr(F_{\mu\nu}F_{\mu\nu})
+{\cal O}(a^2).
\end{align}
This establishes the classical continuum limit of the lattice action for
sufficiently smooth gauge fields. It should not, however, be interpreted as
a configuration-by-configuration correspondence between generic lattice
gauge fields $U_\mu(x)$ and generic gauge potentials in the continuum
$A_\mu(x)$. Typical configurations
contributing to the lattice functional integral are rough, and need not
approach differentiable continuum fields as $a\rightarrow0$. The quantum
continuum limit on the lattice is instead defined through the convergence 
of suitably renormalized observables and correlation functions.

At finite lattice spacing the functional integral is defined by
independent integrations over an $SU(3)$ group element on each link, with
correlations between the link variables generated by the action. The
gauge-field functional measure is the product of the corresponding Haar
measures,
\begin{align}\label{Eq:LatticeMeasure}
{\cal D}U
&=
\textstyle \prod_{\rm links}dU_\mu(x).
\end{align}
No assumption concerning differentiability or the existence of a smooth
continuum gauge connection enters the definition of this lattice functional
integral.

At finite lattice spacing, and before any additional admissibility condition
is imposed, the space of link configurations on a finite lattice with
$N_L$ links is simply
\begin{align}
{\cal C}_{\rm latt}=SU(3)^{N_L}.
\end{align}
Since $SU(3)$ is connected, this configuration space is itself connected.
Thus the unrestricted lattice functional integral does not begin with a
decomposition of the link-configuration space into disconnected sectors
labeled by an integer topological charge. An integer-valued quantity may
nevertheless be assigned to lattice configurations by additional
constructions. Such an assignment should be distinguished from an
\emph{a priori} decomposition of the unrestricted configuration space into
disconnected components.

The notion of topological charge therefore requires additional discussion.
For sufficiently smooth lattice gauge fields satisfying appropriate
admissibility conditions, lattice definitions of the topological charge
can be related to the continuum Pontryagin index and acquire the expected
integer-valued continuum interpretation. 
Furthermore, lattice formulations employing chiral fermions satisfying
the Ginsparg-Wilson relation admit an integer-valued index of the lattice
Dirac operator and an associated lattice index relation. This provides a
natural fermionic definition of topological charge at finite lattice
spacing, with the expected continuum interpretation for sufficiently
regular gauge fields.

In practical lattice calculations, Monte Carlo gauge configurations
typically contain substantial ultraviolet fluctuations. Consequently,
topological charge is often extracted after applying smoothing procedures,
such as cooling or, more systematically, gradient flow, which suppress
short-distance fluctuations and expose the longer-distance structure of
the gauge field. In particular, gradient-flow definitions of topological
observables possess a controlled continuum limit and have become standard
tools in modern lattice studies of topology.

For the purposes of the present review, the important observation is that
lattice QCD begins with a well-defined regularized functional integral over
group-valued link variables, without assuming an \emph{a priori}
decomposition into smooth topological sectors. 
The identification and characterization of topological structure on the
lattice therefore involve additional definitions or regularity conditions
beyond the bare specification of the unrestricted lattice functional
integral, and the relation of these constructions to
continuum topology must be examined carefully.
These issues are discussed further in the following subsections.

The correspondence between lattice QCD and continuum QCD in the continuum
limit is a correspondence between the resulting quantum theories, as
expressed through renormalized physical observables and correlation
functions. It is not established through a configuration-by-configuration 
correspondence between generic rough lattice fields and generic 
rough continuum gauge fields. The familiar explicit correspondence 
between link variables and continuum gauge potentials applies only 
to sufficiently smooth fields.

\subsection{Definitions of topological charge}
\label{Sec:LatticeTopology}

For sufficiently smooth gauge fields in the continuum limit, the
topological classification on the four-torus $T^4$ may be formulated
in terms of principal $SU(3)$ bundles. As discussed previously, such
a bundle describes how copies of the gauge group $SU(3)$ attached to
different spacetime points are connected together globally. 
Different bundles may be locally indistinguishable, since every bundle 
is locally trivial, while differing globally in how these local 
descriptions are patched together by transition functions.

For a nontrivial bundle, the gauge connection need not be representable
by a single globally defined gauge potential $A_\mu(x)$ over all of
$T^4$. Rather, it is described by gauge potentials on overlapping
coordinate patches related by transition functions. It is precisely
this global patching information that permits the second Chern number
to be nonzero on a manifold without boundary.

The relevant mathematical quantity characterizing this global structure
in four dimensions is the second Chern class,
\begin{align}
c_2\in H^4(T^4,\mathbb Z).
\end{align}
Here $H^4(T^4,\mathbb Z)$ is the fourth cohomology group of $T^4$.
For principal $SU(3)$ bundles over the four-dimensional base $T^4$,
the relevant topological information is encoded in the second Chern
class. Since
\begin{align}
H^4(T^4,\mathbb Z)\simeq\mathbb Z,
\end{align}
the second Chern class is specified by a single integer, whose evaluation
on the fundamental four-cycle gives the second Chern number.

More concretely, the whole four-torus $T^4$ constitutes a closed
four-dimensional cycle, conventionally called the fundamental
four-cycle. Evaluating the second Chern class on this four-cycle means,
in its differential-form representation, integrating the corresponding
local curvature density over the complete four-dimensional manifold.
The resulting integer is called the second Chern number. With the
conventional normalization this integer is identified with the gluonic
topological charge,
\begin{align}
Q_{\rm gluon}
= \textstyle [g^2/(16\pi^2)]
\int\!d^4x\,
\tr(F_{\mu\nu}\tilde F_{\mu\nu}) .
\end{align}
Thus, in physical terms, the second Chern number measures a global
topological property of how the gauge bundle is arranged over the
entire four-dimensional spacetime, while the above integral provides
its representation in terms of the local gauge-field curvature.

The way in which this integer is represented on $T^4$ differs from the
familiar description for smooth finite-action fields on $\mathbb R^4$.
On $\mathbb R^4$, the usual construction assumes sufficiently regular
asymptotic behavior such that the gauge field approaches a pure gauge at
infinity. The asymptotic gauge transformation then defines a map
$U:S^3_\infty\to SU(3)$ whose winding number gives the integer topological
charge. Under the additional regularity conditions required to identify all
directions at infinity with a single added point, with any residual
direction dependence accommodated by the gauge transition function,
$\mathbb R^4$ may be compactified to $S^4$, and the same integer can
equivalently be viewed as the second Chern number of the corresponding
bundle over $S^4$.
In this description the asymptotic map $U:S^3\to SU(3)$ appears as the
transition function between gauge patches on $S^4$. The four-torus has
no boundary or spacetime infinity, so this asymptotic construction is
unavailable; instead, the global bundle structure over $T^4$ is described
directly by its second Chern class. Under the appropriate regularity 
conditions, both constructions nevertheless give an integer represented
by the  spacetime integral of the local density $\mathrm{tr}(F\tilde F)$.

The situation for generic lattice gauge configurations is different.
At finite lattice spacing a straightforward discretization of
$F_{\mu\nu}\tilde F_{\mu\nu}$ gives a gluonic lattice topological
charge that is not, in general, an integer on typical Monte Carlo
configurations. Such configurations contain substantial ultraviolet
fluctuations and need not be interpretable as discretizations of smooth
continuum gauge fields. As emphasized in Sec.~\ref{Sec:Lattice}, the
continuum limit of lattice QCD is defined through the continuum behavior
of renormalized observables and correlation functions rather than through
a configuration-by-configuration convergence of generic rough lattice
fields to differentiable continuum gauge potentials. Consequently, the
integer topological classification of smooth continuum fields cannot
simply be transferred configuration by configuration to the unrestricted
lattice ensemble by applying a naive discretization of $F\tilde F$.

There are, however, well-developed procedures for assigning or extracting
an integer topological charge from lattice gauge configurations. Two
particularly important approaches are smoothing by gradient flow and
fermionic definitions based upon the index of an appropriate lattice
Dirac operator.

\emph{Gradient flow and gluonic topological charge:}
Under gradient flow the original link variables are evolved continuously
in an auxiliary flow time $t$. The flow suppresses ultraviolet
fluctuations over a characteristic physical distance of order
\begin{align}
r_{\rm sm}\simeq\sqrt{8t},
\end{align}
while preserving gauge covariance. For positive flow time the resulting
gauge field is substantially smoother than the original field, and
gluonic definitions of the local topological charge density become
well behaved. One may then define
\begin{align}
Q_{\rm gluon}(t) = \textstyle \int\!d^4x\,q_t(x),
\end{align}
where $q_t(x)$ is constructed from the field strength of the flowed
gauge field.

For sufficiently smooth configurations, and with lattice artifacts
under control, $Q_{\rm gluon}(t)$ approaches values close to integers.
Gradient flow therefore provides a particularly useful means of exposing
the topological structure associated with a rough lattice configuration
\citep{Luscher:1981zq,Luscher2010iy,Bruno2014,L_scher_2010,L_scher_2011}.
In practical calculations the flowed gluonic charge becomes increasingly
concentrated near integer values as lattice artifacts are reduced. When an
integer label is required for an individual configuration, one may assign
the nearby integer after sufficient flow. For observables such as the
topological susceptibility, however, the flowed charge itself can be used
and the continuum limit taken without requiring such a rounding procedure.

It is important to distinguish this procedure from the assertion that
the original unflowed lattice configuration possessed a smooth continuum
topological charge. Gradient flow defines a map from the original rough
configuration to a smoother configuration on which the familiar gluonic
notion of topological charge becomes applicable. The resulting construction
is extremely useful, but the smoothing procedure constitutes additional
structure beyond the definition of the unrestricted lattice functional
integral itself.

\emph{Fermionic definition and the index:}
A second approach is provided by the relation between topology and
fermion zero modes. For a sufficiently smooth continuum gauge field,
the Atiyah-Singer index theorem gives
\begin{align}\label{Eq:AS_index}
Q_{\rm gluon}
=
\operatorname{index}(D)
=
n_- - n_+
\in\mathbb Z ,
\end{align}
where $D$ is the Dirac operator in the background gauge field and
$n_-$ and $n_+$ denote the numbers of negative- and positive-chirality
zero modes, respectively, with the sign convention correlated with
that adopted for $Q_{\rm gluon}$.

The index theorem fixes the difference $n_--n_+$; it does not, for a
general smooth gauge field, imply that zero modes of only one chirality
must occur. Stronger vanishing results apply for special backgrounds,
in particular self-dual or anti-self-dual instanton configurations,
for which zero modes of one of the two chiralities are excluded
\citep{tHooft:1976snw,JackiwRebbi:1977}.

On the lattice, Dirac operators satisfying the Ginsparg-Wilson relation
provide an exact lattice realization of chiral symmetry. In particular,
the overlap Dirac operator possesses exact chiral zero modes and an
integer-valued index, allowing a fermionic topological charge to be
defined at finite lattice spacing,
\begin{align}
Q_{\rm index}
\equiv
\operatorname{index}(D_{\rm GW})
=
n_- - n_+
\in\mathbb Z .
\end{align}
This provides an integer-valued definition even when a straightforward
gluonic discretization of $F\tilde F$ on the same unflowed configuration
is not close to an integer
\citep{Neuberger:1998,Hasenfratz:1998,Luscher:1998}.
The existence of this integer index does not imply that the unrestricted
link-configuration space $SU(3)^{N_L}$ has become disconnected. As the
links are varied continuously, the index is constant within regions in
which the overlap operator remains well defined and local, but it can
change when an exceptional configuration is encountered at which an
eigenvalue of the kernel passes through zero. Admissibility or related
regularity conditions exclude such configurations and thereby separate
the allowed configuration space into sectors of fixed index.

The existence of an integer lattice index should nevertheless be
distinguished from the continuum Atiyah-Singer theorem applied directly
to a generic rough continuum gauge potential. For a sufficiently
irregular continuum field, the regularity properties required for the
usual continuum Dirac operator and its index theorem cannot simply be
assumed. The lattice construction instead provides its own well-defined
fermionic definition at finite lattice spacing.

For sufficiently smooth lattice configurations the gluonic and fermionic
definitions agree with the expected continuum topological charge.
Numerical studies have verified that, after sufficient smoothing, the
gluonic charge becomes close to an integer and agrees with the index
obtained from suitable chiral lattice Dirac operators
\citep{Luscher2010iy,Bruno2014,L_scher_2010,L_scher_2011,
Alexandrou2020,Leinweber2004,Zhang2002}.
This agreement provides an important connection between the gluonic,
fermionic, and continuum descriptions of topology.
 
At positive flow time $t$, the flow suppresses short-distance
fluctuations over the radius $(8t)^{1/2}$ and permits well-defined
renormalized composite quantities to be constructed from the flowed
field. One may take the continuum limit at fixed positive physical
flow time. For local flowed observables one may subsequently study the
small-flow-time expansion, while the integrated continuum topological
charge is flow-time independent under the usual boundary conditions.
Alternatively, in considering the removal of lattice-scale fluctuations,
one may consider flow times satisfying
\begin{align}
a\ll\sqrt{8t}\ll L_{\rm phys},
\end{align}
where $L_{\rm phys}$ denotes a relevant fixed physical length scale.
The existence of such a separation of scales allows lattice-scale
roughness to be suppressed without requiring smoothing over macroscopic
physical distances.

This should not be interpreted as replacing the original QCD functional
integral by a functional integral over smooth fields. The flowed fields
are constructed from configurations distributed according to the original
QCD measure, and observables at positive flow time are flowed observables
whose relation to the original theory is controlled by the continuum and
small-flow-time limits. Gradient flow therefore provides a controlled
means of defining and studying topological quantities without requiring
the original configurations contributing to the lattice functional
integral to be smooth.

The important point for the present discussion is that the unrestricted
finite-lattice configuration space described in Sec.~\ref{Sec:Lattice}
is connected and is not intrinsically decomposed into disconnected
integer-valued topological sectors. Integer topological charge can
nevertheless be defined through additional constructions, notably
regularity or admissibility conditions, smoothing procedures such as
gradient flow, or the index of an appropriate lattice Dirac operator.
For sufficiently smooth fields these constructions reproduce the familiar
continuum topological classification. For generic rough lattice
configurations they provide lattice definitions whose relation to the
continuum topological charge becomes unambiguous in the appropriate
continuum and regularity limits.

Thus lattice QCD gives a precise and nonperturbative framework in which
topological observables can be defined and calculated, but it is important
to distinguish this statement from the stronger assertion that every raw
gauge configuration appearing in the unrestricted lattice functional
integral possesses, without additional construction, the same intrinsic
integer topological classification as a smooth continuum gauge field.
The distinction between these statements will be important in what
follows. 

\emph{Note:} It is important to distinguish the addition of a term proportional to
$Q_{\rm gluon}$ to the action from a decomposition of the functional
integral into integer-valued topological sectors. The distinction between
these two constructions is displayed explicitly in
\Eq{Eq:theta_actions_simeq}. The two constructions are directly equivalent
only when the class of gauge fields under consideration is such that
$Q_{\rm gluon}$ is integer valued and provides a corresponding topological
classification of configuration space. This is the case for sufficiently
smooth gauge fields satisfying the appropriate global conditions. For
generic rough configurations in the unrestricted lattice functional
integral, the expression involving $Q_{\rm gluon}$ does not, by itself,
imply a decomposition into sectors labeled by $Q\in\mathbb Z$; an additional
construction is required to assign such an integer topological charge.

\subsection{Physical interpretation}
\label{Sec:LatticePhysical}

The local topological charge density $q(x)$, as well as correlation 
functions constructed from it, play an important role in nonperturbative 
lattice studies of QCD.
In particular, lattice methods provide well-defined procedures for
calculating the topological susceptibility and related quantities associated
with anomalous chiral symmetry. These results constitute important
quantitative tests of our understanding of nonperturbative QCD.

The most widely studied quantity is the topological susceptibility, which
is associated with the correlation function of the local topological charge
density,
\begin{align}\label{Eq:LatticeChi}
\chi
&=
\textstyle \int\!d^4x\,
\langle q(x)q(0)\rangle_{\rm conn}.
\end{align}
Here an appropriate nonperturbative definition of the composite operators
and their short-distance behavior is understood. This qualification is
important because the product of two local topological charge densities has
short-distance singularities at coincident points, so that the expression
above requires a proper nonperturbative prescription.

Modern lattice methods provide such prescriptions. In particular, gradient
flow defines a flowed local density $q_t(x)$ at positive flow time $t$ for
which correlation functions are well behaved. Fermionic constructions based
on lattice Dirac operators satisfying the Ginsparg-Wilson relation provide
another nonperturbative approach to the same physics. The agreement of
different suitable lattice definitions as the continuum limit is approached
provides strong evidence that the resulting topological susceptibility is a
well-defined nonperturbative observable of continuum QCD rather than an
artifact of a particular lattice prescription.

One may also introduce the integrated topological charge
\begin{align}
Q
&=
\textstyle \int\!d^4x\,q(x),
\end{align}
where the same qualifications concerning its nonperturbative definition are
understood. With an appropriate definition of $Q$, translational invariance
allows the susceptibility to be written as
\begin{align}
\chi
&=
\frac{\langle Q^2\rangle-\langle Q\rangle^2}{V}.
\end{align}
At $\theta=0$, if the vacuum preserves $CP$, the expectation value of the
$CP$-odd quantity $Q$ vanishes, $\langle Q\rangle=0$, and hence
\begin{align}
\chi
&=
\frac{\langle Q^2\rangle}{V}.
\end{align}
This representation is extremely useful, but it should be distinguished
from the stronger statement that the unrestricted lattice configuration
space is intrinsically decomposed into disconnected sectors labeled by an
integer $Q$.

The continuum limit is obtained by taking the lattice spacing to zero while
holding appropriate physical quantities fixed. Universality requires that
different lattice regularizations lying in the same universality class
produce the same renormalized continuum observables. The agreement of
independent lattice formulations and different suitable definitions of
topological observables in this limit provides an important nonperturbative
test of this expectation.

It is therefore useful to distinguish the successful nonperturbative
definition and calculation of quantities constructed from $q(x)$ from a
stronger global interpretation of the gauge-field configuration space.
The lattice functional integral is defined directly in terms of group-valued
link variables integrated with the Haar measure. As discussed in the
previous subsection, the unrestricted configuration space at finite lattice
spacing is connected and does not begin with an intrinsic decomposition
into disconnected sectors labeled by an integer topological charge.
Topological charge can nevertheless be defined or characterized through
additional constructions, including regularity or admissibility conditions,
smoothing procedures such as gradient flow, and appropriate fermionic index
definitions. Some of these constructions provide an exactly integer-valued
charge, while flowed gluonic definitions approach the corresponding
integer-valued continuum interpretation as lattice artifacts are removed.

The success of lattice calculations therefore establishes that quantities
such as the topological susceptibility can be given precise
nonperturbative definitions and possess well-defined continuum limits.
It does not, by itself, establish the logically stronger statement that
every raw gauge configuration appearing in the unrestricted functional
integral possesses an intrinsic integer topological charge in precisely
the same sense as a sufficiently smooth continuum gauge field satisfying
the appropriate global conditions.

This distinction is particularly relevant to the conventional
$\theta$-vacuum construction. As emphasized in the previous subsection,
the addition of a term proportional to $Q_{\rm gluon}$ to the action and
the decomposition of the functional integral into sectors labeled by
$Q\in\mathbb Z$ are directly equivalent only when an appropriate
integer-valued topological classification has been established. The
successful lattice calculation of observables constructed from the local
density $q(x)$, including distributions of appropriately defined topological
charge and calculations performed at fixed topology, does not by itself
establish that the unrestricted finite-spacing link-configuration space is
intrinsically decomposed into disconnected integer-valued sectors. Such
constructions employ a particular definition of topological charge or an
additional restriction on the allowed configurations. The existence and
physical usefulness of these constructions should therefore be distinguished
from an \emph{a priori} decomposition of the unrestricted configuration
space itself. The relationship between these statements must therefore be
kept conceptually distinct.


\section{Hamiltonian and Euclidean perspectives}
\label{Sec:H_vs_L}

The Hamiltonian formulation of Yang-Mills theory provides a complementary
description of QCD to the Euclidean functional integral discussed in the
preceding sections. Whereas the functional-integral formulation emphasizes
vacuum expectation values and correlation functions, the canonical approach
focuses on the physical Hilbert space and the implementation of gauge
invariance through operator constraints. It is within this framework that the
conventional construction of the $\theta$ vacuum is most naturally developed.
The discussion presented here follows standard treatments of constrained
Hamiltonian systems; a detailed development of constrained
Hamiltonian systems and an application to electromagnetism
and quantum electrodynamics may be found elsewhere
\citep{Williams:2022bzq}. However, the nonabelian nature of QCD
complicates the situation somewhat.

\subsection{Canonical quantization}
\label{Sec:CanonicalQuantization}

The starting point is the Minkowski-space QCD Lagrangian
\begin{align}
\mathcal L &= \textstyle -\frac14 F_{\mu\nu}^aF^{a\mu\nu}
    + \bar\psi(i\gamma^\mu D_\mu-m)\psi ,
\end{align}
where throughout this discussion we use
\begin{align}
D_\mu &= \partial_\mu-igA_\mu .
\end{align}
Note carefully that the canonical momentum conjugate to the gauge field
$A^{a\mu}$ is $\Pi_\mu^a$, defined by
\begin{align}\label{Eq:CanonicalMomentum}
\Pi_\mu^a &= \frac{\partial\mathcal L}{\partial(\partial_0A^{a\mu})}
    = -F^{a0}{}_\mu .
\end{align}
Thus the canonical pairing is $(A^{a\mu},\Pi_\mu^a)$ and not
$(A^{a\mu},\Pi^{a\mu})$. Equivalently,
\begin{align}
\Pi^{a\mu} &= -F^{a0\mu} = F^{a\mu0}.
\end{align}
Using the notation
\begin{align}
\mathbf E^a &= (E^{a1},E^{a2},E^{a3})
    = (E_x^a,E_y^a,E_z^a),
\end{align}
and similarly for $\mathbf B^a$, the temporal and spatial components give
\begin{align}
\Pi_0^a &= \Pi^{a0} = 0, \qquad
\Pi_i^a = -E^{ai}, \qquad \Pi^{ai} = E^{ai}.
\end{align}
The relation $\Pi_0^a=0$ is a primary constraint rather than an equation
of motion. It arises because the Lagrangian contains no time derivative
of $A^{a0}$, so that $A^{a0}$ is not an independent dynamical degree of
freedom.

The canonical Hamiltonian density obtained by the Legendre transformation
is
\begin{align}\label{Eq:CanonicalHamiltonian}
\mathcal H &= \textstyle \frac12
\left(E^{ai}E^{ai}+B^{ai}B^{ai}\right)
    + \psi^\dagger\left(-i\boldsymbol{\alpha}\cdot\mathbf D
    +\beta m\right)\psi-A_0^aG^a,
\end{align}
where
\begin{align}\label{Eq:GaussGenerator}
G^a&=(D_iE^i)^a+g\psi^\dagger T^a\psi .
\end{align}
The Legendre transformation and the emergence of the primary and
secondary constraints are derived explicitly in
Appendix~\ref{App:H_QCD_deriv}. In particular, $\Pi_0^a=0$ is the
primary constraint, while preservation of this constraint under time
evolution gives the secondary Gauss-law constraint
\begin{align}\label{Eq:GaussLaw}
G^a&=0.
\end{align}
Equivalently,
\begin{align}
(D_iE^i)^a&=-g\psi^\dagger T^a\psi .
\end{align}
Unlike in Maxwell theory, the covariant divergence contains a nonlinear
term. In components,
\begin{align}
(D_iE^i)^a &= \partial_iE^{ai} + gf^{abc}A_i^bE^{ci}
    = \partial_iE^{ai} - gf^{abc}A^{bi}E^{ci},
\end{align}
where in the right-hand side we have used $A_i^a=-A^{ai}$. This nonlinear
term reflects the self-interaction of the nonabelian gauge field.

The Dirac-Bergmann consistency procedure also requires that the secondary
constraint remain satisfied under time evolution. To see this, write
the total Hamiltonian schematically as
\begin{align}
H_T
&=
H_0
-
\int\!d^3x\,A_0^aG^a
+
\int\!d^3x\,\lambda^a\Pi_0^a,
\end{align}
where $H_0$ contains the chromoelectric, chromomagnetic, and fermionic
terms and $\lambda^a$ are arbitrary Lagrange multipliers. The Gauss-law
constraints form a closed algebra,
\begin{align}\label{Eq:GaussAlgebra}
\{G^a(\mathbf x),G^b(\mathbf y)\}_{\rm PB}
&=
gf^{abc}G^c(\mathbf x)
\delta^{(3)}(\mathbf x-\mathbf y),
\end{align}
while $H_0$ is gauge invariant and hence has vanishing Poisson bracket
with the Gauss-law generators. It follows that
\begin{align}
\partial_0G^a(\mathbf x) &= \{G^a(\mathbf x),H_T\}_{\rm PB}
    = -gf^{abc}A_0^b(\mathbf x)G^c(\mathbf x).
\end{align}
Therefore,
\begin{align}
G^a(\mathbf x)=0
\qquad\Longrightarrow\qquad
\partial_0G^a(\mathbf x)=0.
\end{align}
Thus Gauss's law, once imposed on an initial spatial hypersurface, is
preserved by the Hamiltonian time evolution. No additional tertiary
constraint is generated. The Dirac constraint algorithm therefore
closes at this stage.

Canonical quantization now proceeds by promoting the fields and their
conjugate momenta to operators. We do not impose Coulomb gauge or eliminate
the longitudinal gauge-field degrees of freedom at this stage. The
unreduced canonical variables therefore satisfy the ordinary equal-time
commutation relations
\begin{align}
[\hat A^{ai}(\mathbf x),\hat A^{bj}(\mathbf y)] &= 0, \qquad
[\hat\Pi_i^a(\mathbf x),\hat\Pi_j^b(\mathbf y)] = 0, \qquad
[\hat A^{ai}(\mathbf x),\hat\Pi_j^b(\mathbf y)] =
    i\delta^{ab}\delta^i{}_j \delta^{(3)}(\mathbf x-\mathbf y).
\end{align}
Since $\Pi_i^a=-E^{ai}$, these imply
\begin{align}
[\hat E^{ai}(\mathbf x),\hat E^{bj}(\mathbf y)] &= 0, \qquad
[\hat A^{ai}(\mathbf x),\hat E^{bj}(\mathbf y)]
    = -i\delta^{ab}\delta^{ij}\delta^{(3)}(\mathbf x-\mathbf y).
\end{align}
The fermion fields obey the usual corresponding canonical
anticommutation relations, and the fermion and gauge-field canonical
variables commute with one another.

It is worth emphasizing that the ordinary three-dimensional delta
function appears here rather than the transverse delta function familiar
from the reduced Coulomb-gauge quantization of electromagnetism. In the
present formulation Gauss's law has not been solved classically and the
longitudinal gauge-field degrees of freedom have not been eliminated.
Instead, the Gauss-law constraint is retained and imposed directly on the
quantum physical states,
\begin{align}\label{Eq:PhysicalStates}
\hat G^a(\mathbf x)\,|\Psi_{\rm phys}\rangle &= 0 .
\end{align}
This condition implements the local gauge redundancy on the physical
Hilbert space.

To see explicitly how Gauss's law is related to gauge transformations,
introduce the matrix-valued quantities
\begin{align}
A_\mu(\mathbf x)
&=
A_\mu^a(\mathbf x)T^a,
\qquad
\omega(\mathbf x)
=
\omega^a(\mathbf x)T^a,
\end{align}
and consider a local gauge transformation
\begin{align}\label{Eq:InfinitesimalGaugeU}
U(\mathbf x)
&=
\exp[ig\omega(\mathbf x)]
=
1+ig\omega(\mathbf x)+{\cal O}(\omega^2).
\end{align}
The gauge potential transforms as
\begin{align}\label{Eq:FiniteGaugeA}
A_\mu^U &= UA_\mu U^\dagger + (i/g) U\partial_\mu U^\dagger .
\end{align}
Expanding to first order in $\omega$ gives
\begin{align}
UA_\mu U^\dagger &= A_\mu + ig[\omega,A_\mu] +
    {\cal O}(\omega^2)
\qquad\mbox{and}\qquad (i/g)\, U\partial_\mu U^\dagger
    = \partial_\mu\omega + {\cal O}(\omega^2).
\end{align}
We then find
\begin{align}\label{Eq:InfinitesimalGaugeA}
\delta_\omega A_\mu &\equiv A_\mu^U-A_\mu
    = \partial_\mu\omega + ig[\omega,A_\mu]
    = D_\mu\omega .
\end{align}
For an upper spatial component this gives
\begin{align}\label{Eq:Ham_gt_Ai}
\delta_\omega A^i
&=
-(D_i\omega).
\end{align}

We now recover these infinitesimal transformations directly from the
Gauss-law constraint. Since $G^a(\mathbf x)$ is a local constraint, it
is useful to smear it with an arbitrary smooth function
$\omega^a(\mathbf x)$ of compact support,
\begin{align}\label{Eq:SmearedGenerator}
{\cal G}[\omega] &= \textstyle
    \int\!d^3x\, \omega^a(\mathbf x)G^a(\mathbf x).
\end{align}
The terminology ``smeared'' simply means that the local operators
$G^a(\mathbf x)$ are integrated against the ordinary functions
$\omega^a(\mathbf x)$. Each choice of $\omega^a(\mathbf x)$ therefore
defines a single generator ${\cal G}[\omega]$.

Using the canonical commutation and anticommutation relations, the
infinitesimal transformation of a canonical operator $\hat O$ is
generated by
\begin{align}\label{Eq:GaugeGeneratedByG}
\delta_\omega \hat O
&=
-i[\hat{\cal G}[\omega],\hat O].
\end{align}
Direct evaluation gives
\begin{align}
\delta_\omega \hat A^i &= -(\hat D_i\omega) \,,
\qquad
\delta_\omega \hat E^i = ig[\omega,\hat E^i] \, ,
\qquad
\delta_\omega\hat \psi = ig\omega\hat \psi \, ,
\qquad
\delta_\omega\hat \psi^\dagger = -ig\hat \psi^\dagger\omega \, .
\end{align}
The first of these is precisely the infinitesimal transformation obtained
from Eq.~(\ref{Eq:Ham_gt_Ai}), with the spatial index raised. Thus the
Gauss-law constraint is not merely a restriction on the physical states:
its smeared form is the canonical generator of infinitesimal gauge
transformations of the spatial canonical variables on a fixed time
slice.

The corresponding transformation on the Hilbert space is obtained by
exponentiating the generator,
\begin{align}\label{Eq:HilbertGaugeTransformation}
\hat{\cal U}[\omega]
&=
\exp[-i\hat{\cal G}[\omega]].
\end{align}
There are therefore two related objects that should not be confused,
\begin{align}
U(\mathbf x)
&=
\exp[ig\omega^a(\mathbf x)T^a],
\qquad\mbox{and}\qquad
\hat{\cal U}[\omega] = \exp[-i\hat{\cal G}[\omega]].
\end{align}
The first is an $SU(3)$-valued function of position describing the gauge
transformation of the fields, whereas the second is a unitary operator
acting on the quantum Hilbert space. The latter implements the former on
the field operators. For example,
\begin{align}
\hat{\cal U}[\omega]\,\hat A^i(\mathbf x)\,\hat{\cal U}^{-1}[\omega]
    &= \hat A^{iU}(\mathbf x),
\qquad
\hat{\cal U}[\omega]\, \hat \psi(\mathbf x)\, \hat{\cal U}^{-1}[\omega]
    = U(\mathbf x)\hat \psi(\mathbf x).
\end{align}
Thus exponentiation of the smeared Gauss-law generator produces on the
Hilbert space the same finite gauge transformation that is described
classically by the group-valued function $U(\mathbf x)$.

The significance of the physical-state condition now becomes
particularly transparent. From Eq.~(\ref{Eq:PhysicalStates}),
\begin{align}
\hat{\cal G}[\omega]\,|\Psi_{\rm phys}\rangle
&=
0
\end{align}
for every allowed $\omega^a(\mathbf x)$, and hence
\begin{align}
\hat{\cal U}[\omega]\,|\Psi_{\rm phys}\rangle
&=
|\Psi_{\rm phys}\rangle .
\end{align}
Physical states are therefore invariant under finite gauge transformations
that can be obtained by continuously exponentiating the infinitesimal
transformations generated by Gauss's law.

This last qualification is important. The Gauss-law constraint establishes
invariance under the component of the admissible gauge transformation group
that is continuously connected to the identity. It does not, by itself,
establish the existence or physical significance of additional disconnected
components. As discussed in Sec.~\ref{Sec:ThetaVacuum}, the conventional
distinction between small and large gauge transformations arises only after
imposing additional global conditions on the admissible gauge
transformations. In particular, if gauge transformations are required to
approach a fixed, direction-independent value at spatial infinity $U_\infty$,
$U(\mathbf x) \to U_\infty$ as $|\mathbf x|\to\infty$,
then the spatial $\mathbb R^3$ may be one-point compactified to $S^3$.
The admissible gauge transformations may consequently be regarded as maps
$U:S^3_{\rm space}\to SU(3)$,
which are classified by $\pi_3(SU(3)) = \mathbb Z$.
A transformation in the class of zero winding number can be continuously
deformed to the identity while preserving the imposed asymptotic condition
and is conventionally called a small gauge transformation. Such
transformations are generated continuously from the identity by the local
Gauss-law generators.
By contrast, a transformation having nonzero winding number cannot be
continuously deformed to the identity while remaining within this class of
admissible gauge transformations. Such a transformation
cannot be obtained by continuously exponentiating the
infinitesimal transformations generated by the local Gauss-law constraint
while remaining within the prescribed class of admissible transformations.
It is conventionally called a large gauge transformation.

The disconnected components are therefore components of the space of
admissible maps into $SU(3)$, not disconnected components of $SU(3)$
itself, which is a connected group. It is this additional global structure,
rather than the local Gauss-law constraint alone, that gives rise to the
conventional distinction between small and large gauge transformations and
provides the starting point for the Hamiltonian construction of the
$\theta$ vacuum. We examine this construction more closely in the following
subsection.

No gauge fixing has been imposed in the construction above and $A_0^a$
remains a Lagrange multiplier. The spatial gauge fields $\mathbf A^a$ and
their conjugate momenta therefore still contain gauge redundancy. An ideal
gauge fixing would eliminate this redundancy by selecting exactly one
representative from each gauge orbit, thereby providing a description in
which distinct canonical configurations correspond to distinct physical
configurations.

This question of gauge fixing is separate from the Dirac-Bergmann treatment
of constrained dynamics used above. In that treatment the primary and
secondary constraints are identified and their consistency under time
evolution is established. After quantization, Gauss's law is imposed as a
constraint on physical states,
\begin{align}
\hat G^a(\mathbf{x})|\Psi_{\rm phys}\rangle &= 0 .
\end{align}
None of these steps fixes the gauge or selects a unique representative from
each gauge orbit.

An alternative route, often used in canonical treatments of Yang-Mills
theory, is to begin by imposing temporal, or Weyl, gauge,
\begin{align}
A_0^a(\mathbf{x},t)&=0.
\end{align}
This removes the nondynamical temporal component but does not completely
fix the gauge. Time-independent gauge transformations remain, and Gauss's
law must still be imposed as a constraint on physical states. Thus temporal
gauge should not be confused with a complete reduction to the physical
degrees of freedom.

To obtain a reduced Hamiltonian description one must also fix the residual
spatial gauge freedom. For example, one may impose Coulomb gauge,
\begin{align}
\nabla\cdot \mathbf A^a = \partial_iA^{ai}&=0.
\end{align}
In the Dirac-Bergmann formulation this gauge condition must itself be
preserved under time evolution. Since
\begin{align}
\dot A^{ai}&=-E^{ai}-(D_iA_0)^a,
\end{align}
preservation of Coulomb gauge requires
\begin{align}
0
&=
\partial_i\dot A^{ai}
=
-\partial_iE^{ai}-\partial_i(D_iA_0)^a.
\end{align}
Thus, once Coulomb gauge has been imposed, $A_0^a$ is no longer arbitrary
but must be chosen so that
\begin{align}
\partial_i(D_iA_0)^a&=-\partial_iE^{ai}.
\end{align}
Introducing the Coulomb-gauge Faddeev-Popov operator
\begin{align}
{\cal M}^{ab}[A]&=-\partial_iD_i^{ab}[A],
\end{align}
this condition becomes
\begin{align}
{\cal M}^{ab}[A]A_0^b&=\partial_iE^{ai}.
\end{align}
Subject to appropriate boundary conditions and provided that ${\cal M}$
has no relevant zero modes, this equation determines $A_0^a$. Gauss's law
may then be solved for the longitudinal chromoelectric field, thereby
eliminating the corresponding redundant canonical momentum.

Alternatively, if one begins in temporal gauge $A_0^a=0$, Gauss's law
remains as a constraint and a subsequent spatial gauge condition fixes the
residual gauge freedom. Where both procedures are well defined they lead
to the same reduced gauge-fixed Hamiltonian structure. The inverse of
${\cal M}^{ab}[A]$ enters both the solution of Gauss's law and the resulting
nonlocal color-Coulomb interaction.

The further difficulty in a nonabelian theory is that a gauge condition
such as Coulomb gauge does not in general provide the ideal global gauge
fixing just described. Distinct gauge-equivalent configurations can satisfy
the same condition $\nabla\cdot\mathbf A^a=0$. These are Gribov copies,
and the associated Faddeev-Popov operator can develop zero modes at the
Gribov horizon. More generally, the Gribov-Singer obstruction prevents a
globally unique smooth gauge choice of the usual type over the full
nonabelian configuration space. Thus, although a gauge condition can
generally be imposed locally in configuration space, the construction of a
globally valid reduced Hamiltonian description in terms of a unique
representative of each gauge orbit is obstructed in the nonabelian theory.

For discussions and reviews of the Hamiltonian formulation of Yang-Mills
theory and QCD, including Coulomb-gauge quantization, the resolution of
Gauss's law, the nonabelian Coulomb interaction, and the associated
gauge-fixing issues, see for example
Refs.~\cite{Henneaux:1992,Christ:1980ku,Zwanziger:1997mk,
Heinzl:1995jn,Reinhardt:2018}.

\subsection{Large gauge transformations}
\label{Sec:LargeGauge}

The discussion in the previous subsection established that, after
quantization, the Gauss-law constraint becomes the operator condition
\begin{align}
\hat G^a(\mathbf{x})\,|\Psi_{\rm phys}\rangle &= 0 .
\end{align}
The corresponding smeared operator,
\begin{align}
\hat{\cal G}[\omega] &= \textstyle 
    \int\!d^3x\,\omega^a(\mathbf{x})\hat G^a(\mathbf{x}),
\end{align}
generates infinitesimal gauge transformations of the canonical field
operators on a fixed time slice. Exponentiation gives the Hilbert-space
operator
\begin{align}
\hat{\cal U}[\omega] &= \exp[-i\hat{\cal G}[\omega]],
\end{align}
which implements the corresponding finite gauge transformation
$U(\mathbf{x})$ on the field operators. The physical-state condition
therefore implies
\begin{align}
\hat{\cal U}[\omega]\,|\Psi_{\rm phys}\rangle &= |\Psi_{\rm phys}\rangle
\end{align}
for transformations continuously connected to the identity within the
admissible gauge transformation group. These form its identity component
and are conventionally referred to as small gauge transformations.

The important point is that the local Gauss-law constraint determines the
action of this identity component on physical states, but says nothing by
itself about whether the admissible gauge transformation group has
additional disconnected components. The existence and classification of
such components depend on the global definition of the admissible gauge
transformations.

As discussed in Sec.~\ref{Sec:ThetaVacuum}, the conventional construction
restricts sufficiently smooth gauge transformations on a spatial
hypersurface by imposing
\begin{align}
U(\mathbf{x}) &\to U_\infty,
\qquad
|\mathbf{x}| \to \infty .
\end{align}
With this additional asymptotic condition, $\mathbb{R}^3$ may be
one-point compactified for the purpose of classifying the admissible gauge
transformations, and these may be regarded as based maps
$U: S^3 \to SU(3)$. Since $\pi_3(SU(3)) = \mathbb{Z}$,
the admissible gauge transformation group then has disconnected components
labelled by an integer winding number. The component containing the
identity has winding number zero, while transformations belonging to the
remaining components are conventionally referred to as large gauge
transformations.

It is important to distinguish this global statement from the local
canonical constraint. A large gauge transformation $U_m(\mathbf{x})$ of
nonzero winding number cannot be connected to the identity by a continuous
path lying entirely within the assumed class of admissible gauge
transformations. Its action on the Hilbert space therefore cannot be
obtained by continuously integrating the infinitesimal transformations
generated by the Gauss-law operators from the identity component. In
particular, the local physical-state condition does not by itself require
\begin{align}
\hat{\cal U}_m\,|\Psi_{\rm phys}\rangle &= |\Psi_{\rm phys}\rangle ,
\end{align}
where $\hat{\cal U}_m$ denotes the Hilbert-space operator implementing the
classical transformation $U_m(\mathbf{x})$.
Related questions concerning the implementation of large gauge
transformations and the definition of the physical Hilbert space have
recently been reconsidered in the canonical formulation
\citep{Ai:2024vfa}.

Within the conventional construction, one assumes that the disconnected
components are represented on the physical Hilbert space and allows
physical states to transform according to a one-dimensional unitary
representation of the group of connected components. Since the winding 
numbers add under composition, the corresponding Hilbert-space operators
satisfy
\begin{align}
\hat{\cal U}_m\hat{\cal U}_n &= \hat{\cal U}_{m+n}.
\end{align}
A physical state may therefore be chosen to satisfy
\begin{align}\label{Eq:LargeGaugeThetaPhase}
\hat{\cal U}_m\,|\Psi_\theta\rangle &= e^{-im\theta}\,|\Psi_\theta\rangle ,
\end{align}
where $\theta$ is defined modulo $2\pi$ in the usual way.
The phase factors form a one-dimensional unitary representation of
$\mathbb{Z}$. Within the conventional Hamiltonian construction, $\theta$
may therefore be viewed as labelling the possible representations of the
disconnected components of the assumed gauge transformation group on the
physical Hilbert space.
Further discussion of large gauge transformations
can be found in Appendix \ref{App:ChernSimons}.

The corresponding construction in terms of the classical Chern-Simons
functional $N_{\rm CS}[A]$, the states $|n\rangle$, instanton transitions,
and the superposition
\begin{align}
|\theta\rangle &= \textstyle \sum_{n=-\infty}^{\infty} e^{in\theta}|n\rangle
\end{align}
was described in Sec.~\ref{Sec:ThetaVacuum} and need not be repeated here.
The additional insight supplied by the canonical analysis is that the
$\theta$ construction contains global information that is not implied by
the local Gauss-law constraint. The condition
\begin{align}
\hat G^a(\mathbf{x})\,|\Psi_{\rm phys}\rangle &= 0
\end{align}
fixes the behavior of physical states under the identity component of the
admissible gauge transformation group. Extending the construction to
disconnected components requires additional assumptions concerning the
global class of admissible gauge transformations and their behavior at
spatial infinity.

Thus, if the conventional asymptotic and regularity conditions are adopted,
the admissible gauge transformations separate into disconnected components
classified by $\pi_3(SU(3))=\mathbb{Z}$, and their one-dimensional
representations lead to the parameter $\theta$. This construction depends
both on the global definition of the admissible gauge transformation group
and on sufficient smoothness for the winding number and associated
Chern-Simons construction to be well defined and to possess their usual
integer-valued topological interpretation. These are additional assumptions
beyond the local canonical structure of Yang-Mills theory and do not follow
from the Gauss-law constraint itself.

\subsection{Comparison with the Euclidean picture}
\label{Sec:HamiltonianEuclidean}

The Hamiltonian and Euclidean formulations provide complementary
descriptions of quantum Yang-Mills theory. The Hamiltonian formulation
emphasizes the physical Hilbert space, the canonical constraints and their
implementation on physical states, whereas the Euclidean functional
integral expresses correlation functions as averages over gauge-field
configurations. Subject to the usual assumptions required for analytic
continuation and reconstruction, the two formulations are expected to
describe the same quantum theory. In the conventional treatment of the
strong $CP$ problem, this correspondence relates the $\theta$ dependence
arising from the action of large gauge transformations in the Hamiltonian
formulation to the $\theta$ dependence introduced through the Euclidean
topological charge.

Formally, the connection between Minkowski and Euclidean correlation
functions is obtained by analytic continuation of the time coordinate,
\begin{align}
t
&\to
-i\tau ,
\end{align}
under which the Minkowski functional-integral factor
is replaced by the Euclidean weight
\begin{align}
e^{iS_M} \to e^{-S_E} .
\end{align}
This formal continuation underlies the Euclidean functional-integral
description and, after regularization, the lattice formulation of gauge
theory. The precise nonperturbative relation between a Euclidean theory and
a corresponding Hilbert-space quantum theory requires additional
conditions and should not be identified merely with the formal replacement
$t\to -i\tau$.

For QCD with a $\theta$ term, the Euclidean functional integral is
conventionally written as
\begin{align}\label{Eq:ZEthetaAgain}
Z_E(\theta) &= \textstyle \int\!\! {\cal D}A_\mu{\cal D}\bar\psi {\cal D}\psi\,
    \exp\left[-S_E+i\theta Q_{\rm gluon}\right],
\end{align}
where we recall that
\begin{align}
Q_{\rm gluon} &= [g^2/(32\pi^2)]\textstyle \int \! d^4x_E\,
    F_{\mu\nu}^a\tilde F_{\mu\nu}^a .
\end{align}
Under the conventional smoothness and boundary assumptions discussed in
Sec.~\ref{Sec:global_top}, $Q_{\rm gluon}$ may be assigned an integer
topological interpretation, $Q_{\rm gluon}\equiv Q\in\mathbb Z$, and,
when the configuration space is correspondingly classified into such
topological sectors, the functional integral may formally be organized as
\begin{align}
Z_E(\theta) &= \textstyle \sum_{Q\in\mathbb Z} e^{i\theta Q}Z_Q .
\end{align}
Here $Z_Q$ denotes the contribution associated with configurations of
topological charge $Q$. This is the Euclidean counterpart of the
$\theta$ dependence arising in the Hamiltonian construction, but the
integer labels appearing in the two descriptions should not be identified
too literally.

In the Hamiltonian picture, the conventional integer structure arises
from the winding numbers of large gauge transformations on individual
spatial hypersurfaces and from the corresponding integer shifts of the
Chern-Simons functional. By contrast, the Euclidean topological charge is
a four-dimensional quantity. For a sufficiently smooth finite-action
configuration interpolating between asymptotic configurations with
Chern-Simons numbers $N_{\rm CS}(\tau_i)$ and $N_{\rm CS}(\tau_f)$, the
standard relation is
\begin{align}
Q
&=
N_{\rm CS}(\tau_f)-N_{\rm CS}(\tau_i),
\end{align}
subject to the appropriate boundary and regularity conditions. Thus an
instanton of charge $Q=1$ is conventionally interpreted in the Hamiltonian
language as an interpolation between configurations whose Chern-Simons
numbers differ by one. The four-dimensional topological charge is therefore
associated with the change in the spatial Chern-Simons quantity rather than
with its value on a single spatial hypersurface.

This distinction is also important for gauge transformations. The local
four-dimensional Euclidean density
\begin{align}
q_E(x) &= [g^2/(32\pi^2)] F_{\mu\nu}^a\tilde F_{\mu\nu}^a
\end{align}
is gauge invariant, and consequently its spacetime integral $Q$ is
unchanged by any globally defined gauge transformation for which the
integral is well defined. By contrast, the three-dimensional Chern-Simons
functional is not gauge invariant and, under the conventional assumptions,
shifts by an integer under a large spatial gauge transformation. There is
therefore no contradiction between the gauge invariance of the
four-dimensional quantity $Q$ and the integer shift of the Chern-Simons
functional on a spatial hypersurface. They are distinct objects playing
different roles in the Euclidean and Hamiltonian constructions.

The conventional correspondence may therefore be summarized more
accurately as
\begin{align}
\begin{array}{ccc}
\mbox{Hamiltonian}
&
\Longleftrightarrow
&
\mbox{Euclidean}
\\[2mm]
\mbox{spatial Chern-Simons structure}
&
\Longleftrightarrow
&
\mbox{four-dimensional topological charge}
\\
\mbox{transition with }\Delta N_{\rm CS}
&
\Longleftrightarrow
&
Q=\Delta N_{\rm CS}
\\
\hat{\cal U}_m|\Psi_\theta\rangle
=
e^{-im\theta}|\Psi_\theta\rangle
&
\Longleftrightarrow
&
e^{i\theta Q}\mbox{ weighting}.
\end{array}
\end{align}
The correspondence is therefore between two related global constructions,
not an identification of the three-dimensional Chern-Simons label with the
four-dimensional topological charge.

It is useful to separate the local ingredients of this correspondence from
the additional global structure. Both formulations are based on the same
local Yang-Mills dynamics, local gauge symmetry and local composite
operators. In particular, the local operator $q(x)$ and the anomalous axial
Ward identity do not depend upon introducing a decomposition into
topological sectors. The additional structure enters when one classifies
the admissible spatial gauge transformations into disconnected homotopy
classes in the Hamiltonian formulation and when, under the corresponding
regularity and global assumptions, one assigns an integer-valued
topological charge to classes of four-dimensional configurations in the
Euclidean formulation.

Although these constructions are closely related, their assumptions should
also be distinguished. In the Hamiltonian construction, the classification
of large gauge transformations requires a specified class of sufficiently
smooth spatial gauge transformations together with asymptotic conditions
that permit their interpretation as maps from $S^3$ into $SU(3)$. The
Chern-Simons functional must likewise be sufficiently well defined for its
transformation under such maps to have the usual integer shift. In the
Euclidean construction, the integer interpretation of $Q$ requires
corresponding smoothness assumptions, finite-action or appropriate boundary 
conditions, and the global structure needed to assign a topological class
to the four-dimensional gauge configuration. These assumptions are related,
but they are not identical and should not be conflated.

Under the conventional assumptions, the Hamiltonian $\theta$ construction
and the Euclidean factor $e^{i\theta Q}$ provide mutually consistent
descriptions of the same $\theta$ dependence. The purpose of emphasizing
the assumptions entering this correspondence is not to question either
formalism, but to distinguish consequences of the local quantum field
theory from conclusions that require additional global and smoothness
conditions. This distinction will be important below when considering the
extent to which the conventional smooth-field topological construction can
be applied to the full nonperturbative quantum theory.


\section{Conceptual foundations}
\label{Sec:foundations}

The Standard Model is formulated as a local relativistic quantum field
theory. Locality and causality are among its most fundamental principles,
underlying both the construction of the Lagrangian and the interpretation of
physical observables. These principles are independent of perturbation
theory and remain central to nonperturbative formulations of quantum field
theory. They therefore provide a useful starting point when examining the
assumptions underlying the strong $CP$ problem.

\subsection{Locality and causality}
\label{Sec:Locality}

The principle of locality states that the dynamics are determined by local
interactions. The Lagrangian density is constructed from fields and a finite
number of their derivatives evaluated at the same spacetime point,
\begin{align}
{\cal L}(x) &= {\cal L}\left(\phi(x),\partial_\mu\phi(x),\ldots\right),
\end{align}
so that the action
\begin{align}
S &= \textstyle \int\!d^4x\,{\cal L}(x)
\end{align}
is obtained by integrating local contributions over spacetime. In gauge
theories the fields enter through combinations consistent with local gauge
invariance, such as covariant derivatives and field strengths, but the
principle remains unchanged: the fundamental interactions are local in
spacetime.

Relativistic causality requires that measurements performed in spacelike
separated regions cannot influence one another. In the operator formulation,
local gauge-invariant bosonic observables satisfy microcausality,
\begin{align}
[\hat O_1(x),\hat O_2(y)]
&=
0,
\qquad
(x-y)^2<0 .
\end{align}
More generally, local quantum fields satisfy graded commutation relations
at spacelike separation, which is to say that bosonic fields commute
and fermionic fields anticommute at spacelike separation.
For physical observables, microcausality ensures the
compatibility of measurements in spacelike separated regions and prevents
the transmission of physical information outside the light cone.

Physical information is encoded in expectation values and correlation
functions of gauge-invariant observables. Of particular importance are
time-ordered correlation functions of the form
\begin{align}
G_n(x_1,\ldots,x_n)
&=
\langle
0|
T\{
\hat O_1(x_1)
\cdots
\hat O_n(x_n)
\}
|0
\rangle .
\end{align}
Within the Euclidean formulation, the corresponding Euclidean correlation
functions are obtained from the functional integral discussed in previous
sections and, subject to the usual reconstruction assumptions, are related
to the correlation functions of the Minkowski theory.

Closely associated with the physical interpretation of locality is the
cluster decomposition property. Consider two gauge-invariant local
operators $\hat O_A$ and $\hat O_B$ whose supports are separated by a large
spacelike distance. In a vacuum satisfying cluster decomposition, their
connected correlation function vanishes in the limit of infinite spacelike
separation,
\begin{align}\label{Eq:Cluster}
\lim_{(x-y)^2\rightarrow-\infty}
\left[
\langle\hat O_A(x)\hat O_B(y)\rangle
-
\langle\hat O_A(x)\rangle
\langle\hat O_B(y)\rangle
\right]
&=
0 .
\end{align}
Equivalently,
\begin{align}
\lim_{(x-y)^2\rightarrow-\infty}
\langle
\hat O_A(x)
\hat O_B(y)
\rangle
&=
\langle
\hat O_A(x)
\rangle
\langle
\hat O_B(y)
\rangle .
\end{align}
Thus correlations between sufficiently distant local observables disappear
in this limit. Cluster decomposition is an important physical property of
the vacuum and plays a central role in relating local quantum field theory
to the statistical independence of widely separated experiments.

Locality and clustering also underlie the emergence of macroscopic
behavior discussed in Sec.~\ref{Sec:ClassicalLimit}. Macroscopic observables
are constructed by averaging local field operators over regions large
compared with the relevant microscopic correlation lengths. When connected
correlations decay sufficiently rapidly with separation, contributions from
widely separated subregions become approximately independent and relative
fluctuations of suitable averaged observables are suppressed. This provides
an important connection between the local quantum dynamics and the
classical or semiclassical behavior of macroscopic observables.

It is important to distinguish locality from global properties of spacetime,
gauge transformations, or gauge-field configuration space. Locality governs
the construction of the Lagrangian, the algebra of local observables, local
operator identities, and the causal propagation of physical information.
Global topological classifications, by contrast, depend on properties of a
configuration as a whole and generally require additional assumptions
concerning boundary conditions, smoothness, or the global domain on which
the fields are defined. The distinction does not imply that topology cannot
be represented locally by a density. For example,
\begin{align}
q(x) &= [g^2/(32\pi^2)]
F_{\mu\nu}^a(x)\tilde F^{a\mu\nu}(x)
\end{align}
is a local gauge-invariant operator, whereas interpreting
\begin{align}
Q_{\rm gluon} &= \textstyle \int\!d^4x\,q(x)
\end{align}
as an integer-valued global topological charge requires additional global
and smoothness assumptions of the kind discussed in previous sections.
Keeping these logically distinct ingredients separate will be important in
the remainder of this review.

\subsection{Global structure}
\label{Sec:GlobalStructure}

The preceding subsection reviewed locality and causality and their relation
to local observables and correlation functions. A quantum field theory may
also depend on global properties of the spacetime and of the fields defined
upon it. Such properties are not determined by the fields in an arbitrarily
small neighborhood of a spacetime point, but depend on their behavior over
extended regions or on the global geometry and topology of the space on
which the theory is defined. It is therefore important to distinguish the
local dynamical structure of a quantum field theory from the additional
global structure entering its formulation.

One source of global structure is the choice of boundary or asymptotic
conditions. For example, integrations by parts in varying the action
\begin{align}
S &= \textstyle \int_M\!d^4x\,{\cal L},
\end{align}
produce surface terms of the general form
\begin{align}
\textstyle \int_M\!d^4x\,\partial_\mu J^\mu
&= \textstyle \int_{\partial M}\!d\Sigma_\mu\,J^\mu .
\end{align}
The local Euler-Lagrange equations may be derived by considering variations
of compact support within the interior of the spacetime manifold $M$, in
which case the surface terms arising from the variation vanish without any
assumption about the behavior of the fields at a distant boundary
$\partial M$ or at infinity. This is the standard justification for
integrations by parts in deriving the local equations of motion in classical
mechanics and quantum field theory. The important point is that this relies
on the compact support of the variations and does not require the imposition
of boundary conditions on the fields themselves. Boundary or asymptotic
conditions constitute additional global information and may subsequently be
required to specify completely the physical or mathematical problem under
consideration.

This distinction is worth emphasizing. In deriving the local equations of
motion, no physical condition on the fields at infinity is required; the
surface contribution can be eliminated simply by restricting the variations
to have compact support. Assumptions concerning the behavior of the fields
at infinity therefore constitute additional global information. They may
of course be required to define particular global solutions or physical
problems, but conclusions that depend upon them do not follow from the local
field equations alone. This distinction becomes particularly important when
asymptotic conditions are used not merely to select solutions, but to
establish a global topological classification of the gauge fields or gauge
transformations.

Quantum field theory is commonly formulated on flat Minkowski spacetime,
\begin{align}
M
&=
\mathbb R^{1,3},
\end{align}
which has no boundary as a spacetime manifold. Nevertheless, asymptotic
conditions on fields and states are frequently imposed at spatial or
temporal infinity in order to define quantities such as conserved charges
and scattering states. These conditions concern the global formulation of
the theory rather than the local equations of motion. Their physical
significance must therefore be considered separately from that of the local
dynamics.

Global structure may also arise from the topology of the underlying
spacetime manifold $M$. Examples other than flat Minkowski spacetime 
include
\begin{align}
M &= S^4,\; T^4,\; S^1\times\mathbb R^3 .
\end{align}
In finite-temperature field
theory, for example, the Euclidean time coordinate is periodic,
\begin{align}
\tau &\sim \tau+\beta,
\qquad \beta = 1/T,
\end{align}
so that the Euclidean spacetime has topology
\begin{align}
S^1\times\mathbb R^3.
\end{align}
Such global properties can affect the allowed field configurations, the
spectrum and correlation functions, and the possible existence of
topological invariants.

Gauge theories possess additional forms of global structure. 
As discussed in Sec.~\ref{Sec:LargeGauge}, if one restricts 
the admissible gauge transformations on a spatial hypersurface 
by requiring $U(\mathbf{x}) \to U_\infty$ at spatial infinity,
then $\mathbb R^3$ may be one-point compactified for the purpose of
classifying these transformations. Sufficiently smooth gauge
transformations may then be regarded as based maps
$U: S^3 \to SU(3)$, whose homotopy classes are classified by
$\pi_3(SU(3)) = \mathbb Z$.
This is a statement about the global structure of the 
admissible gauge transformation group and gives rise 
to the conventional distinction between
small and large gauge transformations. 

Quantum field theory may also be formulated on curved 
spacetime, where the background geometry can influence 
the spectrum of quantum states, the definition of the 
vacuum and the propagation of quantum fields. Familiar
examples include particle production in expanding 
spacetimes and Hawking radiation in black-hole backgrounds. 
These phenomena illustrate more
generally that the same local principles of quantum 
field theory can be implemented in settings having 
different global geometrical structures,
with correspondingly different physical consequences.

The distinction becomes particularly significant when quantum 
field theory is considered in a cosmological setting. 
The actual spacetime on which the Standard Model is 
realized is known not to be globally flat Minkowski
spacetime and is more appropriately described, on 
sufficiently large scales, by an evolving
cosmological geometry. For example, a homogeneous and 
isotropic universe is described by the 
Friedmann-Lema\^itre-Robertson-Walker metric
\begin{align}
ds^2 &= dt^2 - a^2(t)
    \left[\frac{dr^2}{1-kr^2} + r^2d\Omega^2 \right],
\end{align}
where $a(t)$ is the cosmological scale factor and $k=0,\pm1
$ determines the spatial curvature. The local QCD dynamics 
may be formulated on such a background without requiring 
that gauge-field configurations be classified by their 
behavior over an entire spatial hypersurface or throughout the
complete history of the universe. A global topological
classification of this kind therefore introduces information
extending far beyond that required to specify the local 
quantum field theory. This does not imply
that such global classifications are inconsistent or cannot 
have physical consequences, but it emphasizes that their
applicability to the physical QCD vacuum is an additional
assumption rather than a consequence of local
QCD dynamics.

The familiar integer quantization of the gluonic topological 
charge also depends on the global spacetime setting. It is not
restricted to flat Euclidean spacetime. On a compact oriented
Euclidean four-manifold without boundary, a sufficiently smooth
connection on a principal $SU(3)$ bundle defines
\begin{align}
Q_{\rm gluon} &= [g^2/(32\pi^2)]
    \textstyle \int_M\!d^4x\,\sqrt{g}\,
    F_{\mu\nu}^a\tilde F^{a\mu\nu}
    \in \mathbb Z ,
\end{align}
where, with the usual normalization and up to 
convention-dependent signs, $Q_{\rm gluon}$ is the second 
Chern number of the corresponding gauge bundle. Thus 
spacetime curvature does not by itself invalidate the 
integer-valued topological classification. Nor is Euclidean
rather than Lorentzian signature by itself the essential issue,
since the Chern class is a topological property of the gauge bundle.
On noncompact Euclidean $\mathbb R^4$, the corresponding result instead 
requires suitable smoothness, finite-action and asymptotic
conditions. The physical spacetime of an evolving FLRW universe,
however, does not in general possess the compact, boundaryless
global structure assumed above, and there is no general reason for
the integral of $F\tilde F$ over its complete spacetime history
to define an integer. Establishing such a global
classification would require additional assumptions concerning 
the topology of the spacetime, the gauge fields and their
smoothness, and their behavior at the relevant spatial and 
temporal boundaries or asymptotic regions.

These examples illustrate that global structure can enter 
quantum field theory in several logically distinct ways: 
through boundary and asymptotic conditions, the topology of
spacetime, the topology of the admissible gauge 
transformation group, or the global structure of the gauge 
bundle. Such ingredients can have genuine physical 
consequences, but their relevance must be established for 
the physical system under consideration. In particular, 
a classification requiring information about a field
configuration over an entire spatial hypersurface, or over 
an entire four-dimensional spacetime, represents additional 
global information that is not contained in the local 
dynamics. It is therefore important to identify which 
global assumptions are required for a particular conclusion
and whether those assumptions are physically required by QCD 
or have been introduced as part of a particular global 
formulation of the theory.

This distinction is particularly relevant to the strong $CP$
problem. The local structure of QCD determines the local 
dynamics, the gauge-invariant operator 
\begin{align}
q(x) &= [g^2/(32\pi^2)] F_{\mu\nu}^a(x)\tilde F^{a\mu\nu}(x),
\end{align}
and its appearance in the anomalous axial Ward identity. The
conventional topological interpretation of
\begin{align}
Q &= \textstyle \int\!d^4x\,q(x),
\end{align}
however, requires additional global and smoothness assumptions
concerning the gauge fields, their asymptotic behavior and the
space on which they are defined. Similarly, the Hamiltonian
construction of large gauge transformations requires additional
assumptions concerning smoothness and the admissible spatial 
gauge transformations. The question is therefore not whether
global structure can have physical consequences in quantum 
field theory, which it clearly can, but which global 
structures are required by QCD observables and which arise 
only after additional assumptions have been imposed.

\subsection{Open mathematical questions}
\label{Sec:OpenMathematicalQuestions}

Quantum chromodynamics is one of the most successful physical theories ever
constructed. Its predictions have been confirmed over many orders of
magnitude in energy and distance, and lattice gauge theory has provided
impressive quantitative agreement with a wide range of nonperturbative
observables. Nevertheless, the mathematical foundations of four-dimensional
nonabelian quantum field theory remain incomplete. This should not be
regarded as a weakness of QCD itself, but rather as reflecting the difficulty
of constructing interacting quantum field theories rigorously in four
spacetime dimensions.

The Euclidean formulation of QCD is formally represented by the generating
functional
\begin{align}
Z_E &= \textstyle \int \!{\cal D}A_\mu\, {\cal D}\bar\psi\,
    {\cal D}\psi\; e^{-S_E[A,\psi,\bar\psi]},
\end{align}
where the Euclidean action is
\begin{align}
S_E &= \textstyle \int\!d^4x_E\, [ \frac12
    \mathrm{tr}(F_{\mu\nu}F_{\mu\nu}) + \bar\psi
    (\slashed D_E+m)\psi ].
\end{align}
Although expressions of this kind provide the formal starting point for
continuum functional-integral treatments of QCD, the functional measure
\begin{align}
{\cal D}A_\mu\,
{\cal D}\bar\psi\,
{\cal D}\psi
\end{align}
has not been constructed as a mathematically rigorous continuum measure
defining four-dimensional QCD. In practical nonperturbative calculations
the functional integral is instead given a regulated meaning, most
successfully through lattice gauge theory, as discussed in
Sec.~\ref{Sec:Lattice}.

After formally integrating over the Grassmann-valued quark fields one
obtains schematically
\begin{align}
Z_E &= \textstyle \int\! {\cal D}A_\mu\, e^{-S_{\rm YM}[A]}\,
    \det(\slashed D_E+m),
\end{align}
where
\begin{align}
S_{\rm YM} &= \textstyle \frac12 \int\!d^4x_E\, 
    \mathrm{tr}(F_{\mu\nu}F_{\mu\nu})
\end{align}
is the Euclidean Yang-Mills action. The corresponding continuum functional
integral nevertheless remains a formal expression whose complete
mathematical construction is still an open problem.

Constructive quantum field theory seeks to place such expressions on a firm
mathematical foundation by constructing the relevant measures and
demonstrating the existence of the corresponding interacting quantum
theories. In the Euclidean approach, one seeks Schwinger functions satisfying
the Osterwalder-Schrader conditions, including Euclidean covariance,
reflection positivity, symmetry and appropriate regularity properties,
from which a corresponding relativistic quantum field theory satisfying the
Wightman axioms can be reconstructed. Schematically,
\begin{align}
\{\mbox{Osterwalder-Schrader conditions}\}
\quad
\Longrightarrow
\quad
\{\mbox{Wightman QFT}\}.
\end{align}
Constructive field theory has achieved important rigorous results in lower
dimensions. By contrast, no comparable complete construction is presently
known for four-dimensional nonabelian Yang-Mills theory.

A related question concerns the space of gauge-field configurations entering
the continuum functional integral. If ${\cal A}$ denotes the space of gauge
connections and ${\cal G}$ the admissible gauge transformation group, the
physical configuration space is formally associated with the orbit space
\begin{align}
{\cal A}/{\cal G}.
\end{align}
As discussed in detail in Sec.~\ref{Sec:Lattice}, the full nonperturbative
functional integral should not be understood as an ordinary integral over
smooth classical gauge fields. Continuum quantum fields are naturally
distributional rather than ordinary functions, as is well established for
free quantum fields, and asymptotic freedom provides strong reason to expect
analogous short-distance behavior for the QCD gauge field. This expectation
is also consistent with the increasingly rough gauge-field configurations
encountered as the lattice continuum limit is approached. It is therefore
physically natural to expect that any continuum QCD functional measure, if
constructed, would be supported on gauge fields with distributional
short-distance behavior rather than on the sufficiently smooth
configurations for which the usual classical topological constructions are
most directly defined. This is a physical expectation rather than a
presently established mathematical property of a rigorously constructed
four-dimensional QCD measure.

One of the central open problems is the rigorous existence of
four-dimensional quantum Yang-Mills theory itself. The Clay Mathematics
Institute has identified the problem of establishing the existence of
four-dimensional quantum Yang-Mills theory with a nonzero mass gap as one
of the Millennium Prize Problems. This concerns pure Yang-Mills theory
rather than full QCD with dynamical quarks, but it illustrates directly the
unresolved mathematical foundations of four-dimensional nonabelian gauge
theory. The problem includes establishing a mathematically well-defined
quantum theory with
\begin{align}
m_{\rm gap}
&>
0
\end{align}
and the properties required of a consistent relativistic quantum field
theory. Despite overwhelming physical evidence for confinement and a mass
gap in the corresponding gauge dynamics, no complete mathematical proof is
presently known.

These mathematical questions should be distinguished carefully from the
physical validity of QCD. The remarkable phenomenological success of the
theory and the quantitative agreement between lattice calculations and
experiment provide compelling evidence that QCD correctly describes the
strong interaction. At the same time, the absence of a completely rigorous
continuum construction means that care is required when assigning precise
mathematical properties to the continuum functional measure and to the
individual gauge-field configurations that formally constitute its support.
This is particularly relevant to global topological classifications. As
discussed in Sec.~\ref{Sec:Lattice}, the standard classical constructions
are most directly formulated for sufficiently smooth gauge fields, whereas
the gauge fields relevant to the continuum quantum functional integral are
expected to have distributional short-distance behavior. Whether and how
the corresponding global topological classification extends to the full
continuum quantum measure is therefore a separate mathematical question and
should not simply be inferred from its validity for sufficiently smooth
classical configurations.

\section{Perspectives on the strong $CP$ problem}
\label{Sec:Perspectives}

The preceding sections have reviewed the local structure of QCD,
the Hamiltonian and Euclidean formulations of the theory, the
role of global topology, and the mathematical assumptions
underlying the conventional description of the $\theta$ vacuum.
We are now in a position to bring these elements together and
to distinguish clearly between those aspects of the strong $CP$
problem that follow directly from local quantum field theory and
those that rely upon additional assumptions concerning the global
structure of gauge-field configuration space. In particular, it
is important to distinguish the local, gauge-invariant topological
charge density from the global interpretation conventionally
assigned to its spacetime integral.

\subsection{Summary of the conventional viewpoint}
\label{Sec:convent_summary}

The conventional formulation of QCD begins with the most general
renormalizable gauge-invariant local Lagrangian density,
\begin{align}
\mathcal L &= \textstyle -\frac14 F_{\mu\nu}^aF^{a\mu\nu}
 + \bar\psi i\slashed D\psi
 -\bar\psi_L M\psi_R
 -\bar\psi_R M^\dagger\psi_L
 + \theta\,q(x),
\end{align}
where
\begin{align}
q(x) &= \textstyle [g^2/(32\pi^2)]
F_{\mu\nu}^a\tilde F^{a\mu\nu}
\end{align}
is the local topological charge density, as discussed in
Secs.~\ref{Sec:QCD_Lagrangian} and \ref{Sec:TopologicalDensity}.
The term $\theta q(x)$ is allowed by local gauge invariance and
renormalizability and violates $P$ and $CP$ for generic $\theta$, as
discussed in Sec.~\ref{Sec:ThetaDiscrete}. For sufficiently smooth gauge
fields, $q(x)$ may also be written locally as a total four-divergence,
\begin{align}
q(x) &= \partial_\mu K^\mu(x),
\end{align}
where $K^\mu$ is the Chern-Simons current.

The conventional nonperturbative interpretation introduces additional
global structure, as discussed in
Sec.~\ref{Sec:global_top}. For sufficiently smooth
gauge fields satisfying the appropriate finite-action and asymptotic
conditions, the Euclidean spacetime integral
\begin{align}
Q_{\rm gluon} &= \textstyle \int\!d^4x_E\,q_E(x)
\end{align}
is identified with an integer-valued topological charge. The Euclidean
functional integral may then be decomposed into sectors labelled by
$Q_{\rm gluon}\in\mathbb Z$, with the $\theta$ term assigning the phase
\begin{align}
e^{i\theta Q_{\rm gluon}}
\end{align}
to each sector. 
A further element of the conventional argument concerns cluster
decomposition. At fixed finite volume, restricting the functional integral
to a single topological charge imposes a global constraint and produces
long-range finite-volume correlations. Expectation values at fixed topology
differ from those in the corresponding $\theta$ vacuum by corrections that
vanish in the infinite-volume limit, typically as inverse powers of the
four-volume. Thus a fixed-topology formulation can reproduce the same local
infinite-volume physics under the usual assumptions. In the conventional
global construction, however, the $\theta$ vacua provide the natural
cluster-decomposing states before taking this limit and arise from summing
the integer-valued sectors with relative phases $e^{i\theta Q_{\rm gluon}}$.
This argument presupposes the preceding decomposition into well-defined
integer-valued topological sectors; cluster decomposition does not by itself
establish that such a decomposition exists.

Instantons, discussed in Sec.~\ref{Sec:Instantons}, provide
classical finite-action solutions with nonzero integer topological charge
and give a semiclassical realization of transitions between different
sectors.

The corresponding Hamiltonian construction was discussed in
Secs.~\ref{Sec:CanonicalQuantization} and \ref{Sec:LargeGauge}. It begins
by imposing Gauss's law on physical states. Gauge transformations
continuously connected to the identity act trivially on these states.
With the conventional asymptotic conditions imposed on the admissible
gauge transformations, however, the gauge transformation group possesses
disconnected components classified by
\begin{align}
\pi_3(SU(3))
&=
\mathbb Z.
\end{align}
These are the large gauge transformations. The classical Chern-Simons
functional changes by an integer under such transformations, leading to
the conventional description in terms of vacuum-type states $|n\rangle$
related by large gauge transformations. The physical vacuum is then taken
to be the coherent superposition
\begin{align}
|\theta\rangle &= \textstyle \sum_{n=-\infty}^{\infty}
    e^{in\theta}|n\rangle,
\end{align}
which transforms by a phase under a large gauge transformation. As
discussed in Sec.~\ref{Sec:HamiltonianEuclidean}, the Hamiltonian
$\theta$ vacuum and the weighting of integer topological sectors by
$e^{i\theta Q_{\rm gluon}}$ in the Euclidean functional integral are
conventionally regarded as equivalent descriptions of the same
nonperturbative vacuum structure.

The axial anomaly provides the connection between this gluonic
$\theta$ dependence and the quark mass matrix, as discussed in
Secs.~\ref{Sec:AxialAnomaly} and \ref{Sec:thetabar}. The divergence of
the flavor-singlet axial current contains the local operator $q(x)$,
\begin{align}
\partial_\mu J_5^\mu
&=
2i\bar\psi M\gamma_5\psi
+
2N_f q(x),
\end{align}
with the appropriate flavor-matrix interpretation of the mass term. An
anomalous singlet chiral transformation changes the phase of the quark
mass matrix while simultaneously shifting the coefficient of $q(x)$.
The invariant combination is therefore
\begin{align}
\bar\theta
&=
\theta+\arg\det M.
\end{align}
Equivalently, the $CP$-violating phase may be represented entirely by the
gluonic $\theta$ term, entirely by a pseudoscalar component of the quark
mass matrix, or distributed between the two; physical quantities depend
only on $\bar\theta$.

For nonzero $\bar\theta$, QCD therefore conventionally predicts
$CP$-violating observables, most notably a neutron electric dipole moment.
As discussed in Sec.~\ref{Sec:Experimental}, the experimental upper limit
on the neutron electric dipole moment implies approximately
\begin{align}
|\bar\theta|
&\lesssim
10^{-10}.
\end{align}
The absence of an apparent symmetry or other mechanism within conventional
QCD requiring this dimensionless parameter to be so small constitutes the
strong $CP$ problem.

Thus the conventional strong $CP$ construction contains several logically
distinct ingredients: the locally allowed operator $q(x)$ and the axial
anomaly; the global classification of sufficiently smooth gauge fields and
gauge transformations; the integer topological charge and Chern-Simons
number; the Hamiltonian $\theta$ vacuum and the corresponding Euclidean
topological-sector decomposition; the invariant combination
$\bar\theta=\theta+\arg\det M$; and the experimental requirement that
$|\bar\theta|$ be extremely small. The following discussion examines which
of these ingredients follow directly from the local structure of QCD and
which depend upon the additional global assumptions entering the
conventional construction.

\subsection{Alternative perspectives}
\label{Sec:AlternativePerspectives}

The conventional picture summarized above treats $\bar\theta$ as a physical
coupling of QCD and interprets its extremely small observed value as requiring
an explanation. A logically prior perspective is to ask which ingredients of
this construction follow from the established local structure and observables
of QCD, and which depend upon the additional smoothness, asymptotic and global
assumptions required for the conventional topological classification.

The local structure of QCD is unaffected by this distinction. The QCD
Lagrangian, local gauge invariance, the functional integral, the algebra of
local gauge-invariant operators and the axial anomaly are defined without
first imposing a global topological classification of gauge-field
configurations, as reviewed in Secs.~\ref{Sec:local_QCD} and
\ref{Sec:AxialAnomaly}. In particular, the local gauge-invariant operator
\begin{align}
q(x) &= \textstyle [g^2/(32\pi^2)] F_{\mu\nu}^a \tilde F^{a\mu\nu}
\end{align}
continues to exist independently of such a classification, as discussed in
Sec.~\ref{Sec:TopologicalDensity}. Its correlation functions therefore
remain well-defined physical quantities once the usual ultraviolet
regularization and renormalization of the composite operator are understood.

An important example is the topological susceptibility,
\begin{align}
\chi &= \textstyle \int\!d^4x_E\, \langle \hat q_E(x_E)\hat q_E(0) \rangle_{\theta=0},
\end{align}
discussed in Secs.~\ref{Sec:top_suscept}. It is
determined by correlation functions of the local operator $\hat q(x)$ and
does not by itself require the exact functional integral to be decomposed
configuration by configuration into sectors of integer-valued topological
charge. Equivalently, one may simply regard $\theta$ as a source 
parameter in the generating functional with no decomposition of the
gauge functional integral,
\begin{align}
Z_E(\theta) &= \textstyle \int\! 
    \mathcal{D}A_\mu \mathcal{D}\bar{\psi}  \mathcal{D}\psi\, 
    \exp \left[-S_E + i\theta\int\!d^4x_E\,q(x_E) \right],
\end{align}
and obtain moments and correlation functions of $q(x)$ by differentiation
with respect to this source. At this stage $\theta$ need not be interpreted
as a physical vacuum angle. In particular, a nonzero susceptibility at
$\theta=0$ does not by itself imply the existence of a nonzero, or even an
independently required, physical vacuum angle.

The axial anomaly leads to a closely related distinction. As discussed in
Secs.~\ref{Sec:AxialAnomaly} and \ref{Sec:thetabar}, an anomalous singlet
chiral transformation relates the coefficient of $q(x)$ to the flavor-singlet
phase of the quark mass matrix and identifies the invariant combination
\begin{align}
\bar\theta
&=
\theta+\arg\det M.
\end{align}
This equivalence follows from the local anomaly and the transformation of
the fermion functional measure; it does not require a global classification
of gauge fields into integer-valued topological sectors. However, the
equivalence between two possible parameterizations of an independent
$CP$-violating coupling does not by itself require that such an independent
coupling be present. If one is introduced, whether as a gluonic $\theta$
term, a flavor-singlet pseudoscalar quark mass term, or a combination of the
two, then the usual strong $CP$ phenomenology and the corresponding
fine-tuning problem follow.

Nor does abandoning the assumption that the exact functional integral is
globally decomposed into integer-valued sectors remove the established
semiclassical role of instantons. As discussed in Sec.~\ref{Sec:Instantons},
instantons are sufficiently smooth finite-action stationary points of the
Euclidean action and remain valid semiclassical saddle points. Expanding
about such configurations gives the familiar instanton physics, including
fermion zero modes and the 't Hooft effective interaction. The existence
and usefulness of these saddle points do not, however, by themselves
establish that every configuration contributing to the exact functional
integral must possess the same smoothness and admit the corresponding
integer-valued global classification.

The same observation applies to other established nonperturbative
consequences of the topological charge density. As reviewed in
Sec.~\ref{Sec:app_top_suscept}, the Witten-Veneziano relation depends upon
the anomalous singlet axial Ward identity, the nonzero pure Yang-Mills
topological susceptibility and the large-$N_c$ expansion. The
Leutwyler-Smilga relation follows from the anomalous Ward identity and
low-energy chiral dynamics. In both cases the relevant quantities can be
expressed in terms of correlation functions of the local operator $q(x)$
and their low-energy consequences. Their phenomenological success
therefore establishes the physical importance of local topological charge
density fluctuations, but does not by itself require a
configuration-by-configuration decomposition of the exact functional
integral into integer-valued topological sectors.

Lattice QCD leads to a similar distinction, as discussed in
Secs.~\ref{Sec:Lattice}-\ref{Sec:LatticePhysical}. Lattice calculations
provide compelling nonperturbative evidence for topological charge-density
fluctuations and successfully determine quantities such as the topological
susceptibility. Topological charge can also be characterized using smoothing
procedures such as gradient flow, for which the flowed gluonic charge
approaches the integer-valued continuum interpretation as lattice artifacts
are removed, or through exactly integer-valued index constructions involving
suitable lattice Dirac operators. These are powerful and well-defined
constructions. Nevertheless, the roughness of typical configurations as the
continuum limit is approached means that their existence does not by itself
establish that the original continuum functional integral is fundamentally
restricted to sufficiently smooth gauge fields for which
\begin{align}
Q_{\rm gluon}
&=
\int\!d^4x\,q(x)
\end{align}
is configuration by configuration an integer-valued topological invariant.

If the smoothness and global assumptions discussed in
Secs.~\ref{Sec:FiniteAction}-\ref{Sec:Top_sectors} are not imposed, the
spacetime integral of the local density $q(x)$ may still be considered,
but its interpretation as an integer-valued topological charge does not
follow from the usual smooth-field argument. Consequently, neither the
decomposition
\begin{align}
Z_E &= \textstyle \sum_{Q\in\mathbb Z} Z_Q
\end{align}
nor the associated relative weighting $e^{i\theta Q}$ is forced by the
local formulation alone. The conventional $2\pi$ periodicity of $\theta$
likewise follows from integer-valued $Q$ and is therefore part of this
additional global construction rather than a consequence of the local
operator $q(x)$ by itself.

The conventional cluster-decomposition argument given in \Sec{Sec:Locality}
does not alter this conclusion. Once a
decomposition into sectors of fixed integer topological charge has been
assumed, restricting the theory to a single such sector imposes a global
constraint and, at finite volume, leads to violations of cluster
decomposition. This provides an additional motivation for combining the
sectors into the conventional $\theta$ vacuum. The argument is conditional,
however, on the prior existence of the sector decomposition. If the
smoothness and global assumptions required to define configuration-by-
configuration integer topological charge are not imposed, there is no
fundamental restriction to a fixed-$Q$ sector and hence no corresponding
fixed-sector obstruction to cluster decomposition. Cluster decomposition
remains a requirement on correlation functions of local physical
observables, but it does not by itself require either an integer
topological-sector decomposition or a $\theta$ vacuum.

The corresponding conclusion in the Hamiltonian formulation was discussed
in Secs.~\ref{Sec:CanonicalQuantization} and \ref{Sec:LargeGauge}.
Gauss's law and the local canonical constraint structure remain unchanged.
The further decomposition of the admissible gauge transformation group into
disconnected components classified by
\begin{align}
\pi_3(SU(3)) &= \mathbb Z
\end{align}
requires additional assumptions concerning smoothness and the behavior of
the admissible gauge transformations at spatial infinity. Without those
assumptions, the identity-connected gauge transformations generated by the
Gauss-law constraint remain, but the additional large gauge transformations
and their representation on the physical Hilbert space do not follow from
the local canonical structure. The corresponding Hamiltonian $\theta$
vacuum is therefore likewise an additional global construction rather than
a consequence of Gauss's law alone.

It is useful to give a name to the resulting formulation. For the
purposes of this review, it is convenient to refer to
the theory obtained by retaining the established
local structure of QCD without additionally imposing the global
topological classification described above as 
\emph{minimal QCD}. Its Lagrangian may be written
with a real quark mass matrix and no independent $CP$-violating
strong interaction term,
\begin{align} {\cal L}_{\rm min}
&= \textstyle -\frac14 F_{\mu\nu}^aF^{a\mu\nu} + \bar\psi(i\slashed D-M)\psi ,
\end{align}
where the quark masses may be chosen real and positive. No physical
$\theta q(x)$ term is introduced and, equivalently through the anomalous
singlet chiral transformation, no independent flavor-singlet pseudoscalar
mass term is introduced. This does not amount to imposing
$\bar\theta=0$ on a pre-existing one-parameter family of physically
inequivalent theories. Rather, the additional physical parameter
$\bar\theta$ has not been introduced in the first place.

Minimal QCD still contains the local operator $q(x)$, the axial anomaly
and its associated Ward identities, and the full nonperturbative dynamics
of the gauge and quark fields. A source $\theta$ may therefore still be
coupled to $q(x)$ when required to generate its correlation functions,
without interpreting that source as an additional physical coupling of
the theory. What is omitted is specifically the additional global
classification and the associated independent strong-interaction
$CP$-violating parameter, not the local physics associated with
topological charge-density fluctuations.

On the analysis presented in this review, minimal QCD appears sufficient
to reproduce all presently established QCD phenomenology considered here.
The
$\theta$ term contributes neither to the classical equations of motion nor
to ordinary perturbation theory about the trivial vacuum, as discussed in
Secs.~\ref{Sec:theta_noclass} and \ref{Sec:theta_perturb}. The axial anomaly,
topological susceptibility, instanton saddle points, the 't Hooft
interaction, chiral symmetry breaking, the Witten-Veneziano and
Leutwyler-Smilga relations, lattice QCD and the established hadronic
phenomenology remain intact. In particular, the successful resolution of
the $U(1)_A$ problem and the description of the $\eta'$ mass require
nontrivial topological charge-density correlations and the anomalous Ward
identity, but do not by themselves establish that the exact functional
integral must possess a fundamental decomposition into integer-valued
topological sectors.

The essential issue is therefore not whether a term proportional to $q(x)$
is permitted in the QCD Lagrangian. It clearly is: such a term is compatible
with locality, Lorentz invariance, gauge invariance and renormalizability.
The logically prior question is whether established QCD phenomenology
requires its coefficient to be introduced as an independent physical
parameter. In the conventional formulation this physical interpretation is
supplied by the additional global construction: integer-valued topological
sectors admit relative weights $e^{i\theta Q}$, and $\theta$ consequently
acquires the interpretation of a vacuum angle. If the assumptions required
for this global classification are not imposed, this argument for promoting
$\theta$ to a physical coupling is absent.

The axial anomaly does not alter this conclusion. It establishes that, if
an independent flavor-singlet $CP$-violating parameter is introduced, a
gluonic $\theta$ term and a pseudoscalar phase in the quark mass matrix are
equivalent parameterizations, with the physical combination
\begin{align}
\bar\theta
&=
\theta+\arg\det M.
\end{align}
This equivalence does not, however, require either parameterization to be
present as an independent coupling. We are not aware of any established
QCD observable that requires an independent flavor-singlet pseudoscalar
mass term any more than it requires the additional global assumptions
leading to a physical vacuum angle.

Thus, in the absence of a physical requirement for either the global
topological construction or an independent flavor-singlet $CP$-violating
mass term, there is no corresponding requirement to regard $\theta$ as a
physical parameter of the QCD Lagrangian. The parameter $\theta$ may still
be introduced as a source in the generating functional in order to generate
correlation functions of the local operator $q(x)$ and quantities such as
the topological susceptibility. Its use for this purpose does not by itself
give it the status of a physical vacuum angle. In such a formulation the
QCD Lagrangian may be taken to have $\theta=0$, all established QCD
phenomenology discussed above remains intact, and the conventional strong
$CP$ fine-tuning problem does not arise.

\subsection{Local versus global assumptions}

One of the principal themes of this review has been the distinction between
local properties of QCD and additional assumptions and constructions
concerning its global structure.

The local structure includes the Yang-Mills Lagrangian density, local gauge
invariance, the axial anomaly, local composite operators and Ward identities.
The corresponding unrestricted functional-integral formulation and
correlation functions can be defined without first imposing the additional
global topological classification discussed below. These ingredients
determine the local dynamics of QCD and are common to the conventional and
minimal formulations discussed above.

By contrast, the conventional discussion of the $\theta$ vacuum introduces
additional global assumptions and constructions, including finite-action
and asymptotic conditions, compactification of spatial infinity,
classification of admissible gauge transformations by
$\pi_3(SU(3))$, decomposition of the functional integral into integer
topological sectors, and the identification of
$Q_{\rm gluon} = \int\!d^4x\,q(x)$
with a globally quantized topological charge for the relevant class of
gauge fields.

The conventional formulation incorporates this additional global structure,
whereas minimal QCD does not. As discussed in the preceding subsection,
all presently established QCD phenomenology considered in this review is
consistent with the common local and nonperturbative structure retained in
both formulations; no known observable has been identified that requires
the additional global classification. Thus, as far as presently established
QCD phenomenology is concerned, both formulations are sufficient. The
remaining question is whether the additional global structure is required
by some independent theoretical principle or by physics not captured by
the presently established observables. This question should be kept
distinct from the demonstrated usefulness and consistency of the
conventional construction within the settings in which it is employed.

\subsection{Current understanding}

At present there is no experimental evidence for strong $CP$ violation.
Existing measurements are entirely consistent with
$\bar\theta=0$
within experimental uncertainties.

Theoretical studies involving instantons, the axial anomaly, large-$N_c$
arguments, lattice gauge theory and effective field theory have provided
powerful and mutually consistent insights into the nonperturbative structure
of QCD. At the same time, the rigorous mathematical foundations of
four-dimensional nonabelian quantum field theory remain incomplete, and
several aspects of the continuum functional integral continue to be the
subject of active mathematical research.

From this perspective, the strong $CP$ problem may be viewed at two
complementary levels. Within the conventional framework, the central
question is why the physical parameter $\bar\theta$ is so extraordinarily
small, motivating mechanisms such as the Peccei-Quinn symmetry and the
associated axion. At a more foundational level, one may instead ask which
additional mathematical assumptions are required before $\theta$ acquires
the status of an independent physical coupling of the theory, and whether
any presently established QCD observable requires those assumptions.

The purpose of this review has not been to advocate one viewpoint over the
other, but rather to clarify the logical structure underlying the different
approaches. On the analysis presented in this review, the conventional
formulation and the minimal formulation described above are both consistent
with the presently established QCD phenomenology considered here, while
differing in the additional global structure that is assumed. By
distinguishing carefully between the local and nonperturbative physics common
to both and the additional global constructions entering the conventional
formulation, one can identify more precisely what is presently established,
what is additionally assumed, and which questions remain open.

\section{Open questions and outlook}
\label{Sec:Outlook}

The strong $CP$ problem has occupied a central position in theoretical
particle physics for almost five decades and has profoundly influenced the
development of modern nonperturbative quantum field theory. During this
period a powerful theoretical framework has emerged connecting the axial
anomaly, instantons, topological charge-density fluctuations, large-$N_c$
methods, lattice gauge theory, effective field theory and precision searches
for $CP$ violation. At the same time, no experimental evidence for strong
$CP$ violation has been found. Within the conventional formulation, the
physical parameter
\begin{align}
\bar\theta &= \theta+\arg\det M
\end{align}
is therefore constrained to be extraordinarily small, and explaining this
smallness constitutes the strong $CP$ problem.

One of the principal aims of this review has been to distinguish clearly
between those aspects of this framework that follow from the established
local and nonperturbative structure of QCD and those that involve additional
assumptions and constructions concerning the global organization of
gauge-field configuration space. This distinction does not challenge the
phenomenological success of QCD. Rather, it allows the logical ingredients
entering the conventional formulation of the strong $CP$ problem to be
identified more precisely.

The local structure of QCD is well established. The Yang-Mills Lagrangian
density, local gauge invariance, causal locality, the operator algebra, the axial
anomaly, Ward identities and the local topological charge density
\begin{align}
q(x) &= [g^2/(32\pi^2)] F_{\mu\nu}^a \tilde F^{a\mu\nu}
\end{align}
are common ingredients of the formulations considered in this review.
Likewise, there is compelling theoretical and numerical evidence for
nontrivial fluctuations and correlations of $q(x)$. Quantities such as the
topological susceptibility,
\begin{align}
\chi &= \textstyle \int\!d^4x_E\,
    \langle \hat q_E(x_E)\hat q_E(0) \rangle_E ,
\end{align}
are routinely determined in lattice simulations and play an important role
in our understanding of nonperturbative QCD, including the physics underlying
the Witten-Veneziano relation for the $\eta'$ mass. None of these established
results requires $\theta$ itself to be nonzero.

It is also useful to recall why possible physical effects of a constant
$\theta$ lie outside ordinary perturbation theory. For sufficiently smooth
gauge fields the local density $q(x)$ may be written as a total
four-divergence,
\begin{align}
q(x) &= \partial_\mu K^\mu(x),
\end{align}
where $K^\mu$ is the Chern-Simons current. A constant $\theta$ term therefore
does not modify the classical equations of motion under the usual local
variational assumptions. Ordinary perturbation theory about the trivial
vacuum likewise produces no dependence upon a constant $\theta$. The
possible physical significance of $\theta$ is consequently a
nonperturbative question. This fact should be distinguished from the
separate question of whether the full nonperturbative theory requires a
global classification of gauge-field configurations.

In the conventional formulation, additional global structure is introduced.
For sufficiently smooth fields satisfying appropriate finite-action and
asymptotic conditions, the spacetime integral
\begin{align}
Q_{\rm gluon} &= \textstyle \int\!d^4x_E\,q_E(x)
\end{align}
is identified with an integer-valued topological charge. Correspondingly,
admissible gauge transformations on a spatial hypersurface may, with the
usual asymptotic conditions, be classified by
\begin{align}
\pi_3(SU(3)) &= \mathbb Z.
\end{align}
These constructions lead respectively to the decomposition of the Euclidean
functional integral into integer-labelled topological sectors and to the
Hamiltonian $\theta$ vacuum. Within this framework $\theta$ is interpreted
as a physical vacuum angle and, together with the phase of the quark mass
matrix, gives the invariant parameter $\bar\theta$.

\begin{figure}
\centering
\includegraphics[width=0.95\columnwidth]{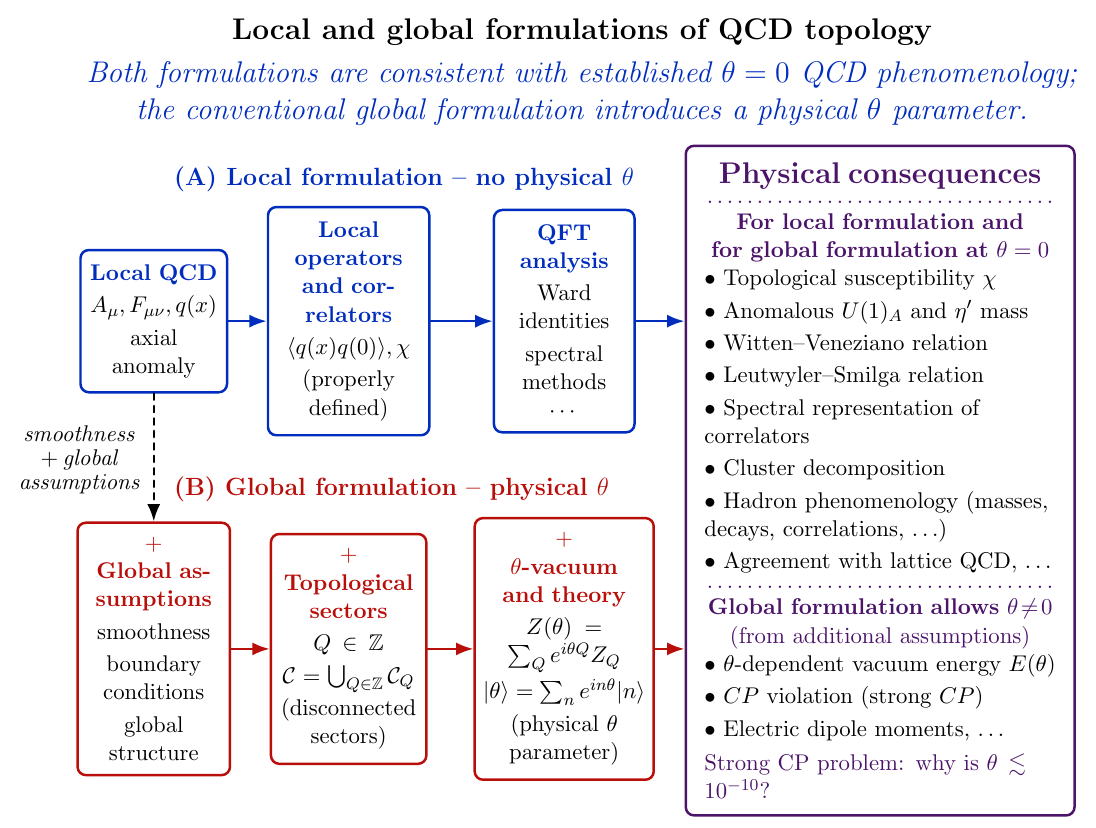}
\caption{Schematic comparison of the local and global formulations of
QCD topology discussed in this review. In the local formulation, shown
in the top row, the gauge fields, the local topological charge density $q(x)$,
the axial anomaly, and appropriately defined correlation functions are
treated using the usual local methods of quantum field theory, without
introducing a physical $\theta$ parameter. In the conventional global
formulation, shown in the bottom row, additional assumptions concerning 
smoothness, boundary conditions, and global structure permit the 
configuration space to be decomposed into integer-valued topological 
sectors and lead to the construction of the $\theta$ vacuum and a physical 
$\theta$ parameter. The $+$ signs emphasize that these global ingredients
supplement, rather than replace, the underlying local QCD structure.
Both routes reproduce the established $\theta=0$ consequences shown on
the right, including the topological susceptibility, anomalous
$U(1)_A$ physics, the Witten--Veneziano and Leutwyler--Smilga relations,
spectral results, cluster decomposition, lattice and hadron
phenomenology. The interpretation of the $\theta$-dependent
generating-functional density as the physical vacuum energy of a family
of theories, together with $CP$ violation, electric dipole moments, and
hence the conventional strong $CP$ problem, arises only when the physical
$\theta$ parameter of the global construction is introduced.}
\label{Fig:local_vs_global}
\end{figure}

The alternative perspective considered in this review asks what remains if
the smoothness, asymptotic and global assumptions required for this
classification are not imposed on the full quantum theory. The local
operator $q(x)$, its correlation functions, the axial anomaly and its Ward
identities, the topological susceptibility, instanton saddle points and the
associated semiclassical physics all remain. A parameter $\theta$ may also
be introduced as an auxiliary source coupled to $q(x)$ in order to generate
its correlation functions, without thereby interpreting $\theta$ as an
independent physical vacuum angle. In the minimal QCD formulation described
in Sec.~\ref{Sec:AlternativePerspectives}, neither a physical gluonic
$\theta$ term nor, equivalently, an independent flavor-singlet pseudoscalar
quark mass term is introduced.

As emphasized in Sec.~\ref{Sec:Perspectives}, the conventional and minimal
formulations are both consistent with the presently established QCD
phenomenology considered in this review. No known QCD observable discussed
here has been identified that requires the additional global classification
of the exact continuum gauge-field configuration space. Conversely, this
does not demonstrate that such global structure is absent or physically
irrelevant. It leaves open the question of whether some deeper theoretical
principle, presently unknown observable, or more complete mathematical
construction of the theory requires it.

The Witten-Veneziano relation provides an instructive example of this
distinction. The pure Yang-Mills topological susceptibility entering the
relation can be expressed through the correlation function of the local
topological charge density,
\begin{align}
\chi_{\rm YM}
&=
\textstyle \int\!d^4x_E\,
\langle\hat q_E(x_E)\hat q_E(0)\rangle_{\rm YM}.
\end{align}
The successful description of the $\eta'$ mass therefore provides compelling
evidence for physically important topological charge-density correlations.
It does not by itself establish that the exact continuum functional integral
must admit a configuration-by-configuration decomposition into sectors of
integer-valued $Q_{\rm gluon}$. Similar distinctions apply to the other
nonperturbative results reviewed in the preceding sections.

A related point concerns the Vafa-Witten theorem
\citep{Vafa:1984xg}. Within its domain of applicability, the theorem
establishes important properties of vector-like gauge theories at
$\theta=0$, including the absence of spontaneous parity breaking. Its
application to $\theta$ dependence begins within a formulation in which
$\theta$ has already been introduced into the Euclidean action. The
foundational question considered here is logically prior: under what
conditions must this source parameter be promoted to an independent
physical coupling? The two questions should therefore be distinguished.

The comparison with QED is also instructive. In both QED and QCD the
corresponding $F\tilde F$ density may be written locally as a total
four-divergence, while its physical interpretation depends upon the
admissible fields, boundary conditions and global structure of the theory.
The Abelian and nonabelian cases nevertheless differ substantially in
their gauge-group topology and nonperturbative dynamics. The comparison is
discussed in Appendix~\ref{App:QEDvsQCD}. Its purpose is not to argue that
QED and QCD must be treated identically, but to illustrate more generally
the need to distinguish a locally allowed total-derivative operator from
the additional global structure that may give its coefficient physical
significance.

Several important conceptual questions therefore remain open. Perhaps the
most fundamental concerns the mathematical structure of the continuum
functional integral itself. As discussed in Sec.~\ref{Sec:OpenMathematicalQuestions},
even free quantum fields are naturally described by distribution-valued
configurations, and asymptotic freedom provides reason to expect analogous
short-distance behavior in QCD. How the global topological constructions
defined for sufficiently smooth classical gauge fields should be related
rigorously to the configuration space and measure of the full continuum
quantum theory remains incompletely understood.

Closely related questions concern the relationship between the Hamiltonian
and Euclidean formulations. The local canonical Gauss-law constraint is
well defined independently of the additional classification of admissible
gauge transformations into disconnected components. Similarly, the local
Euclidean operator $q(x)$ and its correlation functions exist independently
of a decomposition into sectors of integer-valued $Q_{\rm gluon}$. A more
complete mathematical understanding of how these local structures are
related to the conventional global constructions in the full
four-dimensional quantum theory would be valuable.

There is also a basic operational aspect of this question. Any realizable 
experiment probes fields and correlations only within a finite region of
spacetime and over a finite interval of time. Conditions imposed at 
spatial or spacetime infinity, and global classifications that depend 
upon them, therefore cannot be tested directly. This does not imply that
global structure is physically irrelevant: global properties may have
observable consequences within finite regions. It does imply, however,
that the physical significance of such structure must ultimately be 
established through consequences for observables accessible within finite 
spacetime regions. The anomalous singlet chiral transformation provides
an instructive example. Once a physical $\bar\theta$ is admitted, the
$CP$-violating phase conventionally associated with the $\theta$ term may
be transferred to a local pseudoscalar component of the quark mass matrix,
whose observable consequences can then be described through local
operators and nonperturbative QCD dynamics. Thus the fact that a physical
$\bar\theta$ can produce local observable effects does not by itself
establish that the additional global structure conventionally used to
introduce the vacuum angle is required. From this perspective, an
important question is whether any observable accessible within a finite
spacetime region requires that additional global structure, rather than
only the local operators, correlations, and nonperturbative dynamics of
QCD once a $CP$-violating parameter has been introduced.

Another open question concerns the precise physical status of $\theta$.
There is a clear mathematical distinction between introducing $\theta$ as
an auxiliary source parameter,
\begin{align}
Z_E(\theta) &= \textstyle \int\! 
    \mathcal{D}A_\mu \mathcal{D}\bar{\psi}  \mathcal{D}\psi\,
    \exp \left[-S_E + i\theta\int\!d^4x_E\,q_E(x_E)
\right],
\end{align}
and interpreting a constant $\theta$ as an independent physical coupling.
The former is sufficient to generate moments and correlation functions of
$q(x)$, including the topological susceptibility. The latter is the
additional step that leads, through the anomaly and the quark mass matrix,
to a physical $\bar\theta$ and hence to the conventional strong $CP$
problem. Determining whether some independent physical or mathematical
principle requires this latter interpretation remains an important
foundational question.

Future progress may come from several complementary directions. On the
mathematical side, advances in constructive quantum field theory, the
rigorous definition of continuum Yang-Mills theory and the analysis of
functional-integral measures may clarify the relationship between rough
quantum gauge fields and the global topology familiar from smooth classical
configurations. On the computational side, increasingly precise lattice
studies employing gradient flow, chiral lattice fermions and controlled
continuum extrapolations will continue to deepen our understanding of
topological observables. It would be particularly valuable to identify an
observable, if one exists, whose description genuinely requires the global
topological classification rather than only the local correlations of
$q(x)$ and their nonperturbative dynamics.

Experimental searches remain equally important. Increasingly sensitive
measurements of the neutron electric dipole moment provide direct tests for
strong $CP$ violation and, more generally, powerful probes of new sources of
$CP$ violation beyond the Standard Model. Searches for axions and axion-like
particles are likewise of major importance. The foundational questions
examined in this review concerning the conventional formulation of the
strong $CP$ problem do not weaken the scientific case for these searches.

Indeed, axions and axion-like particles constitute a broad and independently
well-motivated class of physics beyond the Standard Model. The QCD axion
provides an elegant dynamical solution of the strong $CP$ problem within its
conventional formulation, but the motivation for light pseudoscalar degrees
of freedom extends considerably beyond this role. Axions and axion-like
particles arise naturally in many extensions of the Standard Model,
including theories with spontaneously broken approximate global symmetries
and ultraviolet constructions containing multiple pseudoscalar fields. They
provide compelling dark-matter candidates over a wide range of masses and
production mechanisms, can play important roles in early-universe cosmology,
and lead to distinctive astrophysical and laboratory signatures through
their couplings to photons, fermions and gauge fields. More broadly, searches
for such particles provide sensitive probes of extremely weakly coupled
hidden sectors and of new physics at energy scales far beyond those directly
accessible to accelerators. The continuing experimental program of
laboratory, astrophysical and cosmological searches for axions and axion-like
particles is therefore strongly motivated independently of the foundational
status of the strong $CP$ problem considered in this review.

We conclude by returning to the principal theme of this review. The
conventional formulation provides a consistent framework in which the
additional global classification leads to a physical parameter
$\bar\theta$, whose observed smallness constitutes the strong $CP$ problem.
A minimal formulation retaining the established local and nonperturbative
structure of QCD without imposing this additional global classification is,
as far as presently known, also consistent with established QCD
phenomenology and need not contain an independent strong-$CP$ parameter.
Present knowledge does not by itself determine which of these formulations
provides the more fundamental description.

Clarifying this distinction does not diminish the remarkable success of
QCD or the importance of the extensive theoretical work based upon its
conventional nonperturbative formulation. Rather, it identifies more
precisely where additional physical or mathematical input is required.
Whether future progress comes from experiment, lattice gauge theory,
continuum methods or a more rigorous understanding of four-dimensional
quantum gauge theory, the strong $CP$ problem will continue to provide an
important setting in which to explore the relationship between locality,
symmetry, nonperturbative dynamics and global topology in quantum field
theory.

\subsection*{Acknowledgments}
AGW is funded by the ARC Centre of Excellence for Dark Matter Particle Physics
(CDMPP) CE200100008 and further supported by the Centre for the Subatomic 
Structure of Matter (CSSM). AGW gratefully acknowledges discussions with 
Bj\"orn Garbrecht, Martin L\"uscher, Andreas Ringwald, Francesco Sannino 
and Gerrit Schierholz on various issues addressed in this work.


\appendix

\section{Fujikawa's derivation of the axial anomaly}
\label{App:Fujikawa}

One of the most illuminating derivations of the axial anomaly is due to
Fujikawa~\citep{Fujikawa:1979ay,Fujikawa:1980eg,Williams:2022bzq}. In this
approach the anomaly is understood directly from the fermionic functional
integral. A chiral transformation that is a symmetry of the classical
massless action fails to leave the quantum functional measure invariant once
that measure is defined using a gauge-invariant ultraviolet regulator. This
formulation is particularly useful for the present review because the
derivation is local and makes no prior assumption concerning instantons,
integer-valued topological sectors, or the global classification of gauge
fields.

We work in Euclidean space using the conventions introduced in Sec.~2.1. The
Euclidean Dirac operator is
\begin{align}\label{Eq:DiracOperator}
\slashed D_E&=\gamma_{E\mu}D_\mu,\qquad D_\mu=\partial_\mu-igA_\mu .
\end{align}
Since the Euclidean gamma matrices are Hermitian, while $D_\mu$ is
anti-Hermitian, $\slashed D_E$ is anti-Hermitian. It is therefore convenient
to introduce the Hermitian operator
\begin{align}
{\cal H}&=i\slashed D_E .
\end{align}
Let its orthonormal eigenfunctions satisfy
\begin{align}\label{Eq:EigenvalueEq}
{\cal H}\phi_n(x)&=\lambda_n\phi_n(x),
\end{align}
where the eigenvalues $\lambda_n$ are real and
\begin{align}
\int d^4x_E\,\phi_m^\dagger(x)\phi_n(x)&=\delta_{mn}.
\end{align}
For a single quark flavor the fermion fields may then be expanded as
\begin{align}\label{Eq:ModeExpansion}
\psi(x)&=\textstyle \sum_n a_n\phi_n(x),\qquad
    \bar\psi(x)= \sum_n\bar b_n\phi_n^\dagger(x),
\end{align}
so that the fermionic functional measure is defined by
\begin{align}\label{Eq:MeasureModes}
{\cal D}\bar\psi\,{\cal D}\psi
&=
\prod_n d\bar b_n\,da_n .
\end{align}
Thus the functional measure is defined through the Grassmann coefficients
appearing in an expansion of the fermion fields in a complete orthonormal
basis.

Consider now an infinitesimal local axial transformation,
\begin{align}\label{Eq:ChiralTransform}
\psi(x)&\to\psi'(x)= [1+i\alpha(x)\gamma_5]\psi(x), \qquad
    \bar\psi(x) \to \bar\psi'(x)=
    \bar\psi(x)[1+i\alpha(x)\gamma_5],
\end{align}
where $\alpha(x)$ is an arbitrary smooth function of compact support. The use
of a local parameter allows the corresponding local Ward identity to be
obtained directly.

For the Euclidean fermion action
\begin{align}\label{Eq:FermionAction}
S_F &= \textstyle \int\! d^4x_E\,\bar\psi(\slashed D_E+m)\psi ,
\end{align}
the infinitesimal variation is
\begin{align}\label{Eq:ClassicalVariationBeforeIBP}
\delta S_F &= \textstyle \int\! d^4x_E\, \left[
    i(\partial_\mu\alpha)J_{5\mu}+2im\alpha\,\bar\psi\gamma_5\psi
    \right],
\end{align}
where
\begin{align}
J_{5\mu}(x) &= \bar\psi(x)\gamma_{E\mu}\gamma_5\psi(x).
\end{align}
Integrating the first term by parts gives
\begin{align}\label{Eq:ClassicalVariation}
\delta S_F&= \textstyle i\int \! d^4x_E\,\alpha(x)
    \left[-\partial_\mu J_{5\mu}(x)+
    2m\bar\psi(x)\gamma_5\psi(x)\right],
\end{align}
where no boundary term occurs because $\alpha(x)$ has compact support.

If the fermionic measure were invariant under the transformation in
Eq.~(\ref{Eq:ChiralTransform}), the massless theory would possess a conserved
singlet axial current. Fujikawa's essential observation is that the
regularized fermionic measure is not invariant.

Under the transformation, the mode coefficients in
Eq.~(\ref{Eq:ModeExpansion}) transform according to
\begin{align}
a_n'&=\textstyle \sum_m\left[\delta_{nm}+
    i\int d^4x_E\,\phi_n^\dagger(x)\alpha(x)\gamma_5\phi_m(x)
    \right]a_m ,
\end{align}
and similarly for $\bar b_n$. Since the coefficients are Grassmann
variables, their integration measure transforms with the inverse determinant.
The transformations of $\psi$ and $\bar\psi$ give identical contributions,
and hence for one Dirac flavor
\begin{align}\label{Eq:JacobianFormal}
{\cal D}\bar\psi'\,{\cal D}\psi'
    &={\cal D}\bar\psi\,{\cal D}\psi\,J[\alpha],
\end{align}
where formally
\begin{align}\label{Eq:FormalJacobian}
J[\alpha]&=\textstyle \exp\left[-2i\int\! d^4x_E\,\alpha(x)
    \sum_n\phi_n^\dagger(x)\gamma_5\phi_n(x)\right].
\end{align}
The factor of two arises because both $\psi$ and $\bar\psi$ contribute to
the Jacobian.

The quantity
\begin{align}
\textstyle \sum_n \phi_n^\dagger(x)\gamma_5\phi_n(x)
\end{align}
is ultraviolet divergent. Formally it is a coincident-point trace and cannot
consistently be set to zero simply by using $\mathrm{tr}(\gamma_5)=0$,
because the coincident-point completeness relation contains the divergent
quantity $\delta^{(4)}(0)$. The transformation of the functional measure must
therefore be defined only after regularization.

Fujikawa's essential step is to regulate this expression in a
gauge-invariant manner. Since ${\cal H}=i\slashed D_E$ transforms covariantly
under gauge transformations, one may define
\begin{align}\label{Eq:RegulatedAnomaly}
{\cal A}(x) &\equiv \textstyle \lim_{\Lambda\rightarrow\infty}
    \sum_n\phi_n^\dagger(x)\gamma_5e^{-\lambda_n^2/\Lambda^2}\phi_n(x)
    =\lim_{\Lambda\rightarrow\infty}\mathrm{tr}\left[\gamma_5\langle x|
    e^{-{\cal H}^2/\Lambda^2}|x\rangle\right],
\end{align}
where the trace acts on spinor and color indices. The regulator suppresses
eigenmodes with $|\lambda_n|\gg\Lambda$ and is removed only after the
regulated expression has been evaluated.

We now evaluate the regulated trace. From the definition of the field
strength used throughout this review,
\begin{align}
[D_\mu,D_\nu]&=-igF_{\mu\nu}.
\end{align}
Consequently,
\begin{align}
\slashed D_E^{\,2}&=\gamma_{E\mu}\gamma_{E\nu}D_\mu D_\nu
    =D^2+\frac14[\gamma_{E\mu},\gamma_{E\nu}][D_\mu,D_\nu]
    =D^2-\frac{ig}{4}[\gamma_{E\mu},\gamma_{E\nu}]F_{\mu\nu},
\end{align}
and hence
\begin{align}\label{Eq:Hsquare}
{\cal H}^2&=-D^2+\frac{ig}{4}[\gamma_{E\mu},\gamma_{E\nu}]F_{\mu\nu}.
\end{align}
The regulated trace may now be evaluated using the short-time heat-kernel
expansion. The trace containing $\gamma_5$ vanishes unless four Euclidean
gamma matrices are present. Therefore the terms containing zero or one
power of $F_{\mu\nu}$ vanish, and the first nonzero contribution is
quadratic in the field strength.

To the order that survives as $\Lambda\rightarrow\infty$,
\begin{align}\label{Eq:AnomalyExpansion}
{\cal A}(x)&=\lim_{\Lambda\rightarrow\infty}\mathrm{tr}
    \left[\gamma_5\frac{1}{2}\left(-\frac{ig}{4\Lambda^2}
    [\gamma_{E\mu},\gamma_{E\nu}]F_{\mu\nu}\right)^2\right]
    \langle x|e^{D^2/\Lambda^2}|x\rangle+{\cal O}(\Lambda^{-2}).
\end{align}
At leading order in the short-time expansion,
\begin{align}\label{Eq:HeatKernelCoincident}
\langle x|e^{D^2/\Lambda^2}|x\rangle&=\int\frac{d^4p_E}{(2\pi)^4}
    e^{-p_E^2/\Lambda^2}+{\cal O}(\Lambda^2)
    =\frac{\Lambda^4}{16\pi^2}+{\cal O}(\Lambda^2).
\end{align}
The terms represented by ${\cal O}(\Lambda^2)$ contain gauge-field-dependent
contributions to the heat kernel, but when multiplied by the explicit
$\Lambda^{-4}$ in Eq.~(\ref{Eq:AnomalyExpansion}) they do not contribute to
the limit.

Using the conventions of Sec.~2.1,
\begin{align}
\gamma_5&=-\gamma_{E1}\gamma_{E2}\gamma_{E3}\gamma_{E4},
    \qquad\epsilon_{1234}=+1,
\end{align}
and therefore
\begin{align}\label{Eq:Gamma5Trace}
\mathrm{tr}\left(\gamma_5\gamma_{E\mu}\gamma_{E\nu}
    \gamma_{E\rho}\gamma_{E\sigma}\right)
    &=-4\epsilon_{\mu\nu\rho\sigma}.
\end{align}
Since $F_{\mu\nu}$ is antisymmetric,
\begin{align}
[\gamma_{E\mu},\gamma_{E\nu}]F_{\mu\nu}
    &=2\gamma_{E\mu}\gamma_{E\nu}F_{\mu\nu},
\end{align}
so that
\begin{align}
&\mathrm{tr}\left[\gamma_5[\gamma_{E\mu},\gamma_{E\nu}]
    [\gamma_{E\rho},\gamma_{E\sigma}]\right]F_{\mu\nu}^aF_{\rho\sigma}^b
    \mathrm{tr}(T^aT^b)
    = 4\, \mathrm{tr} \left(\gamma_5\gamma_{E\mu}\gamma_{E\nu}
    \gamma_{E\rho}\gamma_{E\sigma}\right)
    F_{\mu\nu}^aF_{\rho\sigma}^b \frac12\delta^{ab}
    = -8\epsilon_{\mu\nu\rho\sigma}
    F_{\mu\nu}^aF_{\rho\sigma}^a.
\end{align}
Substitution into Eq.~(\ref{Eq:AnomalyExpansion}) then gives
\begin{align}
{\cal A}(x)&=\left(-\frac{g^2}{32\Lambda^4}\right)\left(
    -8\epsilon_{\mu\nu\rho\sigma}F_{\mu\nu}^aF_{\rho\sigma}^a\right)
    \frac{\Lambda^4}{16\pi^2}=\frac{g^2}{64\pi^2}\epsilon_{\mu\nu\rho\sigma}
    F_{\mu\nu}^aF_{\rho\sigma}^a
    =\frac{g^2}{32\pi^2}F_{\mu\nu}^a\tilde F_{\mu\nu}^a
    \nonumber\\
    &=\frac{g^2}{16\pi^2}\mathrm{tr}\left(F_{\mu\nu}\tilde F_{\mu\nu}
    \right) = q_E(x).
\label{Eq:FujikawaLocalResult}
\end{align}
Here $q_E(x)$ denotes the Euclidean continuation of the local topological
charge density defined in Sec.~2.1. As shown explicitly in
Appendix~\ref{App:Wick_rot}, its spacetime integral continues
to the corresponding Minkowski quantity with the conventions used in this
review.

For one quark flavor, the regulated Jacobian is therefore
\begin{align}\label{Eq:FinalJacobianOneFlavor}
\mathcal J[\alpha]&=\exp\left[-2i \textstyle \int\! d^4x_E\,
    \alpha(x)q_E(x) \right].
\end{align}
For a common singlet axial transformation of $N_f$ quark flavors, each
flavor contributes equally, giving
\begin{align}\label{Eq:FinalJacobian}
\mathcal J_{N_f}[\alpha]&=\exp\left[-2iN_f\int d^4x_E\,
    \alpha(x)q_E(x)\right].
\end{align}
Thus the fermionic functional measure is not invariant under the singlet
axial transformation.

The functional integral itself is unchanged by a change of integration
variables. Combining the variation of the action with the nontrivial
Jacobian therefore gives the local anomalous Ward identity. After
continuation to Minkowski space this becomes
\begin{align}\label{Eq:AnomalousWard}
\partial_\mu J_5^\mu &= 2i\left(\bar\psi_LM\psi_R
    -\bar\psi_RM^\dagger\psi_L\right)+2N_fq(x)
    =2i\bar\psi M_S\gamma_5\psi-2\bar\psi M_P\psi+2N_fq(x).
\end{align}
In the special case of equal real quark masses this reduces to
\begin{align}
\partial_\mu J_5^\mu&=2im\bar\psi\gamma_5\psi+2N_fq(x).
\end{align}
The first term represents explicit breaking of the axial symmetry by the
quark masses, while the second is the quantum anomaly arising from the
non-invariance of the regulated fermionic functional measure.

Several features of this derivation are particularly important for the
discussion in this review.

First, the anomaly is a local quantum-field-theoretic statement. It arises
from the ultraviolet regularization of the fermionic functional measure and
is expressed entirely in terms of the local gauge-invariant operator $q(x)$.
The derivation does not require instantons or any classification of gauge
fields by winding number or integer-valued topological charge.

Second, no assumption has been made that
\begin{align}
Q_{\rm gluon} &= \textstyle \int\! d^4x\,q(x)
\end{align}
is integer valued, nor has the functional integral been decomposed into
topological sectors. Such global constructions require additional
assumptions concerning the smoothness, boundary conditions and global
behavior of the gauge fields. Instantons provide an important semiclassical
realization of the anomaly through the zero modes of the Dirac operator, but
they are not required for the local anomalous Ward identity itself.

Third, the same Jacobian determines the transformation under a constant
singlet axial rotation. Setting $\alpha(x)=\alpha$ gives
\begin{align}\label{Eq:chiral_Jacobian}
\mathcal J_{N_f}(\alpha) &= \exp\left[-2iN_f\alpha Q \right],
\end{align}
where at this stage $Q$ denotes simply the spacetime integral of the local
density $q(x)$; no assumption that it is integer valued is required. The
axial rotation therefore changes the coefficient multiplying $q(x)$ while
simultaneously changing the flavor-singlet phase of the quark mass matrix.
This is the origin of the invariant combination
\begin{align}
\bar\theta &= \theta+\arg\det M
\end{align}
discussed in \Sec{Sec:thetabar}.

The Fujikawa derivation therefore establishes the anomalous relation between
a flavor-singlet pseudoscalar phase of the quark mass matrix and the
coefficient multiplying the local operator $q(x)$. If an independent
$CP$-violating parameter of this type is introduced, the two
parameterizations are physically equivalent and are described by the
invariant combination $\bar\theta$. The derivation does not by itself,
however, establish that such an independent $CP$-violating parameter must
be introduced. That question is logically distinct from the local anomaly
and forms part of the broader discussion of the conventional and alternative
perspectives developed in the main text.

\section{On the reality of the Euclidean functional integral}
\label{App:real_Z_E}

As discussed in \Eqs{Eq:theta_actions_simeq} and
(\ref{Eq:L_thetas}), the Euclidean action becomes complex when
$\theta\ne0$,
\begin{align}
S_E(\theta) &= S_E - i\theta \textstyle \int\!d^4x_E\,q_E(x).
\end{align}
The corresponding functional-integral weight therefore contains the phase
\begin{align}
\exp \left[i\theta\textstyle\int\!d^4x_E\,q_E(x)\right].
\end{align}
It might consequently appear that the Euclidean generating functional
$Z_E(\theta)$ should itself be complex. In fact, for real $\theta$ and real
positive quark masses, $Z_E(\theta)$ is real and is an even function of
$\theta$. The reason is the $CP$ invariance of the functional measure at 
$\theta=0$. 

After integrating over the fermion fields, \Eq{Eq:QCD_genZ_theta} may be
written as
\begin{align}\label{Eq:ZEtheta_det}
Z_E(\theta) &= \textstyle
    \int\!\mathcal{D}A_\mu\,e^{-S_{\rm YM}[A]}\det(\slashed D_E+M)
    \exp\left[i\theta\int\!d^4x_E\,q_E(x)\right],
\end{align}
where
\begin{align}
S_{\rm YM}[A] &= \textstyle 
    \frac12\int\!d^4x_E\,\tr(F_{\mu\nu}F_{\mu\nu})
\end{align}
and $M$ denotes the real diagonal quark-mass matrix.

It is useful first to recall why the fermion determinant appearing in
Eq.~(\ref{Eq:ZEtheta_det}) is real and nonnegative for positive quark
masses. With the Euclidean conventions used throughout this review,
\begin{align}
\slashed D_E^\dagger &=-\slashed D_E,
\end{align}
and since
\begin{align}
\{\gamma_{E5},\slashed D_E\}&=0,
\end{align}
one has
\begin{align}\label{Eq:gamma5_herm}
\gamma_{E5}\slashed D_E\gamma_{E5}&=-\slashed D_E
    =\slashed D_E^\dagger .
\end{align}
This is the usual $\gamma_5$-hermiticity relation. If
\begin{align}
\slashed D_E\phi_\lambda&=i\lambda\phi_\lambda,
    \qquad\lambda\in\mathbb R,
\end{align}
then
\begin{align}
\slashed D_E(\gamma_{E5}\phi_\lambda)
    &=-i\lambda(\gamma_{E5}\phi_\lambda),
\end{align}
so the nonzero eigenvalues occur in pairs $i\lambda$ and $-i\lambda$.
For a single quark flavor with $m>0$ this gives
\begin{align}
\det(\slashed D_E+m) &= \textstyle m^{n_0} \prod_{\lambda>0}
    (i\lambda+m)(-i\lambda+m) \nonumber\\
    &=\textstyle m^{n_0}\prod_{\lambda>0}(\lambda^2+m^2)\ge 0,
\end{align}
where $n_0$ is the number of zero modes. For several flavors,
\begin{align}
\det(\slashed D_E+M) &= \textstyle \prod_f \det(\slashed D_E+m_f) \ge 0
\end{align}
provided that all $m_f>0$. If $m_f=0$, a configuration possessing zero
modes gives a vanishing determinant rather than a negative one.

We may therefore define the real nonnegative $\theta=0$ functional measure
\begin{align}\label{Eq:positive_measure}
d\mu_0[A]&\equiv\mathcal{D}A_\mu\,e^{-S_{\rm YM}[A]}\det(\slashed D_E+M),
\end{align}
so that
\begin{align}\label{Eq:ZEtheta_measure}
Z_E(\theta) &= \textstyle \int d\mu_0[A]\, e^{i\theta Q[A]},
    \qquad Q[A]\equiv \textstyle \int\!d^4x_E\,q_E(x).
\end{align}
At this point no assumption that $Q[A]$ is integer valued is required.

The $\theta=0$ theory is $CP$ invariant. If $A_\mu^{CP}$ denotes the
$CP$-transformed gauge field, the functional measure satisfies
\begin{align}
d\mu_0[A^{CP}]&=d\mu_0[A],
\end{align}
while the pseudoscalar density is $CP$ odd,
\begin{align}
q_E[A^{CP}](x)&=-q_E[A](x^{CP}),
\end{align}
and hence
\begin{align}
Q[A^{CP}]&=-Q[A].
\end{align}
The functional integration contains every admissible gauge configuration
together with its $CP$ transform. Making the change of integration variable
$A\rightarrow A^{CP}$ in Eq.~(\ref{Eq:ZEtheta_measure}) therefore gives
\begin{align}\label{Eq:ZE_even}
Z_E(\theta) &= \textstyle \int d\mu_0[A^{CP}]\, e^{i\theta Q[A^{CP}]}
    = \int d\mu_0[A]\, e^{-i\theta Q[A]}
    = Z_E(-\theta).
\end{align}
Thus $Z_E(\theta)$ is an even function of $\theta$.

On the other hand, since $d\mu_0[A]$ and $Q[A]$ are real,
\begin{align}
Z_E(\theta)^* &= \textstyle \int d\mu_0[A]\, e^{-i\theta Q[A]}
    = Z_E(-\theta).
\end{align}
Combining this result with Eq.~(\ref{Eq:ZE_even}) gives
\begin{align}
Z_E(\theta)^* &= Z_E(\theta),
    \qquad\Rightarrow\qquad
    Z_E(\theta)\in \mathbb R.
\end{align}
Equivalently, since the sine term is odd under $CP$ and therefore
integrates to zero, one may write
\begin{align}\label{Eq:ZEtheta_real}
Z_E(\theta)
&=
\int d\mu_0[A]\,
\cos\left(\theta Q[A]\right).
\end{align}
This form makes both the reality and the evenness in $\theta$ manifest.

An important point is that this argument does not require a decomposition
of the functional integral into sectors of integer-valued topological
charge. It uses only the reality and $CP$ invariance of the $\theta=0$
measure together with the fact that the local density $q_E(x)$ is $CP$ odd.
Consequently, the result remains valid if $\theta$ is regarded merely as a
source parameter coupled to $q_E(x)$ rather than as a physical vacuum
angle.

Within the conventional global construction one may obtain the same result
by decomposing the functional integral into sectors of integer-valued
topological charge. In the following equations, $Q\in\mathbb Z$ denotes the
integer label of a topological sector, whereas $Q[A]$ continues to denote
the spacetime integral of the local density evaluated on an individual
gauge-field configuration. Within this construction, configurations
belonging to the sector labeled by $Q$ satisfy $Q[A]=Q$. Writing
\begin{align}
Z_E(\theta)&=\textstyle \sum_{Q\in\mathbb Z}e^{i\theta Q}Z_E^Q,
\end{align}
where
\begin{align}
Z_E^Q&= \textstyle \int_{Q[A]=Q}\!\mathcal{D}A_\mu\,
    e^{-S_{\rm YM}[A]}\det(\slashed D_E+M)=\textstyle\int\!\mathcal{D}A_\mu\,
    \delta(Q-Q[A])\,e^{-S_{\rm YM}[A]}\det(\slashed D_E+M),
\end{align}
the $CP$ transformation maps the sector $Q$ into the sector $-Q$. Since the
$\theta=0$ measure is $CP$ invariant,
\begin{align}
Z_E^Q&=Z_E^{-Q}.
\end{align}
Furthermore, each $Z_E^Q$ is real and nonnegative for positive quark masses.
It follows that
\begin{align}\label{Eq:even_in_thetaQ}
Z_E(\theta) &= \textstyle
    Z_E^{Q=0}+\sum_{Q>0}\left(e^{i\theta Q}+e^{-i\theta Q}\right)
    Z_E^Q=Z_E^{Q=0}+2\sum_{Q>0}\cos(\theta Q)Z_E^Q\in\mathbb R .
\end{align}
This second derivation is therefore consistent with the more general
configuration-pairing argument above, but it \emph{additionally}
assumes the global decomposition into integer-$Q$ sectors.

The relative contributions of the different topological sectors within
this conventional construction are discussed in
Appendix~\ref{Sec:ZQ_contrib}. They are not required for the proof of the
reality of $Z_E(\theta)$.

Since $Z_E(\theta)$ is real and even in $\theta$, the corresponding vacuum
energy density is likewise an even function in any region where
$Z_E(\theta)>0$ and the usual real logarithm of the Euclidean generating
functional is well defined. Its derivatives at $\theta=0$, including the
topological susceptibility, consequently have the usual reality properties
discussed in Sec.~\ref{Sec:top_suscept}.

Finally, the positivity of the $\theta=0$ Euclidean measure in
Eq.~(\ref{Eq:positive_measure}) is one of the fundamental properties
underlying lattice QCD at zero chemical potential. It permits the
functional integral to be interpreted as an expectation value with respect
to a positive statistical weight and thereby allows Monte Carlo importance
sampling. It also enters rigorous results for vector-like gauge theories,
including the Vafa-Witten analysis. For real nonzero $\theta$, however, the
factor
\begin{align}
e^{i\theta Q[A]}
\end{align}
is generally complex. Although the complete partition function remains
real by the $CP$ pairing demonstrated above, the configuration-by-
configuration weight is no longer positive, producing the familiar
$\theta$-dependent sign problem.

\section{Derivation of the Leutwyler--Smilga relation}
\label{App:LS_relation}

The Leutwyler--Smilga relation connects the topological susceptibility of
QCD to the light quark masses and the chiral condensate. Its derivation
provides a useful illustration of the interplay between the anomalous axial
Ward identity and low-energy chiral dynamics. The purpose of this appendix
is not to give the most general form of the relation, but rather to derive
its leading-order expression and to identify clearly the assumptions that
enter.

We consider $N_f$ light quark flavors in the regime where chiral effective
theory is applicable. The Euclidean topological susceptibility at
$\theta=0$ may be defined by
\begin{align}
\chi &= {\textstyle\int\!d^4x_E\,\langle\hat q_E(x_E)\hat q_E(0)\rangle}
    =\left.\frac{\partial^2\varepsilon_0(\theta)}{\partial\theta^2}
    \right|_{\theta=0},
\end{align}
where the vacuum energy density, when the usual Hamiltonian interpretation
applies, or more generally the generating-functional density when $\theta$
is regarded simply as a source, is defined by
\begin{align}\label{Eq:vareps_0_again}
\varepsilon_0(\theta)&=
    -\lim_{V\rightarrow\infty}\frac{1}{V}
\ln Z_E(\theta).
\end{align}
Here $\theta$ may simply be regarded as a source parameter coupled to the
local operator $q_E(x)$,
\begin{align}
Z_E(\theta)&=\textstyle \int\!{\cal D}\Phi\,\exp
    \left[-S_E+i\theta\int\!d^4x_E\,q_E(x_E)\right].
\end{align}
No decomposition of the functional integral into sectors of integer-valued
topological charge is required for this definition.

The axial anomaly determines how this source enters the low-energy chiral
theory. As discussed in Sec.~\ref{Sec:thetabar}, a flavor-singlet axial
rotation transfers the coefficient of $q(x)$ into the flavor-singlet phase
of the quark mass matrix. Taking the quark masses initially to be real and 
positive and choosing $\alpha=\theta/(2N_f)$, so that the coefficient of 
the topological charge density $q(x)$ is transformed to zero, 
the $\theta$ dependence may therefore be  represented by
\begin{align}
M &\to M_\theta = M e^{i\theta/N_f}.
\end{align}
This step follows from the anomalous transformation of the fermionic
functional measure and does not require a global classification of the
gauge fields.

At energies well below the characteristic QCD scale, the relevant
Goldstone degrees of freedom are described by the chiral field
\begin{align}
U(x)&\in SU(N_f).
\end{align}
The anomalous flavor-singlet degree of freedom is not included explicitly
in this leading-order $SU(N_f)$ effective theory; its effect relevant here
is encoded through the $\theta$-dependent phase of the mass matrix.
To leading order in the chiral expansion, the mass-dependent part of the
Euclidean effective potential is
\begin{align}\label{Eq:LS_Veff}
{\cal V}_{\rm eff}(U;\theta)&=-\Sigma\,
    \mathrm{Re}\,\mathrm{tr}\left(M e^{i\theta/N_f}U^\dagger\right),
    \qquad\mbox{where}\qquad
    \Sigma\equiv -\langle\bar\psi_f\psi_f\rangle_{m\rightarrow0}>0
\end{align}
denotes the chiral condensate per light flavor in the chiral limit.

In the infinite-volume limit, the vacuum energy density is obtained by
minimizing the chiral effective potential,
\begin{align}
\varepsilon_0(\theta)&=\min_U{\cal V}_{\rm eff}(U;\theta)
    +{\cal O}(m_q^2).
\end{align}
For the vacuum configuration we may take
\begin{align}
U&=\mathrm{diag}\left(e^{i\alpha_1},\ldots,e^{i\alpha_{N_f}}\right),
\end{align}
with
\begin{align}
\textstyle \sum_{f=1}^{N_f}\alpha_f&=0\qquad(\mathrm{mod}\ 2\pi),
\end{align}
as required by $\det U=1$. Equation~(\ref{Eq:LS_Veff}) then becomes
\begin{align}
{\cal V}_{\rm eff}&=\textstyle -\Sigma\sum_{f=1}^{N_f}m_f
    \cos\left(\frac{\theta}{N_f}-\alpha_f\right).
\end{align}
It is convenient to define
\begin{align}
\phi_f&= (\theta/N_f)-\alpha_f.
\end{align}
The $SU(N_f)$ condition then implies
\begin{align}\label{Eq:LS_phase_constraint}
\textstyle \sum_{f=1}^{N_f}\phi_f&=\theta
    \qquad(\mathrm{mod}\ 2\pi).
\end{align}
For the expansion about $\theta=0$ we choose the branch continuously
connected to $\phi_f=0$, so that the constraint is simply
\begin{align}
\textstyle \sum_{f=1}^{N_f}\phi_f&=\theta.
\end{align}

The vacuum energy is therefore obtained by minimizing
\begin{align}\label{Eq:LS_energy_min}
\varepsilon_0(\theta)&=\textstyle -\Sigma\max_{\{\phi_f\}}
    \left[\sum_{f=1}^{N_f}m_f\cos\phi_f\right]
    +{\cal O}(m_q^2),\qquad \mbox{with}\qquad
    \sum_{f=1}^{N_f}\phi_f=\theta.
\end{align}
For sufficiently small $\theta$, all $\phi_f$ are small and
\begin{align}
\cos\phi_f&=\textstyle 1-\frac12\phi_f^2+{\cal O}(\phi_f^4).
\end{align}
This then gives
\begin{align}\label{Eq:LS_energy_quadratic}
\varepsilon_0(\theta)&=\varepsilon_0(0)+(\Sigma/2)
    \textstyle\min_{\{\phi_f\}}
    \left[\sum_{f=1}^{N_f}m_f\phi_f^2\right]+{\cal O}(m_q^2,\theta^4),
    \qquad\mbox{subject to}\qquad \sum_f\phi_f=\theta.
\end{align}

The constrained minimization may be performed using a Lagrange multiplier
$\lambda$,
\begin{align}
F&\equiv\textstyle \sum_{f=1}^{N_f}m_f\phi_f^2-\lambda
    \left(\sum_{f=1}^{N_f}\phi_f-\theta\right).
\end{align}
Stationarity with respect to each $\phi_f$ gives
\begin{align}
\frac{\partial F}{\partial\phi_f} &= 2m_f\phi_f-\lambda=0,
\end{align}
and therefore
\begin{align}
\phi_f&= \lambda/(2m_f).
\end{align}
Using the constraint,
\begin{align}
\theta &= \textstyle \sum_{f=1}^{N_f}\phi_f
    = \frac{1}{2}\lambda\sum_{f=1}^{N_f}(1/m_f),
\end{align}
so that
\begin{align}
\lambda/2 &= \bar m\,\theta,\qquad
    \bar m^{-1} \equiv\textstyle \sum_{f=1}^{N_f}(1/m_f).
\end{align}
So then
\begin{align}\label{Eq:LS_phi_solution}
\phi_f &= (\bar m/m_f)\theta.
\end{align}
Substitution into Eq.~(\ref{Eq:LS_energy_quadratic}) gives
\begin{align}
\textstyle \sum_fm_f\phi_f^2 &=\textstyle
    \bar m^2\theta^2\sum_f(1/m_f) = \bar m\,\theta^2,
\end{align}
and hence
\begin{align}\label{Eq:LS_energy_result}
\varepsilon_0(\theta)
&=
\varepsilon_0(0)
+
\frac12
\Sigma\bar m\,\theta^2
+
{\cal O}(m_q^2,\theta^4).
\end{align}
Taking two derivatives with respect to the source $\theta$ gives the
leading-order Leutwyler--Smilga relation,
\begin{align}\label{Eq:LS_relation}
\chi &= \left.\frac{\partial^2\varepsilon_0(\theta)}
    {\partial\theta^2} \right|_{\theta=0}
    =\Sigma\bar m+{\cal O}(m_q^2)
    =\frac{\Sigma}{\sum_{f=1}^{N_f}(1/m_f)}+{\cal O}(m_q^2).
\end{align}
For $N_f$ degenerate light quarks with $m_f=m$,
\begin{align}
\chi
&=
\frac{m\Sigma}{N_f}
+
{\cal O}(m^2).
\end{align}
An immediate consequence of the leading-order relation is that, as any one
of the light quark masses tends to zero,
\begin{align}
\bar m &\to 0, \qquad\Rightarrow  \qquad \chi \to 0.
\end{align}
This agrees with the exact anomalous Ward-identity result that $\chi=0$
when at least one quark is exactly massless.

The logical ingredients entering this derivation are worth emphasizing.
The relation uses the anomalous singlet axial transformation, spontaneous
chiral symmetry breaking, the low-energy chiral effective theory and the
large-volume limit. It is therefore not a consequence of locality alone.
However, no step requires the exact functional integral to be decomposed
configuration by configuration into sectors of integer-valued topological
charge. The source $\theta$ is introduced through its coupling to the local
operator $q(x)$, and the anomaly determines how this source is represented
in the low-energy quark mass term.

The Leutwyler--Smilga relation therefore provides a concrete example in
which a phenomenologically important result involving the topological
susceptibility follows from the anomalous Ward identities and low-energy
chiral dynamics without requiring a prior global classification of the
gauge-field configuration space. Its success demonstrates the physical
importance of correlations of the local topological charge density, but
does not by itself establish that the exact continuum functional integral
must possess a fundamental decomposition into integer-valued topological
sectors.

\section{Derivation of the Witten-Veneziano relation}
\label{App:WV_relation}

The Witten-Veneziano relation connects the mass generated for the
flavor-singlet pseudoscalar meson to the topological susceptibility of pure
Yang-Mills theory \citep{Witten:1979vv,Veneziano:1979ec}. Its essential
ingredients are the anomalous singlet axial Ward identity, spontaneous
chiral symmetry breaking, and the large-$N_c$ expansion, in which quark-loop
effects are suppressed relative to pure-glue contributions. The purpose of
this appendix is not to give the most general derivation, but to make these
ingredients and their logical roles transparent.

The susceptibility entering the Witten-Veneziano relation is the
topological susceptibility of pure Yang-Mills theory at $\theta=0$,
\begin{align}\label{Eq:chiYM_def}
\chi_{\rm YM}&\equiv\textstyle \int\!d^4x_E\,
    \braket{\hat q_E(x_E)\hat q_E(0)}_{\rm YM},
\qquad\mbox{where}\qquad
\hat q_E(x)=[g^2/(32\pi^2)]\hat F_{\mu\nu}^a\hat{\tilde F}_{\mu\nu}^a .
\end{align}
Equivalently, if a source $\theta$ is coupled to $q_E(x)$ in pure
Yang-Mills theory, then near $\theta=0$ the vacuum energy density has the
expansion
\begin{align}\label{Eq:YM_theta_energy}
\varepsilon_{\rm YM}(\theta)&= \varepsilon_{\rm YM}(0)
    +{\textstyle\frac12}\chi_{\rm YM}\theta^2+{\cal O}(\theta^4),
    \qquad\mbox{where}\qquad
    \chi_{\rm YM}=\left.\frac{\partial^2\varepsilon_{\rm YM}(\theta)}
    {\partial\theta^2}\right|_{\theta=0}.
\end{align}
No decomposition into sectors of integer-valued topological charge is
required in order to define this correlation function or its corresponding
source derivative.

Now include $N_f$ light quark flavors. With the conventional normalization
of the flavor-singlet generator, the singlet axial current in the chiral
limit is
\begin{align}
J_5^\mu(x) &= [1/(2N_f)]^{1/2}\textstyle
\sum_{f=1}^{N_f}
\bar\psi_f(x)\gamma^\mu\gamma_5\psi_f(x),
\end{align}
corresponding to the generator
\begin{align}
T^0
&=  [1/(2N_f)]^{1/2} I, \qquad \tr(T^0T^0)=\textstyle\frac12.
\end{align}
The anomalous Ward identity then gives
\begin{align}\label{Eq:anomaly_chiral}
\partial_\mu J_5^\mu(x) &= (2N_f)^{1/2}\,q(x)
\end{align}
in the chiral limit.

At low energies, spontaneous chiral symmetry breaking gives rise to the
pseudoscalar meson degrees of freedom. For the present purpose it is
important to retain not only the usual flavor-nonsinglet Goldstone fields
but also the flavor-singlet pseudoscalar field associated with the
$U(1)_A$ transformation. We therefore extend the usual $SU(N_f)$ chiral
field to
\begin{align}
U(x)&\in U(N_f).
\end{align}
The field $U(x)$ transforms under the chiral flavor group as
\begin{align}
U(x)&\to V_L U(x)V_R^\dagger ,
\end{align}
and contains both the flavor-nonsinglet pseudoscalar fields and the
flavor-singlet field. For the present purpose it is useful to separate
the singlet phase explicitly by writing
\begin{align}
U(x)&=\exp\left[i\frac{(2/N_f)^{1/2}}{f_0}\eta_0(x)\right]\tilde U(x),
    \qquad\tilde U(x)\in SU(N_f).
\end{align}
Here $f_0$ denotes the singlet decay constant. Since
$\det\tilde U=1$, we then have
\begin{align}
\det U&=\exp\left[i\frac{(2N_f)^{1/2}}{f_0}\eta_0\right].
\end{align}
Thus the phase of $\det U$ isolates precisely the flavor-singlet
pseudoscalar degree of freedom that participates in the anomalous
$U(1)_A$ transformation. This is why the singlet phase, rather than the
flavor-nonsinglet components of $U$, enters the anomalous $\theta$
dependence below.

The singlet decay constant $f_0$ is correspondingly defined, with our
normalization of the singlet axial current and one-particle states, by
\begin{align}
\braket{0|\hat J_5^\mu(0)|\eta_0(k)}&=if_0 k^\mu .
\end{align}
The numerical normalization of $f_0$ is convention dependent, since it
depends on the normalization chosen for both $J_5^\mu$ and the state
$\ket{\eta_0(k)}$. The normalization used here is consistent with
$T^0=[1/(2N_f)]^{1/2}I$ and with the parameterization of $U(x)$ above;
with these conventions the anomalous singlet mass is
$m_0^2=2N_f\chi_{\rm YM}/f_0^2$.

At large $N_c$, quark loops are suppressed, so to leading order in the
$1/N_c$ expansion the anomalous dependence of the low-energy theory is
inherited from the pure Yang-Mills vacuum energy. Under a singlet axial 
transformation the shift of $\theta$ induced by the anomaly is compensated 
by the corresponding transformation of the phase of $\det U$. The 
effective theory can therefore depend on these quantities only through
the invariant combination
\begin{align}
\theta
-
\frac{\sqrt{2N_f}}{f_0}\eta_0 .
\end{align}
The sign of the second term depends only on the convention used for the
singlet field and has no effect on the mass derived below.

Expanding the pure Yang-Mills vacuum energy of
Eq.~(\ref{Eq:YM_theta_energy}) in this invariant combination therefore gives,
to quadratic order,
\begin{align}\label{Eq:WV_effective_potential}
{\cal V}_{\rm anom}(\eta_0;\theta)&=\frac12\chi_{\rm YM}
    \left(\theta-\frac{\sqrt{2N_f}}{f_0}\eta_0\right)^2+\cdots ,
\end{align}
where corrections are suppressed in the large-$N_c$ and low-energy
expansions. The coefficient is $\chi_{\rm YM}$ because setting
$\eta_0=0$ reduces the source dependence to that of the pure Yang-Mills
vacuum energy.

At $\theta=0$, Eq.~(\ref{Eq:WV_effective_potential}) contains the singlet
mass term
\begin{align}
{\cal V}_{\rm anom}(\eta_0;0)&= \frac12\left(\frac{2N_f}{f_0^2}
    \chi_{\rm YM}\right)\eta_0^2+\cdots .
\end{align}
Thus the anomalous contribution to the singlet pseudoscalar mass is
\begin{align}\label{Eq:WV_chiral}
m_0^2&=\frac{2N_f}{f_0^2}\chi_{\rm YM}+\cdots ,
\end{align}
which is the Witten-Veneziano relation in the chiral limit. In that limit
the physical $\eta'$ becomes the flavor-singlet state, so that
\begin{align}
m_{\eta'}^2&=\frac{2N_f}{f_0^2}\chi_{\rm YM}+\cdots .
\end{align}
The ellipses denote corrections from finite quark masses, higher orders in
the chiral expansion and subleading powers of $1/N_c$.

The large-$N_c$ scaling is consistent with this result. One has
\begin{align}
f_0^2&={\cal O}(N_c), \qquad
    \chi_{\rm YM}={\cal O}(1),
\end{align}
and therefore
\begin{align}
m_{\eta'}^2&={\cal O}(1/N_c).
\end{align}
Thus the singlet pseudoscalar becomes massless in the simultaneous chiral
and $N_c\rightarrow\infty$ limits, as expected because the anomalous
contribution to the singlet pseudoscalar mass is suppressed in that limit.

For three light flavors with nonzero quark masses, the ordinary chiral mass
terms generate masses and mixing among the octet and singlet pseudoscalar
states. At leading order, where the singlet and nonsinglet decay constants
may be identified, the anomalous singlet mass contribution may be isolated
through the trace of the $\eta_8$-$\eta_0$ mass matrix, giving
\begin{align}\label{Eq:WV_physical}
m_{\eta'}^2+m_\eta^2-2m_K^2 &= \frac{2N_f}{f_0^2}\chi_{\rm YM} +\cdots ,
\end{align}
where at this order $f_0$ may equivalently be replaced by the common
leading-order pseudoscalar decay constant. The ellipses denote
higher-order chiral and $1/N_c$ corrections, including corrections
associated with the distinction between the singlet and nonsinglet decay
constants. This is the commonly used phenomenological form of the
Witten-Veneziano relation.

It is important to distinguish the pure Yang-Mills susceptibility appearing
in this relation from the susceptibility of full QCD. If at least one quark
is exactly massless, the full QCD vacuum energy is independent of a constant
$\theta$ source and hence
\begin{align}
\chi_{\rm full}&=0.
\end{align}
This result is also reflected in
Eq.~(\ref{Eq:WV_effective_potential}). In the massless effective theory
the singlet field is free to adjust so that
\begin{align}
\frac{\sqrt{2N_f}}{f_0}\eta_0&=\theta,
\end{align}
thereby minimizing the anomalous potential and eliminating its
$\theta$ dependence. In the correlator language, the same result is
expressed by a cancellation between the pure-glue contribution and the
$\eta'$ pole contribution. Schematically,
\begin{align}
\chi_{\rm full}&=\chi_{\rm YM}-\frac{f_0^2m_0^2}{2N_f}+\cdots=0
\end{align}
in the chiral limit, where Eq.~(\ref{Eq:WV_chiral}) has been used. 
Thus one should not identify the full-QCD susceptibility with
$\chi_{\rm YM}$ in the chiral theory. Here the separation into the two
displayed contributions is schematic and is to be understood within the
same consistent prescription for contact and subtraction terms discussed
in Sec.~\ref{Sec:chi_deriv}.

The logical ingredients of the Witten-Veneziano relation are therefore the
anomalous singlet axial Ward identity, low-energy chiral dynamics, a
nonvanishing pure Yang-Mills correlation function $\chi_{\rm YM}$, and the
large-$N_c$ expansion that relates the leading anomalous singlet dynamics to
the pure-glue theory. None of these steps requires a
configuration-by-configuration decomposition of the exact functional
integral into sectors of integer-valued topological charge.

The Witten-Veneziano relation consequently provides strong evidence for
nontrivial correlations of the local topological charge density in pure
Yang-Mills theory and for their physical role in the singlet pseudoscalar
sector. Its phenomenological success does not, by itself, establish that
the full continuum gauge-field configuration space must possess a
fundamental global classification by integer topological charge.

\section{Wick rotation and topological charge}
\label{App:Wick_rot}

The Wick rotation of the topological charge requires some care because the
topological density contains the Levi-Civita tensor and therefore depends
upon the orientation convention adopted for the spacetime manifold. The
familiar rule
\begin{align}
d^4x_M
&\to
-i\,d^4x_E
\end{align}
is a statement about the coordinate Jacobian of the Wick rotation, with
$d^4x_E$ denoting the conventional Euclidean coordinate integration
measure. By itself, however, this rule does not specify how the oriented
four-dimensional volume form is to be identified between Minkowski and
Euclidean space. Keeping these two notions distinct is essential for
understanding why the integrated topological charge is unchanged.

We use the Wick-rotation conventions introduced in Sec.~2.1,
\begin{align}
(x^0)_M &\to -i(x_4)_E \qquad\mbox{and}\qquad (x^i)_M\to (x_i)_E.
\end{align}
Consequently,
\begin{align}
dx^0_M &\to -i\,dx_4 \qquad\mbox{and}\qquad
    d^4x_M=dx^0dx^1dx^2dx^3\to -i\,dx_4dx_1dx_2dx_3 = -i\,d^4x_E,
\end{align}
where the last equality is the usual coordinate-measure convention. We also
have
\begin{align}
(\partial_0)_M &\to i(\partial_4)_E \qquad\mbox{and}\qquad 
    (\partial_i)_M\to (\partial_i)_E.
\end{align}
Requiring the covariant derivative
\begin{align}
D_\mu
&=
\partial_\mu-igA_\mu
\end{align}
to continue consistently gives
\begin{align}\label{Eq:Wick_A}
(A_0)_M &\to i(A_4)_E \qquad\mbox{and}\qquad 
    (A_i)_M \to (A_i)_E.
\end{align}
It follows that the field-strength components transform as
\begin{align}\label{Eq:Wick_F}
(F_{0i})_M &\to i(F_{4i})_E \qquad\mbox{and}\qquad 
    (F_{ij})_M\to (F_{ij})_E.
\end{align}
These relations give the familiar continuation of the Yang-Mills action,
\begin{align}
\textstyle i\int\!d^4x_M\,\left[-\frac12\tr(F_{\mu\nu}F^{\mu\nu})_M\right]
&\to \textstyle -\int\!d^4x_E\,\frac12 \tr(F_{\mu\nu}F_{\mu\nu})_E.
\end{align}

The topological term is subtler because it contains the oriented
four-dimensional Levi-Civita tensor. In Minkowski space we write
\begin{align}
Q_M &= \textstyle[g^2/(16\pi^2)] \int\!d^4x_M\,
    \tr(F_{\mu\nu}\tilde F^{\mu\nu})_M
    = [g^2/(32\pi^2)] \int\!d^4x_M\, (\epsilon^{\mu\nu\rho\sigma})_M
    \tr(F_{\mu\nu}F_{\rho\sigma})_M 
\end{align}
with the convention $\epsilon^{0123}=+1$. The corresponding Euclidean 
quantity is
\begin{align}
Q_E &= \textstyle [g^2/(16\pi^2)] \int\!d^4x_E\,
    \tr(F_{\mu\nu}\tilde F_{\mu\nu})_E
    =[g^2/(32\pi^2)] \int\!d^4x_E\, (\epsilon_{\mu\nu\rho\sigma})_E
    \tr(F_{\mu\nu}F_{\rho\sigma})_E,
\end{align}
with the convention $(\epsilon_{1234})_E=+1$.

The important point is that the coordinate integration measure and the
oriented volume form should not be identified without further thought.
The statement $d^4x_M\rightarrow-i\,d^4x_E$ above refers to the coordinate
Jacobian used in the conventional integration measure. By contrast, the
Minkowski oriented volume form is
\begin{align}
\omega_M &= dx^0\wedge dx^1\wedge dx^2\wedge dx^3.
\end{align}
Under the Wick rotation,
\begin{align}
\omega_M &\to -i\, dx_4\wedge dx_1\wedge dx_2\wedge dx_3.
\end{align}
To compare this with the conventionally oriented Euclidean volume form
\begin{align}
\omega_E
&=
dx_1\wedge dx_2\wedge dx_3\wedge dx_4,
\end{align}
the factor $dx_4$ must be moved through three spatial differentials. Since
each interchange changes the sign of the wedge product,
\begin{align}\label{Eq:Wick_oriented_volume}
dx_4\wedge dx_1\wedge dx_2\wedge dx_3
&=-dx_1\wedge dx_2\wedge dx_3\wedge dx_4
    \qquad\mbox{and so}\qquad 
    \omega_M \to +i\,\omega_E.
\end{align}
There is no contradiction between this result and the coordinate-Jacobian
statement $d^4x_M\to-i\,d^4x_E$. The latter refers to the
conventional coordinate integration measure, whereas
Eq.~(\ref{Eq:Wick_oriented_volume}) refers to the oriented differential
four-form. The Wick rotation maps the Minkowski orientation $(0,1,2,3)$
into $(4,1,2,3)$, which has the opposite orientation to our chosen
Euclidean convention $(1,2,3,4)$. 
Since the Levi-Civita tensor changes sign under reversal of orientation,
expressing the Wick-rotated result in the chosen Euclidean orientation
therefore contributes an additional minus sign beyond the coordinate
Jacobian $d^4x_M\to-i\,d^4x_E$. Equivalently, in carrying out the
component continuation one must also account for
$\epsilon_M\to-\epsilon_E$ when passing from the Wick-rotated orientation
$(4,1,2,3)$ to the chosen Euclidean orientation $(1,2,3,4)$.

Consider, for example,
\begin{align}
&d^4x_M\,(\epsilon^{0123})_M\tr(F_{01}F_{23})_M =
    d^4x_M\,\tr(F_{01}F_{23})_M \\
    \to &(-i\,d^4x_E)\,\tr\!\left[(iF_{41})F_{23}\right]_E
    = d^4x_E\,\tr(F_{41}F_{23})_E = -d^4x_E\,(\epsilon_{4123})_E
    \tr(F_{41}F_{23})_E ,
\end{align}
where in the last expression we have used
$(\epsilon_{4123})_E=-1$. We have \emph{not} yet accounted for the additional
sign change, $\epsilon_M\to -\epsilon_E$, associated with reversing the 
orientation from $(4,1,2,3)$ to the 
chosen Euclidean orientation $(1,2,3,4)$. Including this orientation
sign and summing over all permutations of the indices gives
\begin{align}
d^4x_M\,(\epsilon^{\mu\nu\rho\sigma})_M \tr(F_{\mu\nu}F_{\rho\sigma})_M
    &\to -d^4x_E\,(-\epsilon_{\mu\nu\rho\sigma})_E
    \tr(F_{\mu\nu}F_{\rho\sigma})_E
    = d^4x_E\,(\epsilon_{\mu\nu\rho\sigma})_E
    \tr(F_{\mu\nu}F_{\rho\sigma})_E ,
\end{align}
and hence
\begin{align}
Q_M&\to Q_E.
\end{align}

The same result has a particularly compact geometrical formulation in
terms of differential forms. This also provides a useful independent way
of understanding the orientation bookkeeping in the component derivation.
Writing the gauge connection as the one-form
\begin{align}
A
&=
A_\mu dx^\mu,
\end{align}
Eq.~(\ref{Eq:Wick_A}) gives
\begin{align}
(A_0)_Mdx_M^0
&\to
[i(A_4)_E][-i\,dx_4]
=
(A_4)_Edx_4,
\end{align}
while the spatial components are unchanged. Thus the gauge connection
one-form itself continues smoothly,
\begin{align}
A_M &\to A_E.
\end{align}
Since the field-strength two-form is constructed from the connection,
\begin{align}
F &= dA-igA\wedge A,
\end{align}
it follows directly that
\begin{align}
F_M &\to F_E
\end{align}
as a two-form and therefore
\begin{align}
\tr(F_M\wedge F_M) &\to \tr(F_E\wedge F_E).
\end{align}
With
\begin{align}
F &= \textstyle \frac12 F_{\mu\nu}\,dx^\mu\wedge dx^\nu,
\end{align}
the topological charge can be written as
\begin{align}
Q &= \frac{g^2}{8\pi^2} \int\!\tr(F\wedge F).
\end{align}
The smooth continuation of the four-form $\tr(F\wedge F)$, together with
the consistent continuation of the orientation of the integration
manifold, therefore gives immediately
\begin{align}\label{Eq:EandM_top_Q}
Q_M &\to Q_E.
\end{align}
The differential-form expression does not eliminate the orientation issue.
Rather, it incorporates the orientation directly into the definition of
the integral of the four-form. Equation~(\ref{Eq:Wick_oriented_volume})
makes explicit the same orientation bookkeeping that is otherwise encoded
implicitly in the compact differential-form notation. Thus the component
and differential-form derivations are two equivalent descriptions of the
same analytic continuation.

It is worth emphasizing that Eq.~(\ref{Eq:EandM_top_Q}) follows from the
analytic continuation itself and does not require $Q$ to be integer valued.
For sufficiently smooth gauge fields satisfying the additional boundary,
asymptotic, and global conditions discussed in the main text, the common
value may further be identified as an integer topological invariant,
$Q_M=Q_E\in\mathbb Z$.
That integer quantization is an additional global statement and should not
be confused with the equality of the Minkowski and Euclidean spacetime
integrals under Wick rotation.

\section{Relative contributions of the topological sectors}
\label{Sec:ZQ_contrib}

Within the conventional formulation, the Euclidean functional integral is
decomposed into sectors of integer topological charge,
\begin{align}\label{Eq:ZQ_def}
Z_E(\theta) &= \textstyle \sum_{Q\in\mathbb Z}\, e^{i\theta Q}Z_E^Q,
\end{align}
where
\begin{align}
Z_E^Q
& = \textstyle \int_{Q[\Phi]=Q}\!{\cal D}\Phi\,e^{-S_E[\Phi]}.
\end{align}
Here $\Phi$ denotes the dynamical fields. The purpose of this appendix is to
determine how the relative contributions $Z_E^Q$ are distributed over $Q$
in a large but finite Euclidean four-volume $V$, assuming this integer-sector
decomposition. Relevant discussions may be found, for example, in
Refs.~\cite{DiGiacomo:1992df,Rossi:1980cf,Sharpe:2006vb,Gasser:1987ah}.

At $\theta=0$ we have
\begin{align}
Z_E(0) &= \textstyle \sum_{Q\in\mathbb Z}Z_E^Q,
\end{align}
so that
\begin{align}\label{Eq:PQ_def}
P(Q)&\equiv Z_E^Q/Z_E(0)
\end{align}
is the probability weight associated with the sector of charge $Q$ within
this conventional decomposition.

As discussed in Sec.~\ref{Sec:top_suscept}, $CP$ invariance at $\theta=0$
implies
\begin{align}
\braket{\hat Q} &= 0,
\end{align}
while translational invariance gives
\begin{align}\label{Eq:Q_variance_again}
\braket{\hat Q^2} &= \chi V,
\end{align}
where $\chi$ is the topological susceptibility at $\theta=0$. Equivalently,
the vacuum energy density has the small-$\theta$ expansion
\begin{align}\label{Eq:varepsilon_theta_expand_Q}
\varepsilon_0(\theta) &= \varepsilon_0(0)
    +\frac12\chi\theta^2+{\cal O}(\theta^4).
\end{align}
Since
\begin{align}
Z_E(\theta)
&=
\exp[-V\varepsilon_0(\theta)]
\end{align}
up to an overall $\theta$-independent normalization, it follows that
\begin{align}\label{Eq:Z_gauss_theta}
\frac{Z_E(\theta)}{Z_E(0)}&=\exp\left[-\frac12\chi V\theta^2
    +{\cal O}(V\theta^4)\right].
\end{align}

Because $Q$ is assumed to take integer values, Eq.~(\ref{Eq:ZQ_def}) is a
Fourier series and $Z_E(\theta)$ is correspondingly $2\pi$ periodic within
this conventional construction. The exact inverse relation is therefore
\begin{align}\label{Eq:ZQ_inverse}
Z_E^Q&=\int_{-\pi}^{\pi}\frac{d\theta}{2\pi}\,e^{-i\theta Q}Z_E(\theta).
\end{align}
For $\chi V\gg1$, the factor $Z_E(\theta)/Z_E(0)$ becomes sharply
concentrated near $\theta=0$, with characteristic width
\begin{align}
|\theta| &\sim 1/\sqrt{\chi V}.
\end{align}
The quadratic term in Eq.~(\ref{Eq:Z_gauss_theta}) therefore gives the
leading large-volume contribution in the central region of the
distribution. 
Provided that $\theta=0$ is the unique relevant minimum of the vacuum
energy in the fundamental interval, the integration limits may then be
extended from $[-\pi,\pi]$ to $(-\infty,\infty)$ with exponentially small
error at this order, giving
\begin{align}\label{Eq:ZQ_gaussian}
Z_E^Q&\simeq Z_E(0)\int_{-\infty}^{\infty}\frac{d\theta}{2\pi}\,
    \exp\left[-\frac12\chi V\theta^2-i\theta Q\right]
    =\frac{Z_E(0)}{\sqrt{2\pi\chi V}}\,
    \exp\left[-\frac{Q^2}{2\chi V}\right].
\end{align}
Thus
\begin{align}\label{Eq:PQ_gauss}
P(Q)&=\frac{Z_E^Q}{Z_E(0)}\simeq\frac{1}{\sqrt{2\pi\chi V}}\,
    \exp\left[-\frac{Q^2}{2\chi V}\right],
    \qquad\chi V\gg1.
\end{align}
The normalization is understood in the large-volume approximation in which
the sum over integer $Q$ may be replaced by an integral.

Equation~(\ref{Eq:PQ_gauss}) immediately reproduces
\begin{align}
\braket{\hat Q^2}&=\chi V,
\end{align}
and shows that the characteristic magnitude of the global topological
charge is
\begin{align}\label{Eq:Q_typical}
|Q|_{\rm typical}&\sim\sqrt{\chi V}.
\end{align}
Consequently,
\begin{align}
|Q|_{\rm typical}/V&\sim\sqrt{\chi/V}
    \xrightarrow{V\rightarrow\infty}0.
\end{align}
Thus the distribution of the total charge becomes broader as
$\sqrt{V}$, while the typical charge density tends to zero as
$1/\sqrt{V}$.

At the same time, the probability associated with any fixed finite value
of $Q$ tends to zero as the volume increases. In particular,
\begin{align}
P(0)&\simeq 1/\sqrt{2\pi\chi V}\xrightarrow{V\rightarrow\infty} 0.
\end{align}
This does not mean that the $Q=0$ sector disappears. Rather, an increasing
number of sectors contribute appreciably as the width of the distribution
grows with $\sqrt{V}$.

The Gaussian form is the leading large-volume result in the central region
$Q={\cal O}(\sqrt{V})$. 
The Gaussian form is the leading large-volume result in the central region
$Q={\cal O}(\sqrt{V})$. Indeed, the values of $\theta$ relevant to the
Fourier integral are then of order $V^{-1/2}$, so that a term
$Vc_4\theta^4$ in the exponent is of order $1/V$.
Higher terms in the expansion of the vacuum
energy,
\begin{align}
\varepsilon_0(\theta) &= \textstyle \varepsilon_0(0)
    +\frac12\chi\theta^2+c_4\theta^4+\cdots,
\end{align}
encode higher cumulants of the topological-charge distribution and generate
non-Gaussian corrections. These become relevant particularly away from the
central region of the distribution.

It should finally be emphasized that the entire analysis in this appendix is
conditional on the conventional global decomposition
\begin{align}
Z_E(\theta) &= \textstyle \sum_{Q\in\mathbb Z} \, e^{i\theta Q}Z_E^Q.
\end{align}
The relation $\braket{\hat Q^2}=\chi V$ and the correlation-function
definition of $\chi$ do not themselves require such a sector decomposition.
What has been derived here is therefore the distribution of the relative
sector weights conditional upon the assumption of a global decomposition
into integer-valued topological sectors.

\section{Chiral symmetry breaking and the $U(1)_A$ problem}
\label{App:chiral_symm_U(1)}

Here we review the related issues of chiral symmetry, its explicit and
spontaneous breaking, the $U(1)_A$ problem in the strong interactions,
instantons and topology, and the role of instantons in the conventional
semiclassical description of anomalous $U(1)_A$ dynamics. We do this
because it is important to understand whether the role of topology in this
context has any necessary implications for our understanding of the strong
$CP$ problem.

In QCD, \emph{chiral symmetry} is a symmetry of the QCD Lagrangian density
in the limit in which the relevant quark masses vanish. It is a flavor
symmetry that acts independently on the left-handed and right-handed
components of the quark fields. Consider the QCD Lagrangian density in
Minkowski space in \Eq{Eq:QCDLag}. A Dirac quark field $\psi_f$ with flavor
$f$ can be decomposed into left-handed and right-handed components using
the projectors $P_L$ and $P_R$, respectively,
\begin{align}
\psi_{fL}
&=
P_L\psi_f \equiv\textstyle  \frac12(1-\gamma_5)\psi_f,
\qquad \psi_{fR} = P_R\psi_f \equiv\frac12(1+\gamma_5)\psi_f .
\end{align}
The Minkowski-space QCD Lagrangian density in terms of left- and
right-handed quark fields is
\begin{align}\label{Eq:L_M_LR}
\mathcal{L}_M&=\textstyle -\frac14 F_{\mu\nu}^aF^{a\mu\nu}
    +\sum_{f=1}^{N_f}\left\{\bar{\psi}_{fL}i\gamma^\mu D_\mu\psi_{fL}
    +\bar{\psi}_{fR}i\gamma^\mu D_\mu\psi_{fR}
    -m_f\left(\bar{\psi}_{fL}\psi_{fR}+\bar{\psi}_{fR}\psi_{fL}\right)\right\},
\end{align}
since
\begin{align}
\bar\psi_f\psi_f&=\bar\psi_{fL}\psi_{fR}+\bar\psi_{fR}\psi_{fL}.
\end{align}
We see that only the mass term mixes the left-handed and right-handed quark
fields, while the two chiralities couple to the gluon field in exactly the
same way.

When the $N_f$ quark masses under consideration vanish, we have
\begin{align}\label{Eq:M_L_LR_m=0}
\mathcal{L}_M^{m=0}&=\textstyle -\frac14 F_{\mu\nu}^aF^{a\mu\nu}
    + \sum_{f=1}^{N_f}\left[\bar{\psi}_{fL}i\gamma^\mu D_\mu\psi_{fL}
    +\bar{\psi}_{fR}i\gamma^\mu D_\mu\psi_{fR}\right].
\end{align}
There is then no term mixing $\psi_{fL}$ with $\psi_{fR}$, and the
Lagrangian density is invariant under independent unitary transformations
in flavor space,
\begin{align}\label{Eq:UL_UR}
\psi_L&\to U_L\psi_L,\qquad
    \psi_R\to U_R\psi_R,\qquad U_L,U_R\in U(N_f).
\end{align}
Formally, if all six quark masses vanished one could take $N_f=6$.
For low-energy QCD, however, the useful approximate chiral symmetries
involve the light flavors, in particular $N_f=2$ or $N_f=3$.

The full classical flavor symmetry in the massless theory is therefore
\begin{align}
U(N_f)_L\times U(N_f)_R.
\end{align}
Globally, each $U(N_f)$ factor satisfies
\begin{align}
U(N_f)&\cong\frac{SU(N_f)\times U(1)}{\mathbb Z_{N_f}},
\end{align}
so that
\begin{align}\label{Eq:UxU=etc}
U(N_f)_L\times U(N_f)_R
    &\cong \frac{SU(N_f)_L\times U(1)_L}{\mathbb Z_{N_f}}
    \times\frac{SU(N_f)_R\times U(1)_R}{\mathbb Z_{N_f}}
    \nonumber\\
&\xrightarrow{\mbox{near the identity}}
    SU(N_f)_L\times SU(N_f)_R\times U(1)_V\times U(1)_A .
\end{align}
The symmetry associated with baryon-number conservation is $U(1)_V$,
under which left- and right-handed quarks transform with the same phase.
The singlet axial symmetry $U(1)_A$ instead acts with opposite phases on
the two chiralities. The $U(1)_A$ symmetry is broken in the quantum theory
by the axial anomaly
\citep{Adler:1969er,Adler:1969gk,Bell:1969ts,Fujikawa:1980eg}.
The non-Abelian symmetry
$SU(N_f)_L\times SU(N_f)_R$ is referred to as the chiral flavor symmetry.
These are global symmetries: the corresponding transformations are
spacetime independent and are not gauge transformations.

\emph{Brief explanation:} Consider the defining $N\times N$ matrix
representation of $U(N)$ and $SU(N)$, with $I$ the identity matrix.
Any $U\in U(N)$ can be written as
\begin{align}
U
&=
e^{i\phi}S,
\qquad
S\in SU(N),
\end{align}
where $\phi$ is chosen so that
\begin{align}
e^{iN\phi}
&=
\det U.
\end{align}
The choice of $\phi$ is not unique: it is defined modulo $2\pi/N$. This
nonuniqueness reflects the common subgroup $\mathbb Z_N$ of $SU(N)$ and
$U(1)$ and gives
\begin{align}
U(N)
&\cong
\frac{SU(N)\times U(1)}{\mathbb Z_N}.
\end{align}
Thus the quotient simply avoids double counting the common center
elements.

For the two chiral factors, write the Abelian transformations as
$e^{i\phi_L}I_L\in U(1)_L$ and
$e^{i\phi_R}I_R\in U(1)_R$. Define
\begin{align}
\phi_V&=\textstyle \frac12(\phi_L+\phi_R),
    \qquad \phi_A=\frac12(\phi_L-\phi_R),
\end{align}
or equivalently
\begin{align}
\phi_L
&=
\phi_V+\phi_A,
\qquad
\phi_R
=
\phi_V-\phi_A.
\end{align}
Then the Abelian transformations may be written as a product of a vector
transformation and an axial transformation,
\begin{align}
(e^{i\phi_L}I_L)(e^{i\phi_R}I_R)
&=
\left[
(e^{i\phi_V}I_L)(e^{i\phi_V}I_R)
\right]
\left[
(e^{i\phi_A}I_L)(e^{-i\phi_A}I_R)
\right].
\end{align}
Near the identity the discrete quotient structure plays no role in the
local Lie-algebra analysis, which is why the second line of
\Eq{Eq:UxU=etc} is normally used in perturbative and local discussions.

The presence of quark masses in $\mathcal{L}_M$ in
\Eq{Eq:L_M_LR} does not affect the baryon-number symmetry $U(1)_V$,
because the left- and right-handed fields transform identically under this
symmetry. Transformations acting differently on the two chiralities are
instead broken by the mass term. Thus nonzero quark masses explicitly break
$U(1)_A$ and the non-Abelian chiral flavor symmetry
$SU(N_f)_L\times SU(N_f)_R$. Breaking produced directly by terms in the
Lagrangian density is referred to as \emph{explicit symmetry breaking};
the quark masses therefore produce explicit chiral symmetry breaking.

When electromagnetic and weak interactions are included, the chiral flavor
symmetry is further reduced. The different electromagnetic charges of the
quark flavors explicitly break part of the non-Abelian flavor symmetry,
while singlet axial currents also receive the familiar electromagnetic
anomaly. The weak interactions distinguish left- and right-handed fields
and therefore also break the chiral flavor symmetry. These effects are
small compared with the strong interaction at typical hadronic scales, so
the light-quark sector retains a useful approximate chiral symmetry.

In the physical world the quark masses arise from electroweak symmetry
breaking. After diagonalization of the Yukawa matrices,
\begin{align}
m_f &= v\lambda_f/\sqrt{2},
\end{align}
where $v\simeq246~\mbox{GeV}$ is the Higgs vacuum expectation value and
$\lambda_f$ is the corresponding Yukawa coupling. The $u$ and $d$ quark
masses are only a few MeV in the $\overline{\mbox{MS}}$ scheme at
$\mu=2~\mbox{GeV}$ and are therefore small compared with the characteristic
QCD scale, $\Lambda_{\rm QCD}\sim300~\mbox{MeV}$. Consequently,
$SU(2)_L\times SU(2)_R$ is a good approximate chiral symmetry of the
$(u,d)$ sector. The strange quark mass is of order $100~\mbox{MeV}$ at the
same scale, so that $SU(3)_L\times SU(3)_R$ is a less accurate but still
useful approximate chiral symmetry of the $(u,d,s)$ sector.

In the infrared, QCD spontaneously breaks the non-Abelian chiral symmetry
according to
\begin{align}\label{Eq:SULxSURtoSUV}
SU(N_f)_L\times SU(N_f)_R
&\longrightarrow
SU(N_f)_V
\end{align}
through the formation of a quark condensate,
\begin{align}
\langle\bar\psi\psi\rangle
&\ne
0.
\end{align}
This is referred to as \emph{spontaneous symmetry breaking}, or sometimes
as \emph{dynamical symmetry breaking}, because it arises from the internal
dynamics of the theory rather than from symmetry-violating terms in the
Lagrangian density. Expressed another way, the vacuum state does not possess
a symmetry that is present in the Lagrangian density.

Spontaneous breaking of a continuous global symmetry produces massless
Goldstone bosons. If a continuous symmetry group $G$ is spontaneously
broken to a subgroup $H$, the number of Goldstone bosons is the number of
broken generators,
\begin{align}
N_{\rm GB}
&=
\dim(G)-\dim(H).
\end{align}
In QCD the quark masses explicitly break the chiral symmetry, so the
corresponding states acquire nonzero masses and are therefore
\emph{pseudo-Goldstone bosons}.

\emph{Brief explanation:} The spontaneous formation of a quark condensate
means that
\begin{align}
\braket{\hat{\bar\psi}\hat\psi}
&\equiv
\braket{\Omega|\hat{\bar\psi}\hat\psi|\Omega}
\ne
0,
\end{align}
where $\ket{\Omega}$ is the nonperturbative QCD vacuum state. As we saw
above,
\begin{align}
\bar\psi_f\psi_f
&=
\bar\psi_{fL}\psi_{fR}
+
\bar\psi_{fR}\psi_{fL},
\end{align}
so the condensate correlates left- and right-handed fields.

Let $T_L^a$ and $T_R^a$ denote the generators of
$SU(N_f)_L$ and $SU(N_f)_R$. It is useful to introduce the vector and
axial combinations
\begin{align}
T_V^a
&=
\frac12(T_L^a+T_R^a),
\qquad
T_A^a
=
\frac12(T_L^a-T_R^a).
\end{align}
The vector generators leave the condensate invariant, whereas the axial
generators do not. The axial combinations should not, however, be regarded
as generators of an independent group $SU(N_f)_A$ forming a direct product
with $SU(N_f)_V$: for non-Abelian groups their commutators satisfy
schematically
\begin{align}
[T_A^a,T_A^b]
&\propto
if^{abc}T_V^c,
\end{align}
so the axial generators do not form a closed Lie subalgebra by themselves.
The symmetry-breaking pattern is therefore simply
\begin{align}\label{Eq:SULxSURtoSUV_again}
SU(N_f)_L\times SU(N_f)_R
&\longrightarrow
SU(N_f)_V .
\end{align}
The number of broken generators is
\begin{align}
2(N_f^2-1)-(N_f^2-1)
&=
N_f^2-1,
\end{align}
and therefore the exact chiral limit contains $N_f^2-1$ Goldstone bosons.
Because the broken generators are axial combinations, these Goldstone
bosons are pseudoscalars.

For the three light flavors $(u,d,s)$, approximate
$SU(3)_L\times SU(3)_R$ chiral symmetry therefore gives rise to eight
pseudoscalar pseudo-Goldstone bosons. These are identified with the
pseudoscalar octet
\begin{align}
\{
\pi^+,\pi^0,\pi^-,
K^+,K^0,\bar K^0,K^-,
\eta_8
\}.
\end{align}
For example, $m_{\pi^\pm}\simeq140~\mbox{MeV}$ and
$m_{K^\pm}\simeq494~\mbox{MeV}$. The pions are lighter than the kaons
because the $u$ and $d$ quark masses are much smaller than the strange
quark mass, so explicit chiral symmetry breaking is weaker in the
$(u,d)$ sector. In the exact $SU(3)$ chiral limit the pseudoscalar octet
would become massless and degenerate.

For the light quarks, a typical value of the renormalized condensate is of
order
\begin{align}
\braket{\hat{\bar\psi}_f\hat\psi_f}
&\simeq
-(240~\mbox{MeV})^3
\end{align}
at $\mu=2~\mbox{GeV}$ in the $\overline{\mbox{MS}}$ scheme, with the
precise numerical value depending on the renormalization convention and
scale.

Naively one might then ask why there is not an additional light
flavor-singlet pseudoscalar meson associated with spontaneous breaking of
$U(1)_A$. This is the traditional $U(1)_A$ problem. The essential point is
that $U(1)_A$ is not an exact symmetry of the quantum theory: it is broken
by the axial anomaly. Consequently there is no additional Goldstone boson
associated with $U(1)_A$, even when the quark masses vanish.

For three light flavors, the flavor-octet state $\eta_8$ and the
flavor-singlet state, conventionally denoted $\eta_1$ or $\eta_0$, mix once
flavor symmetry is broken. The resulting physical states are the $\eta$
and $\eta'$, with masses
\begin{align}
m_\eta
&\simeq
548~\mbox{MeV},
\qquad
m_{\eta'}
\simeq
958~\mbox{MeV}.
\end{align}
The anomalous singlet dynamics generates a substantial additional mass for
the flavor-singlet component, while $\eta_8$-$\eta_1$ mixing determines
the composition of the physical $\eta$ and $\eta'$ states. The unusually
large $\eta'$ mass is therefore fundamentally associated with the anomalous
$U(1)_A$ dynamics rather than being generated simply by octet-singlet
mixing. This understanding is supported by lattice QCD studies of the
pseudoscalar meson sector, for example Ref.~\citep{PhysRevD.83.111502}
and references therein.

\subsection*{Instantons in QCD and the $U(1)_A$ problem}
\label{Sec:instantons}

The resolution of the $U(1)_A$ problem is conceptually independent of the
value assigned to the parameter $\theta$. The $U(1)_A$ problem concerns
the absence of an additional light pseudoscalar Goldstone boson associated
with the axial $U(1)_A$ symmetry of the classical massless theory. Its
resolution relies on the fact that $U(1)_A$ is not a symmetry of the
quantum theory. It is broken by the axial anomaly, which is expressed
locally by the anomalous Ward identity and, in the functional-integral
formulation, by the non-invariance of the fermion measure under singlet
axial rotations.

In the conventional semiclassical picture, the anomalous $U(1)_A$
dynamics may be exhibited explicitly using smooth gauge-field
configurations with nonzero topological charge, such as instantons
\citep{SchaferShuryak1998}. In an instanton background the Dirac operator
possesses chiral zero modes whose difference is related to the topological
charge by the index theorem. Integrating over the fermions in such a
background produces an effective multi-fermion interaction, the so-called
{'t}~Hooft vertex, involving $2N_f$ fermion fields and carrying nonzero
axial charge \citep{Belavin:1975fg,tHooft:1976rip,tHooft:1976snw}. 
The {'t}~Hooft interaction
therefore violates $U(1)_A$ while preserving the non-anomalous vector and
chiral flavor symmetries. It provides an important semiclassical
realization of the anomalous singlet dynamics responsible for lifting the
would-be flavor-singlet Goldstone mode.

The same anomalous singlet physics is reflected, from a different
perspective, in the Witten-Veneziano relation
\citep{Witten:1979vv,Veneziano:1979ec}, discussed in
Appendix~\ref{App:WV_relation}. That relation connects the anomalous
singlet pseudoscalar mass to the pure Yang-Mills topological susceptibility
in the large-$N_c$ limit. It should not be interpreted as requiring that
the $\eta'$ mass be generated specifically by a dilute instanton gas; the
Witten-Veneziano relation depends on the nonvanishing pure Yang-Mills
topological susceptibility and anomalous singlet dynamics more generally.

The generation of the $\eta'$ mass therefore arises from nontrivial
singlet dynamics together with the anomalous Ward identity, but it does
not logically require that the exact QCD functional integral be decomposed
configuration by configuration into integer-valued topological sectors.
The phenomenological success of the anomalous $U(1)_A$ mechanism supports
the existence of nontrivial correlations of the local topological charge
density, as encoded in the topological susceptibility
\begin{align}
\chi &= \textstyle \int\!d^4x_E\,\braket{\hat q_E(x)\hat q_E(0)}.
\end{align}
However, this does not by itself establish the stronger global statement
that a minimal definition of the full quantum functional integral must be
fundamentally decomposed into integer-$Q$ sectors.

Crucially, the anomalous $U(1)_A$ dynamics and the resulting singlet
pseudoscalar physics are already present at $\theta=0$. Their existence
therefore does not require a nonzero physical value of $\theta$, nor does
the successful resolution of the $U(1)_A$ problem by itself settle the
separate question of which global assumptions are required in formulating
the strong $CP$ problem.

\section{Large gauge transformations and the Chern-Simons number}
\label{App:ChernSimons}

Large gauge transformations and the associated Chern-Simons number play a
central role in the conventional discussion of the strong $CP$ problem.
However, several distinct concepts are often conflated in the literature.
The purpose of this appendix is therefore to review carefully the
relationship between gauge transformations, the Chern-Simons functional,
the local topological charge density and the topological charge.

Throughout this review we distinguish between gauge transformations defined
on a spatial hypersurface and gauge transformations acting on the full
four-dimensional spacetime. This distinction is essential because the
Chern-Simons functional is intrinsically a three-dimensional quantity,
whereas the topological charge density is a local four-dimensional operator.

The local topological charge density is
\begin{align}\label{Eq:qApp}
q(x) &= [g^2/(16\pi^2)]
\tr(F_{\mu\nu}\tilde F^{\mu\nu}),
\end{align}
which is manifestly gauge invariant since
\begin{align}
F_{\mu\nu}
&\rightarrow
UF_{\mu\nu}U^\dagger .
\end{align}
Consequently,
\begin{align}
q(x)
&\rightarrow
q(x)
\end{align}
pointwise throughout spacetime.

The corresponding spacetime integral is
\begin{align}\label{Eq:QApp}
Q &= \textstyle \int\!d^4x\,q(x).
\end{align}
Since $q(x)$ is pointwise gauge invariant, its spacetime integral $Q$ is
unchanged under any gauge transformation that is well defined on the
integration domain and preserves the boundary or asymptotic conditions
required to define the integral. This conclusion does not require any
reference to winding numbers or homotopy groups. In particular, a globally
defined four-dimensional gauge transformation does not change the value
of $Q$.

A different quantity arises upon writing, for sufficiently smooth gauge
fields,
\begin{align}\label{Eq:Current}
q(x)
&=
\partial_\mu K^\mu(x),
\end{align}
where
\begin{align}\label{Eq:CSCurrent}
K^\mu
&=
\frac{g^2}{16\pi^2}
\epsilon^{\mu\nu\rho\sigma}
\left[
A_\nu^a\partial_\rho A_\sigma^a
+
\frac{g}{3}f^{abc}
A_\nu^aA_\rho^bA_\sigma^c
\right],
\end{align}
is the Chern-Simons current. Unlike $q(x)$, the current $K^\mu$ is not gauge
invariant.

Restricting to a fixed time slice, one defines the Chern-Simons functional
\begin{align}\label{Eq:CSNumber}
N_{\rm CS}(t)
&=
\int\!d^3x\,K^0(\mathbf{x},t).
\end{align}
Under gauge transformations continuously connected to the identity,
$N_{\rm CS}$ changes by a surface term. With the conventional asymptotic
conditions imposed on both the gauge fields and the gauge transformations,
this surface term vanishes.
Under a gauge transformation on the spatial hypersurface having
nontrivial winding number
\begin{align}
n
&\in
\pi_3(SU(3))
=
\mathbb Z,
\end{align}
however, the Chern-Simons functional changes according to
\begin{align}\label{Eq:CSShift}
N_{\rm CS}
&\rightarrow
N_{\rm CS}+n .
\end{align}
Thus the Chern-Simons functional is not itself gauge invariant. Rather, its
value changes by an integer under a large gauge transformation.

This integer shift should not be confused with the behavior of the local
density $q(x)$ or of its four-dimensional spacetime integral $Q$. The
Chern-Simons functional is a gauge-dependent three-dimensional quantity
associated with a particular spatial hypersurface, whereas $Q$ is a
four-dimensional spacetime integral of a local gauge-invariant density.
Thus the integer shift of $N_{\rm CS}$ is a statement about how a
gauge-dependent functional on a single spatial hypersurface transforms.
It is not a statement that the four-dimensional gauge-invariant quantity
$Q$ changes under the corresponding globally defined gauge transformation.

For sufficiently smooth gauge fields satisfying the usual asymptotic
conditions, the spacetime integral of $q(x)$ may nevertheless be related to
the change in the Chern-Simons functional between two spatial hypersurfaces.
Integrating Eq.~(\ref{Eq:Current}) over spacetime gives boundary
contributions from both temporal and spatial infinity. If the spatial
boundary term vanishes under the imposed asymptotic conditions, the
remaining contribution is
\begin{align}\label{Eq:QDifference}
Q
&=
N_{\rm CS}(+\infty)
-
N_{\rm CS}(-\infty).
\end{align}
If the gauge field approaches appropriate pure-gauge configurations at
temporal infinity, then under these additional assumptions the difference
may be identified with an integer winding number, giving the familiar result
\begin{align}
Q &\in \mathbb Z.
\end{align}
The important point is that the integer shift of $N_{\rm CS}$ under a
large gauge transformation on a spatial hypersurface should not be
interpreted as a change of the four-dimensional quantity $Q$ under a gauge
transformation. The former concerns a gauge-dependent three-dimensional
functional, whereas the latter is the spacetime integral of the
gauge-invariant local density $q(x)$. Their relation requires the additional
smoothness and asymptotic conditions stated above. The role and
applicability of these assumptions in the full quantum theory are discussed
in the main text.

\section{Hamiltonian from the Legendre transform in QCD}
\label{App:H_QCD_deriv}

With conventions consistent with Eq.~\eqref{Eq:CanonicalHamiltonian},
the Yang-Mills Lagrangian density may be written as
\begin{align}
\mathcal L_{\rm YM} &= \textstyle \frac12 E^{ai}E^{ai}
    -\frac12 B^{ai}B^{ai},
\end{align}
where
\begin{align}
E^{ai} &= \dot A_i^a-(D_iA_0)^a \qquad\mbox{and}\qquad
   B^{ai} = \frac12\epsilon^{ijk}F_{jk}^a 
   =\epsilon^{ijk}\left(\partial_jA_k^a +\frac{g}{2}f^{abc}A_j^bA_k^c
    \right).
\end{align}
Thus
\begin{align}
\dot A_i^a
&=
E^{ai}+(D_iA_0)^a,
\end{align}
and the momentum conjugate to $A_i^a$ is
\begin{align}
\Pi^{ai}
&=
\frac{\partial\mathcal L}{\partial\dot A_i^a}
=
E^{ai}.
\end{align}
The Yang-Mills contribution to the canonical Hamiltonian density is
therefore
\begin{align}
\mathcal H_{\rm YM}
&=
E^{ai}\dot A_i^a-\mathcal L_{\rm YM}
    =E^{ai}\left[E^{ai}+(D_iA_0)^a\right]-\frac12 E^{ai}E^{ai}
    +\frac12 B^{ai}B^{ai}
    \nonumber\\
    &=
    \frac12\left(E^{ai}E^{ai}+B^{ai}B^{ai}\right)
    +E^{ai}(D_iA_0)^a.
\end{align}

Covariant integration by parts follows directly from the antisymmetry
of the structure constants. In particular,
\begin{align}
E^{ai}(D_iA_0)^a &=\partial_i\left(E^{ai}A_0^a\right)
    -A_0^a(D_iE^i)^a.
\end{align}
Using
\begin{align}
(D_iA_0)^a&=\partial_iA_0^a+gf^{abc}A_i^bA_0^c,
\end{align}
one has
\begin{align}
E^{ai}(D_iA_0)^a&=E^{ai}\partial_iA_0^a+gf^{abc}E^{ai}A_i^bA_0^c
    \nonumber\\
    &=
    \partial_i\left(E^{ai}A_0^a\right)-(\partial_iE^{ai})A_0^a
    +gf^{abc}E^{ai}A_i^bA_0^c.
\end{align}
Interchanging the dummy color indices $a$ and $c$ in the last term
and using the antisymmetry of $f^{abc}$ gives
\begin{align}
gf^{abc}E^{ai}A_i^bA_0^c&=-gf^{abc}A_i^bE^{ci}A_0^a.
\end{align}
Consequently,
\begin{align}
E^{ai}(D_iA_0)^a&=\partial_i\left(E^{ai}A_0^a\right)
    -A_0^a\left[\partial_iE^{ai}+gf^{abc}A_i^bE^{ci}\right]
    \nonumber\\
    &=\partial_i\left(E^{ai}A_0^a\right)-A_0^a(D_iE^i)^a.
\end{align}
Thus, after integration over space,
\begin{align}
\textstyle \int\! d^3x\,E^{ai}(D_iA_0)^a &=\textstyle 
    -\int \! d^3x\,A_0^a(D_iE^i)^a+\int\! dS_i\,E^{ai}A_0^a.
\end{align}
Subject to boundary conditions for which the surface term vanishes,
this gives
\begin{align}
\textstyle\int d^3x\,E^{ai}(D_iA_0)^a &=
    \textstyle-\int d^3x\,A_0^a(D_iE^i)^a.
\end{align}
Thus ``covariant integration by parts'' is simply ordinary integration
by parts together with the antisymmetry of the structure constants.

Hence
\begin{align}
H_{\rm YM}&=\textstyle \int d^3x\left[
\frac12\left(E^{ai}E^{ai}+B^{ai}B^{ai}\right)
    -A_0^a(D_iE^i)^a\right].
\end{align}
The fermion contribution supplies the remaining part of the Gauss
generator. With
\begin{align}
D_0&=\partial_0-igA_0^aT^a,
\end{align}
the fermionic contribution to the Hamiltonian contains
\begin{align}
-gA_0^a\psi^\dagger T^a\psi.
\end{align}
Combining the two $A_0^a$ terms gives
\begin{align}
-A_0^a\left[(D_iE^i)^a+g\psi^\dagger T^a\psi\right]&=-A_0^aG^a,
\end{align}
where
\begin{align}
G^a&=(D_iE^i)^a+g\psi^\dagger T^a\psi.
\end{align}
So the Legendre transformation produces $A_0^a$ multiplying
precisely the Gauss expression.

There is a further important point in the Dirac constraint analysis.
Since the Lagrangian contains no time derivative of $A_0^a$, its
conjugate momentum vanishes,
\begin{align}
\Pi_0^a&=\frac{\partial\mathcal L}{\partial\dot A_0^a}=0,
\end{align}
which is the primary constraint. Gauss's law is the Euler-Lagrange
equation obtained by varying the original action with respect to the
nondynamical field $A_0^a$,
\begin{align}
\frac{\delta S}{\delta A_0^a}=0
\qquad\Longrightarrow\qquad
G^a=0.
\end{align}
Thus Gauss's law is not an additional condition imposed on the theory,
but an equation of motion that appears in the Hamiltonian formulation
as a constraint. Denoting the full canonical Hamiltonian by $H$, the 
total Hamiltonian is defined as
\begin{align}
H_T&=H+\int d^3x\,u^a\Pi_0^a,
\end{align}
where $u^a(\mathbf x)$ is an arbitrary Lagrange multiplier for the primary
constraint $\Pi^a_0(\mathbf x)=0$. One then obtains
\begin{align}
\dot\Pi_0^a&=\{\Pi_0^a,H_T\}=-\frac{\delta H}{\delta A_0^a}
    =G^a\approx0.
\end{align}
Gauss's law therefore emerges as the secondary constraint. Schematically, 
the logic is
\begin{align}
\text{absence of }\dot A_0^a
\quad&\Longrightarrow\quad
\Pi_0^a=0
\qquad\text{(primary constraint)},
\nonumber\\
\text{Legendre transformation}
\quad&\Longrightarrow\quad
H\supset-A_0^aG^a,
\nonumber\\
\dot\Pi_0^a\approx0
\quad&\Longrightarrow\quad
G^a\approx0
\qquad\text{(secondary constraint)}.
\end{align}
None of these steps requires a gauge choice. Gauge fixing is a
separate operation used, if desired, to remove the redundancy
associated with the first-class constraints.

\section{Why is there no analogous strong $CP$ problem in conventional QED?}
\label{App:QEDvsQCD}

A useful perspective on the strong $CP$ problem is obtained by comparing QCD
with quantum electrodynamics. At first sight the two theories appear rather
different, since one is an Abelian gauge theory while the other is
nonabelian. However, from the standpoint of local quantum field theory they
share many common features. Both possess a local gauge symmetry, both admit
an axial anomaly, both allow complex fermion mass terms, and both contain the
local pseudoscalar operator $F_{\mu\nu}\tilde F^{\mu\nu}$. Nevertheless,
there is no conventional strong $CP$ problem in QED. Understanding why
provides useful insight into the assumptions underlying the corresponding
construction in QCD.

The QED Lagrangian is
\begin{align}
\mathcal L_{\rm QED} &= -\textstyle \frac14F_{\mu\nu}F^{\mu\nu}
    + \bar\psi (i\slashed D-m) \psi ,
\end{align}
while the corresponding QCD Lagrangian is
\begin{align}
\mathcal L_{\rm QCD} &= \textstyle -\frac14F_{\mu\nu}^aF^{a\mu\nu}
    + \bar\psi(i\slashed D-m) \psi .
\end{align}
Apart from the nonabelian structure of the gauge field, the local forms of
the two theories are remarkably similar. In QED one may define the local
pseudoscalar density
\begin{align}
q_{\rm photon}(x)
&=
\frac{e^2}{32\pi^2}
F_{\mu\nu}\tilde F^{\mu\nu},
\end{align}
in direct analogy with the corresponding QCD operator
\begin{align}
q_{\rm gluon}(x) &= \frac{g^2}{32\pi^2} F_{\mu\nu}^a\tilde F^{a\mu\nu}.
\end{align}

Likewise, both theories possess an axial anomaly. For a single Dirac fermion
the divergence of the axial current satisfies
\begin{align}
\partial_\mu J_5^\mu &= 2im\bar\psi\gamma_5\psi
    +\frac{e^2}{16\pi^2}F_{\mu\nu}\tilde F^{\mu\nu}
\end{align}
in QED. For $N_f$ quark flavors, the corresponding unnormalized
flavor-singlet axial current in QCD satisfies
\begin{align}
\partial_\mu J_5^\mu
&=
2i\sum_{f=1}^{N_f}m_f\bar\psi_f\gamma_5\psi_f
+
2N_f\frac{g^2}{32\pi^2}
F_{\mu\nu}^a\tilde F^{a\mu\nu}.
\end{align}
Thus the local anomaly equations have the same essential structure. In both
theories an axial rotation of the fermion fields changes the phase of the
fermion mass term while simultaneously shifting the coefficient of the
local operator $F\tilde F$. From the standpoint of local quantum field
theory alone, the two theories therefore possess closely analogous
structures.

The crucial distinction for the conventional $\theta$-vacuum construction
arises when one considers the global topology of the gauge fields and gauge
transformations. In QCD the conventional construction assumes that
sufficiently smooth finite-action gauge fields approach pure gauge
configurations at infinity and imposes corresponding asymptotic conditions
on the admissible gauge transformations. On a fixed spatial hypersurface,
one may then regard the gauge transformations as maps
\begin{align}
U:\; S^3 &\to SU(3).
\end{align}
Since
\begin{align}
\pi_3(SU(3)) &= \mathbb Z ,
\end{align}
such maps possess disconnected homotopy classes labelled by an integer
winding number. This forms the basis for the conventional distinction between
small and large gauge transformations and the Hamiltonian construction of
the $\theta$ vacuum. Under the corresponding smoothness and global 
assumptions, the Euclidean formulation admits the related classification 
of finite-action gauge configurations by integer topological charge.

By contrast, in Abelian gauge theory
\begin{align}
U:\; S^3 &\to U(1),
\end{align}
but now
\begin{align}
\pi_3(U(1)) &= 0.
\end{align}
There are therefore no nontrivial winding sectors of smooth $U(1)$ gauge
transformations of this type on a spatially compactified $\mathbb R^3$.
Correspondingly, for ordinary QED on topologically trivial noncompact
spacetime there is no analogue of the BPST instanton solution and no
Hamiltonian construction of a $\theta$ vacuum analogous to that of QCD.

This conclusion refers to ordinary QED formulated on noncompact
topologically trivial spacetime with the usual asymptotic conditions. On a
compact spacetime, or more generally when nontrivial $U(1)$ bundles are
admitted, additional global structure can arise even in the Abelian theory.
For example, on a compact oriented four-manifold a $U(1)$ gauge field may
define a nontrivial line bundle with nonzero first Chern class. In that
case the appropriately normalized integral of $F\wedge F$ is determined by
the square of the first Chern class and can acquire quantized topological
values. Thus $\pi_3(U(1))=0$ excludes the
particular large-gauge-transformation construction familiar from QCD on a
spatially compactified $\mathbb R^3$, but it does not imply that Abelian
gauge theory is devoid of global topological structure on arbitrary
spacetime manifolds. Such possibilities depend upon additional global
information about the spacetime and the allowed gauge bundles and are not
part of ordinary QED on Minkowski spacetime.

The local operator
\begin{align}
F_{\mu\nu}\tilde F^{\mu\nu} &= \partial_\mu K^\mu
\end{align}
may nevertheless be written in QED exactly as in QCD. Likewise, the axial
anomaly implies that a chiral rotation of the fermion fields formally shifts
the coefficient multiplying this operator. The absence of a conventional
strong $CP$ problem in ordinary QED therefore does not arise because the
local operator or the anomaly is absent. Rather, the additional global
topological construction that conventionally gives the corresponding
coefficient the interpretation of a physical vacuum angle in QCD is absent
in ordinary QED on topologically trivial Minkowski spacetime.

It is useful to summarize the comparison.

\begin{center}
\begin{tabular}{lcc}
\hline
Feature & QED & QCD\\
\hline
Local gauge symmetry & Yes & Yes\\
Axial anomaly & Yes & Yes\\
Complex fermion masses & Yes & Yes\\
Local operator $F\tilde F$ & Yes & Yes\\
Nontrivial $\pi_3(G)$ & No & Yes\\
BPST instantons on Euclidean $\mathbb R^4$ & No & Yes\\
Conventional $\theta$ vacuum construction & No & Yes\\
Conventional strong $CP$ problem & No & Yes\\
\hline
\end{tabular}
\end{center}

The comparison highlights the central conceptual distinction developed
throughout this review. The relevant local structures of QED and QCD are
closely analogous, while the conventional $\theta$-vacuum construction in
QCD depends upon additional global topological structure that is absent in
ordinary QED on topologically trivial Minkowski spacetime. The absence of a
conventional strong $CP$ problem in QED is therefore not attributable to
the absence of the local anomaly or of the local operator
$F_{\mu\nu}\tilde F^{\mu\nu}$.

More generally, the QED comparison illustrates that the existence of a
locally allowed total-divergence operator does not by itself determine
whether its coefficient represents an independent physical parameter.
That question can depend upon additional global information concerning the
spacetime, the admissible gauge fields and gauge transformations, and the
allowed gauge bundles. QED and QCD need not have identical global
structures, but their comparison provides a useful illustration of the
distinction between local and global ingredients that has been emphasized
throughout this review.



\bibliographystyle{cas-model2-names}
\bibliography{theta_refs}

@article{Baker2006ts,
  author = {Baker, C. A. and others},
  title = {An Improved experimental limit on the electric dipole moment of the neutron},
  journal = {Phys. Rev. Lett.},
  volume = {97},
  pages = {131801},
  year = {2006},
  eprint = {hep-ex/0602020},
  archivePrefix = {arXiv}
}

@article{Abel2020gbr,
  author = {Abel, C. and others},
  title = {Measurement of the Permanent Electric Dipole Moment of the Neutron},
  journal = {Phys. Rev. Lett.},
  volume = {124},
  pages = {081803},
  year = {2020},
  eprint = {2001.11966},
  archivePrefix = {arXiv},
  primaryClass = {hep-ex}
}

@article{Benabou2025,
  author = {Benabou, J. N. and Hook, A. and Manzari, C. A. and Murayama, H. and Safdi, B. R.},
  title = {Clearing up the strong CP problem},
  journal = {arXiv},
  eprint = {2510.18951},
  archivePrefix = {arXiv},
  primaryClass = {hep-ph},
  year = {2025}
}

@article{Kaplan:2025JHEP050,
  author       = {David E. Kaplan and Tom Melia and Surjeet Rajendran},
  title        = {What can solve the strong CP problem?},
  journal      = {J. High Energy Phys.},
  year         = {2025},
  month        = {Aug},
  pages        = {050},
  doi          = {10.1007/JHEP08(2025)050},
  note         = {Also available as arXiv:2505.08358 [hep-ph]}
}

@article{Dvali2025,
  author = {Dvali, G. and Komisel, L. and Sakhelashvili, O. and Wachowitz, A.},
  title = {The strong CP problem and its gauge axion solution as evidence for fundamental strings},
  journal = {arXiv},
  eprint = {2512.08834},
  archivePrefix = {arXiv},
  primaryClass = {hep-th},
  year = {2025}
}

@article{Strumia2025,
  author = {Strumia, A.},
  title = {Solving the strong CP problem},
  journal = {arXiv},
  eprint = {2501.16427},
  archivePrefix = {arXiv},
  primaryClass = {hep-ph},
  year = {2025}
}

@article{Witten:1979vv,
    author = "Witten, Edward",
    title = "{Current Algebra Theorems for the U(1) Goldstone Boson}",
    reportNumber = "HUTP-79/A014",
    doi = "10.1016/0550-3213(79)90031-2",
    journal = "Nucl. Phys. B",
    volume = "156",
    pages = "269--283",
    year = "1979"
}

@article{Veneziano:1979ec,
    author = "Veneziano, G.",
    title = "{U(1) Without Instantons}",
    reportNumber = "CERN-TH-2651",
    doi = "10.1016/0550-3213(79)90332-8",
    journal = "Nucl. Phys. B",
    volume = "159",
    pages = "213--224",
    year = "1979"
}

@book{Williams:2022bzq,
  author = {Williams, A. G.},
  title = {Introduction to Quantum Field Theory},
  publisher = {Cambridge University Press},
  year = {2022},
  isbn = {9781108470902},
  doi = {10.1017/9781108585286}
}

@book{Creutz1983,
  author = {Creutz, M.},
  title = {Quarks, Gluons and Lattices},
  publisher = {Cambridge University Press},
  year = {1983}
}

@book{Rothe2012,
  author = {Rothe, H. J.},
  title = {Lattice Gauge Theories: An Introduction},
  edition = {4},
  publisher = {World Scientific},
  year = {2012}
}

@book{MontvayMunster,
  author = {Montvay, I. and Muenster, G.},
  title = {Quantum Fields on a Lattice},
  publisher = {Cambridge University Press},
  year = {1994}
}

@book{GattringerLang,
  author = {Gattringer, C. and Lang, C. B.},
  title = {Quantum Chromodynamics on the Lattice},
  publisher = {Springer},
  year = {2010}
}

@book{Smit,
  author = {Smit, J.},
  title = {Introduction to Quantum Fields on a Lattice},
  publisher = {Cambridge University Press},
  year = {2002}
}

@article{Leinweber2004,
  author = {Leinweber, D. B. and Williams, A. G. and Zhang, J. and Lee, F. X.},
  title = {Topological charge barrier in the Markov chain of QCD},
  journal = {Phys. Lett. B},
  volume = {585},
  pages = {187--191},
  year = {2004},
  eprint = {hep-lat/0312035}
}

@article{Alexandrou2020,
  author = {Alexandrou, C. and others},
  title = {Comparison of topological charge definitions in Lattice QCD},
  journal = {Eur. Phys. J. C},
  volume = {80},
  pages = {424},
  year = {2020},
  eprint = {1708.00696}
}

@article{Zhang2002,
  author = {Zhang, J. B. and others},
  title = {Numerical study of lattice index theorem using improved cooling and overlap fermions},
  journal = {Phys. Rev. D},
  volume = {65},
  pages = {074510},
  year = {2002},
  eprint = {hep-lat/0111060}
}

@article{Luscher2010iy,
  author = {Luscher, M.},
  title = {Properties and uses of the Wilson flow in lattice QCD},
  journal = {JHEP},
  volume = {08},
  pages = {071},
  year = {2010},
  eprint = {1006.4518}
}

@article{Bruno2014,
   title={Topological susceptibility and the sampling of field space in N f = 2 lattice QCD simulations},
   volume={2014},
   ISSN={1029-8479},
   url={http://dx.doi.org/10.1007/JHEP08(2014)150},
   DOI={10.1007/jhep08(2014)150},
   number={8},
   journal={Journal of High Energy Physics},
   publisher={Springer Science and Business Media LLC},
   author={Bruno, Mattia and Sch{\"{a}}fer, Stefan and Sommer, Rainer},
   year={2014},
   month=aug }

@article{L_scher_2010,
   title={Universality of the topological susceptibility in the SU(3) gauge theory},
   volume={2010},
   ISSN={1029-8479},
   url={http://dx.doi.org/10.1007/JHEP09(2010)110},
   DOI={10.1007/jhep09(2010)110},
   number={9},
   journal={Journal of High Energy Physics},
   publisher={Springer Science and Business Media LLC},
   author={Lüscher, Martin and Palombi, Filippo},
   year={2010},
   month=sep }

@article{L_scher_2011,
   title={Perturbative analysis of the gradient flow in non-abelian gauge theories},
   volume={2011},
   ISSN={1029-8479},
   url={http://dx.doi.org/10.1007/JHEP02(2011)051},
   DOI={10.1007/jhep02(2011)051},
   number={2},
   journal={Journal of High Energy Physics},
   publisher={Springer Science and Business Media LLC},
   author={Lüscher, Martin and Weisz, Peter},
   year={2011},
   month=feb }

@book{ZinnJustin2002,
  author = {Zinn-Justin, J.},
  title = {Quantum Field Theory and Critical Phenomena},
  edition = {4},
  publisher = {Oxford University Press},
  year = {2002}
}

@book{Varadhan1984,
  author = {Varadhan, S. R. S.},
  title = {Large Deviations and Applications},
  publisher = {SIAM},
  year = {1984}
}

@article{PhysRevD.83.111502,
  title = {Isoscalar meson spectroscopy from lattice QCD},
  author = {Dudek, Jozef J. and Edwards, Robert G. and Jo\'o, B\'alint and Peardon, Michael J. and Richards, David G. and Thomas, Christopher E.},
  collaboration = {Hadron Spectrum Collaboration},
  journal = {Phys. Rev. D},
  volume = {83},
  issue = {11},
  pages = {111502},
  numpages = {5},
  year = {2011},
  month = {Jun},
  publisher = {American Physical Society},
  doi = {10.1103/PhysRevD.83.111502},
  url = {https://link.aps.org/doi/10.1103/PhysRevD.83.111502}
}

@article{SchaferShuryak1998,
  author = {Schafer, T. and Shuryak, E. V.},
  title = {Instantons in QCD},
  journal = {Rev. Mod. Phys.},
  volume = {70},
  pages = {323--426},
  year = {1998},
  eprint = {hep-ph/9610451}
}

@article{Crewther1979,
  author = {Crewther, R. J. and Di Vecchia, P. and Veneziano, G. and Witten, E.},
  title = {Chiral Estimate of the Electric Dipole Moment of the Neutron in Quantum Chromodynamics},
  journal = {Phys. Lett. B},
  volume = {88},
  pages = {123},
  year = {1979}
}

@article{DiGiacomo:1992df,
  author       = {Di Giacomo, A. and Panagopoulos, H.},
  title        = {Topological charge in lattice gauge theory},
  journal      = {Phys.\ Lett.\ B},
  volume       = {285},
  pages        = {133--136},
  year         = {1992},
}

@article{Rossi:1980cf,
  author       = {Rossi, P. and Testa, M.},
  title        = {The cluster expansion and the topological susceptibility in the lattice CP(N) model},
  journal      = {Nucl.\ Phys.\ B},
  volume       = {163},
  pages        = {109--125},
  year         = {1980},
}

@article{Sharpe:2006vb,
  author       = {Sharpe, S.~R.},
  title        = {An introduction to partially quenched chiral perturbation theory},
  journal      = {arXiv:hep-lat/0607016},
  year         = {2006},
}

@article{Gasser:1987ah,
  author       = {Gasser, J. and Leutwyler, H.},
  title        = {Thermodynamics of chiral symmetry},
  journal      = {Phys.\ Lett.\ B},
  volume       = {188},
  pages        = {477--481},
  year         = {1987},
}

@article{Schierholz:2024var,
    author = "Schierholz, Gerrit",
    title = "{Absence of strong CP violation}",
    eprint = "2403.13508",
    archivePrefix = "arXiv",
    primaryClass = "hep-ph",
    reportNumber = "DESY-24-038",
    doi = "10.1088/1361-6471/adc31d",
    journal = "J. Phys. G",
    volume = "52",
    number = "4",
    pages = "04LT01",
    year = "2025"
}

@article{Nakamura:2021meh,
    author = "Nakamura, Y. and Schierholz, G.",
    title = "{The strong CP problem solved by itself due to long-distance vacuum effects}",
    eprint = "2106.11369",
    archivePrefix = "arXiv",
    primaryClass = "hep-ph",
    reportNumber = "DESY 21-078, DESY-21-078",
    doi = "10.1016/j.nuclphysb.2022.116063",
    journal = "Nucl. Phys. B",
    volume = "986",
    pages = "116063",
    year = "2023"
}

@misc{bonanno2025strongcpproblemtheta,
      title={Strong CP problem, theta term and QCD topological properties}, 
      author={Claudio Bonanno and Claudio Bonati and Massimo D'Elia},
      year={2025},
      eprint={2510.03059},
      archivePrefix={arXiv},
      primaryClass={hep-lat},
      url={https://arxiv.org/abs/2510.03059}, 
}

@article{Kotov_2025,
   title={Topological observables and $\theta$ dependence in high temperature QCD from lattice simulations},
   volume={2025},
   ISSN={1029-8479},
   url={http://dx.doi.org/10.1007/JHEP09(2025)045},
   DOI={10.1007/jhep09(2025)045},
   number={9},
   journal={Journal of High Energy Physics},
   publisher={Springer Science and Business Media LLC},
   author={Kotov, A. Yu. and Lombardo, M. P. and Trunin, A.},
   year={2025},
   month=sep }

@article{Iida:2024irv,
    author = "Iida, Kei and Itou, Etsuko and Murakami, Kotaro and Suenaga, Daiki",
    title = "{Lattice study on finite density QC$_{2}$D towards zero temperature}",
    eprint = "2405.20566",
    archivePrefix = "arXiv",
    primaryClass = "hep-lat",
    reportNumber = "YITP-24-68, RIKEN-iTHEMS-Report-24",
    doi = "10.1007/JHEP10(2024)022",
    journal = "JHEP",
    volume = "10",
    pages = "022",
    year = "2024"
}

@article{Gattringer:2020mbf,
    author = "Gattringer, Christof and Orasch, Oliver",
    title = "{Density of states approach for lattice gauge theory with a $\theta$-term}",
    eprint = "2004.03837",
    archivePrefix = "arXiv",
    primaryClass = "hep-lat",
    doi = "10.1016/j.nuclphysb.2020.115097",
    journal = "Nucl. Phys. B",
    volume = "957",
    pages = "115097",
    year = "2020"
}

@article{Sannino:2026wgx,
    author = "Sannino, Francesco",
    title = "{Strong CP and the QCD axion lecture notes via effective field theory}",
    eprint = "2601.19735",
    archivePrefix = "arXiv",
    primaryClass = "hep-ph",
    doi = "10.1140/epjc/s10052-026-15612-4",
    journal = "Eur. Phys. J. C",
    volume = "86",
    number = "5",
    pages = "465",
    year = "2026"
}

@article{Vafa:1984xg,
title = {Restrictions on symmetry breaking in vector-like gauge theories},
journal = {Nuclear Physics B},
volume = {234},
number = {1},
pages = {173-188},
year = {1984},
issn = {0550-3213},
doi = {https://doi.org/10.1016/0550-3213(84)90230-X},
url = {https://www.sciencedirect.com/science/article/pii/055032138490230X},
author = {C. Vafa and E. Witten}
}

@book{Glimm:1987ylb,
    author = "Glimm, James and Jaffe, Arthur",
    title = "{Quantum Physics: A Functional Integral Point of View}",
    doi = "10.1007/978-1-4612-4728-9",
    isbn = "978-0-387-96477-5, 978-1-4612-4728-9",
    publisher = "Springer",
    year = "1987"
}

@book{Simon:2015spl,
    author = "Simon, Barry",
    title = "{The P ($\Phi$)$_2$ euclidean (Quantum) Field Theory}",
    isbn = "978-0-691-61849-4",
    publisher = "Princeton University Press",
    month = "3",
    year = "2015"
}

@article{Neuberger:1998,
   title={Exactly massless quarks on the lattice},
   volume={417},
   ISSN={0370-2693},
   url={http://dx.doi.org/10.1016/S0370-2693(97)01368-3},
   DOI={10.1016/s0370-2693(97)01368-3},
   number={1–2},
   journal={Physics Letters B},
   publisher={Elsevier BV},
   author={Neuberger, Herbert},
   year={1998},
   month=jan, pages={141–144} }

@article{Hasenfratz:1998,
   title={The index theorem in QCD with a finite cut-off},
   volume={427},
   ISSN={0370-2693},
   url={http://dx.doi.org/10.1016/S0370-2693(98)00315-3},
   DOI={10.1016/s0370-2693(98)00315-3},
   number={1–2},
   journal={Physics Letters B},
   publisher={Elsevier BV},
   author={Hasenfratz, Peter and Laliena, Victor and Niedermayer, Ferenc},
   year={1998},
   month=may, pages={125–131} }

@article{Luscher:1998,
   title={Exact chiral symmetry on the lattice and the Ginsparg-Wilson relation},
   volume={428},
   ISSN={0370-2693},
   url={http://dx.doi.org/10.1016/S0370-2693(98)00423-7},
   DOI={10.1016/s0370-2693(98)00423-7},
   number={3–4},
   journal={Physics Letters B},
   publisher={Elsevier BV},
   author={Lüscher, Martin},
   year={1998},
   month=jun, pages={342–345} }

@article{JackiwRebbi:1977,
  title = {Spinor analysis of Yang-Mills theory},
  author = {Jackiw, R. and Rebbi, C.},
  journal = {Phys. Rev. D},
  volume = {16},
  issue = {4},
  pages = {1052--1060},
  numpages = {0},
  year = {1977},
  month = {Aug},
  publisher = {American Physical Society},
  doi = {10.1103/PhysRevD.16.1052},
  url = {https://link.aps.org/doi/10.1103/PhysRevD.16.1052}
}

@article{Gross:1973id,
  author = {Gross, David J. and Wilczek, Frank},
  title = {Ultraviolet Behavior of Non-Abelian Gauge Theories},
  journal = {Phys. Rev. Lett.},
  volume = {30},
  pages = {1343--1346},
  year = {1973},
  doi = {10.1103/PhysRevLett.30.1343}
}

@article{Politzer:1973fx,
  author = {Politzer, H. David},
  title = {Reliable Perturbative Results for Strong Interactions?},
  journal = {Phys. Rev. Lett.},
  volume = {30},
  pages = {1346--1349},
  year = {1973},
  doi = {10.1103/PhysRevLett.30.1346}
}

@article{Wilson:1974sk,
  author = {Wilson, Kenneth G.},
  title = {Confinement of Quarks},
  journal = {Phys. Rev. D},
  volume = {10},
  pages = {2445--2459},
  year = {1974},
  doi = {10.1103/PhysRevD.10.2445}
}

@article{Leutwyler:1992yt,
  author = {Leutwyler, H. and Smilga, A. V.},
  title = {Spectrum of Dirac Operator and Role of Winding Number in QCD},
  journal = {Phys. Rev. D},
  volume = {46},
  pages = {5607--5632},
  year = {1992},
  doi = {10.1103/PhysRevD.46.5607}
}

@article{Smilga:1998dh,
  author = {Smilga, A. V.},
  title = {Physics of Theta Parameter},
  journal = {Lect. Notes Phys.},
  volume = {521},
  pages = {247--266},
  year = {1999},
  eprint = {hep-ph/9805214},
  archivePrefix = {arXiv}
}

@article{Kim:1979if,
  author = {Kim, Jihn E.},
  title = {Weak Interaction Singlet and Strong CP Invariance},
  journal = {Phys. Rev. Lett.},
  volume = {43},
  pages = {103},
  year = {1979},
  doi = {10.1103/PhysRevLett.43.103}
}

@article{Shifman:1979if,
  author = {Shifman, M. A. and Vainshtein, A. I. and Zakharov, V. I.},
  title = {Can Confinement Ensure Natural CP Invariance of Strong Interactions?},
  journal = {Nucl. Phys. B},
  volume = {166},
  pages = {493--506},
  year = {1980},
  doi = {10.1016/0550-3213(80)90209-6}
}

@article{Dine:1981rt,
  author = {Dine, Michael and Fischler, Willy and Srednicki, Mark},
  title = {A Simple Solution to the Strong CP Problem with a Harmless Axion},
  journal = {Phys. Lett. B},
  volume = {104},
  pages = {199--202},
  year = {1981},
  doi = {10.1016/0370-2693(81)90590-6}
}

@article{Zhitnitsky:1980tq,
  author = {Zhitnitsky, A. R.},
  title = {On Possible Suppression of the Axion Hadron Interactions},
  journal = {Sov. J. Nucl. Phys.},
  volume = {31},
  pages = {260},
  year = {1980}
}

@article{Belavin:1975fg,
    author = "Belavin, A. A. and Polyakov, Alexander M. and Schwartz, A. S. and Tyupkin, Yu. S.",
    editor = "Taylor, J. C.",
    title = "{Pseudoparticle Solutions of the Yang-Mills Equations}",
    doi = "10.1016/0370-2693(75)90163-X",
    journal = "Phys. Lett. B",
    volume = "59",
    pages = "85--87",
    year = "1975"
}

@article{tHooft:1976rip,
    author = "'t Hooft, Gerard",
    editor = "Shifman, Mikhail A.",
    title = "{Symmetry Breaking Through Bell-Jackiw Anomalies}",
    reportNumber = "PRINT-76-0254 (HARVARD)",
    doi = "10.1103/PhysRevLett.37.8",
    journal = "Phys. Rev. Lett.",
    volume = "37",
    pages = "8--11",
    year = "1976"
}

@article{tHooft:1976snw,
    author = "'t Hooft, Gerard",
    editor = "Shifman, Mikhail A.",
    title = "{Computation of the Quantum Effects Due to a Four-Dimensional Pseudoparticle}",
    reportNumber = "PRINT-76-0551 (HARVARD)",
    doi = "10.1103/PhysRevD.14.3432",
    journal = "Phys. Rev. D",
    volume = "14",
    pages = "3432--3450",
    year = "1976",
    note = "[Erratum: Phys.Rev.D 18, 2199 (1978)]"
}

@article{Jackiw:1976pf,
    author = "Jackiw, R. and Rebbi, C.",
    editor = "Taylor, J. C.",
    title = "{Vacuum Periodicity in a Yang-Mills Quantum Theory}",
    reportNumber = "MIT-CTP-548",
    doi = "10.1103/PhysRevLett.37.172",
    journal = "Phys. Rev. Lett.",
    volume = "37",
    pages = "172--175",
    year = "1976"
}

@article{Callan:1976je,
    author = "Callan, Jr., Curtis G. and Dashen, R. F. and Gross, David J.",
    editor = "Taylor, J. C.",
    title = "{The Structure of the Gauge Theory Vacuum}",
    reportNumber = "COO-2220-75",
    doi = "10.1016/0370-2693(76)90277-X",
    journal = "Phys. Lett. B",
    volume = "63",
    pages = "334--340",
    year = "1976"
}

@article{Adler:1969gk,
    author = "Adler, Stephen L.",
    title = "{Axial vector vertex in spinor electrodynamics}",
    doi = "10.1103/PhysRev.177.2426",
    journal = "Phys. Rev.",
    volume = "177",
    pages = "2426--2438",
    year = "1969"
}

@article{Adler:1969er,
    author = "Adler, Stephen L. and Bardeen, William A.",
    title = "{Absence of higher order corrections in the anomalous axial vector divergence equation}",
    doi = "10.1103/PhysRev.182.1517",
    journal = "Phys. Rev.",
    volume = "182",
    pages = "1517--1536",
    year = "1969"
}

@article{Bell:1969ts,
    author = "Bell, J. S. and Jackiw, R.",
    title = "{A PCAC puzzle: $\pi^0 \to \gamma \gamma$ in the $\sigma$ model}",
    doi = "10.1007/BF02823296",
    journal = "Nuovo Cim. A",
    volume = "60",
    pages = "47--61",
    year = "1969"
}

@article{Fujikawa:1979ay,
    author = "Fujikawa, Kazuo",
    title = "{Path Integral Measure for Gauge Invariant Fermion Theories}",
    reportNumber = "INS-328",
    doi = "10.1103/PhysRevLett.42.1195",
    journal = "Phys. Rev. Lett.",
    volume = "42",
    pages = "1195--1198",
    year = "1979"
}

@article{Fujikawa:1980eg,
  author = {Fujikawa, Kazuo},
  title = {Path Integral Measure for Gauge Invariant Fermion Theories},
  journal = {Phys. Rev. Lett.},
  volume = {44},
  pages = {1733--1736},
  year = {1980},
  doi = {10.1103/PhysRevLett.44.1733}
}

@article{Weinberg:1977ma,
    author = "Weinberg, Steven",
    title = "{A New Light Boson?}",
    reportNumber = "HUTP-77/A074",
    doi = "10.1103/PhysRevLett.40.223",
    journal = "Phys. Rev. Lett.",
    volume = "40",
    pages = "223--226",
    year = "1978"
}

@article{Wilczek:1977pj,
    author = "Wilczek, Frank",
    title = "{Problem of Strong  $P$  and  $T$  Invariance in the Presence of Instantons}",
    reportNumber = "Print-77-0939 (COLUMBIA)",
    doi = "10.1103/PhysRevLett.40.279",
    journal = "Phys. Rev. Lett.",
    volume = "40",
    pages = "279--282",
    year = "1978"
}

@article{Williams:2026tuw,
    author = "Williams, Anthony G.",
    title = "{On the Origins of the Strong CP Problem}",
    eprint = "2607.10272",
    archivePrefix = "arXiv",
    primaryClass = "hep-ph",
    reportNumber = "ADP-26-12/T1309",
    month = "7",
    year = "2026"
}

@article{Peccei:1977hh,
    author = "Peccei, R. D. and Quinn, Helen R.",
    title = "{CP Conservation in the Presence of Instantons}",
    reportNumber = "ITP-568-STANFORD",
    doi = "10.1103/PhysRevLett.38.1440",
    journal = "Phys. Rev. Lett.",
    volume = "38",
    pages = "1440--1443",
    year = "1977"
}

@article{Peccei:1977ur,
    author = "Peccei, R. D. and Quinn, Helen R.",
    title = "{Constraints Imposed by CP Conservation in the Presence of Instantons}",
    reportNumber = "ITP-572-STANFORD",
    doi = "10.1103/PhysRevD.16.1791",
    journal = "Phys. Rev. D",
    volume = "16",
    pages = "1791--1797",
    year = "1977"
}

@article{Nelson:1983zb,
    author = "Nelson, Ann E.",
    title = "{Naturally Weak CP Violation}",
    reportNumber = "HUTP-83/A080",
    doi = "10.1016/0370-2693(84)92025-2",
    journal = "Phys. Lett. B",
    volume = "136",
    pages = "387--391",
    year = "1984"
}

@article{Barr:1984qx,
    author = "Barr, Stephen M.",
    title = "{Solving the Strong CP Problem Without the Peccei-Quinn Symmetry}",
    reportNumber = "DOE-ER-40048-08 P4",
    doi = "10.1103/PhysRevLett.53.329",
    journal = "Phys. Rev. Lett.",
    volume = "53",
    pages = "329",
    year = "1984"
}

@article{Mohapatra:1978fy,
    author = "Mohapatra, Rabindra N. and Senjanovic, G.",
    title = "{Natural Suppression of Strong p and t Noninvariance}",
    reportNumber = "CCNY-HEP-78-12",
    doi = "10.1016/0370-2693(78)90243-5",
    journal = "Phys. Lett. B",
    volume = "79",
    pages = "283--286",
    year = "1978"
}

@article{Senjanovic:1978ev,
    author = "Senjanovic, Goran",
    title = "{Spontaneous Breakdown of Parity in a Class of Gauge Theories}",
    reportNumber = "CCNY-HEP-78/20",
    doi = "10.1016/0550-3213(79)90604-7",
    journal = "Nucl. Phys. B",
    volume = "153",
    pages = "334--364",
    year = "1979"
}

@article{Pospelov:2005pr,
    author = "Pospelov, Maxim and Ritz, Adam",
    title = "{Electric dipole moments as probes of new physics}",
    eprint = "hep-ph/0504231",
    archivePrefix = "arXiv",
    doi = "10.1016/j.aop.2005.04.002",
    journal = "Annals Phys.",
    volume = "318",
    pages = "119--169",
    year = "2005"
}

@article{Guo:2015tla,
    author = "Guo, F. -K. and Horsley, R. and Meissner, U. -G. and Nakamura, Y. and Perlt, H. and Rakow, P. E. L. and Schierholz, G. and Schiller, A. and Zanotti, J. M.",
    title = "{The electric dipole moment of the neutron from 2+1 flavor lattice QCD}",
    eprint = "1502.02295",
    archivePrefix = "arXiv",
    primaryClass = "hep-lat",
    reportNumber = "ADP-15-6-T908, DESY-15-018, EDINBURGH-2015-01, LIVERPOOL-LTH-1034",
    doi = "10.1103/PhysRevLett.115.062001",
    journal = "Phys. Rev. Lett.",
    volume = "115",
    number = "6",
    pages = "062001",
    year = "2015"
}

@article{Dragos:2019oxn,
    author = "Dragos, Jack and Luu, Thomas and Shindler, Andrea and de Vries, Jordy and Yousif, Ahmed",
    title = "{Confirming the Existence of the strong CP Problem in Lattice QCD with the Gradient Flow}",
    eprint = "1902.03254",
    archivePrefix = "arXiv",
    primaryClass = "hep-lat",
    doi = "10.1103/PhysRevC.103.015202",
    journal = "Phys. Rev. C",
    volume = "103",
    number = "1",
    pages = "015202",
    year = "2021"
}

@article{Peccei:2006as,
    author = "Peccei, R. D.",
    editor = "Kuster, Markus and Raffelt, Georg and Beltran, Berta",
    title = "{The Strong CP problem and axions}",
    eprint = "hep-ph/0607268",
    archivePrefix = "arXiv",
    doi = "10.1007/978-3-540-73518-2_1",
    journal = "Lect. Notes Phys.",
    volume = "741",
    pages = "3--17",
    year = "2008"
}

@article{Faddeev:1967fc,
    author = "Faddeev, L. D. and Popov, V. N.",
    editor = "Hsu, Jong-Ping and Fine, D.",
    title = "{Feynman Diagrams for the Yang-Mills Field}",
    doi = "10.1016/0370-2693(67)90067-6",
    journal = "Phys. Lett. B",
    volume = "25",
    pages = "29--30",
    year = "1967"
}

@article{Becchi:1975nq,
    author = "Becchi, C. and Rouet, A. and Stora, R.",
    title = "{Renormalization of Gauge Theories}",
    reportNumber = "CPT-75-P.723-MARSEILLE",
    doi = "10.1016/0003-4916(76)90156-1",
    journal = "Annals Phys.",
    volume = "98",
    pages = "287--321",
    year = "1976"
}

@article{Tyutin:1975qk,
    author = "Tyutin, I. V.",
    title = "{Gauge Invariance in Field Theory and Statistical Physics in Operator Formalism}",
    eprint = "0812.0580",
    archivePrefix = "arXiv",
    primaryClass = "hep-th",
    reportNumber = "LEBEDEV-75-39",
    year = "1975"
}

@article{Gribov:1977wm,
    author = "Gribov, V. N.",
    editor = "Nyiri, J.",
    title = "{Quantization of Nonabelian Gauge Theories}",
    reportNumber = "LENINGRAD-77-367",
    doi = "10.1016/0550-3213(78)90175-X",
    journal = "Nucl. Phys. B",
    volume = "139",
    pages = "1",
    year = "1978"
}

@article{Yang:1954ek,
    author = "Yang, Chen-Ning and Mills, Robert L.",
    editor = "Hsu, Jong-Ping and Fine, D.",
    title = "{Conservation of Isotopic Spin and Isotopic Gauge Invariance}",
    doi = "10.1103/PhysRev.96.191",
    journal = "Phys. Rev.",
    volume = "96",
    pages = "191--195",
    year = "1954"
}

@article{Engel:2013lsa,
    author = "Engel, Jonathan and Ramsey-Musolf, Michael J. and van Kolck, U.",
    title = "{Electric Dipole Moments of Nucleons, Nuclei, and Atoms: The Standard Model and Beyond}",
    eprint = "1303.2371",
    archivePrefix = "arXiv",
    primaryClass = "nucl-th",
    doi = "10.1016/j.ppnp.2013.03.003",
    journal = "Prog. Part. Nucl. Phys.",
    volume = "71",
    pages = "21--74",
    year = "2013"
}

@incollection{colella1973sample,
  author    = {Colella, Phillip and Lanford, III, Oscar E.},
  title     = {Appendix: Sample Field Behavior for the Free Markov Random Field},
  booktitle = {Constructive Quantum Field Theory},
  editor    = {Velo, Giorgio and Wightman, Arthur S.},
  series    = {Lecture Notes in Physics},
  volume    = {25},
  pages     = {44--70},
  year      = {1973},
  publisher = {Springer-Verlag},
  address   = {Berlin, Heidelberg},
  doi       = {10.1007/BFb0113082}
}

@article{Atiyah:1978ri,
    author = "Atiyah, M. F. and Hitchin, Nigel J. and Drinfeld, V. G. and Manin, Yu. I.",
    editor = "Shifman, Mikhail A.",
    title = "{Construction of Instantons}",
    doi = "10.1016/0375-9601(78)90141-X",
    journal = "Phys. Lett. A",
    volume = "65",
    pages = "185--187",
    year = "1978"
}

@book{Henneaux:1992,
    author = "Henneaux, Marc and Teitelboim, Claudio",
    title = "{Quantization of Gauge Systems}",
    publisher = "Princeton University Press",
    address = "Princeton, New Jersey",
    year = "1992",
    isbn = "978-0-691-08775-7"
}

@article{Christ:1980ku,
    author = "Christ, Norman H. and Lee, T. D.",
    title = "{Operator Ordering and Feynman Rules in Gauge Theories}",
    journal = "Phys. Rev. D",
    volume = "22",
    pages = "939--958",
    year = "1980",
    doi = "10.1103/PhysRevD.22.939"
}

@inproceedings{Zwanziger:1997mk,
    author = "Zwanziger, Daniel",
    title = "{Continuum and lattice Coulomb gauge Hamiltonian}",
    booktitle = "{NATO Advanced Study Institute on Confinement, Duality and Nonperturbative Aspects of QCD}",
    eprint = "hep-th/9710157",
    archivePrefix = "arXiv",
    reportNumber = "NYU-TH-PH-97-22",
    pages = "145--160",
    month = "10",
    year = "1997"
}

@inproceedings{Heinzl:1995jn,
    author = "Heinzl, T.",
    title = "{Hamiltonian formulations of Yang-Mills quantum theory and the Gribov problem}",
    booktitle = "{5th Meeting on Light Cone Quantization and Nonperturbative QCD}",
    eprint = "hep-th/9604018",
    archivePrefix = "arXiv",
    reportNumber = "TPR-95-30",
    month = "6",
    year = "1995"
}

@article{Reinhardt:2018,
    author = "Reinhardt, Hugo and Burgio, Giuseppe and
        Campagnari, Davide R. and Ebadati, Ehsan and
        Heffner, Jan and Quandt, Markus and Vastag, Peter
        and Vogt, Hannes",
    title = "{Hamiltonian Approach to QCD in Coulomb Gauge:
        A Survey of Recent Results}",
    journal = "Adv. High Energy Phys.",
    volume = "2018",
    pages = "2312498",
    year = "2018",
    eprint = "1706.02702",
    archivePrefix = "arXiv",
    primaryClass = "hep-th",
    doi = "10.1155/2018/2312498"
}

@article{Luscher:2004fu,
    author = "Luscher, Martin",
    title = "{Topological effects in QCD and the problem of short distance singularities}",
    eprint = "hep-th/0404034",
    archivePrefix = "arXiv",
    reportNumber = "CERN-PH-TH-2004-062",
    doi = "10.1016/j.physletb.2004.04.076",
    journal = "Phys. Lett. B",
    volume = "593",
    pages = "296--301",
    year = "2004"
}

@article{Luscher:2021ygc,
    author = {L{\"u}scher, Martin},
    title = "{On the chiral anomaly and the Yang{\textendash}Mills gradient flow}",
    eprint = "2109.07965",
    archivePrefix = "arXiv",
    primaryClass = "hep-lat",
    reportNumber = "CERN-TH-2021-132",
    doi = "10.1016/j.physletb.2021.136725",
    journal = "Phys. Lett. B",
    volume = "823",
    pages = "136725",
    year = "2021"
}

@article{Ai:2020ptm,
    author = {Ai, Wen-Yuan and Cruz, Juan S. and Garbrecht, Bj{\"o}rn and Tamarit, Carlos},
    title = "{Consequences of the order of the limit of infinite spacetime volume and the sum over topological sectors for CP violation in the strong interactions}",
    eprint = "2001.07152",
    archivePrefix = "arXiv",
    primaryClass = "hep-th",
    reportNumber = "TUM-HEP-1249/20, CP3-20-02",
    doi = "10.1016/j.physletb.2021.136616",
    journal = "Phys. Lett. B",
    volume = "822",
    pages = "136616",
    year = "2021"
}

@article{Ai:2024vfa,
    author = "Ai, Wen-Yuan and Garbrecht, Bjorn and Tamarit, Carlos",
    title = "{The QCD theta-parameter in canonical quantization}",
    eprint = "2403.00747",
    archivePrefix = "arXiv",
    primaryClass = "hep-th",
    reportNumber = "KCL-PH-TH/2024-13, TUM-HEP-1499/24, MITP-24-031",
    month = "3",
    year = "2024"
}

@article{Strocchi:2024tis,
    author = "Strocchi, F.",
    title = "{The strong CP problem revisited and solved by the gauge group topology}",
    eprint = "2404.19400",
    archivePrefix = "arXiv",
    primaryClass = "hep-th",
    month = "4",
    year = "2024"
}

@article{Albandea:2024fui,
    author = "Albandea, David and Catumba, Guilherme and Ramos, Alberto",
    title = "{Strong CP problem in the quantum rotor}",
    eprint = "2402.17518",
    archivePrefix = "arXiv",
    primaryClass = "hep-lat",
    doi = "10.1103/PhysRevD.110.094512",
    journal = "Phys. Rev. D",
    volume = "110",
    number = "9",
    pages = "094512",
    year = "2024"
}

@article{Khoze:2025auv,
    author = "Khoze, Valentin V.",
    title = "{A note on instantons, {\ensuremath{\theta}}-dependence and strong CP}",
    eprint = "2512.06827",
    archivePrefix = "arXiv",
    primaryClass = "hep-ph",
    reportNumber = "IPPP-25-XXX",
    doi = "10.1007/JHEP08(2026)063",
    journal = "JHEP",
    volume = "08",
    pages = "063",
    year = "2026"
}

@article{Bhattacharya:2025qsk,
    author = "Bhattacharya, Tanmoy",
    title = "{Comment on the claim of physical irrelevance of the topological term}",
    eprint = "2512.10127",
    archivePrefix = "arXiv",
    primaryClass = "hep-lat",
    reportNumber = "LA-UR-22-29996",
    month = "12",
    year = "2025"
}

@article{Luscher:1981zq,
    author = "Luscher, M.",
    title = "{Topology of Lattice Gauge Fields}",
    reportNumber = "BUTP-10/1981-BERN",
    doi = "10.1007/BF02029132",
    journal = "Commun. Math. Phys.",
    volume = "85",
    pages = "39",
    year = "1982"
}

\end{document}